\documentclass[hidelinks,11pt]{book}
\usepackage[square, numbers, comma, compress, nonamebreak ]{natbib}
\usepackage{geometry}
\usepackage[utf8]{inputenc}
\usepackage[T1]{fontenc}
\usepackage[english]{babel}
\usepackage{a4wide}
\usepackage{amsfonts}
\usepackage{amssymb}
\usepackage{graphics,amsmath}
\usepackage{graphicx}
\usepackage{physics}
\usepackage{amsthm,textcomp}
\usepackage{mathrsfs}
\usepackage{bm,bbm,bbold}
\usepackage{mathtools}
\usepackage{fancyhdr}
\usepackage[Sonny]{fncychap}
\usepackage{appendix}
\usepackage{pdfpages}
\usepackage{afterpage}
\usepackage{dsfont}
\usepackage{subcaption}
\usepackage{tikz}
\usetikzlibrary{positioning}

\newgeometry{top=3.5cm,bottom=2.5cm,outer=2.5cm,inner=3.5cm}

\usepackage[normalem]{ulem} 
\usepackage{xcolor}

\usepackage[linktocpage=true]{hyperref}
\usepackage[font=small,labelfont=bf]{caption}

\graphicspath{{Figs/}}
\usepackage{newpxtext,newpxmath}

\newcommand{\mathH}{\hat{\mathcal{H}}}
\newcommand{\bfk}{\textbf{k}}
\newcommand{\bfx}{\textbf{x}}

\newcommand{\fidq}{{}^o\!q}
\newcommand{\fide}{{}^o\!e}
\addto{\captionsenglish}{}


\newenvironment{dedication}{\phantom{}\vfill\begin{flushright}\begin{minipage}{0.5\textwidth}\raggedleft}{\end{minipage}\end{flushright}\vfill}
\begin{document}

\pagestyle{plain}

\begin{titlepage}
\newgeometry{top=3.5cm,bottom=2.5cm,outer=2cm,inner=3cm}
\vfill
\eject

\begin{center}

{\Large \bf \textsc{Universidad Complutense de Madrid}}
\vskip .12in
{\large \textsc{Facultad de Ciencias F\'isicas}}
\vskip .4in
\begin{center}
    \includegraphics[height=5cm]{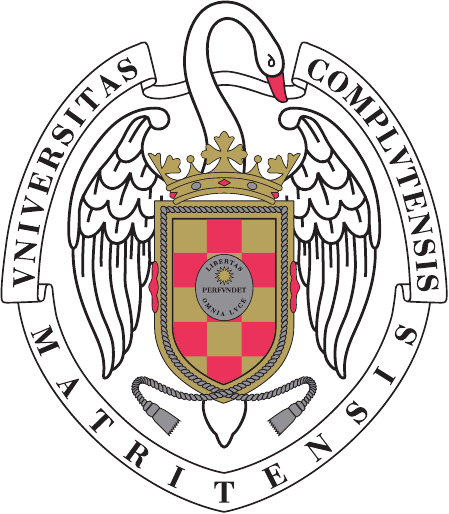}
\end{center}
\vskip .3in
{\Large \bf \textsc{Tesis doctoral}}
\vskip .4in
{\LARGE \bf Observational imprints from Loop Quantum Cosmology\\}
\rule{.6667\linewidth}{.4pt}
\vskip 0.15in
{\LARGE\bf Vestigios observacionales de Cosmologia Cu\'antica de Lazos}

\vskip 0.4in
Memoria para optar al grado de Doctora por
\vskip 0.4in
          
{\Large \bf Rita Barcelos Guerreiro Gon\c{c}alves Neves} 
\vskip 1.3in
\end{center}

\begin{minipage}{.45\linewidth}
\begin{flushleft}
Directores:\\
Dr. Mercedes Martín Benito\\
Dr. Javier Antonio Olmedo Nieto
\end{flushleft}
\end{minipage}
\hfill
\begin{minipage}{.45\linewidth}
\begin{flushright}
\ \\
\ \\
Madrid, Octubre 2023
\end{flushright}
\end{minipage}

\vskip .4in
\noindent
\end{titlepage}
\newgeometry{top=3.5cm,bottom=2.5cm,outer=2.5cm,inner=3.5cm}
\vfill
\eject

\def\baselinestretch{1.2}

\vfill
\eject

\def\baselinestretch{1.2}


\baselineskip 20pt

\null 

\null 
\thispagestyle{empty}
\newpage

\clearpage
\begin{dedication}
\textit{To my parents Isabel and Paulo}
\end{dedication}
\thispagestyle{empty}
\clearpage
\thispagestyle{empty}
\chapter*{Acknowledgements}

Firstly I would like to thank my supervisors Merce and Javi for their unconditional support and confidence in my abilities. Thank you for the patience and understanding in the lows and for the words of appreciation in the highs. I was very lucky to have you as mentors. I also want to extend my gratitude to all the members of the research group that supported me throughout the four years.

I would like to thank all my colleagues in the department for the breaks that I enjoyed very much and greatly improved my vitamin D levels that would otherwise be in the trash, in particular to Mercè, Álvaro, Sara and Lucía in order of appearance in my life. Friends really make life great and without you I am sure these would not have been such wonderful years. Thanks for all the coffee and tea. And beers. And burritos.

A big thanks to my friends further away, Sofia, Elsa, Isma, Almeida and Taanit, for all the emotional support. A special thanks to Elsa without whom I would still be stuck with the plik errors in Montepython. I would have suffered way longer without you. An extra special thanks to Francisco Bioucas, for the emotional but also practical support that made my life easier many times.

Finally a huge thanks to my family, your faith in me has always helped me through the toughest moments, your unwavering support is a great source of comfort. I feel much better doing anything because I know you are there for me mom, dad and mana.

This PhD was financially supported by Funda\c{c}\~ao para a Ci\^encia e a Tecnologia (FCT) through the research grant SFRH/BD/143525/2019
\thispagestyle{empty}
\newpage
\null 
\thispagestyle{empty}
\newpage

\setcounter{page}{0}
\pagenumbering{roman}
\setcounter{tocdepth}{2}
\tableofcontents
\listoffigures
\listoftables

\chapter*{Publications}
\addcontentsline{toc}{chapter}{Publications}
\vspace*{-15pt}The work developed for this thesis has led to five publications and one article that has been accepted for publication in \textit{Phys. Rev. D}:
\begin{itemize}
    \item Mercedes Martín-Benito and Rita B. Neves, ``The Effect of a positive cosmological constant on the bounce of Loop Quantum Cosmology'', \textit{Mathematics}, \textbf{8} no. 2, (2020) 186.
    \item Mercedes Martín-Benito, Rita B. Neves, and Javier Olmedo, ``States of Low Energy in bouncing inflationary scenarios in Loop Quantum Cosmology'', \textit{Phys. Rev. D}, \textbf{103} (2021) 123524.
    \item Mercedes Martín-Benito, Rita B. Neves, and Javier Olmedo, ``Non-Oscillatory Power Spectrum From States of Low Energy in Kinetically Dominated Early Universes'', \textit{Front. Astron. Space Sci}. \textbf{0} (2021) 133.
    \item Álvaro Álvarez-Domínguez, Luis J. Garay, Mercedes Martín-Benito, and Rita B. Neves, ``States of low energy in the Schwinger effect'', \textit{JHEP}, \textbf{06} (2023) 093.
    \item Mercedes Martín-Benito, Rita B. Neves, and Javier Olmedo, ``Alleviation of anomalies from the non-oscillatory vacuum in loop quantum cosmology'', \href{https://arxiv.org/abs/2305.09599}{\texttt{arXiv:2305.09599 [gr-qc]}}.
\end{itemize}
Furthermore, some results are based on an earlier publication:
\begin{itemize}
    \item Mercedes Martín-Benito and Rita B. Neves, ``Solvable Loop Quantum Cosmology: domain of the volume observable and semiclassical states'', \textit{Phys. Rev. D} \textbf{99} no. 4, (2019) 043525.
\end{itemize}

\chapter*{Resumen}
\addcontentsline{toc}{chapter}{Resumen}

El modelo estándar de cosmología supone un universo homogéneo e isótropo que sufre un periodo de expansión exponencial muy temprano denominado inflación.  Las fluctuaciones cuánticas sufren un estiramiento desde el inicio de inflación hasta escalas cosmológicas, dando lugar a las anisotropías de temperatura, polarización y materia que observamos. Sin embargo, este paradigma ignora la física preinflacionaria, ya que poco antes de inflación se produce la singularidad inicial del big-bang. La Cosmología Cuántica de Lazos (LQC por su sigla en inglés) es una teoría prometedora de cosmología cuántica. Su resultado más importante es la resolución de la singularidad inicial clásica en términos de un rebote cuántico que conecta una rama del Universo en contracción con otra en expansión. En consecuencia, proporciona una dinámica preinflacionaria libre de singularidades. En este contexto, ya no está justificado considerar que cada modo de las perturbaciones cosmológicas alcanza el  inicio de inflación en el estado de vacío natural de la cosmología estándar. De hecho, algunos modos podrían alcanzarlo en un estado excitado, lo que podría dejar huellas en sus espectros de potencia al final de inflación y, por tanto, en las observaciones del Fondo Cósmico de Microondas (CMB). El objetivo de esta tesis es buscar tales huellas de LQC en el CMB. Para ello trabajamos en el contexto  de la llamada LQC híbrida.

Primeramente, exploramos la dinámica del fondo más allá de la dinámica efectiva de LQC que se suele adoptar. Aplicamos un tratamiento perturbativo para extraer las principales contribuciones de un potencial constante de campo escalar, considerado como una perturbación del modelo cosmológico con un campo escalar sin masa. Para estados genéricos, el potencial produce asimetrías en el rebote, dando lugar a términos antisimétricos y asimétricos en el valor esperado del volumen. Sin embargo, notablemente, en el caso de estados semiclásicos con un perfil espectral gaussiano, el rebote vuelve a ser simétrico. El efecto es desplazarlo en una cantidad que depende del estado. Entonces, para estos estados la dinámica efectiva describe bien la dinámica cuántica.

Para estudiar las perturbaciones cosmológicas, partimos de ecuaciones de movimiento corregidas cuánticamente obtenidas mediante la LQC híbrida, que se basan en la dinámica efectiva del fondo. Evolucionarlas requiere establecer condiciones iniciales, o equivalentemente elegir un estado de vacío para las perturbaciones. En esta tesis proponemos Estados de Baja Energía (SLEs) como vacíos viables de perturbaciones cosmológicas en LQC. Estos son estados que minimizan la densidad de energía cuando se promedian a lo largo de un periodo dado mediante una función test. Se ha demostrado que los SLEs son estados Hadamard en modelos cosmológicos, y vacíos adecuados de perturbaciones cosmológicas en modelos con un periodo de dominio cinético antes de inflación, que es lo que ocurre en LQC. Encontramos que hay dos opciones naturales para la función test dentro de LQC. Los SLE correspondientes y los espectros de potencia resultantes al final de inflación son bastante insensibles a la forma exacta y al soporte de la función test, siempre que incluya el régimen de alta curvatura (es decir, el rebote). Estos estados producen espectros de potencia con supresión exponencial en los modos infrarrojos, así como aumento de potencia y oscilaciones en escalas intermedias. Si se desplaza el soporte de la función test fuera de este régimen, se obtienen espectros de potencia primordiales en los que el aumento de potencia y las oscilaciones desaparecen. Esto se aproxima más a los espectros de potencia del vacío No Oscilatorio (NO), otra opción viable introducida previamente en LQC. Podemos concluir entonces que las oscilaciones y el aumento de potencia son vestigios claros del rebote en los espectros de potencia primordiales en el contexto de los SLE.

En última instancia, la comparación de observaciones con predicciones se centra en el CMB. Comenzamos con el espectro de potencia más simple obtenido con SLEs en LQC, es decir, el que es equivalente al espectro de potencia del vacío NO. Parametrizándolo y dejando libre la escala a la que se produce la supresión de potencia, realizamos un análisis de  inferencia bayesiana. Esto nos permite restringir la escala libre y concluir que los datos muestran una preferencia porque esta escala esté dentro de los modos infrarrojos observables. En otras palabras, las observaciones son más compatibles con un espectro de potencias primordial con esta característica que con el del modelo estándar. Además, descubrimos que esta característica ayuda a aliviar las anomalías de supresión de potencia y <<lensing>> identificadas previamente en los datos. La primera se refiere a la falta de potencia en las escalas angulares más grandes y se aborda directamente a través de la supresión de potencia en el espectro de potencias primordial. La segunda es una prueba de consistencia que el modelo estándar no supera razonablemente. Nuestro modelo alivia esta anomalía desplazando suficientemente las predicciones de los parámetros libres como para mitigar estas inconsistencias. También investigamos la asimetría de paridad que se observa en los datos pero que no predice el modelo estándar. Aunque nuestro modelo introduce una ligera asimetría, concluimos que esta simple característica de supresión de potencia en los espectros primordiales no afecta significativamente a esta anomalía.

Finalmente, extendemos la construcción de SLEs a escenarios homogéneos genéricos. Concretamente, los investigamos dentro del efecto Schwinger: la producción de partículas debido a un campo eléctrico muy intenso en un espaciotiempo plano. Investigamos la elección de la función  test considerando un pulso eléctrico y examinando los SLEs con funciones test de soporte variable alrededor del pulso. Concluimos que estos estados tienen una dependencia no trivial en el soporte cuando este es del orden del pulso. Sin embargo, tanto en el régimen de soportes pequeños como en el de soportes grandes, los SLEs convergen a estados concretos. Mediante este estudio, comprendemos mejor nuestros resultados anteriores en LQC. Allí, el análogo del pulso es el rebote. Habíamos observado el límite de soporte grande, en comparación con la duración del rebote, donde el SLE depende sólo de si el soporte incluye el rebote. También investigamos el espectro de potencias y el número de partículas creadas, que son complementarios en la caracterización del vacío.

\chapter*{Summary}
\addcontentsline{toc}{chapter}{Summary}

The standard model of cosmology assumes a homogeneous and isotropic universe that undergoes a period of exponential expansion very early on, named inflation. This stretches quantum fluctuations from the onset of inflation to cosmological scales, which seed the temperature, polarization and matter anisotropies that we observe. However, this paradigm ignores pre-inflationary physics, as shortly before inflation there is the initial big-bang singularity. Loop Quantum Cosmology (LQC) is a promising approach to quantum cosmology. Its most outstanding result is that of the resolution of the classical initial singularity in terms of a quantum bounce that connects a contracting branch of the Universe with an expanding one. Consequently, it provides singularity-free pre-inflationary dynamics. Within this context, it is no longer justified to consider that every mode of the cosmological perturbations reaches the onset of inflation in the natural vacuum state of standard cosmology. In fact, some modes might reach it in an excited state, which may leave imprints in their power spectra at the end of inflation and therefore in observations of the Cosmic Microwave Background (CMB). The goal of this thesis is to search for such imprints from LQC in the CMB. We work in the hybrid approach to cosmological perturbations in LQC to do so.

In a first instance, we explore the dynamics of the background beyond the effective dynamics of LQC on which one usually relies. We apply a perturbative treatment to extract the main contributions from a constant scalar field potential by considering it as a perturbation of the cosmological model with a minimally coupled massless scalar field, which is analytically solvable. For generic states, the potential produces asymmetries in the bounce, leading to anti-symmetric as well as asymmetric terms in the expectation value of the volume. However, remarkably, in the case of semiclassical states with a Gaussian spectral profile, the bounce becomes symmetric again. The effect is to displace it by a state-dependent amount. Then, for these states the effective dynamics describes well the quantum dynamics.

To study cosmological perturbations, we adopt the quantum corrected equations of motion obtained through the hybrid LQC approach, which rely on the effective dynamics of the background. Evolving them requires setting initial conditions, or equivalently choosing a vacuum state for the perturbations. In this thesis we propose States of Low Energy (SLEs) as viable vacua of cosmological perturbations in LQC. These are states that minimize the energy density when smeared along a given period via a smearing function. They have been shown to be Hadamard states in cosmological settings and suitable vacua of cosmological perturbations in models with a period of kinetic dominance prior to inflation, as LQC. We find that there are two classes of smearing functions that can be seen as natural choices within LQC. The corresponding SLEs and the resulting power spectra at the end of inflation are quite insensitive to the exact shape and support of the smearing function, as long as it includes the high curvature regime (i.e. the bounce). They produce power spectra with exponential suppression at infrared modes as well as power enhancement and oscillations at intermediate scales. Shifting the support of the smearing function away from this regime results in primordial power spectra where the enhancement and oscillations essentially vanish. This is closer to the power spectra of the so-called Non-oscillatory (NO) vacuum, another viable choice of initial conditions previously introduced in LQC. We can conclude then that the oscillations and enhancement are clear imprints of the bounce in the primordial power spectra in the context of SLEs.

Ultimately, the comparison of observations with predictions focuses on the CMB. We start with the simplest power spectrum obtained with SLEs in LQC, i.e. the one that is equivalent to the power spectrum of the NO vacuum. By parametrizing it and leaving free the scale at which the power suppression occurs, we perform a Bayesian analysis of the model. This allows us to constrain the free scale and conclude that the data shows a preference for this scale to be within the observable infrared modes. In other words, observations are more compatible with a primordial power spectrum with this feature than with that of the standard model. Furthermore, we find that this characteristic mitigates the previously identified power suppression and lensing anomalies in the data. The first relates to a lack of power in the largest angular scales and is addressed directly through the power suppression in the primordial power spectrum. The second is a consistency check that the standard model does not reasonably pass. Our model alleviates this anomaly by shifting the overall predictions for the free parameters enough so that these inconsistencies are mitigated. We also investigate the parity asymmetry that is observed in the data but not predicted by the standard model. Even though our model introduces a slight asymmetry, we conclude that this simple feature of power suppression in the power spectra does not significantly affect this anomaly one way or another.

Finally, we extend the construction of SLEs to generic homogeneous scenarios, possibly anisotropic. Concretely, we investigate them within the Schwinger effect: the particle pair production due to a very strong electric field in a flat spacetime. We investigate the choice of smearing function by considering an electric pulse and examining SLEs with smearing functions of varying support around the pulse. We conclude that these states have a non-trivial dependence on the size of the support when it is of the order of the pulse. However, both in the small and large-support regimes they converge to concrete states. Through this study, we understand better our previous results in LQC. There, the analogous of the pulse is the bounce. We had observed the large support limit, in comparison to the typical time-scale of the bounce, where the SLE depends only on whether the support includes the bounce. We also investigate the power spectrum and number of created particles, which are complementary in characterising the vacua.

\newpage
\null
\newpage
\pagestyle{fancy}
\setcounter{page}{0}
\pagenumbering{arabic}


\chapter*{Motivation and structure}
\addcontentsline{toc}{chapter}{Motivation and structure}

The standard model of cosmology considers a homogeneous and isotropic universe undergoing a finite and nearly exponential expansion at very early times, which is called inflation. Within this paradigm, due to the rapid expansion, inhomogeneities are stretched out, resulting in the statistically isotropic and homogeneous Universe that we observe today. Additionally, this mechanism stretches quantum fluctuations from the onset of inflation to cosmological scales. These induce energy density fluctuations which seed temperature, polarization and matter anisotropies, sourcing all of the structure that we observe.

The Cosmic Microwave Background (CMB) is an excellent test bench for the study of the physics of the large scale structure of our Universe, especially after the recent high-precision observations reported by the Planck Collaboration \cite{Planck2018_parameters,Planck2018_inflation}. There is considerable consensus that the origin of this large scale structure is primordial and naturally explained by the paradigm of cosmological inflation. In fact, the standard model of cosmology predicts a distribution of temperature and polarization anisotropies in the CMB with properties that are remarkably compatible with current observations for a large range of scales. However, some anomalies have been detected for the largest angular scales and persist in the most recent observations. It is thought that these may be hints of non-standard processes occurring in the very early Universe. Naturally, this has captured the attention of researchers in the field of quantum gravity, as it may open an observational window to the quantum nature of spacetime.

Despite the success of this paradigm, which combines simplicity and predictability, inflationary scenarios in General Relativity (GR) usually ignore the pre-inflationary dynamics, since soon before the onset of inflation one encounters the classical big-bang singularity, and the model loses predictability. A singularity-free model that provides well-defined pre-inflationary dynamics may source a different evolution of the perturbations that could leave imprints in the CMB. It is natural to think that the solution to the issue of the singularity could lie in a theory of quantum gravity. GR describes classically the interaction between gravity (as geometry) and matter. In this spirit, the fundamentally quantum nature of matter hints at a fundamentally quantum description of geometry. In contrast, primordial quantum fluctuations, are usually analyzed within the scheme of Quantum Field Theory (QFT) in curved spacetimes, where they are treated as quantum fields propagating on a classical fixed background. A more consistent picture would be to describe both gravity and matter as quantum fields, specially in regimes of extremely high curvature where the right description of gravitational phenomena might come from a non-perturbative approach. A first step in understanding the ramifications of such a theory could be to consider corrections to the classical background originating from an underlying theory of quantum gravity. The efforts in the field of quantum cosmology have mainly been in the spirit of finding imprints of such corrections in predictions of the CMB, motivated to some level by the increasingly more accurate observations.

Loop Quantum Cosmology (LQC) is one of the most promising approaches to quantum cosmology in the literature \cite{LQCreview_Bojowald2005,LQCreview_Ashtekar2011,LQCreview_Banerjee2012,LQCreview_Agullo2016}. It applies the non-perturbative and background independent quantization program of Loop Quantum Gravity (LQG) to cosmological models. It has been successfully applied to Friedmann-Lema\^itre-Robertson-Walker (FLRW) spacetimes \cite{APS_PRL,APS_extended,ACS,Bentivegna2008,Kaminski2009,Pawlowski2012} through its improved dynamics prescription \cite{APS_extended}, leading to a resolution of the big-bang singularity in terms of a quantum bounce that connects a contracting epoch of the Universe with an expanding one. It has also been applied to models with non-zero cosmological constant \cite{Bentivegna2008,Kaminski2009,Pawlowski2012}, spatially compact models \cite{K=1Cosmologies,closedFRWcosmologies} and Bianchi models \cite{Merce_bianchiI,Ashtekar_bianchiI,Ashtekar_bianchiII,WilsonEwing_bianchiIX}. The robust consequence seems to be that when spacetime curvature approaches a critical scale (of Planck order) the quantum effects render gravity repulsive and the bounce occurs. Away from this regime, the quantum dynamics agrees with classical trajectories of a contracting or expanding universe. The main effect can be captured in the effective dynamics obtained rigorously in \cite{Taveras_2008} for the case of a flat FLRW spacetime minimally coupled to a massless scalar field. This procedure results in modified classical equations that accurately describe the quantum dynamics of semiclassical states. It has since been assumed that the method to find these classical equations can be extended to more intricate models, and indeed these effective equations accurately reproduce the corresponding quantum dynamics, that were previously obtained numerically. We will assume this holds also in this work.

Two main strategies have been followed to introduce perturbations in this formalism: the hybrid approach \cite{FernandezMendez2012,FernandezMendez2013,FernandezMendez2014,CastelloGomar2014,CastelloGomar2015,CastelloGomar2016,hybrid_Martinez2016,hyb-vs-dress} and the dressed metric approach \cite{Ashtekar:2009mb,dressedmetric_Agullo2012,dressedmetric_Agullo_PRD_2013,dressedmetric_Agullo_CQG_2013}. Both combine a polymeric quantization based on the techniques of LQG for the geometric sector with a standard Fock quantization of QFT in curved spacetimes for the perturbations. This is based on the assumption that there exists a regime of interest where the main quantum effects are those affecting the global degrees of freedom (homogeneous sector), and was actually first introduced to deal with inhomogeneities in Gowdy models \cite{Merce_2008Hybrid,Garay_2008hybrid,Merce_2010hybrid}. Most works pursuing observational consequences of LQC have been done within the dressed metric approach. In this thesis we will adopt the hybrid approach instead.  In both formalisms the equations of motion of the perturbations are obtained in the form of corrected semiclassical equations. In the case of hybrid LQC, the equations of motion of the perturbations are all hyperbolic at the bounce, unlike in the dressed metric approach. Besides including explicit quantum corrections, these equations of motion rely on the background obtained through the effective dynamics. In a first instance, we will consider in this work further corrections beyond the effective dynamics, by resorting to the procedure outlined in \cite{CastelloGomar2016}.

Independently of the concrete pre-inflationary dynamics  under consideration, cosmological perturbations will in general reach the onset of inflation in an excited quantum state with respect to the Bunch-Davies vacuum of standard cosmology, affecting the power spectra at the end of the exponential expansion and, therefore, potentially leaving imprints in the CMB. Although this is true in general, the predictability of these scenarios clashes with the lack of unique criteria for the choice of vacuum state. Indeed, while during an exponential expansion the natural vacuum state is the Bunch-Davies vacuum, selected by the symmetries of the spacetime, under other evolutions the symmetries are not enough to select a preferred vacuum state. In other words, in general cosmological spacetimes there is no unique notion of particle.

This question has received some attention. Actually, there are several criteria proposing candidates of vacuum states in cosmological settings based on natural requirements and applicable in quite general situations. The most popular prescription \cite{parker1969}, based on a WKB approximation, yields the so-called adiabatic states. Other proposals have explored a Hamiltonian diagonalization \cite{fulling1979,Fahn:2018,Elizaga2019}, the minimization of the renormalized stress-energy tensor \cite{agullo2015,handley2016}, and the minimization of the uncertainty relations \cite{danielsson2002,Ashtekar:2016}. Other choices minimizing physical quantities but smeared along time-like curves have also been considered.

A recent application in kinetically dominated (bouncing) cosmologies consists of minimizing the oscillations of the power spectrum within the whole kinetic dominated era \cite{nonosc,hyb-obs,menava,Navascues:2021}. In these lines, one of the most interesting proposals is that of the States of Low Energy (SLEs) defined in \cite{Olbermann2007}, which minimize the energy density smeared along a time-like curve. These had been overlooked as suitable vacuum states for cosmological perturbations before that work, even though they present some appealing properties. The first is that they are proven to be Hadamard states in cosmological spacetimes. This guarantees certain mathematical properties that allow for computations such as that of the renormalized stress-energy tensor to be well defined, and thus are deemed necessary for states to be physically relevant. Furthermore, more recently they have been shown to source the qualitatively correct asymptotic infrared and ultraviolet behaviors of the primordial power spectra of cosmological perturbations to agree with observations in models where a period of kinetic dominance precedes inflation \cite{Niedermaier2020}. Since this is precisely the case in some bouncing inflationary scenarios, such as Loop Quantum Cosmology (LQC), in this thesis we will explore the consequences of SLEs on these quantum cosmology settings.

In most LQC models, the bounce occurs in a kinetically dominated epoch of the Universe and connects a contracting branch with an expanding one. In contrast to standard cosmology, slow-roll inflation occurs naturally in this scenario \cite{AshtekarSloan_probinflation}. Soon after the bounce, the Universe goes through a period of decelerated expansion, before the potential of the scalar field begins to dominate and standard slow-roll inflation begins. This affects the evolution of primordial perturbations, as some modes cross out of and back into the horizon before the onset of inflation. It is no longer reasonable to assume that they reach it in the Bunch-Davies vacuum of standard cosmology. This may then have an effect in the primordial power spectrum. In the standard $\Lambda$CDM model, this is a near-scale-invariant power spectrum, with a slight red tilt. In the case of LQC, it is affected in the infrared, where the particular deviations from near scale invariance depend on details of the quantization and on the vacuum choice \cite{Li2021,Agullo2023}. 

Previous works within LQC \cite{Ashtekar:2020,Ashtekar:2021,Agullo:2020b,Agullo2021} have shown that such departures from near scale invariance may alleviate some anomalies in observations of large scales. In \cite{Ashtekar:2020,Ashtekar:2021}, adopting the equations of motion of the dressed metric approach \cite{Ashtekar:2009mb,dressedmetric_Agullo2012,dressedmetric_Agullo_PRD_2013,dressedmetric_Agullo_CQG_2013} and a particular vacuum state for the cosmological perturbations \cite{Ashtekar:2016}, it was shown that both the power suppression anomaly and the lensing anomaly may be alleviated. The first is related to a lack of power in the CMB for large multipoles, which is a consequence of the fact that the temperature-temperature correlation function is consistent with zero for large angular scales \cite{Copi2008,Copi2013,Copi2018}. This corresponds to a very unlikely realization of a $\Lambda$CDM universe, which predicts a much larger value of the corresponding estimator than what is observed. The effects of the LQC model considered in references \cite{Ashtekar:2020,Ashtekar:2021} were able to alleviate the anomaly in the concrete sense that the expected value of this estimator is lower than in $\Lambda$CDM. However, one could argue that this alone is not enough to conclude an alleviation of the anomaly, and a computation of the distribution of the estimator would be necessary, so that the p-value of the observation may be found. The second anomaly was found as a consequence of a consistency test in $\Lambda$CDM. A phenomenological parameter is introduced to quantify how the CMB is lensed from the surface of last scattering until today. It is found to be incompatible with the prediction from $\Lambda$CDM at $\sim$2-$\sigma$ level. The LQC model of \cite{Ashtekar:2020,Ashtekar:2021} is able to alleviate this anomaly, by affecting the statistics of other parameters. It does not affect post-inflationary physics that relate to lensing, rather it shifts predictions enough that the inconsistencies no longer present as strongly as in the standard model. References \cite{Agullo:2020b,Agullo2021} consider primordial power spectra with power enhancement of different slopes.  These may be motivated by different models, of which LQC is an example. Here, non-Gaussianities become a key ingredient. They provide a mechanism that correlates the largest wavelength modes of the CMB and super-horizon (non-observable) modes. The effect is to modify the variance of the perturbations, even if the mean remains unaltered. Then, certain features become more likely in this scenario than in standard cosmology. This alleviates the aforementioned anomalies as well as the parity asymmetry anomaly, which refers to the fact that more power is observed in odd multipoles than in even ones, which is not predicted by $\Lambda$CDM.

The main goal of this research is to contribute to a falsifying picture of LQC, by exploring underlying ambiguities and comparing predictions with observations through a rigorous statistical analysis. The structure of this thesis is as follows.

In chapter \ref{chap:intro} we introduce the formalism and review the concepts and tools that we will rely on throughout the thesis. We start in section \ref{sec:earlyUniverse_obs} by reviewing the standard inflationary model of cosmology and introducing cosmological perturbations. We review the ambiguity in the choice of vacuum and introduce the notion of power spectrum, from which we will be able to compare predictions of LQC to those of standard cosmology. We also introduce the formalism of the CMB predictions and review three anomalies that have been identified in the data and that are relevant for the remaining work. In section \ref{sec:introLQC} we review the quantization procedure of LQC, focusing on the solvable prescription which renders the model with a massless scalar field analytically solvable. Afterwards, we introduce a perturbative procedure to compute corrections due to the introduction of a scalar field potential. Furthermore we review the effective dynamics that are capable of capturing the features of the quantum trajectories on semiclassical states. Finally, the results of introducing cosmological perturbations in this context through the hybrid approach are summarized via the quantum corrected equations of motion. Lastly, in section \ref{sec:intro_Bayes} we review the basics of Bayesian analysis and the computational methods that we have used to perform the statistical analyses of this thesis.

In chapter \ref{chap:cosmologicalconstant} we apply the aforementioned perturbative procedure to obtain corrections to the expectation value of the volume of the massless scalar field model from the introduction of a constant scalar field potential. This is equivalent to considering a model with a massless scalar field and a cosmological constant. The goal of this chapter is to explore the quantum corrections to the expectation value of the volume beyond the effective dynamics usually employed in the literature. We find that for generic states the bounce is actually asymmetric. However, for semiclassical states the asymmetric contributions vanish and the bounce becomes symmetric as in the free model. In this case, our results agree with the effective dynamics.

In chapter \ref{chap:SLEsinLQC} we investigate the implications of choosing SLEs as the vacuum of cosmological perturbations in LQC. We focus on the consequences of different supports of the smearing function. First we compare the SLE associated to a smearing function with a wide-enough support around the bounce with one whose support begins at the bounce and ends far enough into the future. Then we explore SLEs associated to smearing functions with a support that starts gradually further away from the bounce. A comparison of the primordial power spectra reveals that the bounce has direct effects at intermediate scales, where it introduces power enhancement and oscillations. When these are dampened by not including the bounce in the support, the power spectra show a power suppression at infrared modes, which coincides with the power spectra obtained with the Non-Oscillatory (NO) vacuum previously proposed in LQC. Through a comparison of the ultraviolet behaviors of the SLEs with that of the NO vacuum we conclude that the latter is also a Hadamard state.

Chapter \ref{chap:anomalies} is dedicated to the Bayesian analysis of the LQC model with the NO state as the vacuum of cosmological perturbations. We are able to constrain the scale at which the power suppression occurs and determine whether observations show a preference for power suppression at observable scales. Furthermore we investigate the potential impact of such a feature on the alleviation of the three aforementioned anomalies.

In chapter \ref{chap:Schwinger} we apply the construction of SLEs in the Schwinger effect. This is the particle creation effect due to a very strong electric field on a flat spacetime. The electric field introduces an anisotropy, though  we consider a homogeneous configuration. We investigate the behavior of SLEs in this context and compare with the cosmological scenario, regarding asymptotic behaviors and the Hadamard property. We focus particularly on the choice of the smearing function and especially on its support. We also analyse the spectral properties of the power spectrum and the number of created particles, which are complementary in characterising the vacuum, and investigate the multipolar contributions coming from the anisotropies.

Chapter \ref{chap:conclusions} is dedicated to conclusions and final remarks.

Throughout we adopt $c=\hbar=1$ and for numerical computations we also take $G = 1$.
\chapter{Introduction}\label{chap:intro}


This chapter is dedicated to an introduction on the subjects on which this thesis fundamentally relies. We have divided them into three main topics: standard cosmology, Loop Quantum Cosmology (LQC) and statistical Bayesian analysis.

Throughout this work, we refer to predictions from standard cosmology, by which we mean the $\Lambda$CDM model. This is a model of cosmology where the Universe is described by a cosmological (FLRW) spacetime which contains a cosmological constant $\Lambda$, that seeds late-time acceleration, cold dark matter (CDM) and ordinary matter. It is capable of depicting the Universe that we observe reasonably well, although some tensions with data have been identified. My work focuses on physics of the very early Universe. Therefore, the first section of this chapter aims at describing the standard model's dynamics of the background and of cosmological perturbations during the inflationary epoch, as well as the observed Cosmic Microwave Background (CMB).

The second section briefly reviews the quantization procedure of LQC. We focus mostly on the effective dynamics, which capture the main effects of the quantization on the expectation values of observables on semiclassical states. This results in modified classical equations that we rely on to describe both the background and the evolution of cosmological perturbations.

Finally, in the third section we introduce the statistical tools that we make use of when comparing predictions with observations. As is usual in cosmological studies, we follow the Bayesian philosophy, and resort to Markov Chain Monte Carlo (MCMC) methods to aid in the exploration of the large parameter space.

\section{Early Universe and observations}\label{sec:earlyUniverse_obs}

We will start by reviewing the dynamics of the early Universe in the standard inflationary model, leaving some notation general when appropriate to pave the way for later endeavors in non-standard scenarios. This will include a review of the background dynamics, as well as the dynamics of cosmological perturbations, and an introduction to the ambiguity in their choice of vacuum along with two particular constructions that we will rely on later. We also provide an overview of the statistical properties of the observed CMB and three anomalies in the data that will be addressed later in the main work.

\subsection{FLRW background}

Let us begin with the background dynamics. For further details see, e.g. \cite{Weinberg2008}. In full GR, the action is given by
\begin{equation}\label{eq:SGR}
    S = \frac{1}{16\pi G}\int d^4x \sqrt{-g}\mathcal{R}+ \int d^4x\sqrt{-g}\mathcal{L}_m,
\end{equation}
where the $\mathcal{L}_m$ stands for the Lagrangian density of the matter component which we write generically for now, $G$ is Newton's gravitational constant, $g$ is the determinant of the metric $g_{\mu\nu}$ and $\mathcal{R}$ is the Ricci scalar. Through the variation of the metric, one obtains the Einstein equations
\begin{equation}\label{eq:Einstein}
    \mathcal{R}_{\mu\nu}-\frac{1}{2}g_{\mu\nu}\mathcal{R} = 8\pi G T_{\mu\nu},
\end{equation}
where $\mathcal{R}_{\mu\nu}$ is the Ricci tensor and $T_{\mu\nu}$ is the energy-momentum tensor of all present fields (e.g. radiation, matter, the inflaton). However, we are particularly interested in flat cosmological spacetimes, i.e., ones that are homogeneous and isotropic. These are described through the flat FLRW metric with line element
\begin{equation}\label{eq:linelementFLRW}
    ds^2 = -N^2(t) dt^2 + a^2(t)\left(dr^2+r^2d\theta^2+r^2\sin^2\theta d\phi^2\right),
\end{equation}
where $(t,r,\theta,\phi)$ are comoving coordinates, $N(t)$ is the lapse function, and $a(t)$ is the cosmological scale factor, which is dimensionless so that $r$ has dimensions of length. Note that only ratios of $a(t)$ are physical. To speak of $a(t)$ we need to choose a reference point in time where the value of this scale is known. This choice of reference is arbitrary and is usually motivated by convenience: in standard cosmology, where the Universe is typically studied from today into the past or the future, $a(t)$ is usually set to 1 today, whereas in LQC it is set to 1 at the bounce, the natural reference point in that case. When comparing the two models, this needs to be taken into account. Concretely, when faced with two scales $a(t)$ and $\tilde{a}(t)$ such that $a(t_0) = 1$ and $\tilde{a}(t_1) = 1$, we can relate them by equating the physical quantity: $a(t)/a(t_{\star}) = \tilde{a}(t)/\tilde{a}(t_{\star})$. In particular, it is most useful to consider $t_{\star} = t_0$ so that we see that $a(t) = \tilde{a}(t)/\tilde{a}(t_0)$, or $t_{\star} = t_1$ so that we see that $\tilde{a}(t) = a(t)/a(t_1)$.

Throughout this work, we will use cosmological time $t$, fixing the lapse $N(t) = 1$, or conformal time $\eta$, corresponding to $N(t) = a(t)$. Derivatives with respect to conformal time will be denoted with prime, whereas the dot represents derivatives with respect to cosmological time.\footnote{Note that conformal and cosmological time can be related by $dt = a d\eta$, which translates to $\dot{f}=f^{\prime}/a$ for any time-dependent function $f$.}

The dynamics is obtained through the Einstein equations \eqref{eq:Einstein}. Following the symmetries of the metric, $T_{\mu\nu}$ must be diagonal, and given isotropy its spatial components must all be equal. For simplicity, we will consider the energy-momentum tensor of a perfect fluid with energy density $\rho(t)$ and pressure $P(t)$: $T_{\mu\nu} = \text{diag}\left(-\rho,P,P,P\right)$. Thus one obtains the Friedmann equations
\begin{align}
    H^2 &= \frac{8\pi G}{3}\rho,\label{eq:FriedmannH}\\
    \frac{\Ddot{a}}{a} &=-\frac{4\pi G}{3}(\rho+3P),\label{eq:Friedmanndda}
\end{align}
where $H = \dot{a}/a$ is the Hubble parameter. These govern the dynamics of the background of a generic FLRW spacetime. However, standard cosmology describes a universe that undergoes a period of inflation. By definition, inflation is a period of accelerated expansion, and so:
\begin{equation}
    \Ddot{a} >0  \qquad \Leftrightarrow \qquad \rho + 3 P < 0,
\end{equation}
from \eqref{eq:Friedmanndda}. This can be driven by a minimally coupled scalar field $\phi$ subject to an appropriate potential $V(\phi)$, which is aptly called the inflaton field. The corresponding action is:
\begin{equation}
    \mathcal{S}_{\phi} = \int d^4x \sqrt{-g}\left[-\frac{1}{2}\partial_{\mu}\phi\partial^{\mu}\phi-V(\phi)\right].
\end{equation}
From this action, through the Euler-Lagrange equations, we obtain the equations of motion of the field: 
\begin{equation}\label{eq:eomPhi}
    \Ddot{\phi}+3H\dot{\phi}-\frac{\boldsymbol{\nabla}^2\phi}{a^2}+V_{,\phi}(\phi) = 0,
\end{equation}
where $\boldsymbol{\nabla}^2$ is the three dimensional Laplacian in flat space and $V_{,\phi}$ denotes the first derivative of $V$ with respect to $\phi$. 

In cosmology, this field is taken to be mostly homogeneous, with inhomogeneities arising from quantum fluctuations, which are much smaller than the expectation value:
\begin{equation}
    \phi(t,\bfx) = \phi_0(t) + \delta\phi(t,\bfx).
\end{equation}
Let us deal with $\phi_0(t)$ first and consider the consequences of the perturbations $\delta\phi(t,\bfx)$ shortly. The homogeneous part of the field is governed by the equation of motion
\begin{equation}\label{eq:eomPhi0}
    \Ddot{\phi}_0+3H\dot{\phi}_0+V_{,\phi_0}(\phi_0) = 0,
\end{equation}
and has energy and pressure:
\begin{equation}
    \rho_{\phi_0} = \frac{\dot{\phi}_0}{2} + V(\phi_0), \qquad P_{\phi_0} = \frac{\dot{\phi}_0}{2} - V(\phi_0).
\end{equation}
If the kinetic term $\dot{\phi}_0/2$ dominates over the potential, we have that $P_{\phi_0}\approx \rho_{\phi_0}$, which is incapable of driving inflation. On the other hand, if the potential dominates, i.e. $\dot{\phi}_0 \ll V(\phi_0)$, then $P_{\phi_0}\approx -\rho_{\phi_0}$ which is a sufficient condition for inflation to occur.\footnote{Noticeably, this is not a necessary condition for $\rho + 3 P < 0$, and so it is not the only possibility in order to obtain inflation. However, it is an appealingly simple scenario, and the one adopted in the standard model.} In this case, the field is rolling slowly, which is why this is called slow-roll inflation. This translates to conditions on the form of the potential. A simple case of such a potential that is widely used for simplicity of the calculations is that of a quadratic potential:
\begin{equation}\label{eq:quadV}
    V_{\text{q}}(\phi) = \frac{1}{2}m^2 \phi^2.
\end{equation}
In fact, this model is all but ruled out in the context of standard $\Lambda$CDM due to observations of the tensor-to-scalar ratio (which we will introduce shortly). However, in the context of non-standard cosmologies, as the ones that we will consider, there is no reason to assume a priori that this is the case.\footnote{Besides, most of our investigations in this work depend very little on the choice of the potential, as it is mostly relevant once tensor modes are fully investigated.} For example, in the context of warm inflation \cite{Bastero-Gil:2009sdq} the quadratic model is not ruled out. Nevertheless, we will also consider the Starobinsky potential (which is more favored by observations in standard cosmologies) in future chapters:
\begin{equation}\label{eq:starV}
    V_{\text{s}}(\phi)= V_0 \left(1-e^{-\sqrt{\frac{16\pi G}{3}}\phi}\right)^2.
\end{equation}

If we further impose that $\rho_{\phi_0}$ is approximately constant, then inflation becomes a quasi-de Sitter epoch, where $H$ is also approximately constant and
\begin{equation}
    a^{\text{dS}}(t) = e^{H(t-t_0)} \qquad \Leftrightarrow \qquad a^{\text{dS}}(\eta) = -\frac{1}{H\eta},
\end{equation}
where we have fixed without loss of generality $a(t_0) = 1$.

\subsection{Cosmological perturbations}

The background exposed in the previous section generates a universe that goes through an inflationary period but that is completely homogeneous and isotropic. As such, it is not capable of describing the structures that we observe in our Universe. For that, and to fully harness the power of the inflationary paradigm, we need to consider perturbations to the background. For further details we refer the reader to e.g. \cite{Riotto:2002yw}. Let us expand the perturbations $\delta \phi(t,\bfx)$ in Fourier modes
\begin{equation}\label{eq:Fourierexpansion}
    \delta\phi(t,\bfx) = \frac{1}{(2\pi)^{3/2}} \int d^3 \bfk\,\delta\phi_{\bfk}(t)e^{i\bfk\bfx}.
\end{equation}
In this case, the spatial derivative term of \eqref{eq:eomPhi} will carry over. For concreteness, let us consider the quadratic potential \eqref{eq:quadV}. In this case, the Fourier modes of the perturbations of the field obey the equations of motion:
\begin{equation}\label{eq:eomdPhik}
    \delta\Ddot{\phi}_{\bfk}(t)+3H\delta\dot{\phi}_{\bfk}(t)+\left(m^2 +\frac{k^2}{a^2(t)}\right) \delta\phi_{\bfk}(t)= 0,
\end{equation}
where $k = |\bfk|$. Perturbations of this field translate into perturbations of the energy-momentum tensor, which in turn induce perturbations of the metric, through the Einstein equations. The perturbations in the metric take the form of scalar and tensor perturbations (vector perturbations are not physical, as they can be absorbed by a choice of gauge). These are referred to as cosmological perturbations. Inflation then stretches these perturbations to cosmological scales, seeding the matter and temperature anisotropies that now populate the Universe. They provide a remarkable opportunity to relate predictions from theoretical models with observations of the CMB.

Cosmological perturbations are commonly described by scalar and tensor gauge-invariant perturbations of the metric, typically denoted by $Q$ and ${\cal T}^I$ (where $I$ encodes the two possible polarizations of the tensor modes) \cite{Langlois:2010xc}. Let us refer to both generically as $T$. Expanding in Fourier modes $T_{\bfk}(t)$ (similarly to \eqref{eq:Fourierexpansion}) we find that they obey the equations of motion of minimally coupled fields:
\begin{equation}\label{eq:eomT}
    \Ddot{T}_{\bfk}(t)+3H\dot{T}_{\bfk}(t)+\left(\omega_{\bfk}^{(i)}(t)\right)^2 T_{\bfk}(t)= 0,
\end{equation}
where $i = s, t$ denotes scalar or tensor modes, respectively. For now, we have written $\omega_{\bfk}(t)$ generically to accommodate different models easily. In the particular case of the classical FLRW model we have \cite{Weinberg2008}:
\begin{equation}
    \left(\omega_{\bfk}^{(s)}(t) \right)^2 = \frac{1}{a^2(t)} \left[k^2+\frac{a^{\prime\prime}(t)}{a(t)}-\frac{z^{\prime\prime}(t)}{z(t)}\right],\qquad \left(\omega_{\bfk}^{(t)}(t) \right)^2 = \frac{k^2}{a^2(t)},
\end{equation}
where $z(t)=a(t)\dot{\phi}(t)/H(t)$. Notice that the equation of motion for tensor modes in this case is the same as that of the perturbations of a minimally coupled massless scalar field, i.e. \eqref{eq:eomdPhik} with $m = 0$, whereas for scalar modes it is more complicated. As will be discussed in section \ref{sec:LQCperts},  $\omega_{\bfk}^{(i)}(t)$  take more complicated forms in LQC. However, $\omega_{\bfk}^{(s)}(t)$ and $\omega_{\bfk}^{(t)}(t)$ are very similar both in the classical and the quantum theory for kinetically dominated universes, where $z^{\prime\prime}/z=a^{\prime\prime}/a$, and also during slow-roll inflation.

Already we can discern two kinds of behaviors of the perturbations depending on the magnitude of $\omega_{\bfk}^{(i)}(t)$ with respect to $H$. This defines a critical wavelength $1/H$ that separates the two regimes and that is named the Hubble radius or Hubble horizon. In the case of kinetic dominance and inflation, when $\left(\omega_{\bfk}^{(i)}(t)\right)^2 = k^2/a^2$ both for tensor and scalar modes, the perturbations have two different kinds of behaviors depending on whether $k/a \gg H$ (or equivalently $k|\eta| \gg 1$ in de Sitter) or $k/a \ll H$ (or $k|\eta| \ll 1$ in de Sitter). The first are said to be inside the Hubble horizon, as their wavelength ($\lambda=2\pi/k$) is smaller than the Hubble radius, whereas the latter are said to be outside the horizon. Notice that this condition is time dependent. Perhaps a more intuitive way of understanding this is to think of the comoving Hubble radius $1/(aH)$. During inflation, this radius shrinks, and therefore the range of modes with wavenumber $k$ that is inside this radius is also shrinking. Conversely, in a decelerated period this radius grows and the range of modes inside the radius grows as well. During inflation a great range of modes will start inside the Hubble radius and gradually cross out, each mode at a different time.

Sub-Hubble modes are (approximately) governed by an equation of motion of a ``harmonic'' oscillator with time-dependent frequency $\omega_{\bfk}^{(i)}(t)$:
\begin{equation}
    \Ddot{T}_{\bfk}(t)+\left(\omega_{\bfk}^{(i)}(t)\right)^2 T_{\bfk}(t) \approx 0,
\end{equation}
and so they oscillate in time. Conversely, for super-Hubble modes:
\begin{equation}
    \Ddot{T}_{\bfk}(t)+3H\dot{T}_{\bfk}(t) \approx 0.
\end{equation}
In standard cosmology, where $H$ is approximately constant, this equation has the general solution $T_{\bfk}(t) = c_1 e^{-3Ht}+c_2$, where $c_1$ and $c_2$ are integration constants. Noticeably, this solution is not oscillatory and quickly tends to a constant. In summary, modes that are inside the Hubble radius have an oscillatory behavior, but shortly after they cross it, they ``freeze-out''. They will remain frozen for as long as they are super-Hubble.

The equations of motion can be cast in a simpler form if we work in conformal time $\eta$ and perform a redefinition of the fields to the Mukhanov-Sasaki variables $u(\eta,\bfx) = a(\eta) Q(\eta,\bfx)$ and $\mu^I(\eta,\bfx)=a(\eta) {\cal T}^I(\eta,\bfx)$. Denoting both generically as $v$ we find that the Fourier modes obey the Mukhanov-Sasaki equation \cite{Mukhanov:1988jd}:
\begin{equation}
    v^{\prime\prime}_{\bfk}(\eta) + \left(\Omega^{(i)}_{\bfk}(\eta)\right)^2 v_{\bfk}(\eta) = 0,\label{eq:eom_uk}
\end{equation}
where $\Omega^{(i)}_{\bfk}(\eta) = k^2+s^{(i)}(\eta)$, and $s^{(i)}(\eta)$ are the time-dependent masses of scalar ($i=s$) and tensor modes ($i=t$). This notation can be related to that of \eqref{eq:eomT} through
\begin{equation}
    \left(\Omega^{(i)}_{\bfk}(\eta)\right)^2 = a^2(\eta)\left(\omega_{\bfk}^{(i)}(t)\right)^2-\frac{a^{\prime\prime}(\eta)}{a(\eta)}.
\end{equation}
In the particular case of the classical FLRW model we have
\begin{equation}
    s^{(s)}(\eta)=-\frac{z^{\prime\prime}(\eta)}{z(\eta)},\qquad s^{(t)}(\eta)=-\frac{a^{\prime\prime}(\eta)}{a(\eta)}.
\end{equation}
For future reference, let us note that in FLRW models the latter can be written as
\begin{equation}\label{eq:stFLRW}
    s^{(t)}(\eta)=-\frac{a^{\prime\prime}(\eta)}{a(\eta)} = -\frac{4\pi G}{3}a^2(\eta)\left(\rho-3P\right).
\end{equation}

To proceed with the study of cosmological perturbations, we have to choose initial conditions $\left(v_{\bfk}(\eta_0), v_{\bfk}^{\prime}(\eta_0)\right)$ for each mode so that we find solutions to their equations of motion. Generally, this is subject to ambiguities.

\subsection{Choice of vacuum}\label{sec:introvacuum}

The ambiguity in the choice of vacuum arises in generic time-dependent backgrounds (such as curved spacetimes) in QFT. The underlying ambiguity is that there are infinite possible quantizations of the same classical model when the spacetime is not maximally symmetric (see e.g. \cite{parker1969}). To canonically quantize the Mukhanov-Sasaki field $v(\eta,\bfx)$, following the usual strategy for the Fock quantization, we choose a basis of solutions $\{\nu_{\bfk}(\eta)\}_{\bfk}$ to the equations of motion \eqref{eq:eom_uk}. This basis is normalized with respect to the Klein-Gordon product:
\begin{equation}
\nu_{\bfk}(\eta)\left(\nu_{\bfk}^{\prime}(\eta)\right)^{*}-\left(\nu_{\bfk}(\eta)\right)^{*} \nu_{\bfk}^{\prime}(\eta)=i,
\end{equation}
such that it preserves the Poisson bracket structure. This way, $\nu_{\bfk}(\eta)$ span the positive norm sector, and their complex conjugate $\nu_{\bfk}^{*}(\eta)$ the negative norm sector of the theory. The fact that $v(\eta,\bfx)$ is a real field implies that $\nu_{\bfk}^{*}(\eta) = \nu_{-\bfk}(\eta)$.

To quantize this real field we define the operator that represents it as
\begin{align}
    \hat{v}(\eta,\bfx) &= \frac{1}{(2\pi)^{3/2}} \int d^3\bfk\, e^{i\bfk\bfx} \hat{v}_{\bfk}(\eta),\\
    \hat{v}_{\bfk}(\eta) &= \hat{a}_{\bfk} \nu_{\bfk}(\eta) + \hat{a}^\dagger_{-\bfk} \nu^*_{-\bfk}(\eta),
\end{align}
where the creation and annihilation operators obey the usual commutation relations
\begin{equation}
    \left[\hat{a}_{\bfk},\hat{a}_{\bfk'}^{\dagger}\right] =  \delta\left(\bfk-\bfk'\right).
\end{equation}

Different choices of solutions $\nu_{\bfk}(t)$ translate into different annihilation and creation operators $\hat{a}_{\bfk}$ and $\hat{a}^{\dagger}_{\bfk}$, and vice versa. The quantum vacuum $\ket{0}$ associated with a given choice is defined as the state annihilated by all the annihilation operators; i.e., $\hat{a}_{\bfk}\ket{0}=0$, for all~$\bfk$. Thus, constructing a particular quantum theory is equivalent to choosing one solution $\nu_{\bfk}(\eta)$ to the equation of motion for each $\bfk$. In turn, this is equivalent to choosing initial conditions $(\nu_{\bfk}(\eta_0),\nu^{\prime}_{\bfk}(\eta_0))$ to that equation at a certain time $\eta_0$.

Here lies the ambiguity: if each choice defines a different quantum theory, each leading to different theoretical predictions, which one should we select? Depending on the particular system that we are studying, one can find in the literature many vacuum proposals. For example, in the simplest case in which we have a matter field in flat spacetime where no external agent breaking Poincar\'e symmetry is present, the frequency $\Omega_{\bfk}(\eta)$ is constant and the Minkowski vacuum is the only one invariant under such a symmetry group. This well-known vacuum is defined by a basis $\{\nu_{\bfk}(\eta)\}_{\bfk}$ which are positive-frequency plane waves, i.e., $\nu_{\bfk} = e^{-ik\eta}/\sqrt{2 k}$. However, in general when we introduce a time dependency in the frequency by means of curvature or another time-dependent external agent, such as an electromagnetic field, Poincaré invariance is broken and the classical group of symmetries is severely reduced. Then, the criterion of preservation of the classical symmetries in the quantum theory is not strong enough to select one unique vacuum. One needs further criteria, either mathematical or physical, to define such a state.

Strictly speaking, this is also the case in standard cosmology. In a de Sitter Universe, where the expansion is uniform ($H$ is constant), there is not a unique vacuum which is invariant under the symmetries of the spacetime. The equations of motion of the modes of the perturbations are equivalent to \eqref{eq:eomdPhik} with $m = 0$ for both tensor and scalar modes (although the latter only approximately). In fact there is a family of solutions for the equations of motion of the Fourier modes of a minimally coupled massless scalar field of the type:
\begin{equation}\label{eq:dSvacuum}
    \nu_{\bfk}^{\text{dS}}(\eta) = \alpha e^{-ik\eta}\left(1-\frac{i}{k\eta}\right)+\beta e^{ik\eta}\left(1+\frac{i}{k\eta}\right).
\end{equation}
However, physical intuition dictates that modes with very small wavelengths (i.e. large $k$) should not be affected by curvature and so the solution should tend to the unique vacuum in flat spacetime, i.e. the Minkowski vacuum. These can be expressed as the modes whose wavelengths are much smaller than the Hubble radius, i.e. , when $k|\eta|\gg 1$. Equivalently, one can also argue that in the distant past, $\eta\rightarrow -\infty$, the vacuum should agree with the Minkowski vacuum. This leads us to choose $\beta = 0$ and $\alpha = 1/\sqrt{2k}$ for proper normalization, which is known as the Bunch-Davies vacuum \cite{Allen:1985ux}:
\begin{equation}\label{eq:v(eta)BD}
    \nu_{\bfk}^{\text{BD}}(\eta) = \frac{1}{\sqrt{2k}}e^{-ik\eta}\left(1-\frac{i}{k\eta}\right).
\end{equation}
Standard inflation is not strictly de Sitter, rather it is a quasi-de Sitter epoch, where $H$ is only approximately constant. Nevertheless, the usual approach is to consider this approximation to be sufficient. Furthermore, the scalar modes of cosmological perturbations do not have exactly the same time-dependent mass function as massless minimally coupled fields. However, standard inflation happens during the slow-roll period of the inflaton, where essentially $\dot{\phi}$ is sufficiently small, and so the scalar and tensor modes of perturbations obey approximately the same equations of motion, i.e., those of a minimally coupled massless scalar field with solutions \eqref{eq:dSvacuum}. The same reasoning that we have just described applies and thus the Bunch-Davies state is chosen as the vacuum of perturbations in standard cosmology.

In more general scenarios, however, there are several proposals for vacua, all equally valid a priori. We will now review the construction of two of them that we will rely on in later chapters. Before we do, it is useful to note that, using the normalization condition, we can parameterize the different possible choices (up to an irrelevant phase) as follows~\cite{Mottola2000}:
\begin{equation}\label{eq:CkDk}
    \nu_{\bfk}(\eta_0) = \frac{1}{\sqrt{2 D_{\bfk}(\eta_0)}}, \qquad \nu_{\bfk}^{\prime}(\eta_0) = \sqrt{\frac{D_{\bfk}(\eta_0)}{2}}\left[C_{\bfk}(\eta_0)-i\right],
\end{equation}
where $D_{\bfk}(t_0)>0$ and $C_{\bfk}(t_0)$ are independent real quantities which determine the particular vacuum that we select. For example, the Minkowski vacuum can be parametrized as
\begin{equation} \label{eq:planewave}
    D_{\bfk}(\eta_0)=\Omega_{\bfk}, \qquad C_{\bfk}(\eta_0)=0.
\end{equation}

\subsubsection{Adiabatic states}\label{sec:adiabatic}

Among the possible candidates for initial conditions, adiabatic vacua are one of the most popular choices for selecting a set of positive frequency solutions.\footnote{They were originally proposed as approximated solutions to the equations of motion although nowadays it is common to consider them for the selection of initial conditions for cosmological perturbations.} To define them, one first considers the ansatz:
\begin{equation}
    \nu_{\bfk}(\eta) = \frac{1}{\sqrt{2W_{\bfk}(\eta)}}e^{-i\int^{\eta}W_{\bfk}(\bar{\eta}){\rm d}\bar{\eta}},
\end{equation}
which is plugged into the equations of motion of the perturbations, yielding:
\begin{equation}\label{eq:W_adiabatic}
    W_{\bfk}^2(\eta) = \Omega_{\bfk}(\eta) -\frac{1}{2}\frac{W^{\prime\prime}_{\bfk}(\eta)}{W_{\bfk}(\eta)}+\frac{3}{4}\left(\frac{W^{\prime}_{\bfk}(\eta)}{W_{\bfk}(\eta)}\right)^2.
\end{equation}

A solution of order $n$ denoted by $W_{\bfk}^{(n)}(\eta)$ is defined as an approximated solution that converges to $W_{\bfk}(\eta)$ in the limit of large $k$ at least as $\mathcal{O}\left(k^{n-\frac{1}{2}}\right)$.

There is not a unique procedure to obtain the functions $W_{\bfk}^{(n)}(\eta)$. We will adopt the following one: the adiabatic solution of order $n+2$, namely $W_{\bfk}^{(n+2)}(\eta)$, is obtained by inserting in the right-hand side of \eqref{eq:W_adiabatic} the solution $W_{\bfk}^{(n)}(\eta)$:
\begin{equation}\label{eq:ad_n}
    \left(W_{\bfk}^{(n+2)}(\eta)\right)^2 = \Omega_{\bfk}(\eta) -\frac{1}{2}\frac{\left[W_{\bfk}^{(n)}(\eta)\right]^{\prime\prime}}{W^{(n)}_{\bfk}(\eta)}+\frac{3}{4}\left(\frac{\left[W_{\bfk}^{(n)}(\eta)\right]^{\prime}}{W^{(n)}_{\bfk}(\eta)}\right)^2,
\end{equation}
starting with $W_{\bfk}^{(0)}(\eta) = k$. In this work we will rather use adiabatic solutions as initial data that will be evolved numerically with the exact equations of motion. We will therefore refer to the state defined by the set of solutions with initial data 
\begin{equation}
    D_{\bfk}(\eta_0) = W_{\bfk}^{(n)}(\eta_0), \qquad C_{\bfk}(\eta_0) = -\frac{\left[W_{\bfk}^{(n)}(\eta_0)\right]^{\prime}}{2 \left[W_{\bfk}^{(n)}(\eta_0)\right]^2},
\end{equation}
for some initial time $\eta_0$ as an adiabatic vacuum of order $n$.

\subsubsection{States of Low Energy}\label{sec:introSLE}

States of Low Energy (SLEs) have been introduced in \cite{Olbermann2007}. That work is based on the result of \cite{Fewster2000}, where it is shown that, unlike the instantaneous energy density (at one spacetime point), the renormalized energy density smeared along a time-like curve using a point-splitting procedure is bounded from below as a function of the state. Thus, there exists a state that minimizes this quantity. The work of \cite{Olbermann2007} adapts this result to FLRW spacetimes, considering a smearing function supported on the worldline of an isotropic observer and a procedure is developed to obtain such a state.

Let us consider the FLRW model with minimally coupled massless scalar fields with Fourier modes $T_\bfk(t)$ with equations of motion \eqref{eq:eomT}. In this context, we will drop the $(i)$ notation for simplicity. The goal of the procedure is to find the state, or equivalently, the solution $T_{\bfk}(t)$ which has minimal smeared energy density associated to the smearing function $f(t)$, which is called a SLE. This is achieved by minimizing the contribution of each mode ${\bfk}$ to the total smeared energy density. This contribution is given by
\begin{equation}\label{energy}
   E(T_{\bfk}(t)) = \frac{1}{2} \int dt\,f^2(t)\left(\frac{|\pi_{T_{\bfk}}(t)|^2}{a^6(t)} +\omega_{\bfk}^2(t) |T_{\bfk}(t)|^2\right),
\end{equation}
where $\pi_{T_{\bfk}}(t)$ is the conjugate momentum of $T_{\bfk}(t)$, which is related to the velocity by means of $\pi_{T_{\bfk}}(t) = a^3(t)\dot T_{\bfk}(t)$ via Hamilton's equations. As investigated in \cite{Olbermann2007}, we can start by considering a fiducial solution $S_{\bfk}(t)$ to \eqref{eq:eomT}. Then, a generic solution to the equation of motion can be written by a Bogoliubov transformation as
\begin{equation}\label{eq:S_k to T_k}
    T_{\bfk}(t) = \lambda({\bfk}) S_{\bfk}(t) + \mu({\bfk}) \bar{S}_{\bfk}(t),
\end{equation}
where $\lambda({\bfk}),\mu({\bfk}) \in \mathds{C}$ are the Bogoliubov coefficients, thus obeying $|\lambda({\bfk})|^2-|\mu({\bfk})|^2 = 1$. Noting that there is a freedom in the choice of the complex phase of $T_{\bfk}(t)$ (if $T_{\bfk}(t)$ is a solution, then so is $e^{i \delta({\bfk})} T_{\bfk}(t)$, with $\delta({\bfk}) \in \mathds{R}$), we can choose $\mu({\bfk}) \in \mathds{R}^+$ without loss of generality. This leaves us with only the following free parameters: $\mu({\bfk})$ and the complex phase of $\lambda({\bfk})$, which we denote as $\alpha({\bfk})$. Writing $E(T_{\bfk})$ in terms of the reference solution $S_{\bfk}(t)$ yields
\begin{equation}\label{eq:W(T_k)}
    E(T_{\bfk}(t)) = (2 \mu^2({\bfk})+1) E(S_{\bfk}(t)) + 2\mu({\bfk}) \text{Re}[\lambda({\bfk}) c({\bfk})],
\end{equation}
where
\begin{equation}
    c({\bfk}) := \frac{1}{2} \int dt\,f^2(t)\left(\frac{\pi_{S_{\bfk}}^2(t)}{a^6(t)} +\omega_{\bfk}^2(t) S_{\bfk}^2(t)\right),\label{eq:c2}
\end{equation}
with $\pi_{S_{\bfk}}(t)=a^3(t)\dot S_{\bfk}(t)$. Note that for a given fiducial solution $S_{\bfk}(t)$, $E(S_{\bfk}(t))$ and $c({\bfk})$ are fixed. Since $\mu_{\bfk},E(S_{\bfk}(t))\geq0$ and
\begin{equation}
    \text{Re}[\lambda({\bfk}) c({\bfk})] = |\lambda(\bfk)| |c(\bfk)| \cos\left(\alpha(\bfk)+\text{Arg}[c({\bfk})]\right),
\end{equation}it is straightforward to see that a minimum $E(T_{\bfk})$ will require $\alpha({\bfk}) = \pi-\text{Arg}[c({\bfk})]$. Then, minimizing $E(T_{\bfk}(t))$ with respect to $\mu({\bfk})$ yields:
\begin{align}
    \mu({\bfk}) &= \sqrt{\frac{E(S_{\bfk}(t))}{2\sqrt{\left[E(S_{\bfk}(t))\right]^2-|c^2({\bfk})|}}-\frac{1}{2}}\ ,\label{eq:mu}\\
    \lambda({\bfk}) &= -e^{-i\text{Arg}[c({\bfk})]} \sqrt{\frac{E(S_{\bfk}(t))}{2\sqrt{\left[E(S_{\bfk}(t))\right]^2-|c^2({\bfk})|}}+\frac{1}{2}}\ .\label{eq:lambda}
\end{align}

In summary, the SLE associated to a smearing function $f(t)$ is given (up to a phase) by \eqref{eq:S_k to T_k},  starting from an arbitrary fiducial state defined via  the modes $S_{\bfk}(t)$, with $\mu({\bfk})$ and $\lambda({\bfk})$ found by \eqref{eq:mu} and \eqref{eq:lambda}, respectively. 

In \cite{Olbermann2007} it is also shown that for ultrastatic models there exists a state (unique up to a phase) which minimizes the energy density for all test functions, being dubbed the state of \emph{minimal} energy. In non ultrastatic models, however, such a state does not exist, every SLE is associated to a smearing function. This compromises the universality of the proposal and raises questions on the relation between these and the natural vacua of maximally symmetric spacetimes. It is desirable that any procedure aiming at defining a privileged vacuum state should somehow single out the already identified natural vacuum in such spacetimes. In Minkowski, SLEs are trivially found to be identical to the natural Minkowski vacuum, regardless of the choice of the smearing function.\footnote{This is easily proven by checking that \eqref{eq:c2} is identically 0 for the Minkowski vacuum. Then, if one starts with the Minkowski vacuum as the fiducial solution, one obtains that $\mu({\bfk}) = 0$ and $\lambda({\bfk})$ is just a complex phase, indicating that the SLE is in fact the one we started with.} In de Sitter, where  the preferred vacuum is given by the Bunch-Davies vacuum, the question is not trivial. In \cite{DegnerPhD2013} it is shown that, in the massive field case in de Sitter, SLEs converge to the Bunch-Davies vacuum in an appropriate limit of the support of the smearing function. The non-triviality in this case is likely related to the fact that the Bunch-Davies vacuum is not strictly the unique natural vacuum in de Sitter. As mentioned, to obtain it one needs to impose that the vacuum behaves as the Minkowski vacuum in the ultraviolet, or equivalently in the distant past.

Remarkably, SLEs are shown to be of Hadamard type, which guarantees certain mathematical properties that asure that computations such as that of the energy-momentum tensor are well defined. A recent investigation \cite{Niedermaier2020} has explored several additional properties of SLEs. It is shown that this construction is independent on the fiducial solution. Additionally, it has determined that these states admit ultraviolet and infrared expansions. When applying the construction to primordial perturbations, these results allow for an analytical development of the asymptotic behaviors of their power-spectra at the end of inflation, from which it is found that these agree with observations in models where a period of kinetic dominance precedes inflation. As we will detail shortly, this is precisely the case of inflationary LQC models.

We will apply this construction to LQC models in chapter \ref{chap:SLEsinLQC} and to the Schwinger effect in chapter \ref{chap:Schwinger}.

\subsection{Primordial power spectrum}\label{sec:introPPS}

Let us now imagine that we have some preferred notion of vacuum that provides initial conditions for the equations of motion. Finally, we are in a position to evolve the perturbations so that we can compute predictions that we may compare with data. The first step consists of evolving them until the end of inflation, to compute the primordial power spectra. These are obtained through the Wightman function, which for the real field $\hat{v}(t,\bfx)$ is defined as the vacuum expectation value:
\begin{equation}\label{eq:Wightman_cosmo}
    W(t,\bfx;t',\bfx')=\bra{0}\hat{v}(t,\bfx)\hat{v}(t',\bfx')\ket{0}=\int \frac{d^3\bfk}{(2\pi)^3} \ e^{i\bfk\cdot (\bfx-\bfx')} \nu_{\bfk}(t)\nu^*_{\bfk}(t'),
\end{equation}
Due to homogeneity, this function only depends on the position vectors through the difference $\bfx-\bfx'$. Taking the limit of coincidence $t'\rightarrow t$ yields
\begin{equation}
    \lim_{t'\rightarrow t} W(t,\bfx;t',\bfx')=\frac{1}{4\pi}\int \frac{d^3\bfk}{k^3} \ e^{i\bfk\cdot (\bfx-\bfx')} \mathcal{P}(t,\bfk),
\end{equation}
where we defined the power spectrum as
\begin{equation}
\label{eq:PS}
    \mathcal{P}(t,\bfk)=\frac{k^3}{2\pi^2}|\nu_{\bfk}(t)|^2.
\end{equation}
Note that in the equation of motion \eqref{eq:eom_uk} only modes with the same  wave number $\bfk$ are coupled. Therefore, the power spectrum depends only on the modulus $k$ and in the following we will identify the Fourier modes with $u_{k}(\eta)$ and $\mu_{k}^{I}(\eta)$ for all $\bfk$ with the same $k$, both for scalar and tensor perturbations, as defined above equation \eqref{eq:eom_uk}. 

Given any initial conditions, the perturbations can be evolved mode by mode with the equation of motion, until the time $\eta_{\text{end}}$ when the relevant scales have all crossed out the horizon and their evolution has stalled.  After that, the power spectra of the comoving curvature perturbation ${\cal R}_k=u_{k}/z$ and tensor modes will remain frozen until the end of inflation.\footnote{In fact, the perturbations remain frozen for as long as their wavelength is much larger than the Hubble radius, i.e., until they cross back into the horizon. This happens during the period of decelerated expansion which follows inflation.} They are given by
\begin{equation}\label{eq_PSRT}
    \mathcal{P}_{\mathcal{R}}(k) = \frac{k^{3}}{2\pi^{2}}\frac{|u_{k}|^{2}}{z^{2}}\Big|_{\eta = \eta_{\rm end}},\qquad
    \mathcal{P}_{\mathcal{T}}(k) = \frac{32k^{3}}{\pi^2}\frac{|\mu_{k}^{I}|^{2}}{a^{2}}\Big|_{\eta = \eta_{\rm end}},
\end{equation}
respectively.

In standard cosmology, choosing the Bunch-Davies vacuum at the onset of inflation, both for scalar and tensor perturbations, leads to primordial power spectra of the form:
\begin{equation}
    \mathcal{P}_{\mathcal{R}}(k) = A_s \left(\frac{k}{k_*}\right)^{n_s-1},\qquad \mathcal{P}_{\mathcal{T}}(k) = A_t \left(\frac{k}{k_*}\right)^{n_t}.
\end{equation}
Here, $k_*$ is a reference scale, usually called the pivot scale, at which the power spectrum of scalar and tensor modes are $A_s$ and $A_t$ respectively, and $n_s$ and $n_t$ are the spectral tilts of scalar and tensor modes, respectively. Current observations are mostly compatible with a near-scale-invariant scalar power spectrum with $n_s \lesssim 1$, corresponding to a slight red tilt. An important quantity is the tensor-to-scalar ratio $r$, which measures the power of tensor modes with respect to scalar modes:
\begin{equation}\label{eq:rTtoR}
    r = \frac{\mathcal{P}_{\mathcal{T}}(k)}{\mathcal{P}_{\mathcal{R}}(k)}.
\end{equation}
The latest observations constrain this quantity at $k=0.002 \text{Mpc}^{-1}$ to $r_{0.002} < 0.032$ at 95\% confidence, being compatible with zero \cite{Tristram:2021tvh}. This constrain has ruled out several potentials of the inflaton in the context of standard cosmology.

In chapter \ref{chap:SLEsinLQC} we will focus our analysis on the primordial power spectra, even though this is not the quantity that is directly observed. This is because from here we can already determine whether the departures from the predictions of the standard model occur within the observable window. In the case that they do, then we may evolve the perturbations from the end of inflation until the surface of the CMB to compare with observations, as we do in chapter \ref{chap:anomalies}. If they do not occur in the observable window, then the observable predictions will exactly agree with those of the standard model.

\subsection{The observed CMB}\label{sec:anomalies}

As explained in the previous subsection, the perturbations of the inflaton induce perturbations of the energy-momentum tensor, which in particular cause perturbations in the temperature of the background radiation. Before these perturbations reach us today, they experience further variations from gravitational redshifts caused by the potential well at last scattering, and from the evolution along the line-of-sight. These are the perturbations that we can observe today by measuring the temperature map of the CMB. It is remarkably uniform, with an average of $\bar{T} = 2.725 \pm 0.002$ K \cite{Fixsen:1996nj} and fluctuations between different directions $\hat{n}$ of the order of $10^{-5}$ K. The concrete statistical properties of these fluctuations depend not only on the physics of the very early Universe described in the previous subsection, but also on the composition of the Universe from the end of inflation until the surface of last scattering and until today. We do not expect quantum gravity to have an influence on this, besides effects that carry over from pre-inflationary dynamics. For this reason, we will not concern ourselves with these details. For the analyses in this work, we will consider the model of standard cosmology regarding physics of the Universe after inflation. In summary, once we obtain the primordial power spectra, we can make use of publicly available tools like \textit{CLASS}~\cite{Blas_2011} to evolve them and extract predictions for temperature fluctuations today.

These fluctuations can be expanded in spherical harmonics and described in terms of their coefficients $a_{\ell m}$:
\begin{equation}
    \delta T(\hat{n}) = \sum_{\ell m} a_{\ell m}\,Y_{\ell m}(\hat{n}),
\end{equation}
where the sum in $\ell$ is over all positive integers and the sum in $m$ is over all integers from $-\ell$ to $\ell$. Theoretical models, in particular $\Lambda$CDM, can only predict statistical properties of the CMB map. Therefore, we are particularly interested in the moments of the coefficients, rather than their actual values. Furthermore, the $\Lambda$CDM model predicts these fluctuations to be statistically isotropic and Gaussian, thus fully characterized by the mean and the second moment
\begin{equation}
    C(\theta) = \langle \delta T(\hat{n}), \delta T(\hat{n}^{\prime}) \rangle,
\end{equation}
where $\theta$ is the angle between two directions in the sky $\hat{n}$ and $\hat{n}^{\prime}$.
Homogeneity and isotropy imply that the second moments of $a_{\ell m}$ are diagonal and depend only on the multipole $\ell$ and can thus be fully characterized by the angular power-spectrum $C_{\ell}$:\footnote{Note that the angular power spectrum can always be defined as $\langle |a_{\ell m}|^2 \rangle$, but it is only in the case of random Gaussian statistically isotropic temperature fluctuations that it fully describes the second moment of $a_{\ell m}$.}
\begin{equation}
    \langle a_{\ell m} a^*_{\ell' m'}\rangle = C_{\ell} \delta_{\ell \ell'} \delta_{m m'}.
\end{equation}
The angular power-spectrum $C_{\ell}$ is related to the correlations in physical space $C(\theta)$ through:
\begin{equation}\label{eq:ctheta}
    C(\theta) = \frac{1}{4\pi} \sum_{\ell} (2\ell + 1) C_{\ell} P_{\ell}(\cos \theta),
\end{equation}
where $P_{\ell}(\cos \theta)$ are the Legendre polynomials. 

From observations, ideally, one would want to obtain the moments of $a_{\ell m}$ by averaging over several realizations (i.e. several universes) or several patches of the same Universe. Since both are unattainable, they are instead obtained by averaging over all $m$, taking advantage of the $m-$independence of the coefficients. This carries an inescapable intrinsic uncertainty named cosmic variance associated to how many different values $m$ takes: $\Delta C_{\ell}/C_{\ell} = \sqrt{2/(2\ell+1)}$. Thus, lower multipoles will unavoidably have larger cosmic variance, as averages are over less values of $m$. In this work, we will need to consider that $C_{\ell}$ are independent random variables with Gaussian distribution. In this case we will take the variance of this distribution to be the cosmic variance $\Delta C_{\ell}$.

Early observations \cite{COBE} indicated that the fluctuations were consistent with a near-scale-invariant power spectrum. This is precisely the prediction of theoretical models within the inflationary paradigm, as long as inflation lasts long enough. Although the standard $\Lambda$CDM model's predictions are able to fit the data well in general, some anomalies have been identified and persist in more recent observations \cite{PlanckVII}. Let us briefly review three that are relevant in the context of quantum cosmology and bouncing scenarios, and that we will revisit in chapter \ref{chap:anomalies}.

\subsubsection{Power suppression anomaly}

\begin{figure}
    \centering
    \includegraphics[width=0.49\textwidth]{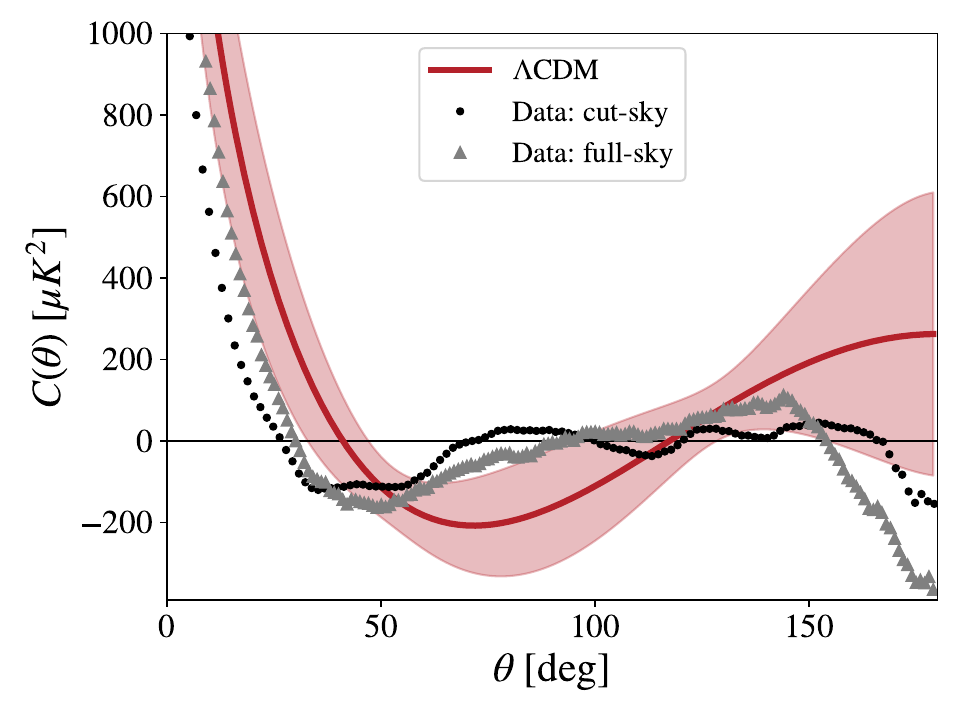}
    \caption[Angular two-point correlation function for $\Lambda$CDM and observed from cut-sky and full-sky data.]{Observed angular two-point correlation function from cut-sky (black dots) and full-sky data (grey triangles), along with the prediction from the $\Lambda$CDM model (red curve) and the corresponding 1-$\sigma$ region due to cosmic variance.}
    \label{fig:CthLCDM}
\end{figure}

Observations show that there is a lack of correlations at large angles (larger than $60^{\circ}$) with respect to the expected behavior for $\Lambda$CDM, as shown in figure \ref{fig:CthLCDM}. This translates to some lack of power in the angular power spectrum for low $\ell$ and is thus commonly called the power suppression anomaly. However, what is most relevant is that the two-point correlation is remarkably consistent with zero for these scales, except for some anti-correlations at $180^{\circ}$. Numerically, the anomaly is best quantified via the estimator
\begin{equation}
    S_{1/2} = \int_{-1}^{1/2} C^2(\theta) d(\cos \theta).
\end{equation}

To simplify calculations and avoid noise coming from $C(\theta)$, this quantity may be computed through $C_{\ell}$ by inserting \eqref{eq:ctheta} as
\begin{equation}\label{eq:S12fromCl}
    S_{1/2} = \sum_{\ell,\ell^{\prime} = 2}^{\ell_{\textrm{max}}} C_{\ell} I_{\ell \ell^{\prime}} C_{\ell^{\prime}},
\end{equation}
where $I_{\ell \ell^{\prime}}$ include the integrals of products of Legendre polynomials and can be found in appendix \ref{sec:appLegendre}. The sum should in principle be over all (available) $\ell$, $\ell^{\prime}$, but it is enough to consider up to $\ell_{\textrm{max}} \sim 100$, as the Legendre polynomials sufficiently suppress higher multipole terms.

From Planck's cut-sky data (where the portion of the sky contaminated by our galactic disk has been removed through a mask), this quantity is found to be around $1200$, the exact value depending on the choice of map and mask \cite{PlanckVII}, whereas for full-sky data (where the contaminated region has been reconstructed) it is around $6700$ \cite{Copi2013}. These correspond to very unlikely realizations of the universe according to the $\Lambda$CDM model, where this quantity is expected to be around $35000$, with the observed values corresponding to p-values of $\sim 0.1 \%$ and $\sim 5\%$, respectively.\footnote{Here we define the p-values as the probability of finding values of $S_{1/2}$ at least as low as that observed in a random realization, given cosmic variance. As such, a low p-value indicates a very unlikely realization of the model.}

\subsubsection{Lensing anomaly}

\begin{figure}
    \centering
    \includegraphics[width=0.49\textwidth]{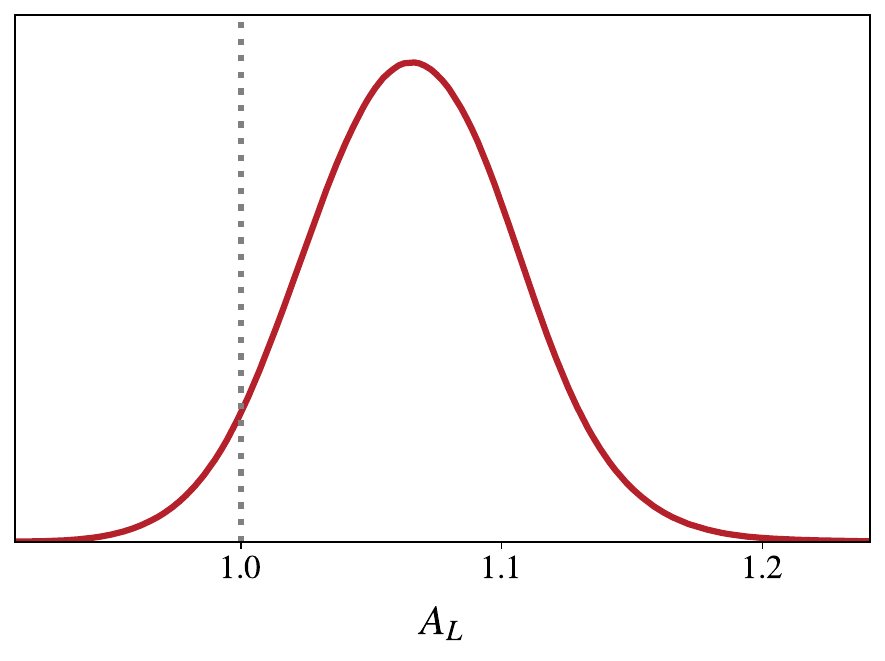}
    \caption[Posterior probability of the lensing amplitude parameter for the $\Lambda$CDM$+A_L$ model given Planck 2018 data with lensing.]{Posterior probability of the lensing amplitude parameter for the $\Lambda$CDM+$A_L$ model given Planck 2018 data with lensing. Also indicated is the value $A_L = 1$, which would correspond to a $\Lambda$CDM universe.}
    \label{fig:ALLCDM}
\end{figure}

As a consistency check, a phenomenological parameter, $A_L$, can be introduced to control how much or how little the CMB is lensed from the surface of last scattering until today, such that $A_L = 0$ means it is not at all lensed, and $A_L = 1$ corresponds to the prediction of the standard model. The anomaly is manifest in a Bayesian analysis of the $\Lambda$CDM+$A_L$ model, as it shows that the data prefers $A_L > 1$ at a 2-sigma level, and with large improvements of $\chi^2$.\footnote{The significance of this anomaly depends on the set of data. For the Planck 2018 data without lensing, this significance reaches about 3-$\sigma$, whereas for the data with lensing it is lower than 2-$\sigma$. In this work we will consider the data with lensing, so that our results can be compared with those of other investigations in LQC that rely on the same data, such as \cite{Ashtekar:2020,Ashtekar:2021}.} This can be seen in the marginalized posterior probability of this parameter, as we show in figure \ref{fig:ALLCDM}. As we will explain in the next section, the marginalized posterior is a probability distribution of a single parameter, taking into account the full parameter space of the model. A peak in this distribution indicates a likely value of this parameter given the full model and the data. We can see that the distribution in this case is a Gaussian with an average that is displaced to $A_L > 1$, which means that this is the preference the data shows in a $\Lambda$CDM+$A_L$ model, and that is in tension with the prediction of $\Lambda$CDM.

If one also leaves the curvature of the Universe as a free parameter, then $A_L = 1$ is within the 1-sigma region if we allow the curvature to be negative, corresponding to a closed universe. However, this leads to discrepancies with other sets of data, which has been dubbed a ``crisis'' in cosmology \cite{DiValentino2019}. Concretely, the point of view adopted in \cite{DiValentino2019} is that fixing the curvature of the Universe to be flat hides the inconsistencies in the data that are observed when the curvature is left free. The parameter $A_L$ allows then to quantify these inconsistencies even when they are hidden behind the choice of flat Universe. Our viewpoint is that a resolution of this anomaly must address the overall inconsistencies between predictions and observations, without affecting the lensing physics of the CMB.

\subsubsection{Parity anomaly}\label{sec:anomalies_parity}

\begin{figure}
    \centering
    \includegraphics[width=0.49\textwidth]{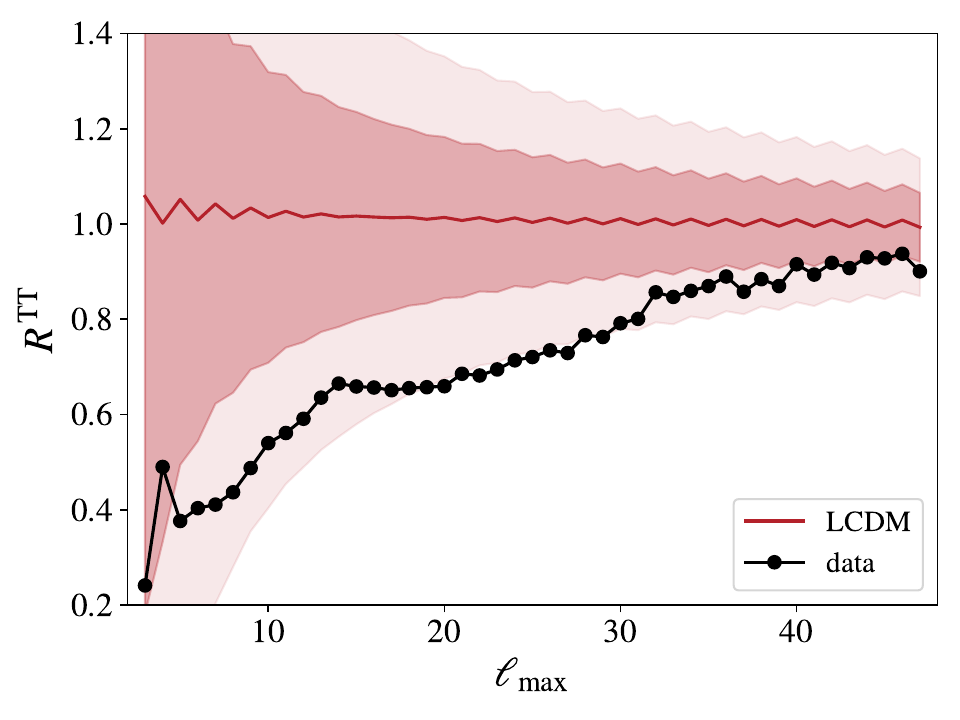}
    \caption[Parity asymmetry observed in Planck 2018 data and that predicted by the $\Lambda$CDM model.]{Parity asymmetry observed in Planck 2018 data (black dots) and that predicted by the $\Lambda$CDM model (red line). Also depicted are the 1 and 2-$\sigma$ regions corresponding to the $\Lambda$CDM prediction considering cosmic variance (shaded red regions).}
    \label{fig:RTT}
\end{figure}

The data also shows an anomalous power excess of odd-$\ell$ multipoles with respect to even ones for large angular scales ($\ell < 30$) in the angular correlation function $C_\ell$. Concretely, the parity asymmetry estimator is taken to be $R^{TT}(\ell_{\textrm{max}}) = D_+(\ell_{\textrm{max}})/D_-(\ell_{\textrm{max}})$, where
\begin{equation}
    D_{\pm}(\ell_{\textrm{max}}) = \frac{1}{\ell^{\pm}_{tot}} \sum_{\ell = 2,\ell_{\textrm{max}}}^{\pm} \frac{\ell(\ell+1)}{2\pi}C_{\ell}
\end{equation}
quantify the mean power contained in even (+) / odd (-) multipoles up to $\ell_{\textrm{max}}$, and $\ell^{\pm}_{tot}$ is the total number of even/ odd multipoles from 2 to $\ell_{\rm max}$. Although the $\Lambda$CDM model predicts neutral parity ($R^{TT} = 1$), it is found that $R^{TT}(\ell_{\textrm{max}}) < 1$ for low multipoles, with a statistical significance $\lesssim 2\sigma$ \cite{PlanckVII}, as we show in figure \ref{fig:RTT}.

\section{Loop Quantum Cosmology}\label{sec:introLQC}

In this section, we review the usual procedures of LQC for the quantization of an FLRW model minimally coupled to a scalar field and focus in particular on the solvable formulation of the case of vanishing potential, as it is essential for the remaining work. For more details on the quantization procedures we refer the reader to \cite{APS_extended, ACS}. We also introduce the so-called effective dynamics which are applied in more general cases. Finally, we consider cosmological perturbations, finding their equations of motion in the hybrid LQC formalism. Note that in this section the potential of the inflaton is denoted by $W(\phi)$ as $V$ is reserved for the volume. In future sections we will make clear what is the notation adopted.

\subsection{Hamiltonian formulation of homogeneous cosmological models}

Let us start with the Hamiltonian formulation of homogeneous and isotropic cosmological models. This will lead us to a formulation of FLRW spacetimes suited to the quantization procedure of LQC. To do so, we will describe the system in terms of an $SU(2)$ connection and a canonically conjugate densitized triad, which are chosen with the goal of describing the system as a gauge theory.

We will start with the action of GR \eqref{eq:SGR} and only later particularize to cosmological models. Applying the ADM formalism \cite{Arnowitt:1962hi}, the 4D metric $g_{\mu\nu}$ is split in the 3-metric $q_{ab}$ induced in the spatial slices that foliate the manifold, the lapse $N$ and the shift vector $N^{a}$, resulting in the line element\footnote{Greek indices range from 0 to 3 whereas latin ones are spatial, ranging from 1 to 3.}
\begin{equation}
    ds^2 = -\left(N^2-q_{ab}N^a N^b\right)dt^2 + 2q_{ab} N^a dt dx^b + q_{ab}dx^a dx^b.
\end{equation}
When writing the action in terms of these variables, we see that $N$ and $N^a$are Lagrange multipliers, and are therefore not physical quantities. Thus, the relevant physical information is encoded in $q_{ab}$ and its canonically conjugate momentum
\begin{equation}
    \pi^{ab} = \sqrt{q}\left(K^{ab}-K q^{ab}\right),
\end{equation}
where $q$ is the determinant of $q_{ab}$, $K_{ab}$ is the extrinsic curvature and $K=K^a_ a$. Following the techniques of LQG \cite{Thiemann_notes}, one considers an $SU(2)$ spinor bundle on the spacetime manifold and introduces the associated triad $e_i^a$ and co-triad $e_a^i$ such that\footnote{Here, indices $i,j$,... are $SU(2)$ Lie Algebra indices, whereas $a,b$... are spatial ones. Both range from 1 to 3.}
\begin{equation}
    q_{ab} = e_a^i e_b^j \delta_{ij}, \qquad K_a^i = K_{ab} e_j^b\delta^{ij}, \qquad e_i^a e_b^j = \delta_b^a \delta_i^j,
\end{equation}
where $\delta_{ij}$ is the Kronecker delta. Then, we describe the system in terms of the densitized triad $E_i^a$ and its canonically conjugate pair, the Ashtekar-Barbero gauge connection $A_a^i$:
\begin{equation}
    E_i^a = \sqrt{q}e_i^a, \qquad A^i_a = \Gamma_a^i + \gamma K_a^i,
\end{equation}
where $\gamma$ is an arbitrary real number named the Immirzi parameter, and $\Gamma_a^i$ is the Levi-Civita spin connection ruling the parallel transport on the spinor bundle.

In the case of flat FLRW models, due to homogeneity, some integrals in the Hamiltonian framework diverge. This divergence is spurious and simply due to the fact that the spatial slices are non-compact. To deal with this, we restrict calculations to a fixed fiducial cell $\mathcal{V}$, which acts as an infrared regulator. Homogeneity implies that the dynamics on $\mathcal{V}$ reproduce those of the entire Universe. Indeed, at the end of the calculations, the volume of the fiducial cell can be taken to infinity, and in fact physical quantities do not depend on the choice of this cell. We fix a fiducial Euclidean metric $\fidq_{ab}$, with respect to which the fiducial cell has a volume $V_0$. In this setup, due to homogeneity and isotropy, we can also parametrize the connection and triad by only one spatially constant parameter for each \cite{Ashtekar_Bojowald_Lewandowski}:
\begin{equation}
    A_a^i = c(t) V_0^{-1/3}\fide_a^i, \qquad E_i^a = p(t) V_0^{2/3}\sqrt{\fidq}\fide_i^a,
\end{equation}
where $\fidq$ is the determinant of the fiducial metric, with respect to which we define the fiducial co-triad $\fide_a^i$. The system can be described in terms of the new canonical pair $c(t)$ and $p(t)$ with Poisson brackets
\begin{equation}
    \lbrace c(t), p(t) \rbrace = \frac{8\pi G \gamma}{3}.
\end{equation}
These can be related with the scale factor and its momentum through
\begin{equation}
    |p(t)| = a^2(t) V_0^{2/3},\qquad c(t) = \gamma V_0^{-1/3}\dot{a}(t).
\end{equation}
The sign of $p(t)$ represents the relative orientations of the physical and fiducial triads. Finally, in LQC, as in LQG, there is no operator corresponding to the connection itself. The curvature is then defined in terms of holonomies of the connection taken in the fundamental representation of $SU(2)$, and we may describe the system in terms of them and fluxes of the densitized triad. Classically, quantities that are defined in terms of the connection could be obtained by considering holonomies around a loop, and taking the limit where the enclosed area vanishes. In the quantum theory, on the other hand we cannot continuously shrink this area until it vanishes, as this limit is not well defined. Indeed, in LQG the geometry is discrete. The area operator has a discrete spectrum. In LQC this results in a gap between zero and the minimum non-vanishing eigenvalue equal to $\Delta= 4\sqrt{3}\pi G \gamma$ \cite{Ashtekar_bianchiI}, aptly named the area gap. Thus, instead of taking the limit where the area goes to zero, in LQC it is taken as the area goes to $\Delta$. In summary, following the "improved dynamics" prescription \cite{APS_extended}, the holonomies are taken along straight lines which enclose a square of physical area $\Delta$, whereas the fluxes are simply proportional to $p$.

To simplify calculations, we perform another change of variables to a new canonical pair 
\begin{equation}
    v = \text{sign}(p)\frac{|p|^{3/2}}{2\pi G \gamma \sqrt{\Delta}},\qquad b = \sqrt{\frac{\Delta}{|p|}}c,
\end{equation}
with $\lbrace b,v \rbrace = 2$, so that the holonomies produce a constant shift in the new geometric variable $v$. These can be related back to the original pair consisting of the scale factor and its momentum as
\begin{equation}
    |v| = \frac{a^3V_0}{2\pi G \gamma \sqrt{\Delta}}, \qquad b = \gamma \sqrt{\Delta}H,
\end{equation}
such that the physical volume of the fiducial cell $\mathcal{V}$ is
\begin{equation}
    V = a^3V_0 = 2\pi G \gamma \sqrt{\Delta} |v|.
\end{equation}

The simplest cosmological model with non-trivial dynamics includes also a minimally coupled homogeneous massless scalar field $\phi$ and its momentum $\pi_{\phi}$, which form a canonical pair with Poisson brackets $\lbrace \phi, \pi_{\phi} \rbrace = 1$. Let us consider instead the more general case of a scalar field subject to a potential $W(\phi)$, from which we may recover the massless case by taking it to be zero. The Hamiltonian of GR is found to be a linear combination of constraints. Once it is symmetry reduced to a cosmological model, only a global Hamiltonian constraint survives, namely the zero mode of the full Hamiltonian constraint, which generates time reparametrizations:
\begin{equation}\label{eq:constraintVH}
    C(N) = N\left[\frac{\pi_{\phi}^2}{2 V}-\frac{3}{8\pi G} V H^2 + V W(\phi)\right] = 0.
\end{equation}
For convenience, we will fix a harmonic gauge $N = 2 V$ and work with the canonical pair $(v,b)$ such that:
\begin{equation}\label{eq:constraint}
    C = \pi_{\phi}^2 - \frac{3}{4\pi G \gamma^2} \Omega_0^2 + 8\pi^2 G^2 \gamma^2 \Delta v^2 W(\phi) = 0,
\end{equation}
where $\Omega_0 = 2\pi G \gamma b v$. Since the Hamiltonian is a constraint, it does not generate evolution in the time coordinate of the metric, and we are forced to address the concept of evolution in this scenario. This is already true in full GR, where the Hamiltonian is given by a linear combination of constraints. To speak of evolution in the quantum theory, we need to select a variable to undertake the role of internal time. Then we may evolve the remaining variables with respect to this internal clock. In LQC, it is the field $\phi$ that is chosen as the clock variable.

\subsection{Solvable LQC}\label{sec:sLQC}

So far we have described the classical system in our preferred variables. The next step is to promote them (and the constraint \eqref{eq:constraint}) to well-defined operators on a kinematical Hilbert space. The latter will be given by two sectors: one for geometry and another one for matter. The geometrical sector is quantized with a polymeric prescription, which distinguishes LQC from other procedures such as the Wheeler-De Witt approach. On the other hand, for the matter sector a standard Schr\"odinger-like representation is adopted. In a first instance, we are going to consider the case of an FLRW spacetime with a minimally coupled massless scalar field, i.e., corresponding to taking $W(\phi) = 0$. This allows us to cast the system into an analytically solvable formulation. Later we will consider the case with non-vanishing potential.


Mimicking the procedures of LQG, in LQC there is no operator directly representing $b$. Instead, there is one to represent the matrix elements of the holonomies $\widehat{e^{i b/2}}$, and one represents the basic commutation relation  $[\hat{v}, \widehat{e^{i b/2}}]=i \widehat{\{v, e^{i b/2}\}}$. Then, in the polarization where the operator $\hat{v}$ acts by multiplication, the operator $\widehat{e^{i b/2}}$ produces a constant shift on the quantum state $\ket{v}$:
\begin{equation}
    \widehat{e^{i b/2}} \ket{v} = \ket{v+1},\qquad \hat{v}\ket{v} = v\ket{v}.
\end{equation}
In other words, there is no infinitesimal generator of translations in $v$, but finite translations. Accordingly, the inner product for the geometric sector is discrete, given by the Kronecker delta:
\begin{equation}
    \bra{v}\ket{v^{\prime}} = \delta_{v,v^{\prime}}.
\end{equation}
Thus, the basis states $\ket{v}$ are normalizable and provide an orthonormal basis. This is the foundation for the outstanding features of LQC, which set it apart from other quantization procedures. For the matter field, we adopt a standard representation where $\hat{\phi}$ acts by multiplication and $\hat{\pi}_{\phi} = -i \partial_{\phi}$ acts by derivative, so that the Hilbert space for the matter sector is the standard space $L^2(\mathbb{R}, d\phi)$. The total kinematical Hilbert space is the tensor product of both geometric and matter sectors. The quantum counterpart of the constraint \eqref{eq:constraint} is now given by
\begin{equation}
    \hat{C} = -\partial_\phi^2-\hat{\Theta}.
\end{equation}
The geometric part is encoded in $\hat{\Theta} = \frac{3}{4\pi G\gamma^2}\hat{\Omega}_0^2$, where
\begin{equation}
    \hat{\Omega}_0 = \frac{1}{2\sqrt{\Delta}}\hat{V}^{1/2}\left[\widehat{\text{sign}(v)}\widehat{\sin b}+\widehat{\sin b}\widehat{\text{sign}(v)}\right]\hat{V}^{1/2},
\end{equation}
and $\widehat{\sin b} = \left(\widehat{e^{ib}}-\widehat{e^{-ib}}\right)/(2i)$ \cite{Merce_Guillermo_Javier}. Then, $\hat{\Theta}$ is a difference operator of step 4.

Thus the physical states decouple in superlection sectors, spanned by volume eigenstates which differ by multiples of four units between them.

There is an especially useful representation named sLQC \cite{ACS}, in which the free model (with vanishing potential) can be straightforwardly solved analytically. Let us denote as $\Gamma$ the wave functions of the $(v,\phi)$ representation. Taking the sector spanned by the states supported on the volume variable $v$ equal to a multiple of four ($v=4n$, where $n$ is an integer), a discrete Fourier transform is performed from $v$ to $b$, followed by a rescaling of $\Gamma$ by $\chi=\Gamma/(\pi v)$ and the change of variables:
\begin{equation}
	x=\frac{1}{\sqrt{12\pi G}}\ln\left[ \tan\left( \frac{b}{2} \right) \right],
\end{equation}
so that $b=2\tan^{-1} \left( e^{\sqrt{12\pi G}x} \right)$. This way, representing the functions of the connection in terms of holonomy elements $e^{ib/2}$, we obtain a counterpart of $\Omega^2_0$ which is equivalent to replacing $b$ with $\sin b$. Thus, in the particular case of vanishing potential $W(\phi)=0$, the quantum counterpart of the constraint \eqref{eq:constraint} reads
\begin{equation}\label{eq:constraint_x}
    \hat{\pi}_{\phi}^2-\hat{\pi}_x^2=0,
\end{equation}
where $\hat{\pi}_x$ and $\hat{\pi}_{\phi}$ are the two momentum operators, which act by derivative: $\hat{\pi}_x=-i\partial_x$ and $\hat{\pi}_\phi=-i\partial_\phi$. We also have $\hat{\pi}_x = \sqrt{3/(4\pi G\gamma^2)}\hat{\Omega}_0$.

Regarding $\phi$ as (internal) time, the constraint is a Klein-Gordon equation in 1+1 dimensions. To remove double counting of solutions due to time reversal invariance, we take $\hat{\pi}_{\phi}$ to be positive. While in the kinematical Hilbert space the operators $\hat\phi$ and $\hat\pi_\phi$ provide a pair of canonically conjugated essentially self-adjoint operators, note that at the physical level, $\phi$ is the time variable (and it has no associated operator defined in the physical Hilbert space) and $\hat\pi_\phi=\pm|\hat\pi_x|$ is the Hamiltonian generating the evolution of positive and negative frequency solutions respectively. The positive frequency solutions of the constraint \eqref{eq:constraint_x} are then functions of the form $\chi(x_{\pm}) = \chi(\phi \pm x)$, which correspond to left and right moving modes, respectively. Here $\chi$ is any function whose Fourier transform is supported on the positive real line. The operator $\hat{\pi}_{x}=-i\partial_x$ is positive on left-moving modes and negative on right-moving modes. Moreover,
invariance under reversal of the triads imposes the condition that  physical states verify $\chi(x,\phi)=-\chi(-x,\phi)$, which leads to physical states of the form:
\begin{equation}\label{eq:states_sLQC}
	\chi(x,\phi) = \frac{1}{\sqrt{2}}\left[\chi(x_+)-\chi(x_-)\right].
\end{equation}

This way, left and right moving modes are not independent (as would be the case had we performed a standard quantization instead of the polymeric one of LQC), and we can write physical results using only, e.g., left-moving modes. In particular, the (time-independent) inner product on physical states can be written as
\begin{equation}\label{eq:innerprod_sLQC}
    ( \chi_1,\chi_2 )=2 i \int_{-\infty}^\infty dx\ [\partial_x{\chi_1}^*(x_+)] \chi_2(x_+),
\end{equation}
where $*$ denotes complex conjugation.

The dynamics can be integrated straightforwardly: in Schr\"odinger picture, the evolution of positive frequency states from an initial time $\phi_0$, that for simplicity we set equal to 0,  is given by 
\begin{equation}
    \chi(x,\phi)=e^{i|\hat{\pi}_x|\phi}\chi_0(x),
\end{equation}
$\chi_0$ being the initial FLRW state at $\phi_0=0$. In what follows we choose 
\begin{equation}\label{eq:chi0}
    \chi_0(x)=\frac{1}{\sqrt{2}}\left[F(x)-F(-x)\right],
\end{equation}
so that 
$\chi(x,\phi) = \left[F(\phi+x)-F(\phi-x)\right]/{\sqrt{2}}$.
We recall that $F$ is a function with Fourier transform supported on the positive real line as we are restricting to the positive frequency sector.

Moreover, within the Heisenberg picture, $\hat{\pi}_x$ and $\hat{\pi}_{\phi}$ are Dirac observables, preserved by the dynamics, whereas $\hat{x}(\phi)$ is found to be
\begin{equation}\label{eq:xevolution}
 \hat{x}(\phi)= \hat{x}_0 - \phi\, \text{sign}(\hat{\pi}_x).
\end{equation}

Here $\hat{x}_0$ represents $\hat{x}$ in the section of evolution where $\phi=0$.
Furthermore, the operator that represents the volume of the cell $\mathcal V$ is given (in Schr\"odinger picture) by \cite{ACS}:
\begin{equation}
   \hat{V}=\frac{2\pi G\gamma \sqrt{\Delta}}{\sqrt{3\pi G}} \cosh(\sqrt{12\pi G}\hat{x}) |\hat{\pi}_x|.
\end{equation}

The Heisenberg picture counterpart, $\hat{V}(\phi)$, is obtained simply by replacing $\hat{x}$ with $\hat{x}(\phi)$. Then,  as a function of the internal time $\phi$, the expectation value of the volume operator on all physical states turns out to be
\begin{equation}\label{eq:expVsLQC}
	\langle \hat{V}(\phi)\rangle_{\chi_0} = V_+e^{\sqrt{12\pi G}\ \phi}+V_-e^{-\sqrt{12\pi G}\ \phi},
\end{equation}
 where $\langle \hat{V}(\phi)\rangle_{\chi_0}\equiv({\chi_0}, \hat{V}(\phi){\chi_0} )=({\chi}, \hat{V}{\chi} )$ and $V_\pm$ are state-dependent constants defined, in terms of the function $F(x)$, as
\begin{equation}\label{eq:v+-sLQC}
	V_\pm \equiv \frac{2 \pi G \gamma \sqrt{\Delta}}{\sqrt{3\pi G}}\int_{-\infty}^{+\infty} dx \left| \frac{d{F}(x)}{dx} \right|^2 e^{\mp\sqrt{12\pi G}\ x}.
\end{equation}
This can be rewritten as 
\begin{equation}\label{eq:expVsLQCcosh}
    \langle \hat{V}(\phi)\rangle_{\chi_0} = V_B \cosh\left[\sqrt{12\pi G}\left(\phi-\phi_B\right)\right],
\end{equation}
where
\begin{equation}\label{eq:VBphiB_free}
    V_B = 2 \sqrt{V_+V_-},\qquad \phi_B = \frac{1}{12\sqrt{2\pi G}}\ln\left(\frac{V_+}{V_-}\right).
\end{equation}
\begin{figure}
    \centering
    \includegraphics[width=0.5\textwidth]{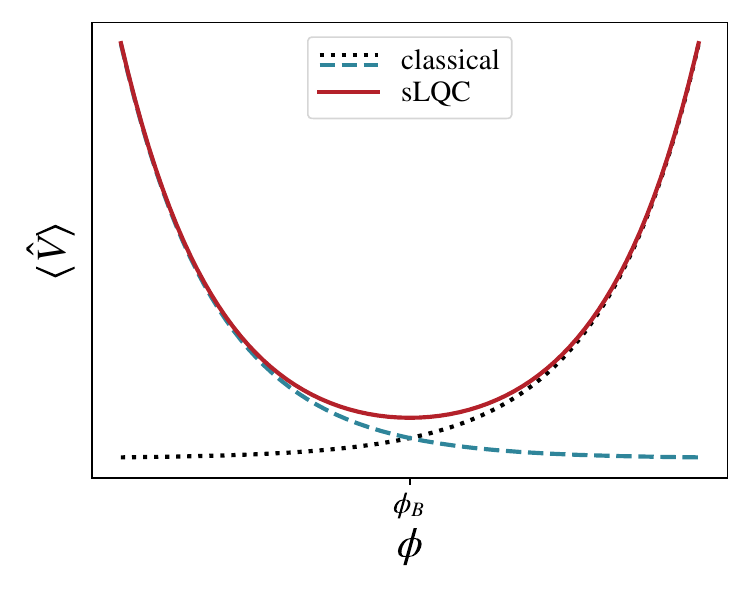}
    \caption[Expectation value of the volume observable in solvable LQC.]{Expectation value of the volume observable in the solvable formulation of LQC (solid red line). Also depicted are the corresponding classical trajectories of an expanding universe (dotted black line) and a contracting one (dashed blue line).}
    \label{fig:bounceSLQC}
\end{figure}
From these it is straightforward to conclude that the expectation value of the volume over any state in its domain\footnote{A state in the domain of the volume operator is one of the form \eqref{eq:chi0} where the Fourier transform of the function $F(x)$ is supported on the positive real line.} decreases for growing $\phi < \phi_B$, reaches a minimum at $\phi=\phi_B$ and increases for increasing $\phi >\phi_B$, as depicted in figure \ref{fig:bounceSLQC}. The point $\phi_B$ where it reaches its minimum $\langle \hat{V}(\phi=\phi_B)\rangle_{\chi_0} = V_B$ is called the bounce. When $\phi \ll \phi_B$ or $\phi \gg \phi_B$ the expectation value agrees with the classical trajectories of a contracting or expanding Universe, respectively.

\subsection{Corrections from a field potential}\label{sec:generic potential}

In this section, we summarize the procedure proposed in \cite{CastelloGomar2016} to obtain further corrections to the quantum dynamics of the model with a generic field potential $W(\phi)$.
The issue is that of the quantum evolution of each FLRW state, as in general, in the presence of a non vanishing field potential, it is determined by a time-dependent quantum Hamiltonian which is not integrable. With this method, the potential is viewed as a kind of geometric interaction. One first takes the free geometric part of the generator of the FLRW dynamics (corresponding to a vanishing potential), and uses it to pass to an interaction picture. The evolution generated by the free geometric operator can be integrated explicitly, and even analytically by using the solvable LQC formulation. Namely, we would obtain the trajectories \eqref{eq:xevolution} and the analytical expression for the states \eqref{eq:states_sLQC} along with the physical inner product~\eqref{eq:innerprod_sLQC}. On the other hand, the potential is regarded as a perturbation to the free dynamics. Its dominant contributions are extracted and passed to a new interaction picture, where the corresponding evolution operator is expanded in powers of the potential, which is truncated at the desired order. If necessary, the remaining evolution can be accounted for with a semiclassical or effective treatment.

The quantum operator that represents the full constraint \eqref{eq:constraint} in the $x$-representation is
\begin{align}
    \hat{C} &= \hat{\pi}_{\phi}^2-\mathH_0^{(2)},\qquad
    \mathH_0^{(2)} = \hat{\pi}_x^2 - 2 W({\hat\phi})\hat{V}^2.
\end{align}

At the physical level, the quantum Hamiltonian is $\mathH_0 = \hat{\pi}_{\phi}$, generating evolution in the time parameter $\phi$. Considering only positive frequency solutions, $\mathH_0$ is given by the positive square root of $\mathH_0^{(2)}$, and is seen as a perturbation of the free system Hamiltonian $\mathH_0^{(F)} = |\hat{\pi}_x|$. In this spirit, we write the states as $\chi(x,\phi) = \hat{U}(x,\phi)\chi_0(x)$, where $\chi_0$ is the initial FLRW state at $\phi=0$ of the free system previously introduced, and $\hat{U}$ is the evolution operator
\begin{equation}
    \hat{U}(x,\phi)=\mathcal{P}\left[\exp\left(i \int_{0}^{\phi} d\tilde{\phi} \mathH_0(x,\tilde{\phi})\right)\right],
\end{equation}
where $\mathcal{P}$ denotes time ordering (with respect to $\phi$). Thus, expectation values are taken on states of the FLRW geometry, with the inner product of sLQC \eqref{eq:innerprod_sLQC}. The computational difficulty lies in the evolution operator $\hat{U}(x,\phi)$. Now, we pass to an interaction picture, where any operator $\hat{A}$ of the original Schr\"odinger picture has its counterpart:
\begin{equation}
    \hat{A}_I(\phi) = e^{-i\mathH_0^{(F)}\phi} \hat{A} e^{i\mathH_0^{(F)}\phi}.
\end{equation}
Note that this is simply the form of the operators of the free system in the Heisenberg picture. Thus, the form of operators of the full system in the interaction picture is simply found by replacing their dependence on $\hat{x}$ by the same dependence on the evolved operator $\hat{x}(\phi)$ defined in equation \eqref{eq:xevolution}. Writing $\mathH_{1} = \mathH_0 - \mathH_0^{(F)}$, the states in this picture can then be described by\footnote{See appendix \ref{sec:appInteractionPicture} for details.}
\begin{align}
    \chi_I(x,\phi) &= \hat{U}_I(x,\phi)\chi_0(x),\\
    \hat{U}_I(x,\phi) &= \mathcal{P}\left[ \exp\left( i \int_{0}^{\phi} d\tilde{\phi} \mathH_{1I}(x,\tilde{\phi})\right) \right].
\end{align}
This leads to the following form for the expectation value of an operator:
\begin{equation}
   \langle \hat{A} \rangle_{\chi}= \langle \hat{A}(\phi) \rangle_{\chi_0} = \langle \hat{U}_I^{\dagger}(\phi) \hat{A}_I(\phi) \hat{U}_I(\phi) \rangle_{\chi_0},
\end{equation}
where the dagger denotes the adjoint. The obstacle is now in the computation of $\hat{U}_I$, which generates the evolution due to $\mathH_{1}$ only, whereas the evolution due to $\mathH_0^{(F)}$ is encoded in $\hat{A}(\phi)$ and is under control. In other words, we have extracted from $\hat{U}(x,\phi)$ the part of the dynamics that is analytically solvable.

Assuming that we can regard the potential as a perturbation of the free system, we can extract from $\hat{U}_I$ the dominant contributions of the potential. Up to first order, and given a suitable factor ordering, it is shown in \cite{CastelloGomar2016} that $\mathH_1$ can be represented approximately as
\begin{equation}\label{eq:H2}
    \mathH_1 \simeq \mathH_2 = -W(\phi) \hat{B},
\end{equation}
where the auxiliary operator $\hat{B}$ is defined as
\begin{equation}\label{eq:B}
    \hat{B} = \frac{4\pi G\Delta \gamma^2}{3}\cosh^2\left(\sqrt{12\pi G}\hat{x}\right)|\hat{\pi}_x|.
\end{equation}

Rigorously, we can write $\mathH_{1I} = \mathH_{2I} + \mathH_{3I}$, where $\mathH_{2I}$ is the operator \eqref{eq:H2} in the interaction picture and $\mathH_{3I}$ is the remaining part of $\mathH_{1I}$, at least of second order in the potential. The dynamics generated by $\mathH_{2I}$ can be obtained by passing to a new interaction picture $J$, for which we introduce:
\begin{equation}
    \hat{U}_{2I}(x,\phi) = \mathcal{P}\left[ \exp\left( i \int_{0}^{\phi} d\tilde{\phi} \mathH_{2I}(x,\tilde{\phi})\right) \right].
\end{equation}
To any operator $\hat{A}_I$ in the original interaction picture corresponds the operator $\hat{A}_J$ of the new interaction picture:
\begin{equation}
    \hat{A}_J(\phi) = \hat{U}_{2I}^{\dagger}(\phi) \hat{A}_I(\phi) \hat{U}_{2I}(\phi). 
\end{equation}

Finally, this leads to the expectation value of any operator $\hat{A}$ of the Schr\"odinger-like picture on states of the FLRW geometry to be given by
\begin{equation}
    \langle \hat{A}(\phi) \rangle_{\chi_0} = \langle \hat{U}_J^{\dagger}(\phi) \hat{A}_J(\phi) \hat{U}_J(\phi) \rangle_{\chi_0},
\end{equation}
where
\begin{equation}
    \hat{U}_J = \mathcal{P}\left[ \exp\left( i \int_{0}^{\phi} d\tilde{\phi} \mathH_{3J}(x,\tilde{\phi})\right) \right].
\end{equation}

Up to this point the treatment has been exact. However, we have only been able to isolate the issues in the computation of the expectation value to the two evolution operators $\hat{U}_{2I}$ and $\hat{U}_J$. To perform explicit computations analytically we need to introduce approximations. Firstly we consider situations for which the evolution generated by $\mathH_3$ is negligible and can be ignored, and thus $\hat{U}_J \simeq \mathds{1}$ and
\begin{equation}
    \langle \hat{A}(\phi) \rangle_{\chi_0} \simeq \langle \hat{A}_J(\phi) \rangle_{\chi_0}.
\end{equation}
Note that this is simply how we formally perform the approximation \eqref{eq:H2}, allowing us to track it rigorously.

Secondly, we expand the time-ordered integral in $\hat{U}_{2I}$ in powers of the potential and truncate at first order:
\begin{equation}
    \hat{U}_{2I} = \mathds{1} + i \int_{0}^{\phi} d\tilde{\phi} \mathH_{2I}(x,\tilde{\phi}) + \mathcal{O}(W^2).
\end{equation}

Keeping only linear contributions of the potential yields
\begin{equation}\label{eq:Aj}
    \hat{A}_J(\phi) \simeq \hat{A}_I(\phi) - i\left[\hat{A}_I(\phi),\int_{0}^{\phi} d\tilde{\phi}\,W(\tilde\phi)\hat{B}_I(\tilde{\phi})\right].
\end{equation}

This way, given the inner product of sLQC \eqref{eq:innerprod_sLQC} and an initial state $\chi_0$, we are ready to compute the expectation value of any operator, to first order in the potential.

We will apply this formalism to the computation of the expectation value of the volume operator in the case of a constant potential in chapter \ref{chap:cosmologicalconstant}.

\subsection{Effective dynamics}\label{sec:effectivedynamics}

 In the previous section we have reviewed a procedure to obtain corrections from a potential of the scalar field to the expectation values of operators in the free model for generic states. However, generally we are particularly interested in the semiclassical sector of the theory, which is represented by quantum states that are sharply peaked on the classical trajectory at low curvatures. In those regimes, general relativity is an excellent approximation. In \cite{Taveras_2008}, effective dynamics are derived for the case of a massless scalar field, and it is found that they may be obtained by a substitution of the variable $b$ in the classical constraint by $\sin b$. In other scenarios, it has been assumed that such a construction is equally valid, as indeed it reproduces the numerical trajectories of expectation values of observables on semiclassical states. 

This way, given a  choice of time gauge, the evolution in the corresponding time is obtained through the Poisson brackets of the variable with the effective Hamiltonian constraint. The dynamics obtained for $V(\phi)$ for the case of vanishing potential match those of $\langle \hat{V}(\phi)\rangle_{\chi_0}$ from the solvable LQC prescription. When the potential is non vanishing, they accurately describe the expectation value of observables on semiclassical states obtained numerically for different models \cite{Singh:2012zc,Diener:2014mia}.

This prescription allows us to work in a classical framework and find quantum corrections to classical equations. Concretely, we can find the Friedmann equation that it produces. Instead of fixing the harmonic gauge, let us fix $N = 1$ so that the Hamiltonian constraint generates evolution in cosmic time. The effective constraint is:
\begin{equation}\label{eq:Ceff}
    C_{\text{eff}} = \frac{\pi_{\phi}^2}{2 V} - \frac{3}{4 \gamma \sqrt{\Delta}} \sin^2b |v| + V W(\phi) = 0,
\end{equation}
where we have used again $V = 2\pi G\gamma\sqrt{\Delta}|v|$. This way, the evolution of the variables is found through the Poisson brackets with the effective constraint:
\begin{align}
    \dot{\phi} &= \lbrace \phi, C_{\text{eff}} \rbrace = \frac{\pi_{\phi}}{V},\\
    \dot{v} &= \lbrace v, C_{\text{eff}} \rbrace = \frac{3}{\gamma\sqrt{\Delta}}|v|\sin b\cos b.
\end{align}
Resorting to $C_{\text{eff}} = 0$ and noting that
\begin{equation}
    \rho = \frac{\dot{\phi}^2}{2}+W(\phi),
\end{equation}
we obtain
\begin{equation}
    \sin b\cos b = \pm \left[\frac{\rho}{\rho_c}\left(1-\frac{\rho}{\rho_c}\right)\right]^{1/2},
\end{equation}
with $\rho_c = 3/(8\pi G \gamma^2\Delta)$. This way we obtain the modified Friedmann equation:
\begin{equation}
    H^2 = \left(\frac{\dot{v}}{3v}\right)^2 = \frac{8\pi G}{3}\rho\left(1-\frac{\rho}{\rho_c}\right).
\end{equation}
This is the classical Friedmann equation with corrections that are of second order in the energy density. When $\rho \ll \rho_c$ the trajectory is that of classical FLRW, whereas when $\rho$ approaches $\rho_c$ the quantum corrections become important causing the bounce to occur at $\rho = \rho_c$.

\subsection{Cosmological perturbations}\label{sec:LQCperts}

There are two main approaches to include perturbations in the formalism of LQC: hybrid LQC \cite{FernandezMendez2012,FernandezMendez2013,FernandezMendez2014,CastelloGomar2014,CastelloGomar2015,CastelloGomar2016,hybrid_Martinez2016,hyb-vs-dress} and the dressed metric approach \cite{Ashtekar:2009mb,dressedmetric_Agullo2012,dressedmetric_Agullo_PRD_2013,dressedmetric_Agullo_CQG_2013}. Both are based on a combination of the polymeric quantization of LQC for the geometry and a standard Fock quantization for the perturbations. The motivation is that there should exist physically interesting regimes where the main quantum gravity effects are those affecting the global degrees of freedom of the model, namely the homogeneous sector.

In the hybrid approach, the background and perturbations are treated as a whole symplectic system. Starting from the Hamiltonian formalism, the action is obtained and truncated at quadratic perturbative order. On the other hand, in the dressed metric approach, the background and perturbations are treated separately since the beginning, assuming from the start that backreaction is negligible. There is no classical Hamiltonian that generates the evolution of the full system at some order of truncation. Instead it is as if there were two Hamiltonians, one is the standard FLRW one, the other generates the evolution of the perturbations when the background is seen as fixed.

In this thesis, we adopt the hybrid formalism. The motivation is two-fold. Firstly, most works on the observational consequences of LQC through comparison with data have been performed following the dressed metric approach. A comparison between the consequences of the two is important when building a falsifiable picture of the theory. Secondly, whereas the equations of motion of the perturbations are hyperbolic at the bounce for all modes in hybrid LQC, this is not the case in the dressed metric approach.

The Fourier modes of gauge-invariant scalar and tensor perturbations satisfy the equations of motion \eqref{eq:eom_uk}, with time-dependent mass terms that depend on the quantization approach adopted (see for example \cite{hyb-vs-dress} for more details). In the case of the hybrid approach, they can be written in terms of the background variables $a$, $\rho$ (inflaton energy density), $P$ (inflaton pressure) and the inflaton potential $V(\phi)$ as follows:
\begin{equation}\label{eq:s(eta)LQC}
    s^{(t)} = -\frac{4\pi G}{3} a^2 \left(\rho - 3P\right),\qquad s^{(s)} =  s^{(t)} + \mathcal{U},
\end{equation}
where
\begin{equation}\label{eq:U_MS}
    \mathcal{U} = a^2\left[V_{,\phi\phi}+48\pi G V(\phi)+ 6 \frac{a^{\prime}\phi^{\prime}}{a^3 \rho} V_{,\phi}-\frac{48\pi G}{\rho} V^2(\phi)\right].
\end{equation}
Note that $s^{(t)}$ depends on $a,\rho$ and $P$ the same way as in classical FLRW \eqref{eq:stFLRW}. But there are two important distinctions: a) the evolution of the background variables in this case is the one of the effective dynamics of LQC, which differs from the classical ones, and b) in this case it is not equal to $a^{\prime\prime}/a$ of the effective dynamics. On the other hand, the scalar modes receive additional explicit corrections through $\mathcal{U}$.
\begin{figure}
    \centering
    \includegraphics[width=0.49\textwidth]{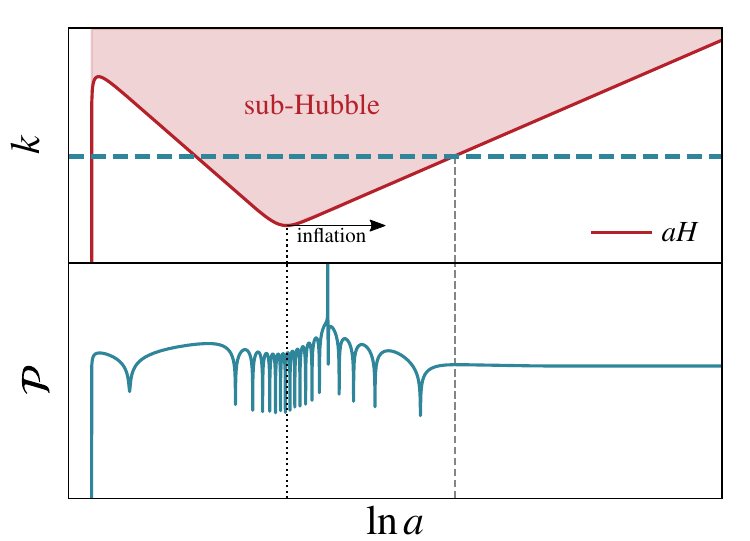}
    \caption[Depiction of the evolution of the inverse of the comoving Hubble radius in a typical LQC model, and that of the power spectrum of a mode that is affected by the dynamics of the background.]{Top: depiction of the evolution of the inverse of the comoving Hubble radius (red line) in a typical LQC model. Modes with wavenumber $k$ above the red line are sub-Hubble, whereas those below the red line are super-Hubble. Also indicated is the onset of inflation, and the moment a particular mode (blue dashed line) crosses the Hubble radius during inflation. Bottom: evolution of the power spectrum of the mode indicated in the above plot (blue line), that is affected by the dynamics of the background. In particular, we note that this mode has an oscillatory behavior at the onset of inflation.}
    \label{fig:aHP}
\end{figure}

Generally, there is no analytical solution to \eqref{eq:eom_uk} with such time-dependent mass terms. These can be solved numerically, given initial conditions $v_k(\eta_0)$, $v^{\prime}_k(\eta_0)$, uniquely specifying a choice of vacuum for perturbations. Unlike in standard cosmology, in this case it is no longer justified that every mode of the perturbation reaches the onset of inflation in the Bunch-Davies vacuum. In fact, pre-inflationary dynamics should affect the evolution of a range of modes. In realistic models of LQC, the bounce occurs in a period of kinetic dominance, i.e., where the kinetic energy of the inflaton dominates over its potential energy. It is followed by a very short period of super inflation and then one of decelerated expansion. Finally slow-roll inflation begins shortly after the potential begins to dominate over the kinetic energy. This can be understood from the evolution of the inverse of the comoving Hubble radius $aH$, as we have depicted in the top panel of figure \ref{fig:aHP} for a typical LQC model. An increase of this quantity signifies accelerated expansion. There we can clearly see the three stages we have mentioned. In the lower panel of the figure we have depicted the power spectrum of a mode that feels the differences in the background during the pre-inflationary dynamics. We see that this mode crosses out of the horizon and back into it before inflation begins. Finally, it crosses back out during inflation and ``freezes''. This may affect the evolution of these modes and their primordial power spectra. In turn, this may leave observable imprints in the anisotropies of the CMB.

We will employ this description in chapters \ref{chap:SLEsinLQC} and \ref{chap:anomalies} for particular choices of vacuum in order to evolve the perturbations until the end of inflation and compare predictions with observations.

\section{Bayesian Analysis}\label{sec:intro_Bayes}

Modern cosmology is at a place where it is capable of developing genuinely falsifiable theories, due to the increasing accuracy of cosmological observations. These observations allow us to constrain parameters of models and update our previous knowledge of them. This is accomplished through thorough statistical analyses that contrast predictions of theoretical models with observations. With this in mind, Bayesian analysis is a very powerful tool that is usually employed in cosmology to help constrain parameters and compare different models, effectively extracting the information that we have obtained from the data. We will use this tool in this work. This section is dedicated to a brief review of Bayesian analysis and the numerical methods used to employ it. For further details see for example \cite{Liddle:2009xe,hobson_2009}.

Let us take a set of data $D$, and a model with parameter values $\theta$. For each fixed model (fixed values $\theta$), we can compute the $\chi^2$ statistic, which is a sum of the distance between the prediction from the model to the measured values. For concreteness, let us consider a model that relates two variables $x$ and $y$ with parameters $\theta$: $y = y(x;\theta)$. The set of data is then given by pairs of $x$ and $y$ and the associated uncertainty in $y$, which we will call $\sigma$: $D=\{x_i,y_i,\sigma_i\}$. The $\chi^2$ is found as
\begin{equation}
    \chi^2(D,\theta) = \sum_i \frac{\left(y_i-y(x_i;\theta)\right)^2}{\sigma_i^2}.
\end{equation}
For each set of parameters $\theta$ we obtain a value of $\chi^2$, which is an indication of goodness of fit: a lower $\chi^2$ indicates that the model fits the data better. The value of $\chi^2$ by itself is not very meaningful, in fact if we have more data points it will increase, which does not mean necessarily that the data is fitted any worse by the model. We may now manipulate this to work in a framework of probabilities or at least of distributions, by computing the likelihood:
\begin{equation}
    \mathcal{L}(D,\theta) = N e^{-\chi^2(D,\theta)/2}.
\end{equation}
Now, a higher value of $\mathcal{L}$ indicates that the model fits the data better. As long as the normalization factor $N$ is chosen appropriately, this can be interpreted as the probability of measuring the data $D$ given the model $\theta$, i.e. $\mathcal{L}(D,\theta) = p(D|\theta)$. Note that in this case this is a probability over the data, it is not normalized over the parameter space $\theta$.

Usually in cosmology, the main goal of statistical analysis is to infer constraints on parameters from the data. In other words, we are interested in obtaining $p(\theta|D)$, i.e. the probability that the model with parameters $\theta$ reproduces the observations $D$. This is called the posterior probability distribution and can be found from the likelihood through Bayes theorem:
\begin{equation}
    p(\theta|D) = \frac{p(D|\theta)p(\theta)}{p(D)} = \frac{\mathcal{L}(D,\theta)p(\theta)}{p(D)}.
\end{equation}
Here, $p(\theta)$ is a total probability of the parameters, transcending the data. This is called the prior, as it offers the possibility of introducing prior knowledge on $\theta$, regardless of the data. For example, physical intuition may dictate that some parameters have to be positive, or previous independent analyses may have constrained $\theta$ in some way. It is in this sense that Bayesian analysis allows us to update prior knowledge. On the other hand, $p(D)$ is an overall probability of measuring the data. However, $p(\theta|D)$ is a probability distribution of the parameters $\theta$. Therefore, since the denominator is independent of them, it can be seen as a normalization constant, and we will not dwell on its significance.

To obtain the posterior, the procedure is to sample the phase space of $\theta$, compute the $\chi^2$ for each point, obtain the likelihood and multiply it by the prior for each point. This leads to a distribution, which can be cast into a proper probability distribution if it is normalized such that $\int p(\theta|D) d\theta = 1$. From this procedure we are able to obtain three important quantities:
\begin{itemize}
    \item the best-fit in the parameter space of $\theta$: the point at which the peak of the posterior is found and which corresponds to the values of the parameters for which the model best fits the data; 
    \item the 1 and 2-$\sigma$ regions in parameter space: portions of the parameter space for which the integration of the posterior yields 68\% and 95\% of the total volume under the curve, respectively. These are usually projected onto 2D contour plots for better readability;
    \item the marginalized posterior probabilities of each parameter: obtained by integrating the posterior over all other parameters, resulting in a probability distribution of one parameter.
\end{itemize}
\begin{figure}
    \centering
    \begin{subfigure}[b]{0.49\textwidth}
    \centering
    \includegraphics[width=\textwidth]{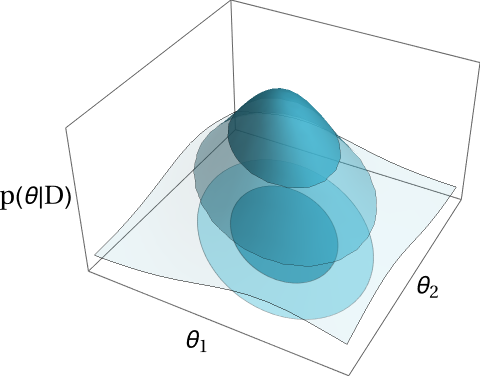}
    \caption{}
    \end{subfigure}
    \begin{subfigure}[b]{0.49\textwidth}
    \centering
    \includegraphics[width=\textwidth]{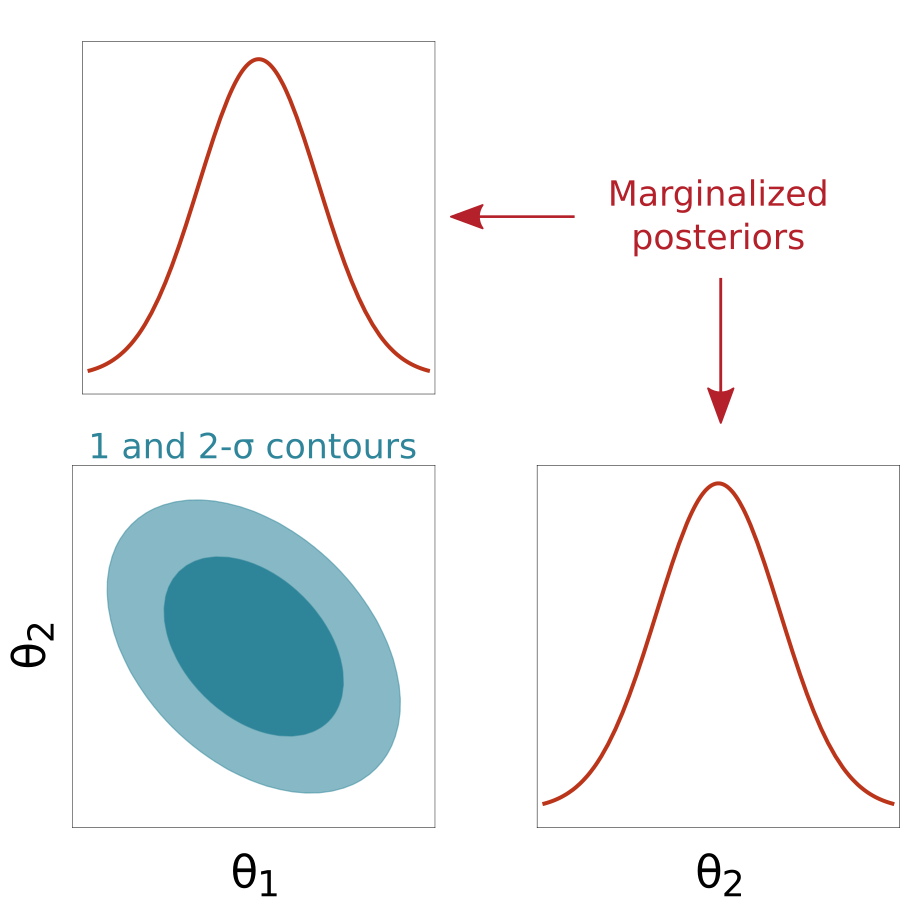}
    \caption{}
    \end{subfigure}
    \caption[Illustration of posterior probability in parameter space, 2D contour plots and marginalized posteriors.]{(a) Depiction of a posterior probability in a 2 parameter model and (b) corresponding 2D contour plot and marginalized posterior probabilities.}
    \label{fig:Bayes}
\end{figure}
These allow us to obtain the constraints on parameters as the best fit $\pm$ the 1-$\sigma$ region when projected onto the parameter in question, as well as the triangle plots which are composed of the 2D contour plots of the 1 and 2-$\sigma$ regions projected onto the 2D plane of each pair of parameters together with the marginalized posterior distribution of each parameter. These concepts are illustrated in figure \ref{fig:Bayes}.

In cosmology, models usually have at least 6 parameters, which means that the parameter space of $\theta$ is at least 6 dimensional. Sampling even just 10 values of each parameter requires the computation of the posterior $10^6$ times. $N$ additional points for each parameter increases this value by $N^6$, which requires great computational power and is in practice not feasible. This requires the clever use of some techniques. In fact, we are not interested in knowing the whole posterior distribution in the same amount of detail. In most cases, a thorough sampling of the region of the peak (or peaks in case the distribution is multimodal) is enough. Furthermore, there are computational tools that can greatly aid in this endeavor. The most common one is that of Markov-Chain-Monte-Carlo (MCMC) methods. Let us briefly explain how they work by breaking them down in their two components.

\begin{itemize}
    \item Monte Carlo methods are ones that are based fundamentally in a random sampling of the phase space. A typical example is the estimation of $\pi$ by randomly sampling the area of a square with $N$ points and counting how many of them ($n$) are inside the circle that is circumscribed by the square. The ratio between the areas of the circle and the square is $\pi/4$. Since the areas are proportional to the number of points sampled within, $\pi$ can be estimated by $4 n/N$, with greater accuracy the more random samples we take.
    \item A Markov chain is a sequence of states that are obtained according to some probabilistic rule from the last state with no memory of previous ones. In other words, it is a sequence $\{X_j\}$ that for any integer $n$ and possible states $\{i_j\}$ of the random variables verifies\\
    $$p(X_n = i_n|X_{n-1} = i_{n-1}) = p(X_n = i_n|X_0 = i_0,X_1 = i_1,...,X_{n-1} = i_{n-1}).$$
\end{itemize}

Thus, MCMC methods are ones that allow us to estimate some quantity based on random sampling to create a Markov chain. Let us consider this quantity to be the posterior probability, which can be computed up to a constant (normalization) by $\mathcal{L}(D,\theta)p(\theta) \equiv f(\theta)$.

The most common algorithm of these methods used in cosmology (and the one that we have employed in the statistical analysis of this work) is the Metropolis-Hastings algorithm. It defines the jumping probability from one point $\theta_n$ in phase space to another $\theta^{\prime}$ as $g(\theta^{\prime}|\theta_n)$. Usually, this is defined as a Gaussian centered in $\theta_n$ with a standard deviation known as the jumping factor. This way, it is more likely to jump to closer values of $\theta_n$, but jumps to points much further away will still occur. This is essential to avoid getting trapped in a local maximum. The manipulation of the jumping factor allows for the efficiency of the algorithm to be improved. A very large jumping factor may cause the algorithm to miss the peak of the distribution, whereas a very small one may cause it to take too long to explore the region of the peak. At each point in the chain, a new candidate point $\theta^{\prime}$ is obtained randomly from the last one $\theta_n$ through $g(\theta^{\prime}|\theta_n)$, and it is accepted as the new point with probability $\text{min}(a,1)$ where $a = f(\theta^{\prime})/f(\theta)$.\footnote{Concretely: a random number $i$ is generated between 0 and 1, if $a \geq i$ the candidate is accepted, otherwise it is rejected.} If it is accepted it becomes the next element of the chain $\theta_{n+1} = \theta^{\prime}$, if it is rejected the previous element is repeated $\theta_{n+1} = \theta_n$. Note that if the candidate point is closer to the peak (in which case $f(\theta^{\prime}) > f(\theta_n))$, it is always accepted as the next element of the chain. On the other hand, even if the candidate is not closer to the peak there is also a chance that it is accepted, allowing for an exploration of the volume around the peak, and avoiding getting trapped in local maxima. 

\begin{figure}
    \centering
    \includegraphics[width=0.4\textwidth]{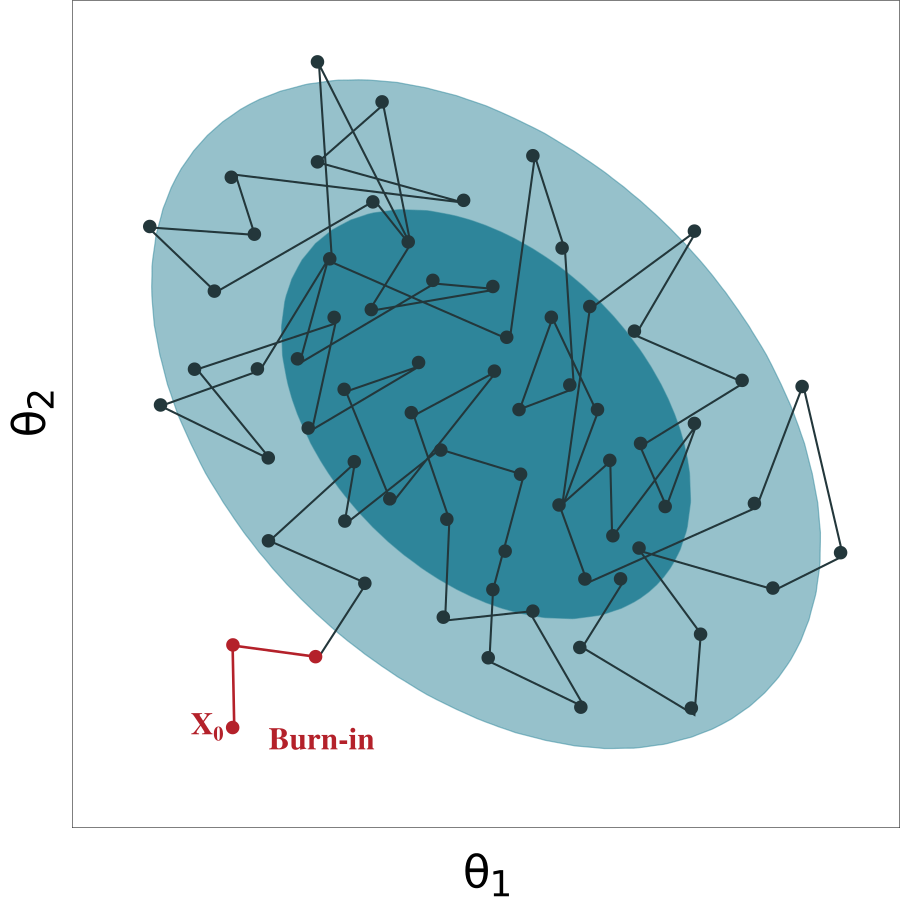}
    \caption{Pictoric illustration of a chain resulting from a MCMC method.}
    \label{fig:chain}
\end{figure}

Figure \ref{fig:chain} illustrates the typical result of one run: a chain that starts somewhere randomly in phase space and explores it moving closer to the peak, while also sometimes exploring regions further away. It is best to run several chains in parallel to decrease dependencies on the initial point and to compute averages of many chains. The beginning phase is usually called burn-in and it is ignored. The decision to stop the chain or to continue computing more points is usually based on the Gelman-Rubin convergence diagnostic $R$, which takes into account the length of the chain as well as the average of the variances of each chain and the variance between the averages of each chain. Convergence is reached when $R-1 \ll 1$. The codes that are implemented in practice employ additional features to increase efficiency and accuracy, but are based on the algorithm that we have described.

\begin{figure}
    \centering
    \begin{subfigure}[b]{0.59\textwidth}
    \centering
    \includegraphics[width=\textwidth]{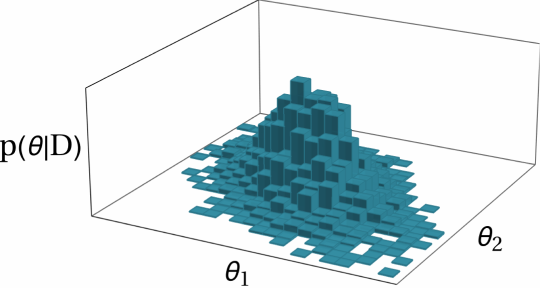}
    \caption{}
    \end{subfigure}
    \begin{subfigure}[b]{0.4\textwidth}
    \centering
    \includegraphics[width=\textwidth]{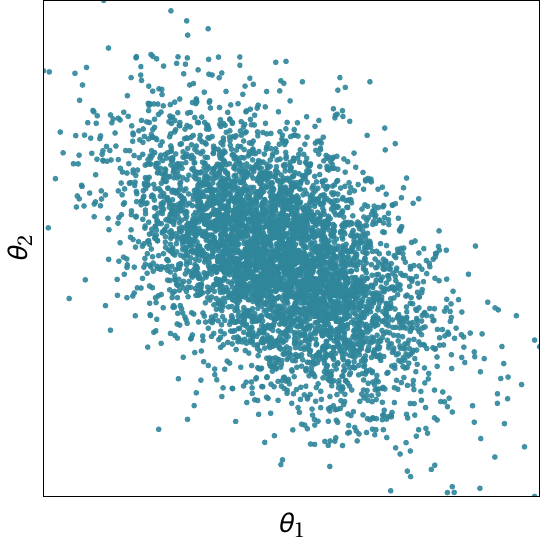}
    \caption{}
    \end{subfigure}
    \caption[Illustration of population of points generated in phase space from a MCMC method.]{(a) Depiction of a histogram of the population of points generated from a MCMC method, which estimates the posterior distribution and (b) corresponding 2D scatter plot that estimates the 2D contour plot of the posterior distribution.}
    \label{fig:scatter}
\end{figure}

This way, as depicted in figure \ref{fig:scatter}, a (properly normalized) histogram of the population of points in phase space adequately estimates the posterior, as the more a point appears in a chain, the more likely it is to be close to the peak, whereas if it is almost never selected it is probably in the tail of the distribution.

We will make use of these techniques in chapter \ref{chap:anomalies}, with the aim of comparing predictions of LQC with observations of the CMB.
\chapter{Beyond the effective dynamics of LQC}\label{chap:cosmologicalconstant}

So far, cosmological perturbations have been studied within LQC by finding corrections to the classical equations of motion after introducing semiclassical or effective approximations for the description of the FLRW geometry. Expectation values on FLRW states are replaced with classical values evolved with the effective dynamics of homogeneous and isotropic LQC introduced in chapter \ref{sec:effectivedynamics}. In \cite{CastelloGomar2016} a procedure is proposed to compute the quantum corrections at the Planck regime to the Mukhanov-Sasaki equations up to the maximum practical extent. The goal is to obtain better quantitative predictions. This way, it is also possible to check that additional quantum corrections are indeed negligible in scenarios where they have been neglected; e.g., when effective dynamics for the description of the FLRW geometry are taken in the place of the full quantum dynamics.

We have reviewed this procedure for a generic potential in chapter \ref{sec:generic potential}. In this chapter, we apply it to the simplest case of a constant positive potential, namely, a positive cosmological constant, and work within linear perturbative order. This has led to the publication \cite{ours_Lambda}. We take the potential to be positive to have a toy model for inflation (even if eternal), although all our considerations apply  for a negative one as well. Even though in this scenario the quantum Hamiltonian is not time dependent, the~corresponding FLRW states do not admit a simple analytically closed form, and heavy computational power is required to generate them numerically, as has been done in \cite{Pawlowski2012}. Hence, this model provides the perfect test of the procedure. Namely, it allows for the development of the required mathematical tools in a simplified scenario, and the final result can be checked against one of the exact (but more time and resource-consuming) treatment. Indeed, the dynamics of this model obtained with our treatment agrees with that of \cite{Pawlowski2012}.

Moreover, we are able to take the analysis further by comparing the evolution of the models with and without cosmological constant. We conclude that, for semiclassical states characterized by a Gaussian profile, the introduction of a positive cosmological constant pushes the bounce to lower volumes and higher values of the scalar field. Furthermore, we are able to uncover interesting behaviors of the dynamics for generic states. In fact, we find that the simple behavior of a symmetric bounce is particular to semiclassical states with Gaussian profiles, but that the dynamics of generic states are more intricate, with terms that are not symmetric. This displays clearly one advantage of this perturbative method: by keeping only the leading order contributions of the potential to the dynamics, we simplify the computations and do not have to particularize the investigation to semiclassical states or to rely on numerical treatments. This allows us to retain more information and uncover effects that might otherwise be ignored.

As our work is seen as a stepping stone to the application of this perturbative method to other models (namely, inflationary ones), we keep the calculations general as much as possible, only particularizing for a constant potential when it is necessary to move forward in the calculations.

In chapter \ref{sec:generic potential}, we have reviewed the procedure proposed in \cite{CastelloGomar2016}. In section \ref{sec:expVgeneric} we use it to compute the expectation value of the volume operator with corrections to first order in the generic potential. In Section \ref{sec:constant potential} we apply the treatment for the case of a constant potential, obtaining the expectation value of the volume, and compare it with classical trajectories. In Section \ref{sec:bounce with potential} we compare the scenarios with and without cosmological constant, and~investigate further the effect of introducing a cosmological constant in the bounce scenario of the solvable model. Section \ref{sec:approx} is devoted to discussing the regime of the validity of our approximation. Lastly, in Section \ref{sec:discussion} we conclude with a discussion of our work.

\section{Expectation value of the volume operator}\label{sec:expVgeneric}

In this section, we apply the procedure proposed in \cite{CastelloGomar2016} to obtain the corrections from a generic field potential $W(\phi)$ to the expectation value of the volume operator. We keep contributions to first order in $W$.

According to the procedure introduced in chapter \ref{sec:generic potential}:
\begin{equation}\label{eq:Vexp1}
    \langle \hat{V}(\phi) \rangle_{\chi_0}\simeq \langle \hat{V}_J(\phi) \rangle_{\chi_0} =2i\int_{-\infty}^\infty dx \frac{dF^*(x)}{dx}\hat{V}_J(\phi)F(x).
\end{equation}

Let us show this calculation with generic potential explicitly, so that this task is simplified when studying particular forms of the potential in the future.
To simplify notation, we define
\begin{equation}\label{eq:I1}
    I_1(x,\phi) \equiv \int_{\phi_0}^{\phi} d\tilde{\phi}\,W(\tilde\phi)\cosh^2\left[\sqrt{12\pi G}(x-\Delta\tilde\phi)\right],
\end{equation}
with $\Delta\tilde\phi=\tilde\phi-\phi_0$. From now on, we will set $\phi_0=0$ for simplicity, as we did it chapters \ref{sec:sLQC} and \ref{sec:generic potential}.
The above integral will be particular to each form of the potential $W(\phi)$. This way, $I_1$ and its derivatives with respect to $x$ are of first order in $W$. Applying \eqref{eq:Aj} for $\hat{V}_J(\phi)$, with $\hat{V}_I(\phi)$ the volume operator of the free model,
and recalling that $\hat{\pi}_x$ is positive for left-moving modes, we obtain
\begin{equation}\label{eq:VexpW}
    \langle \hat{V}(\phi) \rangle_{\chi_0} \simeq \frac{4\pi G\gamma\sqrt{\Delta}}{\sqrt{3\pi G}} \int_{-\infty}^{+\infty} dx \left| \frac{dF(x)}{dx} \right|^2 E(x,\phi),
\end{equation}
to first order in $W$, where
\begin{align}\label{eq:E}
\begin{split}
    E(x,\phi) &= \cosh\left[\sqrt{12\pi G}(x-\phi)\right]+\frac{4\pi G\gamma^2\Delta}{3}\bigg\{2\sqrt{3\pi G} I_1(x,\phi)\sinh\left[\sqrt{12\pi G}(x-\phi)\right] \\
    &- \cosh\left[\sqrt{12\pi G}(x-\phi)\right] \partial_x I_1(x,\phi)\bigg\}.
\end{split}
\end{align}

For specific forms of the potential, it is useful to identify in $\langle \hat{V}(\phi) \rangle_{\chi_0}$ the constants 
\begin{equation}\label{eq:vn}
	V_n \equiv \frac{2\pi G\gamma\sqrt{\Delta}}{\sqrt{3\pi G}}\int_{-\infty}^{+\infty} dx \left| \frac{dF(x)}{dx} \right|^2 e^{-n\sqrt{12\pi G}\ x},
\end{equation}
in order to express $\langle \hat{V}(\phi) \rangle_{\chi_0}$ in a manner analogous to that of \eqref{eq:expVsLQC}. Note that $V_\pm$ defined in \eqref{eq:v+-sLQC} corresponds to taking $V_{\pm 1}$, and so we verify that for $W=0$ ($I_1=0$) we recover the case of sLQC, as we should.

In what follows, to simplify notation, $\langle \hat V \rangle$ means $\langle \hat{V} \rangle_{\chi}=\langle \hat{V}(\phi) \rangle_{\chi_0}$, and likewise for other operators.

\section{Constant Potential}\label{sec:constant potential}

This procedure has been proposed in \cite{CastelloGomar2016}, but so far, it has not been applied to a specific form of the potential. In this section we do so for the simplest form of the potential: a constant positive one. Even though this is a simple scenario, it is already a relevant model for cosmology, as it is equivalent to considering a massless scalar field in the presence of a positive cosmological constant $\Lambda$, having $W = \Lambda / 8\pi G$. It is also relevant as a test model for the procedure, before it is implemented with other more complicated forms of $W$. 

A constant potential provides a time-independent Hamiltonian. This way, it allows for a simplification of the calculations, while still enabling the development of techniques that would be useful for the analysis of other potentials. Additionally, the loop quantization of this model has been rigorously carried out in \cite{Pawlowski2012}, and the quantum dynamics of a family of semiclassical states determined by Gaussian profiles has been obtained by generating numerically the eigenfunctions of the Hamiltonian. This provides a check for the validity of our method. 

Classically, the model presents two types of solutions: universes expanding from a big bang to infinite volume (reached at infinite proper time), and universes contracting from infinite volume to a big crunch (reached at infinite proper time). In internal time $\phi$, the expanding solutions reach infinite volume at a finite value of $\phi$, and one can analytically continue the evolution, matching it with that of the contracting solution.
Moreover, as it happens for vanishing cosmological constant, the loop quantization drastically modifies the dynamics at the Planck regime, replacing the big bang or big crunch singularities with a quantum bounce. This leads to an essentially periodic (in time $\phi$) universe \cite{Pawlowski2012}.

In this work we are interested only on the quantum modifications to the classical dynamics in the Planck regime; namely, in the quantum dynamics around the bounce. Indeed, our approximation method is only valid  when the matter energy density dominates over the energy density associated to the cosmological constant, $\Lambda/(8\pi G)$, so that the latter can be regarded as a perturbation. In consequence, we consider values of $W$ to be far smaller than the Planckian density (as discussed later, in numerical computations we consider values such that the energy density of the cosmological constant at the bounce is four orders of magnitude below that of the scalar field). The main novelty of this work is a comparison between the Planck regimen dynamics of the model with $\Lambda>0$ and that of the free model ($\Lambda=0$). Namely, we analyze explicitly how the introduction of a positive (small) cosmological constant modifies the dynamics of the model without cosmological constant.

With a constant potential, it is straightforward to compute $I_1(x,\phi)$ as defined in \eqref{eq:I1}. This is followed by a manipulation of $E(x,\phi)$ of \eqref{eq:E} in order to identify the $V_n$ functions that appear in $\langle \hat{V} \rangle$ of \eqref{eq:VexpW}. These steps can be found in Appendix \ref{sec:appendix_V}. That way, we find the expectation value of the volume as
\begin{align}\label{eq:expv_W}
\begin{split}
  \langle \hat{V} \rangle \simeq &\left[V_+ +\tilde{W} w_+(\phi)\right] e^{\sqrt{12\pi G}\phi}+ \left[V_- +\tilde{W} w_-(\phi)\right] e^{-\sqrt{12\pi G}\phi}\\
  &+ \tilde{W}\left(V_{+3}e^{3\sqrt{12\pi G}\phi}+ V_{-3}e^{-3\sqrt{12\pi G}\phi}\right),
\end{split}
\end{align}
to first order in the potential, where, to simplify notation we have defined:
\begin{align}
    \tilde{W} &= W\frac{\pi G\gamma^2\Delta}{6},\label{eq:Wtilde}\\
    w_\pm(\phi) &= \left(3\mp 8\sqrt{3\pi G}\phi\right)V_\pm - V_{\pm3}-3V_\mp,
\end{align}
recalling that the constants $V_n$ are state dependent, as defined in \eqref{eq:vn}. 

So far, we have made no restriction to a particular form of the physical state $\chi$ on which we compute the expectation value. However, we are interested in tracking the expectation value of the volume on semiclassical states, by which we mean states that are highly peaked (i.e., with small relative dispersion) on the classical trajectories in the low curvature regime, where quantum effects of the geometry should be negligible. These are determined by fixing adequately the functions $F$ defined in \eqref{eq:chi0}, which in turn define the $V_n$ constants \eqref{eq:vn}. As previously mentioned, $F$ is a function with Fourier transform $\tilde{F}(k)$ supported on the positive real line:
\begin{equation}
    F(x) = \frac{1}{2\pi}\int_0^{+\infty}dk\,\tilde{F}(k)e^{i\sqrt{12\pi G}x}.
\end{equation}

For similarity with previous LQC literature, and in particular \cite{APS_extended,Pawlowski2012}, we choose semiclassical states determined by spectral Gaussian profiles (in the $v$-representation where the volume is diagonal) centered at $k_o \ll -1$ with width $\sigma \ll |k_o|$:
\begin{equation}\label{eq:gaussian profile}
	\tilde{\psi}(k) = \frac{1}{\sqrt{\sigma\sqrt{\pi}}} e^{-\frac{(k-k_o)^2}{2\sigma^2}}.
\end{equation}
These states are chosen such that the expectation value of $\hat\pi_\phi$ and its relative dispersion on them verify
\begin{equation}\label{eq:pi_phi and disp}
    \langle \hat\pi_\phi \rangle=-\sqrt{12 \pi G} k_o,\quad \frac{\langle \Delta\hat\pi_\phi \rangle}{\langle \hat\pi_\phi \rangle}=\frac{\sigma}{\sqrt{2} |k_o|}.
\end{equation}
The condition $k_o \ll -1$ assures that it is centered on a classical trajectory ($\langle \hat\pi_\phi \rangle \gg \hbar$) and $\sigma \ll |k_o|$ assures it is sufficiently peaked on it.

As we investigated in \cite{paper1_domainV}, the relation between the profiles	$\tilde{\psi}(k)$ of the $v$-representation and the profiles $\tilde{F}(k)$ of the solvable representation is given by
\begin{equation}
	\tilde{F}(k) = \frac{1}{\sqrt{k\pi}}\ \tilde{\psi}(-k) \cos\left(\frac{1-2ik}{4}\pi \right)\ \Gamma\left(\frac{1}{2}-ik\right),
\end{equation}
and for the states \eqref{eq:gaussian profile} we obtain \cite{paper1_domainV}
\begin{equation}\label{eq:vn_ko}
    V_n =2\pi G \gamma\sqrt{\Delta} (-k_o)^{1-n}e^{\frac{n^2}{4\sigma^2}}.
\end{equation}

Thus, for given values of $k_o$ and $\sigma$, we can compute $\langle \hat{V} \rangle$ and track its evolution along $\phi$.

\begin{figure}[h]
\centering
    \begin{tikzpicture}
    \tikzstyle{every node}=[font=\normalsize]
    \node (img)  {\includegraphics[width=0.5\textwidth]{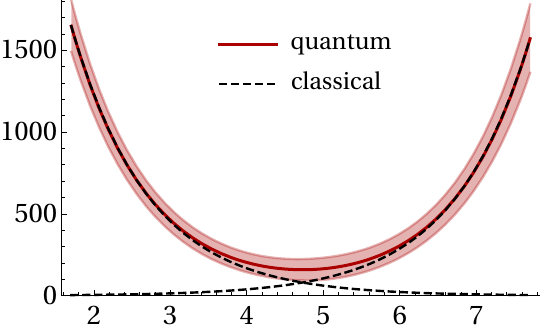}};
    \node[below=of img, node distance=0cm, xshift=.6cm, yshift=1.3cm] {$\sqrt{12\pi G}\phi$};
    \node[left=of img, node distance=0cm, anchor=center,rotate=90,yshift=-0.7cm,] {$V/2\pi G\gamma\sqrt{\Delta}$};
 \end{tikzpicture}
    \caption[Expectation value of volume in model with cosmological constant.]{Expectation value of $\hat{v}$ on a semiclassical state with the profile \eqref{eq:gaussian profile}, along with its dispersion (see~Appendix \ref{sec:appendix_V}), compared with the corresponding classical trajectories. The following values were used for the parameters: $k_o=-100$, $\sigma=10$ ($\langle\Delta\hat{\pi}_{\phi}\rangle/\langle\hat{\pi}_{\phi}\rangle \simeq 0.07$) and $\tilde{W} = 10^{-5}$.}
    \label{fig:dynamics}
\end{figure}

Figure \ref{fig:dynamics} shows the expectation value of $\hat{v}=\hat{V}/2\pi G\gamma\sqrt{\Delta}$ on a semiclassical state determined by \eqref{eq:gaussian profile}, and its dispersion, as  functions of time $\phi$, along with the corresponding classical trajectories. As it should be, this analysis agrees with the results of \cite{Pawlowski2012}.

The classical trajectories are obtained from the classical constraint \eqref{eq:constraint}, by finding the derivatives of $v$ and $\phi$ with respect to harmonic time $\tau$ through the Poisson brackets with the constraint, and finding $dv/d\phi$ through the chain rule. Then, recalling that $V(\phi) = 2\pi G\gamma\sqrt{\Delta}v(\phi)$ one obtains the classical trajectory, which will label with the subscript $c$:
\begin{equation}
        \frac{dV_c(\phi)}{d\phi}= \pm \sqrt{12\pi G}\left(V_c^2(\phi) +\frac{2W}{\pi_{\phi}^2}V_c^4(\phi) \right)^{1/2},
\end{equation}
These differential equations result in the two classical trajectories: an expanding and a contracting solution.

In the low curvature regime, both the classical and quantum trajectories agree. However, close to the Planck regime, the quantum trajectory deviates from the classical ones. As classical trajectories tend to zero volume, the quantum description shows that a bounce occurs, connecting a contracting epoch of the Universe with an expanding one.

In the next section we investigate  the impact of the potential on the dynamics of the system further.

\section{The Effect of the Constant Potential in the Bounce Scenario of sLQC}\label{sec:bounce with potential}

In this section, we compare the trajectory of the volume when the cosmological constant is positive with that of the free case (without cosmological constant). More explicitly, we look for an approximate value of the point $\phi_B$ around which the bounce occurs, and seek an approximate expression for the trajectory around this point, analogously to \eqref{eq:expVsLQCcosh}. 

For simplicity, let us call $V(\phi) = \langle \hat{V} \rangle$. To find $\phi_B$, one would need to solve $V'(\phi) = 0$. Let us refer to $\phi_B$ for the free case as $\phi_B^F$. Recall from \eqref{eq:VBphiB_free}:
\begin{equation}
    \phi_B^F = \frac{1}{2\sqrt{12\pi G}}\ln \left(\frac{V_-}{V_+}\right).
\end{equation}

However, with non vanishing potential, $V'(\phi) = 0$ is a transcendental equation which cannot be solved exactly. To this end, we employ Newton Raphson's method, where, given an initial estimate to $\phi_B$, corrections to it can be found iteratively. This method is usually employed in numerical calculations to find quite accurate roots for transcendental equations. However, we use only one iteration of it analytically, in~order to find $\phi_B$ as a function of the parameters of the system, without the need for numerical substitutions. Indeed, since we are already obtaining $V(\phi)$ from a perturbative treatment in $W$, further corrections to $\phi_B$ would be irrelevant within our truncation scheme.
The initial estimate for $\phi_B$ is given by $\phi_B^F$. The first iteration of Newton Raphson's method reads $\phi_1 = \phi_B^F - V'(\phi_B^F)/V''(\phi_B^F)$. A careful calculation yields
\begin{equation}
    \phi_1 = \phi_B^F+\tilde W\left(4\phi_B^F+\frac{f(V_n)}{4\sqrt{3\pi G}}\right) + \mathcal{O}\left(\tilde{W}^2\right),
\end{equation}
where
\begin{equation}
    f(V_n) = \frac{V_{+3}+3V_-}{V_+}- \frac{V_{-3}+3V_+}{V_-} +3\left( \frac{V_{-3}V_+}{V_-^2} -\frac{V_{+3}V_-}{V_+^2} \right).
\end{equation}

Note that $\phi_1$ already includes all the corrections to first order in the potential. Indeed, the next iteration of the method would provide us with a further improved value for $\phi_B$: $\phi_2 = \phi_1 - V'(\phi_1)/V''(\phi_1)$. However we found that $V'(\phi_1)$ is of $\mathcal{O}\left(\tilde{W}^2\right)$, and so $\phi_B\simeq \phi_1$ to first order in the potential.

It is convenient to write the evolution of the expectation value of the volume in terms of $\Delta\phi\equiv\phi-\phi_B$. Expanding \eqref{eq:expv_W} in powers of $W$ and truncating at first order we find
\begin{equation}\label{eq:Vapprox}
    \langle \hat V \rangle \simeq V_B^F \left\{\cosh(\sqrt{12\pi G}{\Delta}\phi) + \tilde{W} \left[S(\sqrt{12\pi G}{\Delta}\phi)+A(\sqrt{12\pi G}{\Delta}\phi)+g(\sqrt{12\pi G}{\Delta}\phi)\right]\right\},
\end{equation}
where we have used $V_B^F = 2 \sqrt{V_+ V_-}$ to denote value of the volume at the bounce for the free system and
\begin{align}
    S(x) &= \left[3-\frac{1}{2}\left(\frac{V_{-3}+3V_+}{V_-}+\frac{V_{+3}+3V_-}{V_+} \right)\right]\cosh(x)-4x\sinh(x),\\
    A(x) &= \frac{3}{2}\left(\frac{V_{-3}V_+}{V_-^2}-\frac{V_{+3}V_-}{V_+^2}\right)\sinh(x),\\
    g(x) &= \frac{1}{2}\left(\frac{V_{+3}V_-}{V_+^2}e^{3x}+\frac{V_{-3}V_+}{V_-^2}e^{-3x}\right).
\end{align}
The value of the volume at the bounce for the full system, $V_B=V(\phi_B)$, is thus given by
\begin{equation}
    V_B \simeq V_B^F \left[1+\frac{\tilde{W}}{2}\left(6-\frac{V_{+3}+3V_-}{V_+}-\frac{V_{-3}+3V_+}{V_-}+\frac{V_{+3}V_-}{V_+^2}+\frac{V_{-3}V_+}{V_-^2}\right)\right],
\end{equation}
at first order in $W$. Let us remark again that the above expressions are valid for the range of the evolution when the energy density of the matter field dominates over $W=\Lambda/(8\pi G)$, and in particular around the bounce.

Equation \eqref{eq:Vapprox} allows us to analyze in more detail the effect of the potential on the dynamics. We can easily see that there is an impact on the symmetry of the bounce around $\phi_B$. Namely, we identify three different types of state dependent contributions: a symmetric one given by $S(x)$ as $S(-x) = S(x)$, an anti-symmetric one given by $A(x)$ as $A(-x) = -A(x)$, and finally one that is generically asymmetric (neither symmetric nor anti-symmetric) given by $g(x)$, as the weights of the two exponentials are not necessarily even related for generic physical states.

Whereas for generic states, the effect of the potential is somewhat intricate, remarkably, for the specific case of the semiclassical states determined by the Gaussian profiles \eqref{eq:gaussian profile}, we find that
\begin{equation}
    \frac{V_{-3}V_+}{V_-^2}=\frac{V_{+3}V_-}{V_+^2}=e^{2/\sigma^2}.
\end{equation}

Consequently, $A(x) = 0$, i.e. the anti-symmetric term of \eqref{eq:Vapprox} vanishes, and in $g(x)$ the coefficients multiplying the two exponentials become equal, thereby turning this into a symmetric function: $g(x) \propto \cosh(x)$. Therefore, for this family of physical states the bounce remains symmetric, as in the case of the free system.

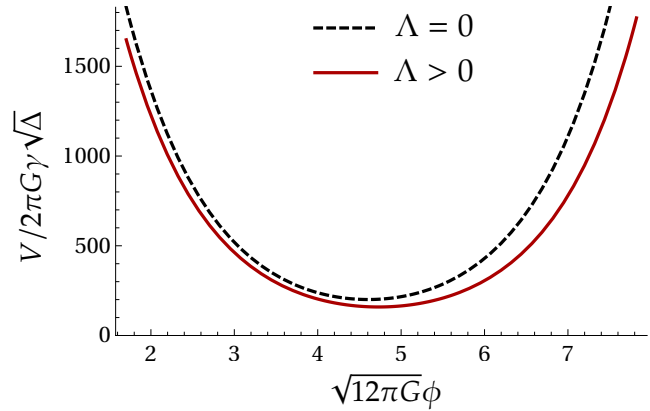
\begin{figure}[h]
\centering
  \begin{tikzpicture}
  \tikzstyle{every node}=[font=\normalsize]
  \node (img) at (0,0) {\includegraphics[width=0.5\textwidth]{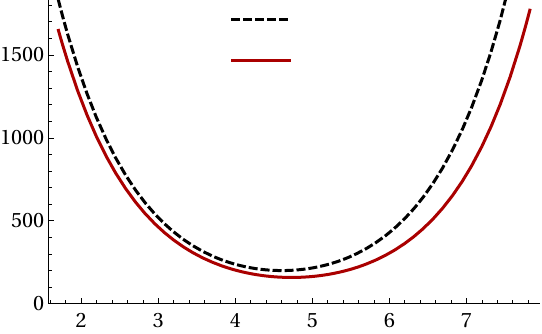}};
  \node[below=of img, node distance=0cm, xshift=.4cm, yshift=1.2cm] {$\sqrt{12\pi G}\phi$};
  \node[left=of img, node distance=0cm, anchor=center,rotate=90,yshift=-0.7cm,] {$V/2\pi G\gamma\sqrt{\Delta}$};
  \node[left] at (1.7,2.15) {\large{$\Lambda=0$}};
  \node[left] at (1.7,1.55) {\large{$\Lambda>0$}};
 \end{tikzpicture}
 \caption[Expectation value of volume with no potential vs with cosmological constant.]{Comparison between the expectation value of $\hat{v}$ on a semiclassical state with the profile \eqref{eq:gaussian profile}, for~the free case ($\tilde{W}=0$), and $\tilde{W} = 10^{-5}$. The parameters of the Gaussian profile were set to $k_o=-100$, $\sigma=10$ ($\langle\Delta\hat{\pi}_{\phi}\rangle/\langle\hat{\pi}_{\phi}\rangle \simeq 0.07$).}
  \label{fig:compare bounces}
\end{figure}

Figure \ref{fig:compare bounces} compares the expectation value of the volume in the two cases. For this analysis, we chose a toy value for $\tilde{W}$ so that its effect in the dynamics is appreciable. Indeed we see that the dynamics of the full system describes a symmetric bounce, which is displaced with respect to that of the free case.

Namely, for our class of states defined by the Gaussian profile \eqref{eq:gaussian profile}, we find that
\begin{equation}
    \sqrt{12\pi G}\phi_B^F=\ln|k_o|, \qquad f(V_n) = \left(3-e^{2/\sigma^2}\right)(k_o^2-k_o^{-2}).
\end{equation}
For $k_o \ll -1$ and $\sigma^{-2} \ll 1$, we obtain $e^{2/\sigma^2} \simeq 1$, and $f(V_n)\simeq 2 k_o^2$, leading to
\begin{equation}\label{eq:phiB}
    \phi_B \simeq \phi_B^F+ \frac{\tilde W}{\sqrt{12\pi G}}\left(4\ln|k_o|+k_o^2\right),
\end{equation}
and
\begin{equation}\label{eq:VB}
    V_B \simeq V_B^F \left(1-2\tilde{W} k_o^2\right).
\end{equation}
Here, $V_B^F=4\pi G\gamma\sqrt{\Delta} |k_o|e^{1/(4\sigma^2)}\simeq 4\pi G\gamma\sqrt{\Delta} |k_o|$.
Thus, we conclude that for our class of states, the effect of the potential is to push the bounce to higher $\phi$ and smaller volumes.

In summary, whereas for the specific case of semiclassical states defined with a Gaussian spectral profile, the introduction of a positive cosmological constant simply displaces the bounce by state-dependent amounts, for generic states it also renders it asymmetrical, in a state-dependent way. As a consequence, the two branches of the Universe (pre and post-bounce) tend, in general, to different classical trajectories. Even though this is the first time that an asymmetric bounce has resulted from this quantization procedure, we note that other prescriptions within the context of LQC have been found to generate an asymmetry between the pre and post-bounce epochs \cite{Assanioussi:2018,Liegener:2019,Liegener:2019_2,Li:2018}.


\section{Regime of Validity of Our Approximation}
\label{sec:approx}

Finally, it should be noted that the perturbative nature of this treatment implies a regime of validity with respect to the parameters of the model. Naturally, the corrections arising from the perturbative treatment should not achieve the order of the leading terms. Although one might at first assume that as long as $\tilde{W} \ll 1$ the perturbation scheme is valid, we can see from \eqref{eq:expv_W} that the corrections arise with the state-dependent variables $V_n$. Thus, there has to be some constraint on the parameters of the chosen state.

In the case of a Gaussian profile such as \eqref{eq:gaussian profile}, making use of \eqref{eq:vn_ko}, the leading contributions to the corrections are found to be of the order of $\tilde{W}k_o^2$ and $\tilde{W}k_o^2 e^{2/\sigma^2}$ with respect to the terms of the free model. Consequently, we need to restrict to the region of parameter space for which $\tilde{W}k_o^2 \ll 1$ and $\tilde{W}k_o^2 e^{2/\sigma^2} \ll 1$.

\begin{figure}[h]
\centering
  \begin{tikzpicture}
  \tikzstyle{every node}=[font=\normalsize]
  \node (img) at (0,0) {\includegraphics[width=0.5\textwidth]{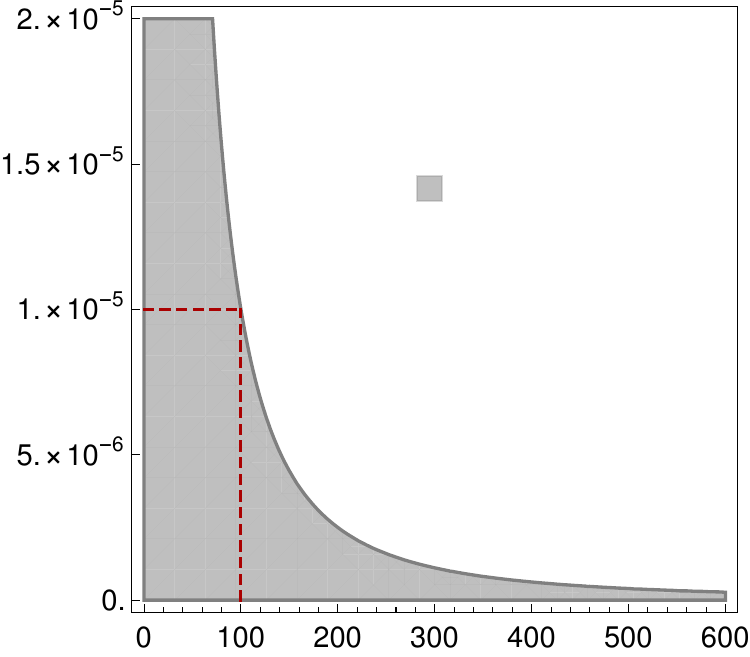}};
  \node[below=of img, node distance=0cm, xshift=.6cm, yshift=1.25cm,] {$|k_o|$};
  \node[left=of img, node distance=0cm, anchor=center,rotate=90,xshift=.25cm,yshift=-0.9cm,] {$\tilde{W}$};
  \node[left] at (3.,1.5) {$\tilde{W}k_o^2 \leq 0.1$};;
 \end{tikzpicture}
 \caption[Allowed region of parameters for validity of approximation.]{Representation of the region of parameters that provides a truncation error $\lesssim 1\%$, when $\sigma = 10$. The red dashed lines represent the point chosen for the graphical representations of the dynamics in this work: $|k_o| = 100$, $\tilde{W} = 10^{-5}$.}
  \label{fig:region of params}
\end{figure}
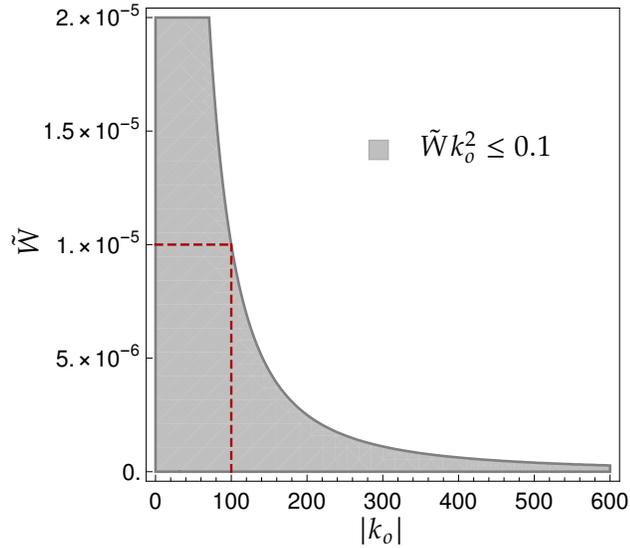

The choice of values for these three parameters requires two types of compromise. Firstly, since we already need $\tilde{W}k_o^2 \ll 1$, the second condition is satisfied if $e^{2/\sigma^2}$ is small. However, note that $e^{2/\sigma^2} \geq 1$, and so we wish to use values for $\sigma$ such that $\sigma^{-2} \ll 1$ (thus $e^{2/\sigma^2} \simeq 1$). However, at the same time, we want to keep the relative dispersion of the states given by \eqref{eq:pi_phi and disp} small, which prevents us from picking high values for $\sigma$, as these would have to be compensated for with $k_o$. This way, we constrain $\sigma$ to an intermediate value $\mathcal{O}(10)$, which allows us in what follows to choose reasonable values for the remaining parameters.
Then, with this choice for $\sigma$, the treatment is valid as long as $\tilde{W}k_o^2 \ll 1$. This is also in agreement with the condition determined in \cite{CastelloGomar2016} for the validity of the treatment to first order in the potential. Consequently, the error associated to truncating the dynamics at first order in the potential is of order of $(\tilde{W} k_o^2)^2$. The values of the parameters used in the previous graphics have been chosen by limiting this error to $1\%$ at the bounce; i.e., by choosing $\tilde{W}$ and $k_o$ such that $\tilde{W} k_o^2 \leq 0.1$. Figure \ref{fig:region of params} shows the corresponding region in parameter space. In it, we highlight the point $(100,10^{-5})$ chosen for the graphical representations presented in this work. This choice was based on the second compromise: on one hand, a~larger value of $\tilde{W}$ makes more obvious the effect of the potential in the dynamics, but on the other hand, $|k_o|$ should be large enough so that the physical state has a small relative dispersion.

One might also wonder how far from the bounce we can trust our quantum evolution truncated at first order in $W$. The agreement of the quantum trajectory with the classical ones displayed in Figure~\ref{fig:dynamics} shows that the approximation remains valid well within the regime in which quantum corrections become negligible. Indeed, from the condition for validity obtained in \cite{CastelloGomar2016}, we are able to determine that the approximation is valid around the bounce for $|\sqrt{12\pi G}(\phi-\phi_B^F)| \leq 1/|\tilde{W}k_o|$, which for our chosen parameters corresponds to $|\sqrt{12\pi G}(\phi-\phi_B^F)| \leq 10^3$.

\section{Conclusions and discussion}\label{sec:discussion}

This work aimed to deepen the analysis of loop quantum gravitational corrections to pre-inflationary dynamics of cosmological models for inflation. To that end, our strategy was to develop the necessary mechanisms to employ the treatment proposed in \cite{CastelloGomar2016} to flat FLRW models with a scalar field subject to a potential that drives inflation. This treatment is based on the dynamics of the model with vanishing potential, obtaining the effect of the potential in the dynamics in a perturbative manner.

In this work, we have applied said method to the simplest possible case: a flat FLRW model with a scalar field subject to a constant potential. However, we showed explicit calculations with generic potential as far as was possible, so as to simplify future applications to other models. For this specific model, equivalent to the model with a massless scalar field in the presence of a cosmological constant, we computed the expectation value of the volume along the evolution. By plotting this quantity taken over semiclassical states along with the classical trajectories, we observed the same qualitative behavior as that of the free case: far from the Planckian regime, where quantum effects are expected to be negligible, the quantum trajectory agrees with the classical ones, whereas close to the Planck scale, the quantum dynamics displays a bounce, connecting a contracting epoch of the Universe with an expanding one. 

However, the main novelty of this work is that the dynamics were obtained perturbatively with respect to that of the free case, making it easy to compare the two scenarios. This comparison revealed that even though the qualitative behavior of the two models (with and without cosmological constant) is the same, the bounce suffers a displacement when a cosmological constant is introduced. Through a more detailed analysis, we were able to write an approximate form of the expectation value of the volume around the new bounce. Focusing on semiclasscial states given by Gaussian profiles peaked on classical trajectories, we were able to determine that a positive cosmological constant displaces the bounce to lower values of the volume and higher values of the scalar field by amounts that are dependent on the state parameters. Remarkably, this simple behavior of a symmetric bounce is particular to such semiclassical states. Indeed, the behavior of the volume for generic states is more intricate, as there are corrections that are not symmetric around the bounce and even ones that are anti-symmetric.

One consequence of our results, whose detailed study we leave for the future, is that the upper bound on the total energy density might be affected by a state-dependent amount. Indeed the maximum of the energy density would surely be affected in such a manner; however, it is not trivial to understand whether the universal upper bound found in the model without cosmological constant \cite{ACS} is altered as well. A~proper analysis of this point requires a study on the different possible factor orderings one may choose for the operator representing the energy density, in accordance with the discussion of \cite{Kaminski:2008td} on the spectrum of the energy density operator.

Finally, it is worth remarking that even though this treatment introduces approximations, it offers several advantages. Firstly, it is less time and resource-consuming than the exact treatment pursued in~\cite{Pawlowski2012}. Secondly, it can be applied to models with non-constant potentials, presumably with little in the way of added difficulties, other than the computation of one integral that depends on the form of the field potential. To this, we add that at least polynomial potentials in the scalar field should not introduce additional challenges. Lastly, the application of this method to a simple model already allowed us to uncover interesting consequences that went unnoticed in the numerical investigations of \cite{Pawlowski2012}. Indeed, even though the treatment of that work was more exact, the form of the physical states was so intricate that it did not provide information on the dynamics for generic states. A particularization to semiclassical states is necessary and some state-dependent effects are naturally ignored.
\chapter{SLE\MakeLowercase{s} in LQC}\label{chap:SLEsinLQC}


In this chapter we propose the SLEs introduced in chapter \ref{sec:introSLE} as candidates for the vacuum of cosmological perturbations in LQC. We will focus our study on the hybrid approach of the theory, though the procedure can be reproduced within the dressed metric approach as well. This has led to the two publications \cite{SLEs_ours,mno-sles}.

The main motivation for this proposal is 2-fold: SLEs provide a minimization of the regularized energy density (of each mode), and they are exact Hadamard states, which guarantees that computations such as that of the expectation value of the renormalized stress-energy tensor are well defined. Furthermore, these states have been shown to provide the correct qualitative asymptotic infrared and ultraviolet behaviors of the primordial power spectra in models with a period of kinetic dominance prior to inflation, which is the case of LQC. However, a comparison with observation is restricted to a finite window of modes. An agreement between predictions and observations relies heavily on the behavior at intermediate scales. In this chapter we explicitly compute primordial power spectra for intermediate scales as well.

The construction of SLEs depends on a choice of smearing (or test) function. Its support determines the window of time in which the regularized energy density is minimized. Given this function, the procedure provides a state that is a global solution. The choice of smearing function can be seen as an ambiguity that is introduced in exchange for the ambiguity in the choice of initial time when to fix the vacuum. This ambiguity will be one of the main focuses of this chapter. In chapter \ref{sec:introSLE} we have shown that in Minkowski the SLEs are trivially found to be the Minskowski vacuum. We have also mentioned that in de Sitter with a massive scalar field this construction has been shown to select the Bunch-Davies vacuum (the natural vacuum in this case) in a given configuration. Concretely, this happens when the support of the smearing function tends to the distant past. For completeness, in section \ref{sec:SLEdeSitter} we prove that in de Sitter the SLE prescription selects the Bunch-Davies vacuum also in the case of a massless scalar field.

Then we explore the construction within LQC in section \ref{sec:SLEinLQC}. We will show that the predictions of power spectra are very insensitive to the choice of smearing function of the SLEs, provided it covers the Planckian regime. By computing the tensor-to-scalar ratio and the spectral index, we also show that this choice of vacuum state can provide at least as good an agreement with the observations of these quantities (which have recently ruled out some models) as the standard cosmological model with the same inflaton potential.

In section \ref{sec:SLEs no bounce}, we investigate the effect of shifting the test function away from the high curvature regime. Firstly, this provides a more complete analysis of the SLEs and the ambiguity of the test function. Secondly, this allows us to distinguish in the primordial power spectra the consequences coming directly from LQC corrections and those related to simply having a period of kinetic dominance prior to inflation, which can also be obtained in a classical scenario. We will show that if the test function ignores the Planckian region, the effect in the resulting SLE is appreciable. Importantly, these effects are translated to the primordial power spectra, where the imprint of the bounce becomes obvious in this context. When the support of the test function includes it, the power spectra show oscillations and power enhancement at intermediate scales, which almost entirely vanish when the support does not include the bounce.

This motivates us to compare our results with those found in the LQC literature that adopts as initial conditions for the perturbations the so-called non-oscillatory (NO) vacuum state \cite{nonosc,hyb-obs} in section \ref{sec:SLEvsNO}. As the name suggests, this state is precisely defined to minimize mode by mode the amplitude of the oscillations of the primordial power spectra in a given time interval. It turns out that this minimization in time is reflected in a minimization of oscillations in the $k$ domain of the power spectra. This NO prescription has been motivated as well as a good candidate for the vacuum of the perturbations \cite{menava}. We show that the same primordial power spectra can be obtained with SLEs when the support of the test function does not include the Planckian regime. Furthermore, one question that so far remained unanswered is whether this NO vacuum is of Hadamard type. In this section, by comparing with SLEs, we show that indeed this is the case. To do so, we resort to their UV expansions, obtained for SLEs in \cite{Niedermaier2020} and for the NO vacuum in \cite{menava}.

Finally, in section \ref{sec:conc} we conclude  summarizing our results and with some closing remarks.

\section{SLE\MakeLowercase{s} in de Sitter}\label{sec:SLEdeSitter}

As we have just seen, SLEs are not unique inasmuch as they depend on the choice of the test function $f$. Then one might wonder whether there is any relation between these states and the natural vacuum present in maximally symmetric spacetimes, namely Minkowski and de Sitter. These spacetimes  admit a unique vacuum invariant under the isometries of the spacetime, the Minkowski vacuum and the Bunch-Davies vacuum respectively. It is therefore desirable that any prescription attempting to define the vacuum, when applied to these spacetimes, singles out such a state.

In Minkowski this question is trivial.  Indeed, if we take as fiducial solution the Minkowski vacuum, it is obvious that the integrand in \eqref{eq:c2} is identically zero for any $f$, and thus the SLE is found to be the Minkowski vacuum (up to a phase). This is not the case in a de Sitter universe. Thus, the question arises whether the SLEs converge to the Bunch-Davies vacuum in some appropriate limit. Reference \cite{DegnerPhD2013} proves that indeed this is the case for the massive field theory.

For the sake of completeness in this discussion, in this section we will extend the analysis to the massless case, which is also common in cosmology. Let us recall that in de Sitter the scale factor evolves as
\begin{equation}
    a^{\text{dS}}(\eta)=-\frac{1}{H\eta},
\end{equation}
where conformal time takes values $\eta\in(-\infty,0]$, and  the Hubble parameter $H$ is constant. The cosmological perturbations are ruled by the equations of motion \eqref{eq:eomT} with $\omega_k^2=k^2/a^2$ for both scalar and tensor modes. In this cosmological chart of de Sitter, even though strictly speaking we cannot define a unique vacuum state invariant under the isometries of the spacetime, cosmologists typically consider as natural vacuum the one defined by \eqref{eq:v(eta)BD}. In $T_{k}$ this translates to:
\begin{align}\label{bdvacuum}
T^{BD}_k(\eta)=-\frac{H}{\sqrt{2k}}e^{-ik\eta}\left( \eta-\frac{i}{k}\right),
\end{align}
that we will keep calling Bunch-Davies state. Let us introduce a test function $f(\eta)$ defining the time-like curve of an isotropic observer in de Sitter,  with support in the interval $[\eta_0,\eta_f]$. Then, the smeared energy density of each mode in the vacuum \eqref{bdvacuum} measured by that observer is 
$E(T^{BD}_k)$ defined in \eqref{energy}. Using conformal time, it explicitly reads
\begin{align}\label{ebd}
  E(T^{BD}_k)=\frac{H^3}{2}\int_{\eta_0}^{\eta_f} d\eta f^2(\eta)|\eta| \left(k\eta^2+\frac{1}{2k} \right).
\end{align}

Let us now consider any other state, defined via $T^\omega_k$.
Such a state will be related to the Bunch-Davies state through a Bogoliubov transformation that, up to an irrelevant global phase, is given by
\begin{align}
T^{\omega}_k=e^{i\alpha(k)} \sqrt{1+\beta^2(k)} T^{BD}_k+\beta(k) \bar{T}^{BD}_k,
\end{align}
with  $\alpha(k)$ and $\beta(k)\neq 0$ real coefficients. 
Then, we can write
\begin{align}
    E(T^{\omega}_k)=E(T^{BD}_k)(1+2\beta^2) +2\beta\sqrt{1+\beta^2}\text{Re}[e^{i\alpha} D(T_k^{BD})],
\end{align}
with
\begin{equation}
   D(T_k) = \frac{1}{2} \int dt\,f^2(t)\left(\dot{T}_k^2 +\omega_k^2 T_k^2\right),
\end{equation}
which implies
\begin{align}
 D(T_k^{BD})=-\frac{H^3}{2}\int_{\eta_0}^{\eta_f} d\eta f^2(\eta)|\eta|e^{-2ik\eta} \left(i\eta+\frac{1}{2k} \right).
\end{align}

Let us now show that, no matter what this state is, it always verifies $E(T^{\omega}_k)\geq E(T^{BD}_k)$ in the limit $\eta_0\rightarrow-\infty$ and for any time-like curve of the observer measuring the state's energy. The  smallest value for $E(T^{\omega}_k)$ is attained by choosing $\alpha(k)$ such that $e^{i\alpha(k)} D(T_k^{BD})=-|D(T_k^{BD})|$ and $\beta(k)>0$. In that case, and defining $\delta(k)=\beta(k)/\sqrt{1+\beta(k)^2}\in(0,1)$, we have the following relation 
\begin{align}
   E(T^{\omega}_k)=\mathcal{R}_k(\eta_f,\eta_0)E(T^{BD}_k),
\end{align}
with
\begin{align}
   \mathcal{R}_k(\eta_f,\eta_0)=\left\{1+\frac{2\delta(k)}{1-\delta(k)^2}\left[\delta(k) -\frac{|D(T_k^{BD})|}{E(T^{BD}_k)}\right]\right\}
\end{align}

Noting that in general for any function $g(x)$ we have that $|\int dx g(x)| \leq \int dx |g(x)|$, it is easy to see that ${|D(T_k^{BD})|}/{E(T^{BD}_k)}$ is bounded by 
\begin{align}\label{ratio}
 \frac{|D(T_k^{BD})|}{E(T^{BD}_k)}\leq \frac{\int_{\eta_0}^{\eta_f} d\eta f^2(\eta)\eta^2\sqrt{1+\frac{1}{(2k\eta)^2} }}{\int_{\eta_0}^{\eta_f} d\eta f^2(\eta)\eta^2\left(k|\eta|+\frac{1}{2k|\eta|} \right)}.
\end{align}
Then, in the limit $\eta_0\rightarrow-\infty$, the denominator in \eqref{ratio} grows faster than the numerator no matter the test function $f$ and we have
\begin{align}
   \lim_{\eta_0\to-\infty}\mathcal{R}_k(\eta_f,\eta_0)= \left[1+\frac{2\delta^2(k)}{1-\delta^2(k)}\right]>1,\quad\forall \,k,\eta_f.
\end{align}

This implies
   \begin{align}
   E(T^{\omega}_k)> E(T^{BD}_k)\,,\quad \forall\,k,
\end{align}
independently of the choice of both $f$ and the final point of its support, provided that the initial point $\eta_0$ tends to the distant past. 

This clarifies the relation between SLEs in the case of the massless field in the cosmological chart of de Sitter and the Bunch-Davies vacuum: they agree whenever the observer starts measuring the energy at the distant past, since then the Bunch-Davies vacuum has the smallest possible energy. This result actually agrees with the conclusion found also in the massive case \cite{DegnerPhD2013}. The condition that the observer starts measuring in the distant past seems somewhat natural, since the Bunch-Davies state results as the natural vacuum after one imposes that the vacuum tends to the Minkowski vacuum for ultraviolet modes or equivalently in the distant past.

\section{SLE\MakeLowercase{s} in bouncing LQC}\label{sec:SLEinLQC}

The work of \cite{Niedermaier2020} has shown that SLEs are an appropriate choice of vacuum for cosmological perturbations in models where a period of kinetic dominance precedes inflation. This is precisely the typical case of some inflationary bouncing cosmologies, like LQC (see chapter \ref{sec:LQCperts}). Here, the Fourier modes of gauge-invariant scalar and tensor perturbations satisfy the equations of motion \eqref{eq:eom_uk} with the time-dependent mass terms $s^{(s)}$ and $s^{(t)}$ of \eqref{eq:s(eta)LQC}, respectively.

To apply the procedure outlined in Sec. \ref{sec:introSLE}, we first obtain a fiducial solution $S_k$, by fixing initial conditions at the bounce, and then we integrate the dynamics numerically. Let us recall that any state can be parametrized by the functions $D_k$ and $C_k$ as in \eqref{eq:CkDk}. Since the fiducial solution chosen is irrelevant, for simplicity we fix $S_k$ at the bounce to be the 0th order adiabatic state, defined in chapter \ref{sec:adiabatic} by choosing $D_k=k$ and $C_k=0$.

Furthermore, for these computations, we have considered a quadratic inflaton potential: $V(\phi)= m^2 \phi^2/2$, with $m=1.2 \times 10^{-6}$ in Planck units. We choose this value of the mass to get agreement with observations, inspired by previous studies in LQC \cite{AshtekarSloan_probinflation}. Finally, we also need to fix the free parameter $\phi_B$, the value of the inflaton field at the bounce. Since we do not intend to do in this chapter a rigorous Bayesian analysis, we choose a toy value for $\phi_B$. We are looking for one that leads to a power spectrum which deviates from that of standard cosmology in the region where observations allow it. Some tuning is necessary: too low values will result in too little inflation, pushing the oscillations of the power spectra to the range where they are excluded by observations, whereas too big values will lead to power spectra that show no deviation from the standard model in the observable scales. In this spirit, we fix $\phi_B = 1.225$ in Planck units, though a range of values around it would still produce the desired qualitative behavior.

Then one has to fix the test function $f^2$. Let us focus initially on the choice of its support. In standard cosmology, it is common to fix the initial time for the vacuum of perturbations at the onset of inflation, since the classical theory breaks down closely before, at the big bang singularity. On the other hand, LQC offers a singularity-free geometry, where curvature never blows up but reaches a maximum Planckian magnitude. Hence, it seems natural to consider initial conditions at this high-curvature region. However, one can also select an initial time in the asymptotic past well before the bounce. These have mainly been the two strategies followed in the literature. Noticeably, for the SLE construction, we do not need to fix an initial time. As long as we have a solution $S_k$ (for all $\eta$), the choice of which is irrelevant, we can construct the solution $T_k$ (for all $\eta$) that provides the minimal smeared energy density. However, the freedom mentioned above is now replaced by the choice of the test function. Initially, one might reckon that, since we are minimising the energy density as measured by an observer, it seems more natural to allow this observer to witness the whole evolution. Concretely, this means choosing $f^2$ with a wide enough support around the bounce (since it should be compactly supported). Notwithstanding, one could argue that this is a naive view, since LQC provides quite simplified scenarios. For instance, cosmological models derived from full LQG provide a quantum bounce that, instead of connecting two low curvature FLRW spacetimes, displays a collapsing branch that is indeed a de Sitter spacetime with a Planckian cosmological constant (see for instance Refs. \cite{Assanioussi:2018,Dapor:2017,Assanioussi:2019,quismondo:2019,Agullo:2018,Li:2019,Li:2020,Olmedo:2018}). In this sense, we might want to remain agnostic about the pre-bounce dynamics. It is therefore  reasonable to explore test functions with support that includes the contracting and expanding branch, as well as those that are supported only on the expanding branch (starting at the bounce).

\begin{figure}[t]
\begin{center}
\includegraphics[width=.6\textwidth]{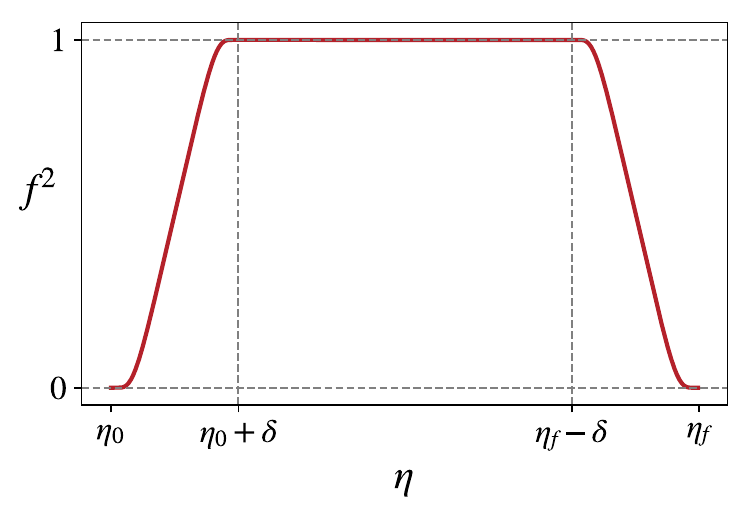}
\end{center}
\caption[Smooth step function used for smearing.]{The smooth step function defined in \eqref{eq:f2step}, represented in terms of its parameters: initial and final points, $\eta_0$ and $\eta_f$ respectively, and $\delta$, which controls the ramping up.}
\label{fig:f2}
\end{figure}

Let us examine these two possibilities separately. For the first class of test functions, where $f^2$ has support on a very wide period from before the bounce to the future, the SLE converges very quickly with the support of $f$. We have also found that the SLE does not depend on the shape of the test function either\footnote{We computed several possibilities, with different test functions, for instance considering the bump function and combining different number of bumps one after the other, even changing their amplitudes, and our computations showed convergence to essentially the same SLE.}. For the second case, we consider test functions whose support begins exactly at the bounce. Again, we find that the SLE converges very quickly with the support. In this case, we are not interested in different shapes for the initial point of the support, as that would dampen or amplify the influence of the Planckian regime over the classical one. Furthermore, the fact that the SLE converges as long as the support is wide enough indicates that different shapes of the test function away from the Planckian regime are irrelevant. Therefore, for the sake of simplicity, in this chapter we will show the results for $f^2$ being a (smooth) window function. The computations were done in conformal time $\eta$, through the change of variables $dt = a d\eta$ in the integrals of $E(S_k)$ defined in \eqref{energy} and of $C(k)$ defined in \eqref{eq:c2}. For this reason, we define the window function in terms of conformal time directly by making use of the auxiliary function
\begin{equation}\label{eq:Sx}
    S(x) = \frac{1-\tanh\left[\cot(x)\right]}{2},
\end{equation}
such that $f^2$, supported in the interval $\eta \in \left[\eta_0,\eta_f\right]$, is defined as:
\begin{equation}\label{eq:f2step}
    f^2(\eta) = 
\begin{cases}
	S\left(\frac{\eta-\eta_0}{\delta}\pi\right) &\eta_0\leq \eta < \eta_0+\delta,\\
	1 & \eta_0+\delta\leq \eta \leq \eta_f-\delta,\\
	S\left(\frac{\eta_f-\eta}{\delta}\pi\right) &\eta_f-\delta < \eta \leq \eta_f,
\end{cases}
\end{equation}
where $\delta$ controls the ramping up, with smaller $\delta$ resulting in a steeper step. The behavior of this function is illustrated in figure \ref{fig:f2}. The first situation that we will consider, which we will name the whole evolution one, is studied by choosing this function with large enough support around the bounce. Concretely, we have determined that for a range around the bounce $\eta_f-\eta_0 = 64$ (Planck seconds) the resulting SLE has already converged. The second situation, which we will refer to as the expanding branch, is investigated by fixing $f^2$ to be this window function starting at the bounce and ending at the onset of inflation, with a very small value of $\delta$, so that the contribution from the dynamics close to the bounce is not dampened. Specifically, we have chosen $\delta \sim 0.06$ (Planck seconds).

\begin{figure}
\centering
    \begin{subfigure}[b]{0.49\textwidth}
    \centering
    \includegraphics[width=\textwidth]{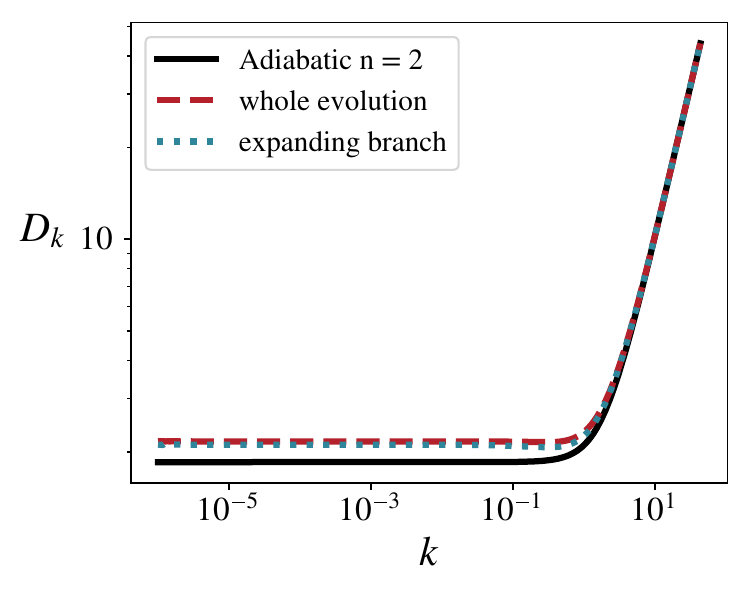}
    \caption{}
    \end{subfigure}
    \begin{subfigure}[b]{0.49\textwidth}
    \centering
    \includegraphics[width=\textwidth]{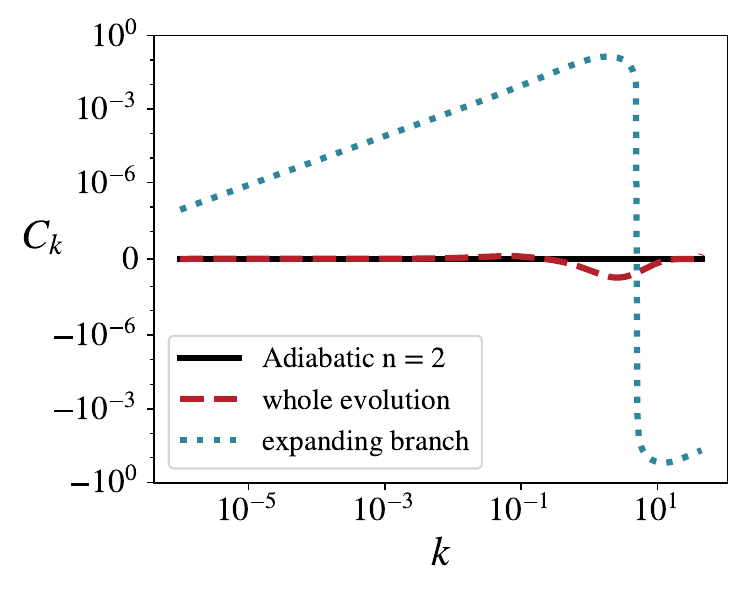}
    \caption{}
    \end{subfigure}
    \caption[Initial conditions for SLEs with test function supported on whole evolution, expanding branch only and adiabatic of order 2.]{Initial conditions for scalar modes at the bounce corresponding to the SLEs obtained with window functions covering the whole evolution (dashed red lines) and only the expanding branch (dotted blue lines) in terms of (a) $D_k$ and (b) $C_k$, as constructed in \eqref{eq:CkDk}. Second order adiabatic initial conditions computed through \eqref{eq:ad_n} are also shown for comparison (solid black line). The scale of $k$ is in Planck units. All computations were performed for a quadratic potential with $m=1.2\times 10^{-6}$, and with $\phi_B = 1.225$. For tensor modes, the resulting SLE at the bounce shows no significant qualitative differences.}
    \label{fig:CD}
\end{figure}

Figure \ref{fig:CD} shows the value at the bounce of the functions $D_k$ and $C_k$ that characterize the SLEs for scalar modes obtained with these two strategies. Though it might seem surprising at first, the procedure gives SLEs with almost exactly the same initial conditions at the bounce for tensor modes, which is why we have omitted that plot. However, it is easy to realize that this has to be the case. Firstly, let us note that, close to the bounce, the time-dependent masses of the two types of perturbations are very similar, as their difference, given by \eqref{eq:U_MS}, is subdominant for a kinetically dominated bounce. As mentioned, the SLE is independent of the fiducial solution $S_k$, and so it is in particular independent of the initial conditions imposed to obtain $S_k$. In this case we are free to choose for $S_k$ the same initial conditions for scalar and tensor modes (as is indeed the case with the 0th order adiabatic ones we have adopted). Then it is easy to see that, given the same initial conditions and very close equations of motion at and around the bounce, the procedure should provide a state that maintains a great similarity between the value of the state at the bounce for scalar and tensor modes.
\begin{figure}
    \centering
    \begin{subfigure}[b]{0.49\textwidth}
    \centering
    \includegraphics[width=\textwidth]{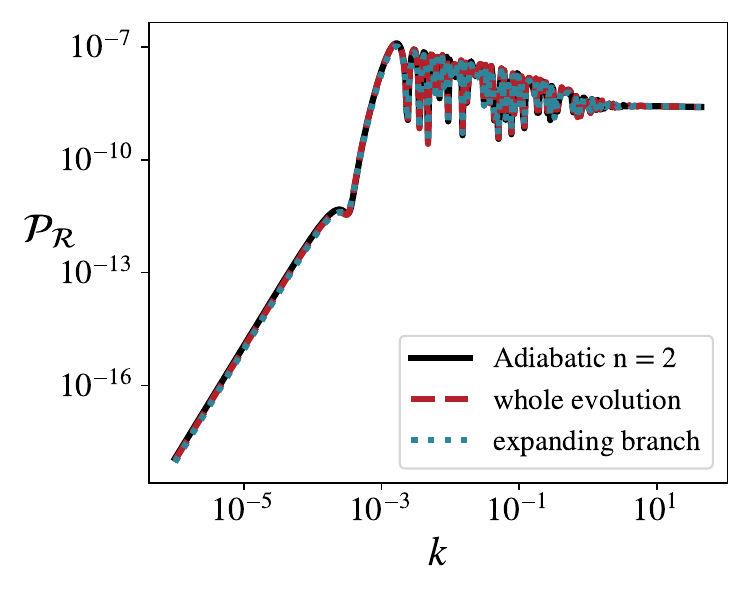}
    \caption{}
    \end{subfigure}
    \begin{subfigure}[b]{0.49\textwidth}
    \centering
    \includegraphics[width=\textwidth]{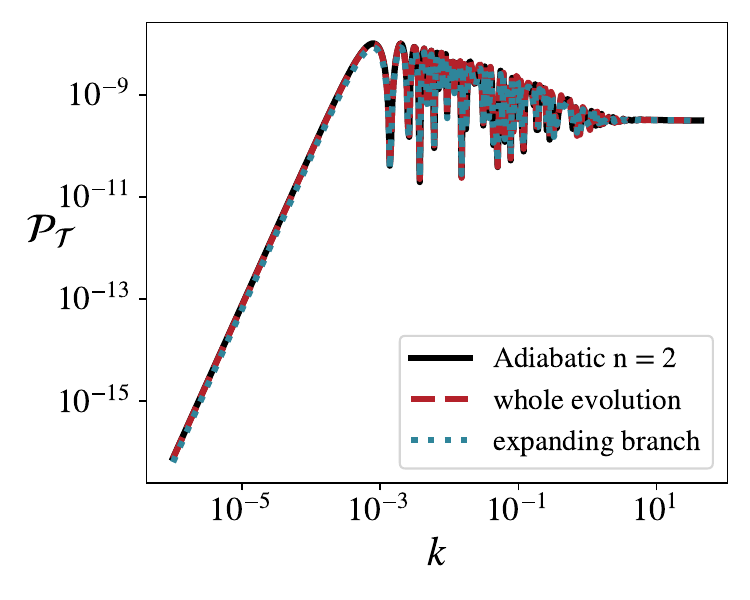}
    \caption{}
    \end{subfigure}
    \caption[Primordial power spectra of SLEs with test function supported on whole evolution, expanding branch only and adiabatic of order 2.]{Power spectra of the (a) comoving curvature perturbation $\mathcal{P}_{\mathcal{R}}$ and (b) tensor perturbation  $\mathcal{P}_{\mathcal{T}}$ corresponding to the SLEs obtained with window functions covering the whole evolution (dashed red lines) and only the expanding branch (dotted blue lines). The ones for second order adiabatic initial conditions at the bounce are also shown for comparison (solid black lines). The scale of $k$ is in Planck units. All computations were performed for a quadratic potential with $m=1.2\times 10^{-6}$, and with $\phi_B = 1.225$.}
    \label{fig:PS}
\end{figure}
Regarding the comparison between the two choices of test function, Figure \ref{fig:CD} shows that $D_k$ has the same kind of behavior in both cases, though it tends to a different value in the infrared. On the other hand, $C_k$ reveals an entirely different behavior between the two scenarios, even though the magnitude in all cases is smaller than unity.

Remarkably, we have found that both scenarios yield extremely similar power spectra at the end of inflation, computed through \eqref{eq_PSRT}, as is shown in Figure \ref{fig:PS}. It seems that the differences observed in initial conditions have no measurable impact in predictions of power spectra. Pairing this to the convergence of the SLE with respect to the form and support of the test function mentioned above, one may say that observational predictions seem to be manifestly insensitive to the choice of test function, within these natural scenarios in LQC.

Interestingly, we find that the power spectra match extremely well with the ones obtained with second order adiabatic initial conditions at the bounce, as computed from \eqref{eq:ad_n}, which are also shown in Figure \ref{fig:PS} for comparison. However, it is important to stress that the SLEs are fundamentally different from adiabatic initial conditions as choices of vacuum. Firstly, SLEs minimize the smeared energy density. Secondly, they are exact Hadamard states, which is not the case for adiabatic states of finite order. Importantly, this allows for the computation of quantities such as the regularized stress-energy tensor. It is worth pointing out that, even if one is not interested in such computations and would therefore be more inclined to use simpler constructions for initial conditions such as adiabatic ones, it would not be necessary to go any further than second order so as to approximate the primordial  power spectra obtained with the SLEs considered here.

\begin{figure}
    \centering
    \includegraphics[width=0.49\linewidth]{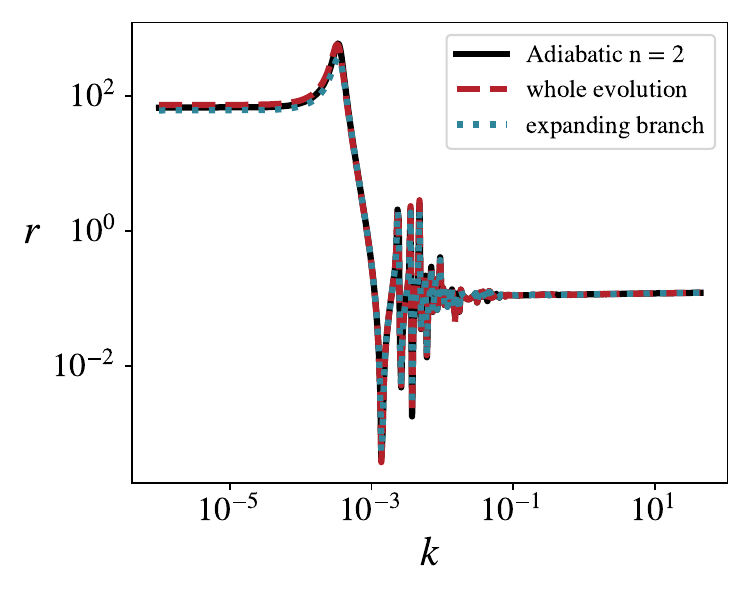}
    \caption[Tensor-to-scalar ratio of SLEs with test function supported on whole evolution, expanding branch only and adiabatic of order 2.]{Tensor-to-scalar ratio corresponding to the SLEs obtained with window functions covering the whole evolution (dashed red line) and only the expanding branch (dotted blue line). The one for second order adiabatic initial conditions at the bounce is also shown for comparison (solid black line). The scale of $k$ is in Planck units. All computations were performed for a quadratic potential with $m=1.2\times 10^{-6}$, and with $\phi_B = 1.225$.}
    \label{fig:r}
\end{figure}

For completeness, in Figure \ref{fig:r} we show the tensor-to-scalar ratio ($r=\mathcal{P}_{\mathcal{T}}/\mathcal{P}_{\mathcal{R}}$ as defined in \eqref{eq:rTtoR} of chapter \ref{sec:introPPS}) as a function of $k$. Scale invariance is achieved for this observable for lower $k$, where the scalar and tensor power spectra still show appreciable departure from scale invariance. This is in agreement with previous results reported in the literature \cite{hyb-obs,dressedmetric_Agullo_CQG_2013}. However, observations of this quantity are usually reported for a particular reference scale.

To contextualize our results in the observations of the CMB, we have computed two quantities: $r_{0.002}$, the value of the tensor-to-scalar ratio at $k = 0.002\ \text{Mpc}^{-1}$, and the spectral index $n_s$. We note that this is only a first investigation with toy values for the model, and that these computations serve merely to show the potential of using SLEs as vacuum states for primordial perturbations. Furthermore, we note that the scale of $k$ shown in the figures of this chapter is in Planck units and corresponds to the usual choice of fixing the scale factor to be $1$ at the bounce in LQC. To relate these to observations of the CMB (namely to identify the scale corresponding to $k = 0.002\ \text{Mpc}^{-1}$ where the tensor-to-scalar ratio is to be computed), which fix the scale factor to be $1$ today, one needs first to identify the correspondence between $k$ of observations, which we will denote as $\tilde{k}$, with that of our own model, which we will keep calling $k$. This is accomplished by identifying the pivot scale $k^{\star}$ that corresponds to the one of observations $\tilde{k}^{\star}$. For greater detail on this matter in LQC see appendix \ref{sec:app_background}. For the pivot scale of the Planck Collaboration of $\tilde{k}^{\star} = 0.05 \text{Mpc}^{-1}$, we find the corresponding one in our model in Planck units to be $k^{\star} = 43.9$ when $\phi_B = 1.225$. Then, it is easy to find that to $\tilde{k} = 0.002\ \text{Mpc}^{-1}$ corresponds $k \simeq 1.76$.

In this work, we will compare our results for $r_{0.002}$ to the ones obtained combining the results of the Planck Collaboration \cite{Planck2018_inflation} and those of BICEP 2018 \cite{BICEP:2021xfz}, which have been reported in \cite{Tristram:2021tvh}: $r_{0.002}< 0.032$ to 95\% confidence level. We find the tensor-to-scalar ratio at this scale for the test function supported along the whole evolution, $r_{0.002}^{\text{w}}$, and for the test function supported on the expanding branch only, $r_{0.002}^{\text{e}}$, as the ratio between the tensor and scalar power spectra at $k \simeq 1.76$:
\begin{equation}
    r_{0.002}^{\text{w}} \simeq 0.118, \qquad r_{0.002}^{\text{e}} \simeq 0.117.
\end{equation}
These values seem to be somewhat disfavored by the observations. However they perfectly agree with the predictions for standard cosmology with the same quadratic potential for the scalar field, also shown in \cite{Planck2018_inflation}. In this sense, the disparity with observations likely stems from the choice of this potential and not from the choice of vacuum for the perturbations.

The computation of the spectral index is more straightforward. It is found by fitting the near-scale-invariant portion of the power spectrum of the comoving curvature perturbation with the function:
\begin{equation}
    \mathcal{P}_{\mathcal{R}} = A_S \left(\frac{k}{k^{\star}}\right)^{n_s-1},
\end{equation}
where $A_S$ is the value of $\mathcal{P}_{\mathcal{R}}$ at $k = k^{\star}$. Concretely, we have computed $n_s$ by fitting the power spectrum with this function in a range in the ultraviolet such that shifting this range to lower or higher values of $k$ made no difference in the final result of $n_s$ to three decimal points. This way we find, for the two possibilities of the test function:
\begin{equation}
    n_s^{\text{w}} \simeq n_s^{\text{e}} \simeq 0.969.
\end{equation}
This value agrees with mean values and error bars from observations from the Planck collaboration: $n_s=0.9649 \pm 0.0042$ at 68\% confidence level.

Finally, it is worth mentioning that we expect the SLEs to lead to different predictions than the standard Bunch-Davies vacuum in GR and previous proposals in LQC. For instance, in \cite{agullo2015}, the resulting vacuum state that yields a vanishing (renormalized) stress-energy tensor at the bounce is a 4th order adiabatic state that produces a stronger enhancement of power at small wave numbers compared to the SLEs. On the other hand, the vacua proposed in \cite{Ashtekar:2016,nonosc} show suppression of power at those scales, rather than enhancement, with respect to the Bunch-Davies power spectrum. We leave for future work a detailed comparison between the predictions coming from our proposal and those of previous ones.

\section{Effect of the bounce in SLE\MakeLowercase{s}}\label{sec:SLEs no bounce}

In the previous section we have shown that there are two families of test functions that can be seen as natural choices for the smearing function within LQC: those that are symmetric and wide enough around the bounce and those that are supported on the expanding branch only, including the high curvature regime. These provide SLEs that are very insensitive to their particular form, as long as they include the high-curvature regime, and which lead to primordial power spectra that are essentially the same for the two choices of test function.

In this section we explore the consequences of not including the bounce (i.e. the high curvature regime) in the support of the test function. This way we will be able to study also the effect of the shape of the test function when supported only on the expanding branch away from the high-curvature regime. This will allow us to provide a comparison with an analogous classical scenario of an FLRW model with a period of kinetic dominance prior to inflation.

\begin{figure}[t]
    \begin{subfigure}[b]{0.49\textwidth}
    \centering
    \includegraphics[width=\textwidth]{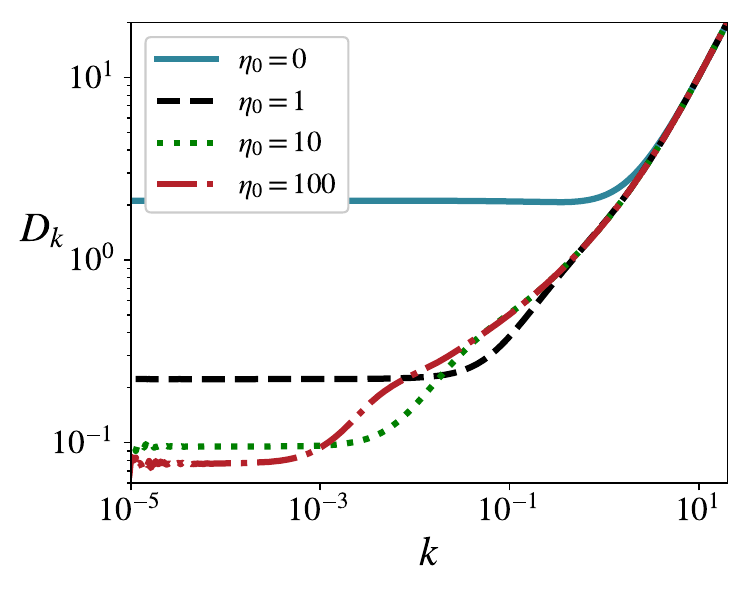}
    \caption{}
    \end{subfigure}
    \begin{subfigure}[b]{0.49\textwidth}
    \centering
    \includegraphics[width=\textwidth]{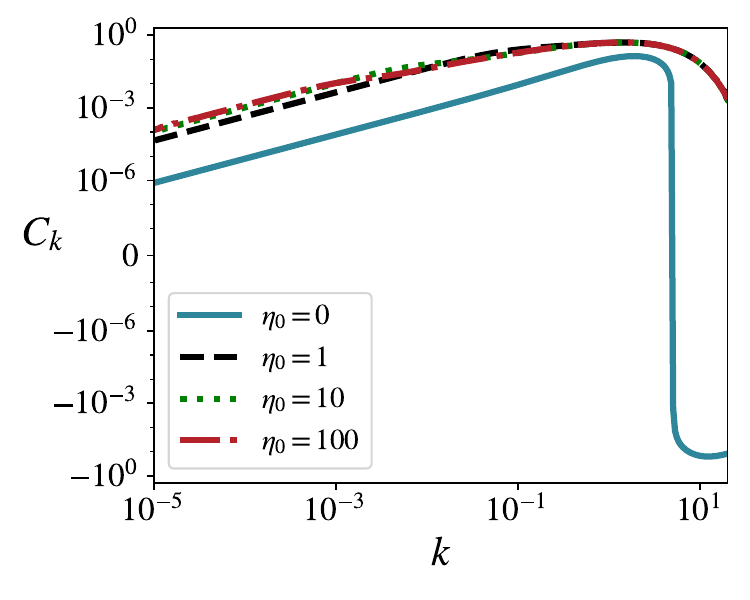}
    \caption{}
    \end{subfigure}
    \caption[Initial conditions of perturbations of SLEs with test function including and gradually excluding the bounce.]{Initial conditions in terms of (a) $D_k$ and (b) $C_k$, as constructed in \eqref{eq:CkDk}, for scalar modes at the bounce corresponding to the SLEs obtained with the window function \eqref{eq:f2step} covering the the expanding branch until the onset of inflation with starting points: $\eta_0=0$ (solid blue line), $\eta_0=1$ (dashed black line), $\eta_0=10$ (dotted green line) and $\eta_0=100$ (dotted-dashed red line). The scale of $k$ is in Planck units. All computations were performed for a quadratic potential $V(\phi) = m^2\phi^2/2$, with $m=1.2\times 10^{-6}$ and with the value of the scalar field at the bounce fixed to $\phi_B = 1.225$ (toy value). For tensor modes, the resulting SLE at the bounce shows no significant qualitative differences.}
\label{fig:CDNO}
\end{figure}

Let us start by considering the smooth step function $f^2$ plotted in figure \ref{fig:f2}, supported in the interval $\eta \in \left[\eta_0,\eta_f\right]$, as defined in \eqref{eq:f2step}. Figure \ref{fig:CDNO} shows the initial conditions, parametrized through \eqref{eq:CkDk}, corresponding to the SLE obtained for scalar perturbations when considering the test function \eqref{eq:f2step}, with $\eta_0=0,1,10$ and $100$ Planck seconds after the bounce, with $\eta_f$ fixed at the onset of inflation, and for a sharp step of $\delta \sim 0.06$. The case of $\eta_0=0$ corresponds to the second one analysed in the previous section, which we had labeled ``expanding branch''. The effect of excluding the bounce is immediately noticed as soon as the support of the test function is moved one Planck second into the expanding branch. If we push the initial time further into the future, the change is gradually decreased, and for $\eta_0 = 100$ we see some convergence. The corresponding figure for tensor modes is omitted since the initial conditions are essentially the same, as discussed previously. Within this family of test functions that exclude the bounce, we have also investigated the consequences of changing their shape. In all these cases, we find that, as the starting point moves further away from the bounce, the SLE becomes more insensitive to the shape of the test function. For this reason,  below, we will focus our comments on the four step functions defined above, as they already show the different qualitative behaviors one may obtain from different test functions in this scenario.

\begin{figure}[t]
\centering
    \begin{subfigure}[b]{0.49\textwidth}
    \centering
    \includegraphics[width=\textwidth]{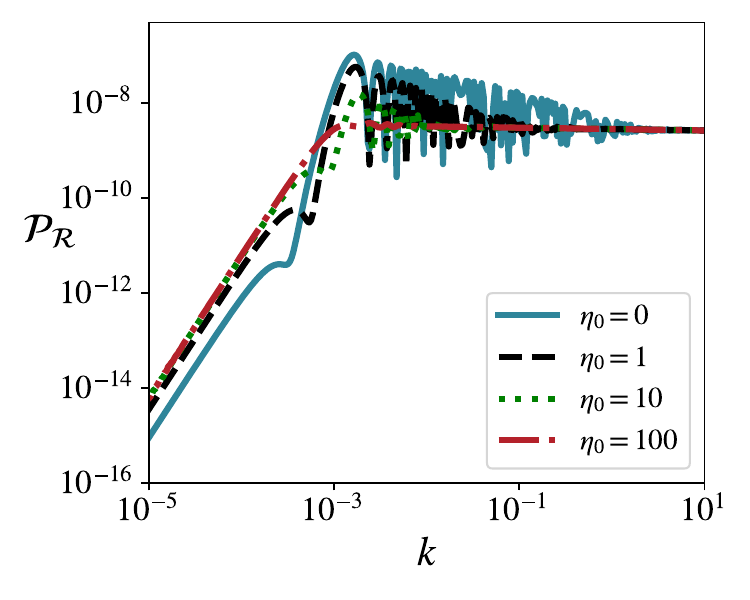}
    \caption{}
    \end{subfigure}
    \begin{subfigure}[b]{0.49\textwidth}
    \centering
    \includegraphics[width=\textwidth]{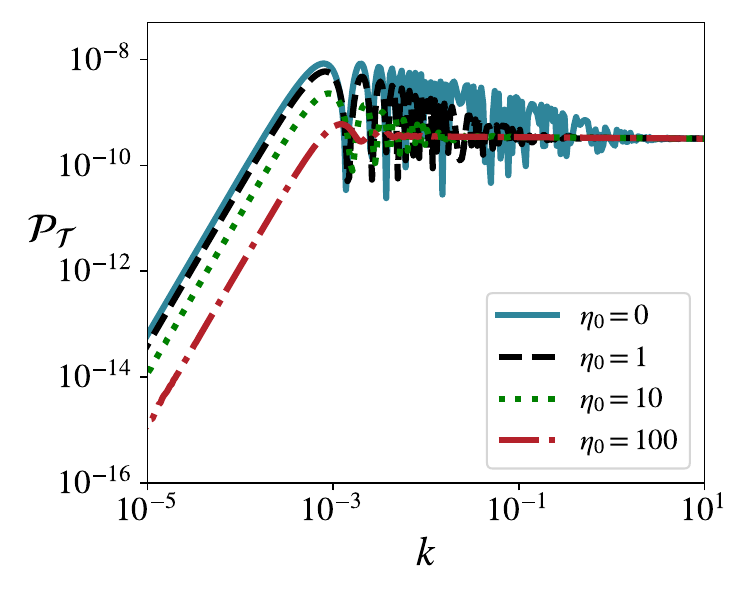}
    \caption{}
    \end{subfigure}
    \caption[Primordial power spectra of SLEs with test function including and gradually excluding the bounce.]{Power spectra of the (a) comoving curvature perturbation $\mathcal{P}_{\mathcal{R}}$ and (b) tensor perturbation  $\mathcal{P}_{\mathcal{T}}$ corresponding to the SLEs obtained with the window function \eqref{eq:f2step} covering the the expanding branch until the onset of inflation with starting points: $\eta_0=0$ (solid blue line), $\eta_0=1$ (dashed black line), $\eta_0=10$ (dotted green line) and $\eta_0=100$ (dotted-dashed red line). The scale of $k$ is in Planck units. All computations were performed for a quadratic potential $V(\phi) = m^2\phi^2/2$, with $m=1.2\times 10^{-6}$ and with the value of the scalar field at the bounce fixed to $\phi_B = 1.225$ (toy value).}
\label{fig:PSNO}
\end{figure}

\begin{figure}[t]
    \centering
    \begin{subfigure}[b]{0.49\textwidth}
    \centering
    \includegraphics[width=\textwidth]{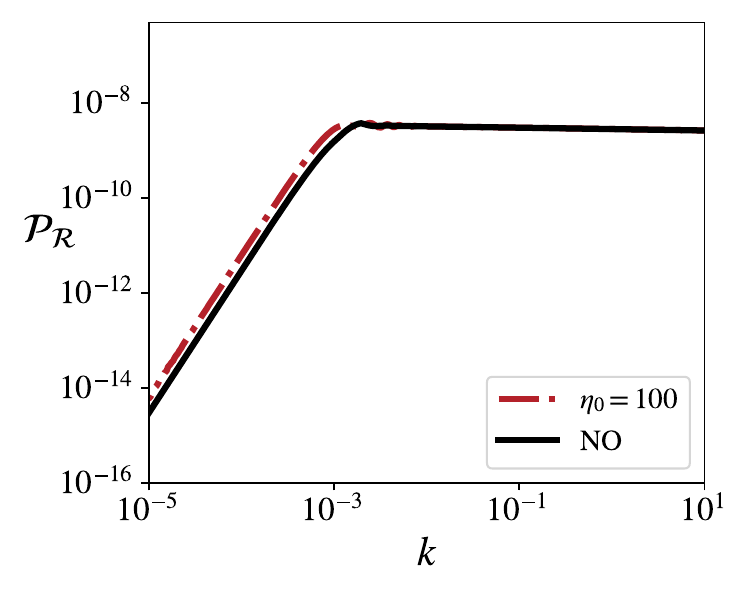}
    \caption{}
    \end{subfigure}
    \begin{subfigure}[b]{0.49\textwidth}
    \centering
    \includegraphics[width=\textwidth]{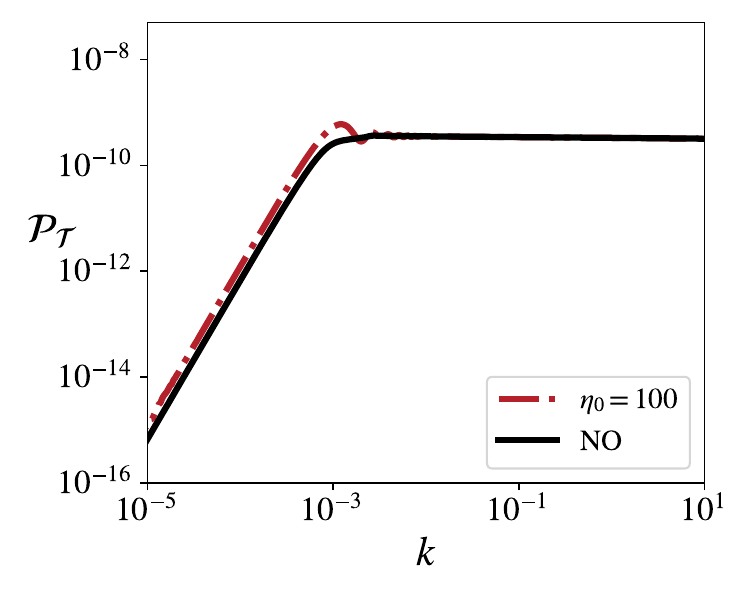}
    \caption{}
    \end{subfigure}
    \caption[Primordial power spectra of SLEs with test function excluding the bounce and of NO vacuum.]{Comparison between the power spectra of the (a) comoving curvature perturbation $\mathcal{P}_{\mathcal{R}}$ and (b) tensor perturbation  $\mathcal{P}_{\mathcal{T}}$ corresponding to the SLE obtained with $\eta_0=100$ (dotted-dashed red line) and to the NO vacuum (solid black line). The scale of $k$ is in Planck units. All computations were performed for a quadratic potential $V(\phi) = m^2\phi^2/2$, with $m=1.2\times 10^{-6}$ and with the value of the scalar field at the bounce fixed to $\phi_B = 1.225$ (toy value).}
\label{fig:PSNO100}
\end{figure}

Figure \ref{fig:PSNO} shows the corresponding primordial power spectra for scalar and tensor perturbations, computed through \eqref{eq_PSRT}. Here, the effect of removing the bounce from the support of the test function is evident. As the support of the test function is pushed further away from the high curvature regime, the oscillations and power enhancement in the power spectra are gradually dampened. It is interesting to note that, in fact,  as Figure \ref{fig:PSNO100} shows, the power spectra are pushed towards those obtained from the non-oscillatory (NO) vacuum state defined in \cite{nonosc}, which is constructed by minimizing the oscillations in time of the power spectrum of perturbations for the whole expanding branch, including the bounce. We further note that the case where the support of the test function starts at $\eta_0=100$ will essentially correspond to that obtained by using the SLE as the vacuum state of primordial perturbations in a classical FLRW model with a period of kinetic dominance prior to inflation. However, for smaller $\eta_0$, SLEs show oscillations in $k$ at and below scales comparable to those of the curvature at that initial time. Then we can conclude that the oscillations that appear in the power spectra when including the high curvature region (for instance the bounce of LQC) in the support of the test function open an interesting observational window into quantum cosmology. 

For completion, we added a appendix \ref{sec:appSLENOKD} where we apply the SLE and NO vacuum prescriptions in a classical universe dominated by the kinetic energy of the scalar field. We discuss the situations in which they agree with the natural choice for vacuum state there, as defined in \cite{Contaldi:2003zv}.

\section{Comparison between SLE and NO vacuum}\label{sec:SLEvsNO}

One remarkable property of the SLEs that is usually not explicitly proven for other vacua proposals is that they are of Hadamard type \cite{Olbermann2007,Niedermaier2020}. This guarantees that computations such as that of the expectation value of the renormalized stress-energy tensor will be well defined. On the other hand, the NO vacuum has only been proven to behave in the ultraviolet (UV) asymptotic regime as a high order adiabatic state, at least of fourth order \cite{nonosc_BeaDaniGuillermo}. Indeed, considering two adiabatic states of orders $n$ and $m$, one can compute the $\beta$ coefficient of the Bogoliubov transformation between the two:
\begin{equation}
\beta = i\left[ u^{n}_k \left(u^{ m}_k\right)^{\prime}-u^{ m}_k \left(u^{ n}_k\right)^{\prime}\right],
\end{equation}
and find that in the UV $|\beta|$ decays with $k^{-l-2}$, where $l = {\rm{min}}(n,m)$. In the case of the comparison of the NO vacuum with an nth-order adiabatic one, it was found that $|\beta|\sim k^{-2-n}$ at least up to $n=4$, which shows that the NO vacuum is the highest order one of the two. As a Hadamard type vacuum is an infinite order adiabatic state, this is an indication that the NO vacuum might be as well, though a stronger proof would be desirable. In this section, we will provide one, through a comparison with the SLEs.

To simplify the comparison, let us write the UV expansions of both the SLE and the NO state as
\begin{equation}\label{eq:uF}
    u_k(\eta) \sim \frac{1}{\sqrt{2 F_k(\eta)}} e^{-i \int d\eta F_k(\eta)},
\end{equation}
where $\sim$ means the behavior in the large $k$ regime. The NO vacuum state has recently been analysed analytically in \cite{menava}. In particular, that work has found that the state admits the UV asymptotic expansion \eqref{eq:uF} with:
\begin{equation}\label{eq:F_NO}
    F_k^{NO}(\eta) = - {\rm Im}(h_k(\eta)),
\end{equation}
where
\begin{equation}
    kh_k^{-1}\sim i\left[1-\frac{1}{2k^2}\sum_{n=0}^{\infty}\left(\frac{-i}{2k}\right)^{n}\gamma_n \right],
\end{equation}
and the $\gamma_n$ coefficients are given by the iterative relation
\begin{equation}\label{eq:gamman}
\gamma_{n+1}=-\gamma_{n}^{\prime}+4s(\eta)\left[\gamma_{n-1}+\sum_{m=0}^{n-3}\gamma_m \gamma_{n-(m+3)}\right]-\sum_{m=0}^{n-1}\gamma_m \gamma_{n-(m+1)},
\end{equation}
with $\gamma_{0}=s(\eta)$ and $\gamma_{-n}=0$ for all $n>0$. With this expansion, we are able to compute the NO state up to any order in $1/k$ easily. Actually, one can check by direct inspection that 
\begin{equation}\label{eq:uFNO}
    F_k^{NO}(\eta) \sim k \left\{ 1 + \sum_{n\geq 1} \frac{(-1)^n}{k^{2n}} G_n(\eta)\right\}^{-1},
\end{equation}
where the $G_n$ are determined recursively by
\begin{equation}\label{eq:Gn}
    G_n(\eta) = \!\!\sum_{m,l\geq 0, m+l =n-1}\! \left\{\frac{1}{4} G_m G_l^{\prime\prime} -\frac{1}{8}G_m^{\prime} G_l^{\prime} +\frac{1}{2}s(\eta)G_m G_l\right\}-\frac{1}{2}\sum_{m,l\geq 1, m+l=n} G_m G_l,
\end{equation}
with $G_0=1$. Remarkably, in \cite{Niedermaier2020}, the SLEs are found to have the same asymptotic expansion \eqref{eq:uFNO}, regardless of the choice of the test function.

Therefore, the $\beta$ coefficients of the Bogoliubov transformation between the SLE and NO vacuum are identically zero in the UV. Thus, we conclude that the NO vacuum is of Hadamard type since it displays exactly the same short-distance structure as the SLEs. 

\section{Conclusions and discussion}\label{sec:conc}

In general cosmological scenarios, SLEs are not free of ambiguities. By definition, they minimize the smeared energy density with respect to a given test function. Hence, one might question their universality as natural vacua for perturbations. To this respect, we have pointed out that the analysis of SLEs in maximally symmetric spacetimes seems to indicate that they do indeed select the preferred vacuum when a clear notion of one already exists. In Minkowski, for example, the situation is trivial and the SLE is immediatly identified as the Minkowski vacuum. In de Sitter, in Ref. \cite{DegnerPhD2013} it was proven that SLEs select the Bunch-Davies vacuum in an appropriate limit of the test functions in the case of the massive theory. For completeness, we have also shown in section \ref{sec:SLEdeSitter} that this is the case in the massless theory as well.

When applied to cosmological perturbations in LQC, we have identified two natural choices for the test function: one supported on a window around the bounce, and one that includes the expanding branch only and is steep so that the bounce contribution is not dampened. In both cases we find that the resulting SLE does not depend on the support of the test function as long as it is wide enough. For the first choice of support, we further find that the SLE does not strongly depend on the shape of the test function. Remarkably, we have found that the SLEs yield power spectra at the end of inflation that do not depend qualitatively on either choice as long as the test function has support on the high-curvature region of the geometry. Even quantitatively this dependence is very weak. Then, the major disadvantage in selecting SLEs as vacuum states for primordial perturbations, namely that of the ambiguity in the test function, effectively disappears at least in the context of hybrid LQC for these natural choices. Furthermore, a preliminary analysis of the tensor-to-scalar ratio and the spectral index shows that these are at least as viable candidates for vacuum states of perturbations as the Bunch-Davies vacuum in standard cosmology regarding their agreement with observations. Additionally, they are proven to be of Hadamard type, which is not the case in other proposals within LQC.

The goal of this part of the work is to present SLEs as suitable candidates for vacua of perturbations in LQC. This has been accomplished by demonstrating that there exists a naturally motivated class of test functions for which the results seem to be insensitive to their particular form and support. A preliminary analysis has also shown that in this case an agreement with observations is attainable. However, to determine this agreement rigorously, we plan to carry out a proper Bayesian analysis, studying not only the possible parameters of the test functions, but also the free parameters coming from the LQC framework, such as the value of the inflaton field at the bounce. 

We have also investigated the effect of pushing the support of the test function away from the bounce and indeed from the high curvature regime. In addition to offering a more complete analysis of SLEs within LQC, this allows us to disentangle the effects coming from quantum corrections to the dynamics, which are important in the high curvature regime, from those that arise from having only a period of kinetic dominance prior to inflation, which can be found also in classical inflationary models, and is not a direct consequence of the quantum nature of geometry.

We have found that whether the support of the test function includes the high curvature regime or not has a greater influence on the resulting SLE than any other parameter of the test function that has been studied previously. As soon as the test function is pushed away from the bounce, the SLE suffers a big shift, which then converges as the test function is pushed further away from the high curvature regime. We have also found that in this case, when convergence with respect to the support is reached, the SLE is again insensitive to the shape of the test function. Furthermore, through the computation of the power spectra of perturbations at the end of inflation, we see that as the test function is shifted away from the high curvature region the oscillations found for intermediate scales (scales comparable to those of the curvature at initial time) are gradually dampened, and the spectra are pushed to those of the NO vacuum state introduced in \cite{nonosc}. Then it is safe to conclude that these oscillations are in fact a consequence of the corrections coming from LQC, which opens an interesting observational window into signatures from LQC in observations of the CMB.  For instance, the enhancement of power at super-Hubble scales in the primordial power spectrum of scalar perturbations has been proposed, together with large scale non-Gaussianities, as a mechanism to explain several anomalies in the CMB \cite{Agullo:2020,Agullo:2020b}. From this perspective, the power spectrum provided by SLEs prescription when including the bounce is physically relevant. On the other hand, the NO-like power spectra show a lack of power at large scales of primordial origin that can alleviate some tensions in the CMB \cite{Ashtekar:2020,Ashtekar:2021}. This is the object of study of the next chapter.

Finally, the fact that SLEs are proven to be Hadamard is a great advantage that most proposals do not enjoy. Typically, this property is difficult to prove explicitly. One strategy, that may be enough for practical purposes, is to compare a state with an adiabatic one of increasing (finite) order, and show, through the $\beta$ coefficients of the Bogoliubov transformation between the two states, that the state in question is always of higher order than the adiabatic one considered. This shows that it is at least a very high order adiabatic state, and, since a Hadamard state is an adiabatic state of infinite order, most likely so is the proposed state. However, we now have a family of states, namely SLEs, that are explicitly Hadamard. Therefore, the $\beta$ coefficients of the transformation between any Hadamard state and any SLE should decrease faster than any power of the wave number. We have applied this reasoning to the NO vacuum state, that had previously been shown to be at least of fourth order \cite{nonosc_BeaDaniGuillermo}.  We find that, in the ultraviolet limit of large wave numbers, the asymptotic expansion that the NO vacuum satisfies (found in \cite{menava}) agrees exactly with that of the SLE \cite{Niedermaier2020} (no matter the test function chosen to define it). As a consequence, the $\beta$ coefficients of the transformation between the two will be identically zero in the ultraviolet. In other words, the NO vacuum has the same short-distance structure than the SLEs, which proves that the NO vacuum is of Hadamard type as well.

\chapter{Contrasting LQC with observations}\label{chap:anomalies}

In the last decades, cosmology has reached a high degree of maturity as a research field thanks in part to the increasingly more accurate measurements of the Cosmic Microwave Background (CMB). For the most part, the observed CMB is well explained by the inflationary paradigm introduced in chapter \ref{sec:earlyUniverse_obs} \cite{Planck2018_parameters,Planck2018_inflation}, where quantum fluctuations in the very early Universe seed the temperature fluctuations observed today. However, as detailed in \ref{sec:anomalies}, some anomalies have been identified in the data with respect to predictions from standard cosmology and persist in recent observations \cite{PlanckVII}. This has captivated researchers in the field of quantum gravity as it can be an indication of non-standard processes occurring in the very early Universe.

Generally, the pre-inflationary dynamics of LQC leave imprints on the primordial power spectra as departures from near scale invariance. In \cite{Ashtekar:2020,Ashtekar:2021,Agullo:2020b,Agullo2021} it was shown that, in the dressed metric approach \cite{Ashtekar:2009mb,dressedmetric_Agullo2012,dressedmetric_Agullo_PRD_2013,dressedmetric_Agullo_CQG_2013} of LQC, such departures may alleviate the anomalies in observations of large scales introduced in chapter \ref{sec:anomalies}. In \cite{Ashtekar:2020,Ashtekar:2021}, a particular choice of vacuum that leads to power suppression for infrared modes of the scalar power spectrum \cite{Ashtekar:2016} is shown to alleviate both the power suppression and lensing anomalies. The first is said to be alleviated since the estimator $S_{1/2}$ is lower in this model than in $\Lambda$CDM. However, we argue that such a conclusion requires the investigation of the p-values of the observation with respect to this model. Furthermore, the lensing anomaly is alleviated by shifting predictions of other parameters enough so that the inconsistencies quantified by the lensing parameter do not present in this model as strongly as in $\Lambda$CDM. On the other hand, the mechanism for alleviation of anomalies in \cite{Agullo:2020b,Agullo2021} is different. In those references, the authors consider several power spectra with power enhancement of different slopes. Various models, including LQC, may serve as motivations for these phenomena. With these power spectra, non-Gaussianities become important. These correlate the largest observable modes and super-Hubble ones. This correlation, though not directly observable, increases the variance of the distribution of perturbations. Thus, certain features, like the anomalies reviewed in chapter \ref{sec:anomalies}, are more likely to occur in this scenario.

In this chapter, we consider LQC but we depart from previous analyses in two ways, both related to the ambiguities present in the construction of the cosmological model at hand. First, we adopt the equations of motion derived from hybrid LQC \cite{CastelloGomar2014,CastelloGomar2015,hyb-vs-dress,Navascues:2021}, which we introduced in \ref{sec:LQCperts}, mainly motivated by the fact that so far no Bayesian analysis comparing predictions with observations has been conducted adopting such prescription. One of its advantages is that the equations of motion for the perturbations are all hyperbolic at the bounce, unlike those of the dressed metric approach \cite{hyb-vs-dress,Iteanu2022}. The other main distinction from previous studies is the choice of initial conditions for perturbations, that we choose to be the Non-Oscillatory (NO) vacuum \cite{nonosc,hyb-obs,nonosc_BeaDaniGuillermo}, introduced in the previous chapter. It minimizes oscillations in the primordial power spectrum, and leads to a power spectrum that is the near-scale-invariant one of $\Lambda$CDM in the ultraviolet, with exponential power suppression in the infrared and some small oscillations in intermediate scales. We have shown that its power spectrum is a particular case of the power spectra obtained from SLEs. It is obtained when the support of the smearing function ignores the Planckian regime. Besides having been motivated as a suitable vacuum in LQC in its own right, it provides the simplest form of the scalar power spectra that we obtain with SLEs within LQC. This means that we may easily parametrize the power spectrum and simplify numerical calculations.

The scale at which these features occur in the primordial power spectrum depends not only on the choice of initial vacua but also on initial conditions of the background at the bounce (see appendix \ref{sec:app_background}). Here, we leave this scale as a free parameter in a first instance. Taking the primordial power spectrum as a starting point, we compute predictions for the angular power spectrum described in chapter $\ref{sec:anomalies}$ and compare with observations. We perform a Bayesian analysis of the model, from which we are able to show that the data prefers some of the effects of LQC to be within the observable range. Guided by this analysis, we are able to fix concrete initial conditions and investigate their effect on the aforementioned anomalies.

The main goal is to investigate the observational consequences of this particular model with this choice of vacuum. On the other hand, this is also relevant for other vacua within LQC which lead to power suppression of infrared modes in the primordial power spectrum \cite{nonosc,Ashtekar:2020,Agullo:2020b,SLEs_ours,ElizagaNavascues2021}. Thus, we contribute to the goal of understanding whether there are some robust features from LQC in predictions that transcend these ambiguities.

We recall that we have reviewed the three aforementioned anomalies in section \ref{sec:anomalies}. Section \ref{sec:results in LQC} is dedicated to the statistical analysis of the model that we are considering within LQC. We present the Bayesian analysis of the model with all parameters free, and then fix initial conditions to investigate possible alleviation of anomalies in concrete cases. Finally, section \ref{sec:conclusions} is dedicated to concluding remarks. We have also included appendix \ref{sec:app_numerics} with some details of our calculations.

\section{Statistical analysis of LQCNO}\label{sec:results in LQC}

The concrete predictions of the primordial power spectrum depend on details of the procedure and, as we showed in the previous chapter, on the choice of vacuum. However, in LQC departures from near scale-invariance will occur always infrared of the characteristic scale of the bounce given by $k_{\text{LQC}}$, which is related to the Ricci scalar (or equivalently the energy density) at the bounce. The question then becomes whether these departures from the near-scale-invariant power spectrum of standard cosmology occur within the observable window. The value of $k_{\text{LQC}}$ also depends on the value of the inflaton field at the bounce, $\phi_B$. A larger $\phi_B$ will generate more e-folds of inflation, washing out the effects of the pre-inflationary dynamics on the power spectrum to more infrared scales, leading to lower $k_{\text{LQC}}$. Although some heuristic arguments may help fix $\phi_B$, we will leave it as a parameter of our model. In short, the pre-inflationary dynamics of LQC, or any bouncing model with a period of kinetic dominance prior to inflation, leads to a primordial power spectrum that agrees very well with the near-scale-invariant one of standard cosmology for $k > k_{\text{LQC}}$, and that departs from it for $k \leq k_{\text{LQC}}$. The value of $k_{\text{LQC}}$ depends on the initial conditions of the background, namely $\phi_B$.

\begin{figure}[t]
    \centering
    \begin{subfigure}[b]{0.48\textwidth}
    \centering
    \includegraphics[width=\textwidth]{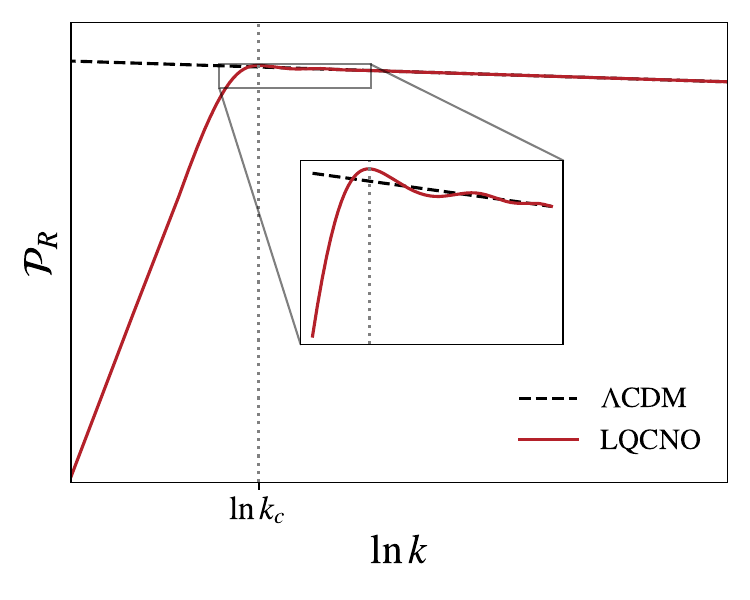}
    \caption{}
    \label{fig:PSNO_pic}
    \end{subfigure}
    \begin{subfigure}[b]{0.5\textwidth}
    \centering
    \includegraphics[width=\textwidth]{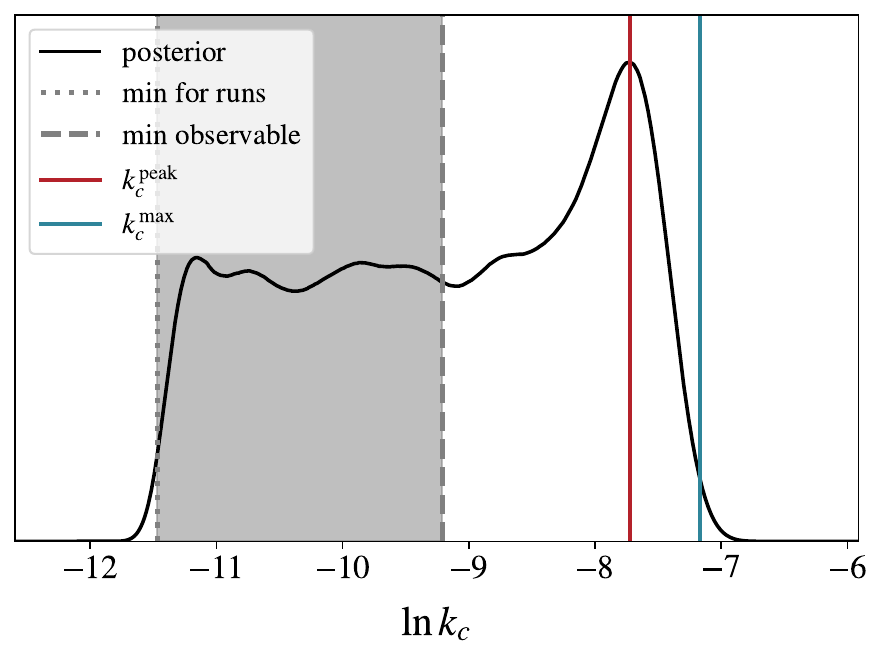}
    \caption{}
    \label{fig:posteriorkc}
    \end{subfigure}
    \caption[Typical primordial power spectrum of LQC with NO vacuum and Marginalized posterior probability o suppression scale.]{(a) Typical primordial power spectrum of scalar perturbations for the LQCNO model (solid red line), as parametrized by \eqref{eq:PSNO}, and the corresponding power spectrum for the $\Lambda$CDM model (dashed black line). (b) Marginalized posterior probability of the parameter $k_c$ of LQCNO resulting from a Bayesian analysis of the model with all its 7 parameters free, and integrating over the remaining 6. The grey region represents the portion of the explored parameter space that leads to a power spectrum where the modifications due to LQCNO are not observable.}
\end{figure}

In this chapter, we will carry out a Bayesian analysis comparing Planck cosmological data with the physical predictions corresponding to LQC when choosing the NO vacuum introduced in the previous chapter. We will refer to this model as LQCNO. Such an analysis, not done so far to the best of our knowledge, is essential to quantify how well the primordial power spectrum corresponding to the NO vacuum agrees with observations. Its power spectrum is exponentially suppressed for $k$ below a certain scale $k_c$, with some minimal oscillations for $k \gtrsim k_c$, as shown in figure \ref{fig:PSNO_pic}. Note that this means that $k_{\text{LQC}}$ is somewhere ultraviolet of $k_c$, though the main modifications to the power spectrum occur infrared of $k_c$, via the power suppression. The two scales are proportional through a factor that depends on the value at the bounce of both the inflaton and the energy density (see the examples in table \ref{tab:kcphiBefolds}). We will consider this energy density scale to be fixed, as it was shown in \cite{nonosc_BeaDaniGuillermo} that the primordial power spectrum arising from the NO vacuum is almost independent of it. To simplify computations, we have parametrized this power spectrum with three free parameters, $k_c$, $A_s$ and $n_s$:
\begin{equation}\label{eq:PSNO}
    \mathcal{P}_{\mathcal{R}}(k) = f(k,k_c)\ \mathcal{P}_{\mathcal{R}}^{\Lambda\textrm{CDM}}(k),
\end{equation}
where
\begin{equation}
    \mathcal{P}_{\mathcal{R}}^{\Lambda\textrm{CDM}}(k) = A_s \left( \frac{k}{k_{\star}} \right)^{n_s-1}
\end{equation}
is the near-scale-invariant power spectrum of the $\Lambda$CDM model, and $f(k,k_c)$ parametrizes the departure from it for LQCNO as described in appendix \ref{sec:app_param} --- see equations \eqref{eq:fkk} and \eqref{eq:fhk}. Then the parameter $k_c$ encodes the freedom particular to the LQCNO model, which relates to the freedom in the choice of $\phi_B$, or equivalently the number of e-folds of inflation. For a more intuitive picture, throughout this work we will cast $k_c$ values into the corresponding (approximate) number of e-folds of inflation in models with quadratic and Starobinsky inflaton potentials given by \eqref{eq:quadV} and \eqref{eq:starV}, respectively. Approximate numerical expressions to relate these quantities can be found in appendix \ref{sec:app_phiB_efolds} and \ref{sec:app_kcefolds}.

\subsection{Bayesian analysis}

Our model comprises of 7 free parameters: $k_c$, which encodes the freedom particular to LQCNO, and the 6 parameters of $\Lambda$CDM.\footnote{One could also consider the energy density scale of the bounce as an extra parameter. However, as explained before, the primordial power spectrum of the NO vacuum is insensitive to this scale, so we will not consider it as a free parameter in this case.} Of these, two refer to the parametrization of the primordial power spectrum, as mentioned in the previous section: $A_s$ and $n_s$. The remaining parameters are the baryonic and cold matter densities $\Omega_b h^2$ and $\Omega_c h^2$, the angular scale of acoustic oscillations $100 \theta_{\textrm{MC}}$, and the optical depth at reioinization $\tau_{\textrm{reio}}$. These four are relevant to characterize the post-inflationary Universe and model the propagation of perturbations from the surface of last scattering until today.

To compute predictions that can be compared with CMB observations (i.e. with the angular power spectra described by $C_{\ell}$ introduced in chapter \ref{sec:anomalies}), we have used the publicly available Boltzmann code \textit{CLASS} \cite{Blas_2011}, to which we have given externally the primordial power spectrum parametrized above. With this strategy there is no need for modifications of the code to accommodate LQC, as it is relevant only inasmuch as the primordial power spectrum is affected. The post-inflationary processes will not be modelled any differently with respect to standard cosmology and thus \textit{CLASS} can be used directly to simulate them. This will allow us to obtain the angular power spectrum $C_{\ell}$ for a given point in parameter space. Furthermore, we have resorted to \textit{MontePython} \cite{Brinckmann2018,Audren2012}, a sampler which is interfaced with \textit{CLASS}, to apply the Markov-Chain-Monte-Carlo (MCMC) method that we have used to explore the parameter space and perform the Bayesian analysis of this work. For these computations we have adopted the convergence criterion $R-1 \leq 0.01$ (for more details on this procedure see chapter \ref{sec:intro_Bayes}). The analysis has been performed using the CMB \textit{Planck} 2018 data with lensing \cite{PlanckV,PlanckVIII}.

We will start by analysing the freedom in $k_c$. To do so, let us first clarify its physical meaning. As explained previously, this scale is closely related with $k_{\text{LQC}}$. Concretely, they are monotonously related and it is always true by construction that $k_{\text{LQC}} \geq k_c$. Then, if $k_c$ is sufficiently smaller than the minimum observable wavenumber, the signatures from LQCNO will not be visible, as the observed portion of the power spectrum will perfectly agree with the standard one. Conversely, if $k_c$ is larger than the minimum observed value, the corresponding power spectrum will be such that the power suppression with respect to near scale-invariance is within the observable window.

With this in mind, let us consider the marginalized posterior probability for $k_c$ shown in figure \ref{fig:posteriorkc}, obtained by analysing the LQCNO model with all 7 parameters free, and integrating over the other 6. The cut-off of the posterior for low $k_c$ is not a reflection of a preference in the data, but rather of the minimum allowed for the runs. Then, the portion of the posterior probability corresponding to $k_c$ lower than the minimum observed is roughly a plateau. This is expected, since for those values of the parameter the theoretical prediction within the observed region is always the same, namely that of $\Lambda$CDM, modulo some small oscillations that may still be within the observable window while $k_c$ is large enough. On the other hand, for $k_c$ larger than the minimum observed wavenumber, the posterior probability displays a maximum followed by a very sharp cut-off. Let us now note that this cut-off is not a reflection of the maximum allowed for the runs, which was set at a much larger value. In this instance it does represent an actual preference in the data for $k_c$ to be below a certain scale. Nevertheless, there is a clear preference for some of the effects of LQCNO to be within the observable window, as the maximum likelihood corresponds to a value of $k_c$ that is observable.\footnote{Usually, one would draw this conclusion from the best-fit instead. The best-fit value for $k_c$ in the 7-dimensional parameter space corresponds to a $k_c$ no longer within the observable window. However, this needs to be interpreted in the context of the corresponding 1-$\sigma$ interval. This depends on the cut-off for low $k_c$ that is artificially imposed, and thus has no physical meaning. We could not have left this parameter with no minimum as the goodness of fit of the model for any $k_c$ lower than the observed minimum is exactly the same. This would lead to an ill-defined posterior ``probability'' distribution for $\ln(k_c)$, with an infinitely large plateau. For this reason, we focus on the conclusions we can draw from the marginalized posterior probability.} We will denote this as $k_c^{\textrm{peak}}$ in the following.

This result agrees qualitatively with that of \cite{Contaldi:2003zv}, where a similar parametrization is used for the power spectrum and that served as inspiration for the parametrization used here. The difference lies essentially in the form of the suppression at infrared scales and in the amplitude of oscillations. Additionally, the motivation for the two cases are different, as in \cite{Contaldi:2003zv} the power spectrum arises from a classical model with a period of kinetic dominance followed by a de Sitter branch, with only two free parameters: $n_s$ and a scale related to our~$k_c$. 

In summary, the data prefers a $k_c$ that is observable over one that is not, but it very strongly constrains it. In other words, the marginalized posterior probability of $k_c$ indicates that some departure from $\Lambda$CDM is preferred. This is further supported by the fact that both the minimum and average of the $\chi^2$ statistic are slightly lower for the LQCNO model with $k_c$ fixed at $k_c^{\rm peak}$ than that for $\Lambda$CDM (with $\Delta {\rm min}\chi^2 \simeq 1.7$, and $\Delta {\rm mean}\chi^2 \simeq 0.5$). This improvement is also discussed in the analyses of references \cite{Agullo:2020b,Agullo2021}. However, references \cite{Ashtekar:2020,Ashtekar:2021} do not report this value.

In the rest of this analysis, we will consider two separate models. The first is LQCNO when fixing $k_c$ to $k_c^{\rm peak}$, the one preferred by observations. The second is LQCNO when $k_c$ is fixed to $k_c^{\rm max}$, which we consider as the maximum possible $k_c$, considering agreement with observations. It is useful to consider such a scenario as it illustrates more appreciably the effect of the power suppression. The concrete values of these scales, as well as the corresponding values of $\phi_B$ and the number of e-folds of inflation $N$ are given in table \ref{tab:kcphiBefolds} for quadratic and Starobinsky inflation. For comparison, table \ref{tab:kcphiBefolds} also displays the corresponding value of the characteristic scale of LQC: $k_{\text{LQC}}\equiv a_\text{B}\sqrt{R_\text{B}/6}$, where $R$ is the scalar curvature and the subscript B denotes evaluation at the bounce.\footnote{In appendix \ref{sec:app_background} we give details on how to translate wave numbers $k$ expressed in natural units to ${\rm Mpc}^{-1}$.}

\begin{table}[t]
    \centering
    \begin{tabular}{c|c|c|c|c|c|c}
         {} & \multicolumn{3}{c|}{quadratic} & \multicolumn{3}{c}{Starobinsky}\\
         $k_c ({\rm Mpc}^{-1})$ & $\phi_B$ & $N$ & $k_{\text LQC} ({\rm Mpc}^{-1})$ & $\phi_B$ & $N$ & $k_{\text LQC}({\rm Mpc}^{-1})$\\
         \hline
         $k_c^{\rm peak} = 4.44\times 10^{-4}$ & $0.940$ & $64.6$ & $0.635$ & $-1.460$ & $61.4$ & $1.12$ \\
         $k_c^{\rm max} = 7.70\times 10^{-4}$ & $0.925$ & $64.1$ & $2.03$ & $-1.462$ & $60.8$ & $3.41$
    \end{tabular}
    \caption[Values of $k_c$ used and corresponding $\phi_B$, $N$ and $k_{\text{LQC}}$ for quadratic and Starobinsky inflation.]{Values of the parameter $k_c$ used in this chapter cast into the corresponding values of $\phi_B$ and number of e-folds of inflation for the quadratic and Starobinsky inflation potentials. We also include the corresponding values of $k_{\text{LQC}}$. More details on these calculations can be found in appendix \ref{sec:app_numerics}.}
    \label{tab:kcphiBefolds}
\end{table}

\begin{figure}[p]
    \centering
    \includegraphics[width=\textwidth]{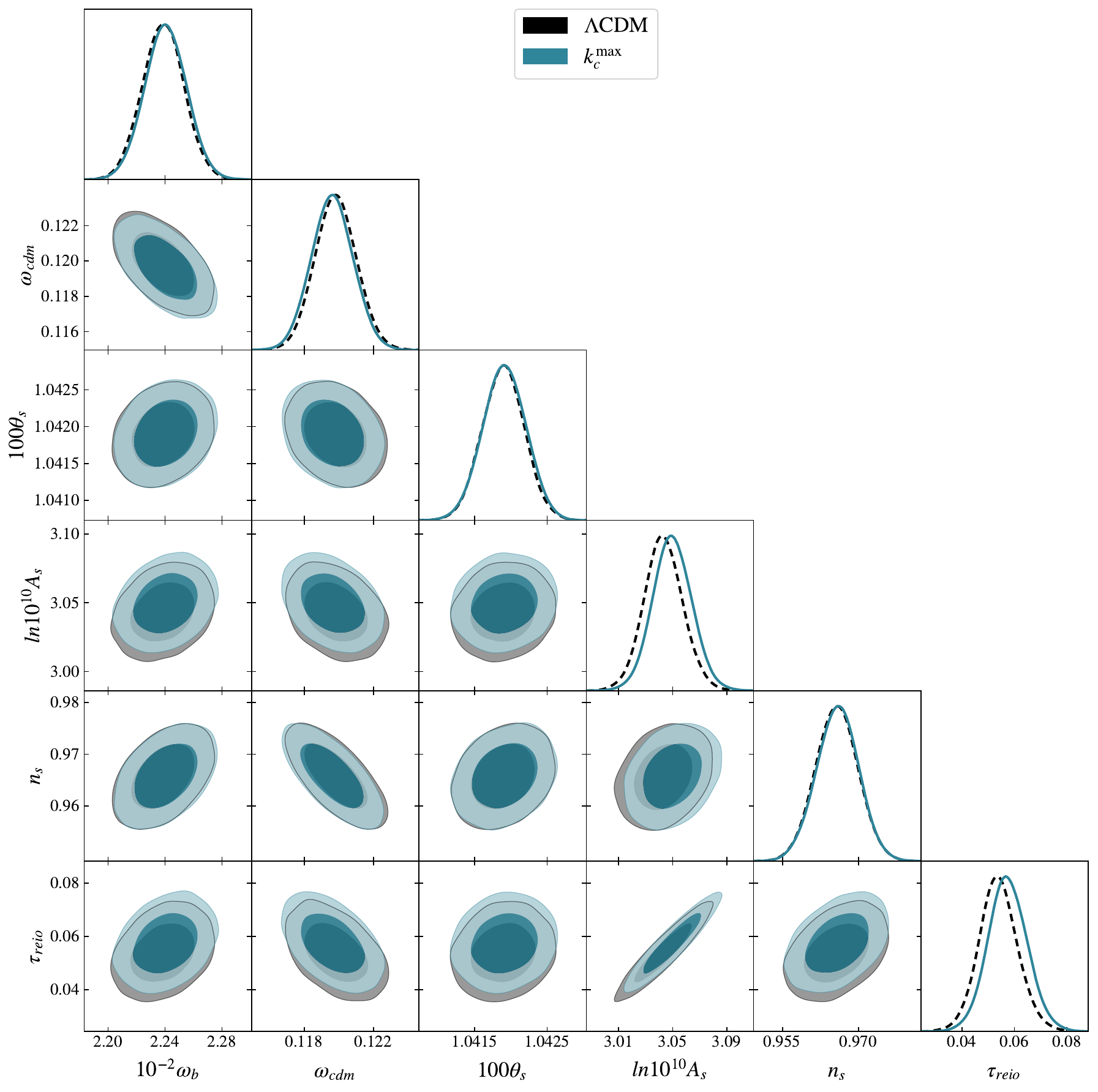}
    \caption[1 and 2-$\sigma$ 2D contours for the 6 $\Lambda$CDM parameters for $\Lambda$CDM and LQCNO with $k_c^{\rm max}$.]{1 and 2-$\sigma$ 2D contours for the 6 $\Lambda$CDM parameters, plus 1D marginalized posteriors, for $\Lambda$CDM (black) and LQCNO with $k_c^{\rm max}$ (blue).}
    \label{fig:triangle}
\end{figure}

Let us now compare the posterior probabilities of the 6 $\Lambda$CDM parameters in the $\Lambda$CDM and LQCNO models, as shown in figure \ref{fig:triangle}. For clarity, in these figures we do not present the case of LQCNO with $k_c^{\rm peak}$, as it sits between $\Lambda$CDM and $k_c^{\rm max}$. Most parameters seem to be mostly unaffected, except for $A_s$ and $\tau_{\textrm{reio}}$, as is evident from the 1-dimensional marginalized posterior distributions. The shifts in the contour plots of figure \ref{fig:triangle} are due to shifts in these two parameters. These are known to be correlated. A more opaque surface of last scattering corresponds to a lower optical depth $\tau$, by construction, and will result in perturbations reaching us with less power, and thus lower $A_s$. It seems natural then that the suppression of infrared modes of the LQCNO power spectrum will be compensated by higher power at the (already ultraviolet) pivot scale $A_s$. Consequently $\tau_{\textrm{reio}}$ also increases. Although this is the least constrained parameter of $\Lambda$CDM, this offers an opportunity for a falsifiable picture of LQCNO in the future, as independent measurements of $\tau_{\textrm{reio}}$ will help to constrain it further and therefore may allow to distinguish between LQCNO and $\Lambda$CDM \cite{Gupt2017}.

\subsection{Alleviation of anomalies}

To investigate the possible alleviation of the anomalies, in this section we will fix $k_c$ to the values of table \ref{tab:kcphiBefolds}: $k_c^{\rm peak}$ and $k_c^{\rm max}$. The goal is to show how much LQC may contribute to the alleviation of anomalies depending on the choice of this parameter.

\subsubsection{Power suppression}

For each value of $k_c$, fixing the remaining parameters to the best-fit values obtained from the Bayesian analysis, we compute the corresponding $C_\ell$ using \textit{CLASS}. From these we are able to compute $S_{1/2}$ through \eqref{eq:S12fromCl}. This is the expected value of $S_{1/2}$ for our model, shown in table \ref{tab:S12_pvalues}, which decreases substantially the higher the $k_c$. This is in agreement with what has been found in other scenarios within LQC \cite{Ashtekar:2021,Agullo2021}.

However, this alone is not enough to infer that the anomaly has been alleviated. Indeed, our model might be capable of lowering the expected value of this quantity, with respect to $\Lambda$CDM, but it might also affect the variance of its distribution such that the observed value still represents a very unlikely realization of a universe according to the model. In other words, it is necessary to compute the p-value of the observed $S_{1/2}$ in the context of our model. To do so, we need to take into account that the previously obtained $C_\ell$'s are random variables with a Gaussian distribution with a variance given by the cosmic variance. Note that $S_{1/2}$ is a sum of products of $C_{\ell}$'s, and therefore its distribution is not Gaussian. It is more straightforward to obtain it numerically, through a Monte Carlo method.

\begin{figure}[t]
    \centering
    \begin{subfigure}[b]{0.49\textwidth}
    \centering
    \includegraphics[width=\textwidth]{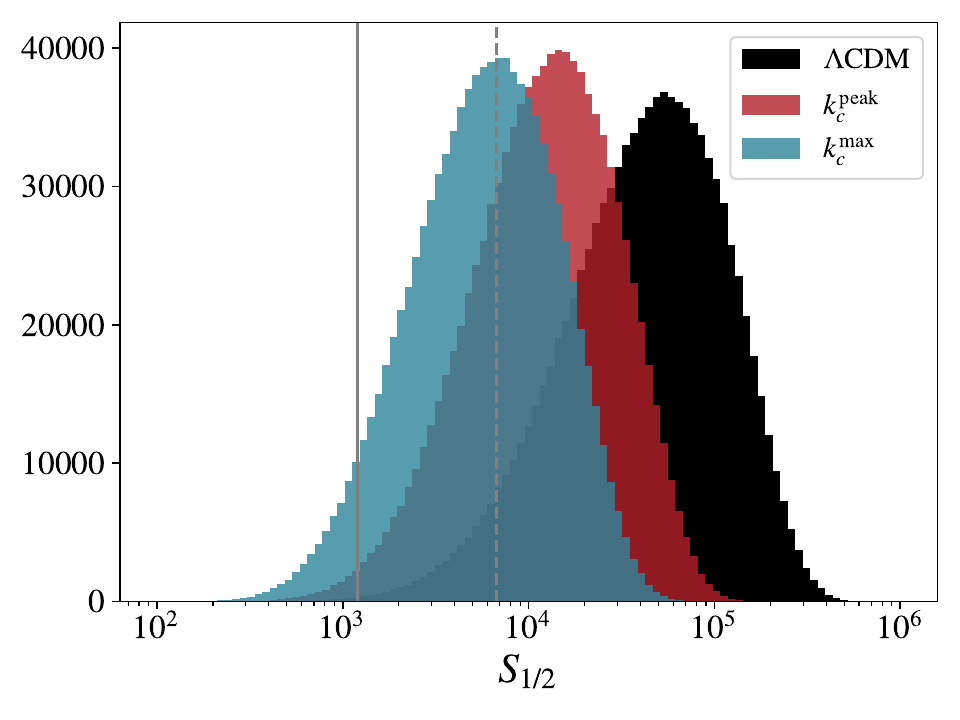}
    \caption{}
    \label{fig:S12}
    \end{subfigure}
    \begin{subfigure}[b]{0.49\textwidth}
    \centering
    \includegraphics[width=\textwidth]{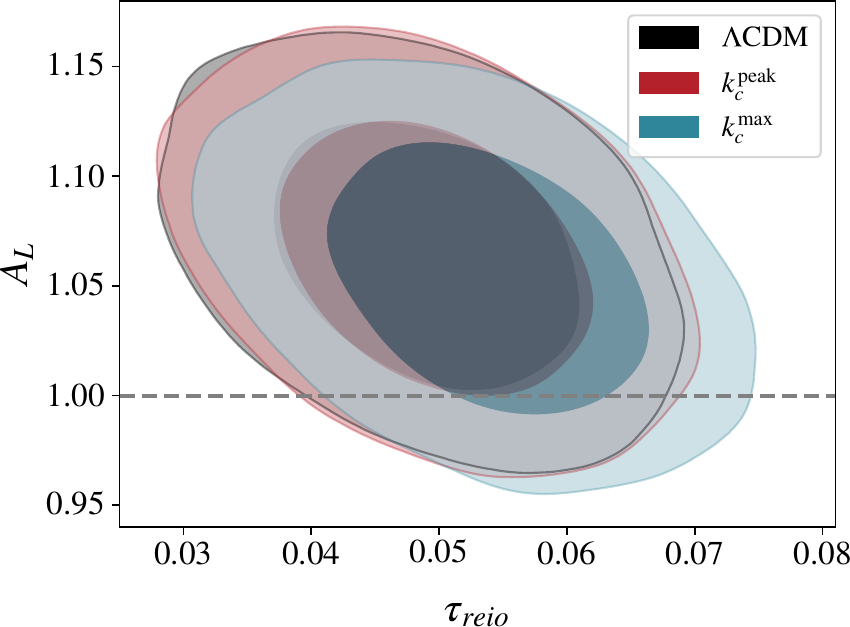}
    \caption{}
    \label{fig:AL}
    \end{subfigure}
    \caption[Distribution of $S_{1/2}$ and 1 and 2-$\sigma$ 2D contour for the parameters $A_L$ and $\tau_{\textrm{reio}}$, for $\Lambda$CDM , LQCNO with $k_c = k_c^{\text{peak}}$ and $k_c = k_c^{\text{max}}$.]{(a) Distribution of $S_{1/2}$ considering a Gaussian distribution of cosmic variance for $C_{\ell}$'s. Also represented are the observed values for cut-sky (solid grey line) and full-sky data (dashed grey line). (b) 1 and 2-$\sigma$ 2D contour for the parameters $A_L$ and $\tau_{\textrm{reio}}$, for $\Lambda$CDM (black), LQCNO with $k_c = k_c^{\text{peak}}$ (red) and $k_c = k_c^{\text{max}}$ (blue).}
\end{figure}

We have sampled the $C_{\ell}$ space randomly, computed $S_{1/2}$ for each point, and thus obtained the corresponding distributions of $S_{1/2}$, shown in figure \ref{fig:S12}. The p-value of the observed $S_{1/2}$ is simply the fraction of points that have resulted in $S_{1/2}$ at least as small as the observed one. As shown in table \ref{tab:S12_pvalues}, the p-value improves substantially for higher $k_c$, both for the cut-sky and full-sky observations. In this manner we are able to say that the LQCNO model is capable of alleviating this anomaly, as the observed values are more likely realizations of the LQCNO model than of $\Lambda$CDM.
\begin{table}[t]
    \centering
    \begin{tabular}{c|c|c|c}
     {} & {} & \multicolumn{2}{c}{p-value} \\
     model & $S_{1/2}$ & cut-sky & full-sky\\
     \hline
     $\Lambda$CDM & 35430 &$\sim 0.1 \%$ & $\sim 5 \%$ \\
     LQCNO $k_c^{\text{peak}}$ & 14557 & $\sim 2 \%$ & $\sim 26 \%$ \\
     LQCNO $k_c^{\text{max}}$ & 7799 & $\sim 5 \%$ & $\sim 55 \%$ 
    \end{tabular}
    \caption[Expected values of $S_{1/2}$ and p-values of observations for $\Lambda$CDM and LQCNO with different choices for $k_c$.]{Expected values of $S_{1/2}$ and corresponding p-values with respect to observations for cut-sky ($S_{1/2} \sim 1200$) and full-sky data ($S_{1/2} \sim 6700$) for $\Lambda$CDM and LQCNO with different choices for $k_c$. Note that we define the p-value as the probability of obtaining a realization with $S_{1/2}$ at least as small as the observed one, according to the model.}
    \label{tab:S12_pvalues}
\end{table}

\subsubsection{Lensing anomaly}

For this analysis, we have included the extra parameter $A_L$ in our model, and performed a Bayesian analysis with the 6 $\Lambda$CDM parameters as well as $A_L$ free. As mentioned, the best fit of almost all $\Lambda$CDM parameters is only sightly affected, except for $\tau_{\textrm{reio}}$ and $A_s$, which are correlated. Looking at the $A_L$ vs. $\tau$ contour plot of the posterior probability in figure \ref{fig:AL}, we can see that it shifts to higher values of $\tau_{\textrm{reio}}$ and lower values of $A_L$ as $k_c$ increases. For an appreciable $k_c$ within the observational window, this shift has pushed the 1-sigma region to include $A_L = 1$. Evidently, the dynamics of LQC does not affect the lensing the CMB goes through before it reaches our telescopes. Instead what we can conclude is that the model no longer presents the tension that is quantified through $A_L$ at the same level as the $\Lambda$CDM model.

Again, this analysis will benefit from future observations. As they constrain further and independently $\tau_{\textrm{reio}}$, they will help constrain $k_c$, and effectively constrain how much the dynamics of LQC may affect predictions in the observable window and contribute to the alleviation of anomalies.

\subsubsection{Parity anomaly}

It has been shown in \cite{Agullo:2020b,Agullo2021} that LQC may be able to alleviate the parity anomaly, when the power spectrum is such that non-Gaussianities become important. In those works, the considered power spectra show a power law suppression for infrared modes, as is the case in the LQCNO model, as well as some power enhancement for intermediate scales. Non-Gaussianities introduce a coupling between (non-observable) super-horizon modes and the largest observable ones. This affects the two-point correlation function, such that the mean value of the perturbations may remain unaltered, but the variance increases. In this sense, the anomalies are alleviated, as observations are more likely realizations in this scenario. Nevertheless, the concrete power spectra considered in those works can also affect the mean value of the parity asymmetry statistic.

\begin{figure}[h]
    \centering
    \begin{subfigure}[b]{0.49\textwidth}
    \centering
    \includegraphics[width=\textwidth]{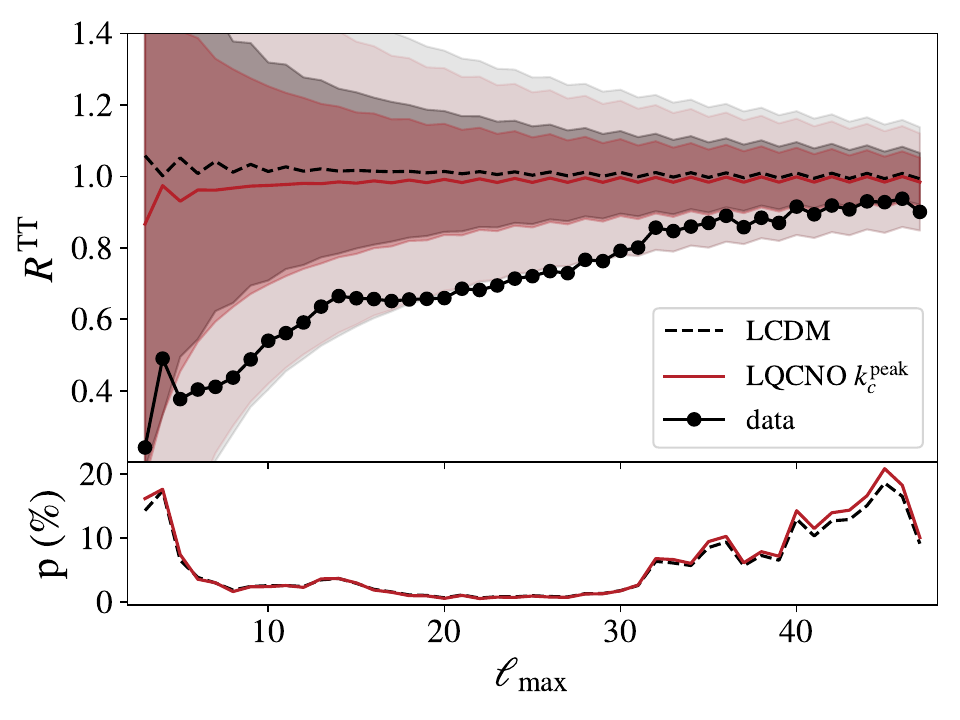}
    \caption{}
    \label{fig:RTT_pvalue}
    \end{subfigure}
    \begin{subfigure}[b]{0.49\textwidth}
    \centering
    \includegraphics[width=\textwidth]{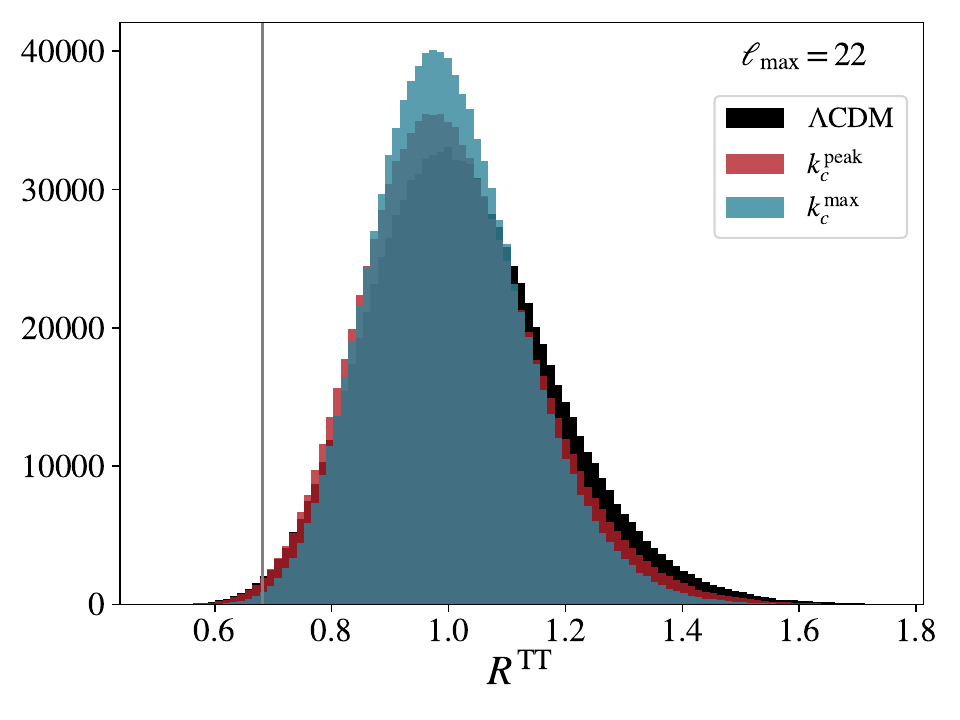}
    \caption{}
    \label{fig:RTTdist}
    \end{subfigure}
    \caption[Parity asymmetry statistic and p-values as function of the maximum multipole and distribution of the statistic for $\Lambda$CDM and LQCNO with $k_c^{\rm max}$ and $k_c^{\rm peak}$]{(a) Upper: parity asymmetry statistic $R^{\rm TT}$ as a function of the maximum multipole $\ell_{\rm max}$ for $\Lambda$CDM (dashed black line) and LQCNO with $k_c^{\rm peak}$ (solid red line) and the corresponding Planck data (black dots). Also represented are the 1 and 2-$\sigma$ regions corresponding to $\Lambda$CDM (grey regions) and LQCNO with $k_c^{\rm max}$ (red regions). Lower: p-value of the observed data with respect to the models. (b) Distribution of $R^{\rm TT}$ for $\ell_{\rm max} = 22$ (lowest p-value) considering a Gaussian distribution of cosmic variance for $C_{\ell}$'s. Also represented is the corresponding observed value (solid grey line).}
\end{figure}

In this work we would like to understand the role of the power suppression of infrared modes in the alleviation of this anomaly. As such, we have computed the parity asymmetry statistic for the LQCNO model with $k_c^{\rm peak}$ as well as $\Lambda$CDM as outlined in section \ref{sec:anomalies_parity}, shown in the upper plot of figure \ref{fig:RTT_pvalue}. For clarity, we do not represent this statistic for LQCNO with $k_c^{\rm max}$, as it sits close to these two. It is not always below that of $k_c^{\rm peak}$, for some $\ell_{\rm max}$ it is between that of $k_c^{\rm peak}$ and $\Lambda$CDM. In other words, there does not seem to be a clear monotonic behavior of this quantity with $k_c$, as we have seen with $S_{1/2}$, $A_L$ and $\tau_{\rm reio}$. In any case, the LQCNO model introduces a small power asymmetry. However, as is the case of the power suppression anomaly, this is not enough to conclude an alleviation of the anomaly. Indeed, the p-values (represented in the lower plot) indicate that it is not the case for most $\ell_{\rm max}$. To compute them, we consider again the $C_{\ell}$ to be Gaussian random variables with cosmic variance, and sample them through a Monte Carlo method, obtaining the corresponding distributions of $R^{\rm TT}$ for each $\ell_{\rm max}$. In figure \ref{fig:RTTdist} we represent these distributions for the case of $\ell_{\rm max} = 22$, which results in the lowest p-value for all three models. Remarkably, from this figure it is clear that actually the distribution of $R^{\rm TT}$ becomes thinner for higher values of $k_c$. The slight shift in the mean is not sufficient to increase the p-value of the observation, which is in the tail of the distribution, and instead it actually decreases it for the particular case of $\ell_{\rm max} = 22$ represented in the figure. This is of course just one case, but it illustrates how the p-values may be lower even when the expected value of the estimator is closer to the data. Evidently for some $\ell_{\max}$ this is not the case and the p-value increases slightly in the case of LQCNO. Overall, the anomaly is essentially as strong as in the standard model. 

\section{Conclusions and discussion}\label{sec:conclusions}

In this chapter, we have aimed to perform a rigorous statistical analysis of the LQCNO model, via a comparison with Planck CMB data. On the one hand, the goal was to find possible signatures from this model in the data. In this spirit, we have first performed a Bayesian analysis with all 7 parameters  of our model free (6 of them coinciding with those of $\Lambda$CDM plus an extra scale $k_c$ inherent to our model). This allowed us to conclude that the data shows a preference for some of the effects particular to LQCNO to be within the observable window. This is quantified in the marginalized posterior probability of the parameter $k_c$, which shows a peak in the observable window, and a sharp cut-off shortly after. It seems that the LQCNO model may adjust the data as well as $\Lambda$CDM, and even slightly better when we fix the free parameter such that LQC effects are observable. We have also found that the LQCNO model with fixed $k_c$ affects appreciably the best-fit of the optical depth at reionization $\tau_{\rm reio}$. This opens the possibility of constraining $k_c$ with future observations of $\tau_{\rm reio}$. A more rigorous comparison between LQCNO and $\Lambda$CDM would require the computation of the Bayesian evidence. Given that this calls for computationally demanding methods to thoroughly explore the parameter space, we believe it is a tool that is most useful when we are able to constrain $k_c$ further from the data, so that we may compare $\Lambda$CDM with a model of LQCNO with fixed initial conditions motivated by observations. 

Furthermore, we have investigated the effect on anomalies of two illustrative models of LQCNO: one with initial conditions corresponding to the peak of the marginalized posterior distribution of $k_c$, and one in the tail of this distribution within the observational window, as a limiting case. We find that, for both the power suppression and the lensing anomalies, the more the effects are within the observational window the more the anomalies are alleviated. We have obtained the distribution of the estimator that quantifies the power suppression anomaly and found that the p-value of the observation improves in the LQCNO models with respect to $\Lambda$CDM. In the case of the lensing anomaly, we found that the 2D posterior probability of the lensing amplitude $A_L$ and optical depth at reionization is shifted, such that the 1-$\sigma$ region is pushed closer to $A_L = 1$. On the other hand, for the parity anomaly we have found that the power suppression that the LQCNO model offers in the primordial power spectrum is not enough to alleviate this tension, as it introduces only a slight asymmetry in the expected value, and in some cases the p-value of the data actually decreases. Additionally, the effect on this asymmetry does not seem to depend on the scale $k_c$ as much as the estimators of the previous anomalies. In this sense, given previous works in the literature \cite{Agullo:2020b,Agullo2021}, we believe that power spectra for which non-Gaussianities are important may be necessary for the alleviation of this anomaly.

On the other hand, this work is part of a larger effort to find robust results that are consequence of LQC in general, regardless of the ambiguities. We have shown that in the context of hybrid LQC, and with the particular choice of the NO vacuum, there is a preference for the effects of LQC to be observable and some alleviation of anomalies is possible. However, a power suppression of infrared modes is common in primordial power spectra of LQC models \cite{nonosc,Ashtekar:2020,Agullo:2020b,SLEs_ours}. Any model that leads to a power spectrum with power suppression in the infrared will have the potential of alleviating these anomalies in the same way. More commonly, some power enhancement is also present for intermediate scales. The natural continuation of this work is to consider such power spectra and perform a similar analysis. We intend to do so for the SLEs with smearing function that does not ignore the Planckian regime investigated in the previous chapter. 

\chapter{SLE\MakeLowercase{s} in the Schwinger effect}\label{chap:Schwinger}


Many particle creation phenomena are an effect of the action of a classical, time-dependent external agent which excites a quantum matter field. This is the case, for instance, in cosmological pair production due to an evolving spacetime geometry \cite{parker1969,Ford1987,Weinberg2008}, and in fermionic pair production resulting from a strong electric field, called the Schwinger effect \cite{Sauter1931,Schwinger1951}.

In this context, it is commonly assumed that the external agents are strong enough so that backreaction of the quantum test fields can be neglected. Then a mean-field approximation is considered. See \cite{Pla2021,Pla2022} for a recent study on the validity of the semiclassical approach in the Schwinger effect. In this case, for strengths of the electric field below the critical Schwinger limit, which is of the order of $10^{18} \ \text{V/m}$, particle production is exponentially suppressed \cite{Schwinger1951,Yakimenko2018}. There is still no consensus on the possibility of reaching such critical strength with intense laser facilities, as this would produce fermionic pair cascades which could deplete the electric field \cite{Fedotov2010,Bulanov2010,Gonoskov2013}. Nevertheless, recent works prove that these obstacles might be overcome, making supercritical electric fields attainable in the future \cite{Ilderton2023,Vincenti2019}. For a recent review see \cite{Fedotov2023}. In addition, the Schwinger effect has recently been observed in an analogue mesoscopic experiment in graphene \cite{Schmitt2023}.

Pair production effects are often explained in the context of quantum field theory in curved spacetimes. The Hamiltonians describing these systems depend on time, and thus when canonically quantizing each classical theory one encounters the possibility of constructing infinitely many different quantum theories. Equivalently, there is freedom in the choice of the annihilation and creation operators, and therefore of the quantum vacuum. This is the ambiguity in the choice of vacuum that we have explored within the context of LQC in the previous chapters. One might then compare two different choices and find that one particular vacuum is excited with respect to another choice of vacuum: particles have been produced.

Different quantum theories predict different expectation values for physical observables. For example, the work in the previous chapters has shown that fixing the vacuum of cosmological perturbations to be the Bunch-Davies state at the onset of inflation results in different predictions than fixing the vacuum to be, e.g. a SLE in LQC. Generally, different choices of vacua provide different power spectra. Since this is a quantity that is directly related with observations, it is considered the relevant magnitude in cosmology. On the other hand, works about the Schwinger effect usually study the number of created particles as the magnitude of physical relevance. As a consequence of the ambiguities, in the literature there is still an open discussion about the physical interpretation of the time evolution of the number of created particles \cite{Alvarez2023,Ilderton2022,Yamada2021}. Here, we argue that we need both the power spectrum and the particle number to completely characterize our choice of vacuum. This is why we introduce the notion of power spectrum in the Schwinger effect. In addition, the study of these magnitudes will allow us to learn about the anisotropies introduced by the electric field. Concretely, we find that anisotropies do not play an important role in the power spectrum neither in the infrared nor in the ultraviolet, whereas at intermediate scale multipoles do contribute. The same applies to the particle number.

As in non-standard cosmology, in the context of the Schwinger effect, many vacua have been proposed, depending on the physical properties that one wants to imprint on the quantum theory. In this chapter, we study SLEs within the context of the Schwinger effect. In particular, we investigate here the role of the support of the smearing function that defines each SLEs, and find asymptotic regimes for sufficiently small and large supports, thus providing physical interpretation to the ambiguities in the choice of this function. This will help to understand the behavior that we have found in chapter \ref{chap:SLEsinLQC}: SLEs are independent of the support of the smearing function in LQC as long as it is wide enough around the bounce, but very sensitive to whether the moment of the bounce is included in the support.

As proven in \cite{Olbermann2007}, SLEs are of Hadamard type in FLRW spacetimes. This relates to the ultraviolet behaviour of the two-point function, and guarantees that computations such as that of the stress-energy tensor are well defined \cite{Fewster2013}. Along these lines, we will see in this chapter that the ultraviolet behaviour of SLEs in the Schwinger effect is compatible with the Hadamard condition, although this property remains to be rigorously proven. 

This work has led to the publication \cite{ours_Schwinger}. The structure of this chapter is as follows. In section \ref{sec:formalismSchwinger} we introduce the formalism that we are working on: a classical charged scalar field in flat spacetime coupled to an external and time-dependent electric field. We extend the construction of SLEs to general homogeneous settings, considering especially the Schwinger effect in section \ref{sec:settings}. In section \ref{sec:sopf} we investigate the dependence on the smearing function. Section \ref{sec:anisotropies} is dedicated to anisotropies. Here, we define the power spectrum in the Schwinger effect and investigate the multipolar contributions. Finally, in section \ref{sec:num} we study the number of created particles for different choices of the smearing function. Section \ref{sec:conclusions_schw} is dedicated to conclusions and closing remarks.

\section{The Schwinger effect}\label{sec:formalismSchwinger}

Let us consider the motion of a classical charged (complex) scalar field $\phi(t,\bfx)$ in flat spacetime coupled to an external spatially-homogeneous time-dependent electric field. The dynamics of this matter field is governed by the Klein-Gordon equation
\begin{equation} \label{eq:KG}
    \left[ (\partial_{\mu}+iqA_{\mu})(\partial^{\mu}+iqA^{\mu})+m^2 \right]\phi(t,\textbf{x})=0,
\end{equation}
where $q$ and $m$ are the charge and the mass of $\phi(t,\bfx)$, and $A_{\mu}(t,\bfx)$ is the electromagnetic potential. We use the temporal gauge $A_{\mu}(t)=(0,\textbf{A}(t))$, as this is the only choice which explicitly translates homogeneity to the equations of motion, and thus, to the quantum theory. Each complex Fourier mode
\begin{equation} \label{eq:Fourier}
    \phi_{\bfk}(t)=\frac{1}{(2\pi)^{3/2}}\int d^3\textbf{x} \ e^{-i\bfk\cdot\textbf{x}}\phi(t,\textbf{x})
\end{equation}
satisfies a harmonic oscillator equation 
\begin{equation} \label{eq:harmonic} 
    \ddot{\phi}_{\bfk}(t)+\omega_{\bfk}(t)^2\phi_{\bfk}(t)=0
\end{equation}
with time-dependent frequency
\begin{equation} \label{eq:omegaSchwinger}
    \omega_{\bfk}(t)=\sqrt{|\bfk+q\textbf{A}(t)|^2+m^2}.
\end{equation}
These equations are decoupled for different wavevectors $\bfk$. In addition, note that the real and imaginary parts of each mode ($\phi_{\bfk}^{\text{R}}(t)$ and $\phi_{\bfk}^{\text{I}}(t)$, respectively) also satisfy the same equation, so we can deal with them independently.

Matter fields coupled to other external time-dependent spatially homogeneous agents different from an electric field are also governed by harmonic oscillator equations of the type \eqref{eq:harmonic}, as we have previously seen for instance in the case of FLRW cosmologies. All the information on the external field is encoded in the time-dependent frequency.

\begin{figure}[t]
    \centering
    \begin{subfigure}[b]{0.49\textwidth}
    \centering
    \includegraphics[width=\textwidth]{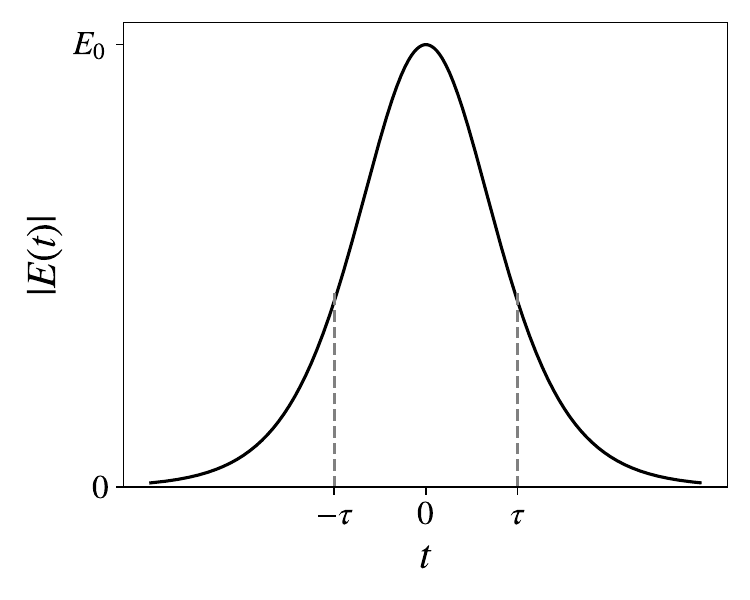}
    \caption{}
    \label{fig:elec}
    \end{subfigure}
    \begin{subfigure}[b]{0.49\textwidth}
    \centering
    \includegraphics[width=\textwidth]{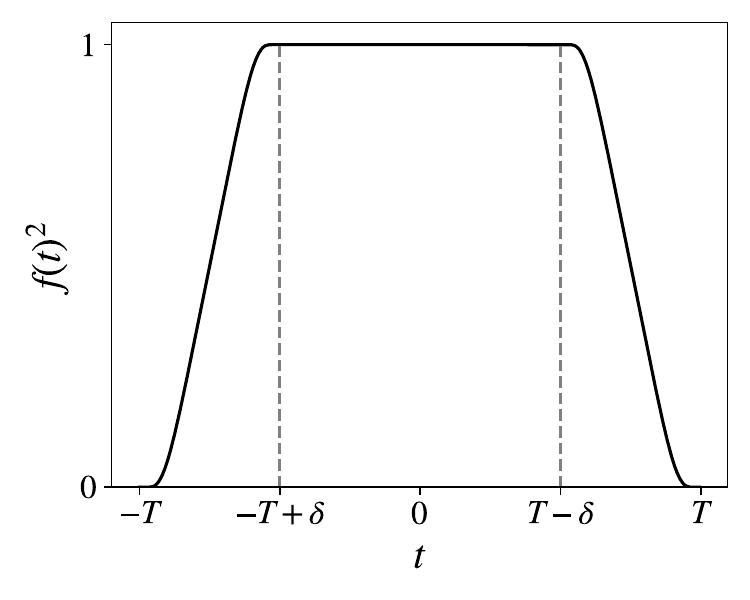}
    \caption{}
    \label{fig:testf}
    \end{subfigure}
    \caption[Illustrative plots of electric field and smearing function.]{Illustrative plots of (a) Sauter-type electric field of time width $\tau$ and maximum amplitude $E_0$ and (b) smearing function \eqref{eq:windowf} of compact support $[-T,T]$ and slope of length $\delta$.}
\end{figure}

In particular, we consider the so-called Sauter-type potential \cite{Sauter1931}, i.e., an electric field potential of the form
\begin{equation}
\label{eq:Sauter}
\textbf{A}(t)=E_0\tau\left[\tanh\left(t/{\tau}\right)+1 \right] \ \textbf{e}_z,
\end{equation}
that, without loss of generality, we have chosen oriented in the $\textbf{e}_z$ direction for simplicity. As it is shown in figure \ref{fig:elec}, it models a P{\"o}schl-Teller electric pulse of maximum amplitude~$E_0$ at time $t=0$ \cite{Poschl1933}. It vanishes asymptotically, and the characteristic width of the pulse is given by~$\tau$. 

In the following sections time will be always expressed in units of the time width~$\tau$ of the Sauter-type pulse, and frequencies in units of~$\tau^{-1}$. On the other hand, in our plots, we fix the value of the mass to $m=\tau^{-1}$ and the maximum amplitude of the electric field to $qE_0=\tau^{-2}$. This corresponds to the critical Schwinger limit $qE_0=m^2$ \cite{Yakimenko2018}. For lower strengths, the probability of pair production is exponentially suppressed. Only for electric fields of this order does the Schwinger effect become physically relevant. In practice, the qualitative behaviour of the system for stronger electric fields is the same, and provides no additional information that is relevant to this work. Furthermore, we are interested in studying the physical differences between choices of vacua. Considering larger intensities than the Schwinger limit makes these differences less clear. Finally, it is worth noting that the Schwinger limit is not attainable experimentally yet, although recent works in laser facilities are optimistic about achieving this in the near future \cite{Vincenti2019,Ilderton2023,Ilderton2013}.

We now wish to canonically quantize the matter field $\phi(t,\bfx)$, as we have done in chapter \ref{sec:introvacuum} for the cosmological perturbations. However in this case we are dealing with a complex field. First, we choose a basis of solutions $\{\varphi_{\bfk}(t)\}_{\bfk}$ to the dynamic equations \eqref{eq:harmonic}. Note that in general we could consider different basis for $\phi_{\bfk}^{\text{R}}(t)$ and $\phi_{\bfk}^{\text{I}}(t)$, but for simplicity we are fixing the same basis for both of them. In order to preserve the Poisson bracket structure, this basis has to be normalized with respect to the Klein-Gordon product:
\begin{equation} \label{eq:normalization}
    \varphi_{\bfk}(t)\dot{\varphi}^*_{\bfk}(t)-\varphi_{\bfk}^*(t)\dot{\varphi}_{\bfk}(t)=i.
\end{equation}
Then, we can define the quantum field operators $\hat{\phi}^{\alpha}_{\bfk}(t)$ ($\alpha=\text{R},\text{I}$) as the linear combination
\begin{equation}
\label{eq:hatphi}
    \hat{\phi}^{\alpha}_{\bfk}(t)=\hat{a}^{\alpha}_{\bfk}\varphi_{\bfk}(t)+\hat{a}^{\alpha\dagger}_{\bfk}\varphi_{\bfk}^*(t).
\end{equation}

Again we are faced with the ambiguity in the choice of vacuum, which translates to an ambiguity in the choice of solutions $\varphi_{\bfk}(t)$, which in turn translate to an ambiguity in the choice of initial conditions $(\varphi_{\bfk}(t_0),\dot\varphi_{\bfk}(t_0))$ at some initial instant $t_0$.

Given the potential \eqref{eq:Sauter}, in the asymptotic past the time-dependent frequency~\eqref{eq:omegaSchwinger} tends to $\omega_{\bfk}^{\text{in}}~=~\sqrt{k^2+m^2}$, where $k=|\bfk|$, whereas in the asymptotic future it tends to $\omega_{\bfk}^{\text{out}}~=~\sqrt{|\bfk+2qE_0\tau \textbf{e}_z|^2+m^2}$. This allows us to find an expression for the in-solution $\varphi_{\bfk}^{\text{in}}(t)$ to equation~\eqref{eq:harmonic}, which asymptotically behaves as a plane wave of frequency $\omega_{\bfk}^{\text{in}}$ for $t\rightarrow -\infty$. According to \cite{BeltranPalau2019}, it can be written in terms of hypergeometric functions \cite{Abramowitz1965}:
\begin{equation}
\label{eq:insolution}
    \varphi_{\bfk}^{\text{in}}(t)=\frac{1}{\sqrt{2\omega_{\bfk}^{\text{in}}}}e^{-i\omega_{\bfk}^{\text{in}}t}\left( 1+e^{2t/\tau} \right)^{(1-id)/2} {}_2F_1\left( \rho_{\bfk}^+,\rho_{\bfk}^-;1-i\tau\omega_{\bfk}^{\text{in}};-e^{-2t/\tau} \right), 
\end{equation}
where 
\begin{equation}
    d=\sqrt{(2qE_0\tau^2)^2-1}, \qquad \rho_{\bfk}^{\pm}=\frac{1}{2}\left[ 1-i\tau\left(\omega_{\bfk}^{\text{in}}\pm \omega_{\bfk}^{\text{out}}\right)-id \right], 
\end{equation}
We may also find the out-solution $\varphi_{\bfk}^{\text{out}}(t)$ to equation~\eqref{eq:harmonic}, which asymptotically behaves as a plane wave of frequency $\omega_{\bfk}^{\text{out}}$ for $t\rightarrow +\infty$. By comparing these two, one finds that the in-state is excited with respect to the out-state. In other words, the number of created particles is not zero. In fact, the number of created particles depends on the choice of vacuum, i.e. on the choice of initial conditions.

The in and out-solutions are natural choices only asymptotically. There is no unique natural vacuum at a finite time $t_0$ in this context. We can again parametrize the initial conditions without loss of generality as in \eqref{eq:CkDk}:
\begin{equation}\label{eq:WY}
    \varphi_{\bfk}(t_0) = \frac{1}{\sqrt{2 D_{\bfk}(t_0)}}, \qquad \dot\varphi_{\bfk}(t_0) = \sqrt{\frac{D_{\bfk}(t_0)}{2}}\left[C_{\bfk}(t_0)-i\right].
\end{equation}

One possibility is to find the vacuum which minimizes the energy density. However, as the Hamiltonian is time dependent, this criterion has ambiguities. One option is to choose a particular time $t_0$ and find the vacuum that instantaneously minimizes the expectation value of the energy density at $t_0$. This prescription defines the so-called instantaneous lowest-energy state (ILES) at $t_0$, given by \cite{Mukhanov2007}
\begin{equation} \label{eq:ILES}
    D_{\bfk}(t_0)=\omega_{\bfk}(t_0), \qquad  C_{\bfk}(t_0)=0.
\end{equation}
It is important to remark that the minimum of the energy density only exists if $\omega_{\bfk}(t_0)^2$ is positive, which is always true in the case of the Schwinger effect \eqref{eq:omegaSchwinger}. Note that at different times $t_0$ we have different notions of instantaneous lowest-energy states as long as the frequency $\omega_{\bfk}(t)$ depends on time. In addition, the ILES has another interesting property: it instantaneously diagonalizes the Hamiltonian in the basis of the vacuum and its excited states.

Another possibility is to minimize the energy density in a finite time interval, instead of at an exact instant of time. This is precisely the goal of the construction of SLEs that we have reviewed in \ref{sec:introSLE} for the case of FLRW models.

\section{SLE\MakeLowercase{s} in general homogeneous settings} \label{sec:settings}

In \ref{sec:introSLE} we have reviewed the procedure introduced in \cite{Olbermann2007} to define and compute the SLEs in FLRW cosmological backgrounds. In this section our aim is to extend this method to generic spatially homogeneous backgrounds, paying special attention to the Schwinger effect.

Here we propose a direct generalisation of Olbermann's procedure \cite{Olbermann2007} to systems characterized by modes $\phi_{\bfk}(t)$ satisfying harmonic oscillator equations with time-dependent frequencies $\omega_{\bfk}(t)$ (generic in this section). Let $f(t)$ be a smearing function of compact support $[t_1,t_2]$. As in the case of cosmological perturbations, it is still true here that each mode $\phi_{\bfk}(t)$ contributes to the total smeared energy as
\begin{equation} \label{eq:Ek}
    E(\phi_{\bfk})=\frac{1}{2}\int dt \ f(t)^2\left[ |\dot{\phi}_{\bfk}(t)|^2+\omega_{\bfk}(t)^2|\phi_{\bfk}(t)|^2 \right].
\end{equation}
This is equivalent to equation \eqref{energy}. Then we can use the same procedure of chapter \ref{sec:introSLE} to obtain the state $T_{\bfk}$ that minimizes this quantity: first we provide a fiducial solution $S_{\bfk}(t)$ to the equation of motion, then we perform the Bogoliubov transformation \eqref{eq:S_k to T_k} with Bogoliubov coefficients that are obtained from the fiducial solution through \eqref{eq:mu} and \eqref{eq:lambda}. We recall that the resulting state of this procedure is independent of the choice of fiducial solution.

For a state of minimal energy to exist in a generic background, the frequencies $\omega_{\bfk}(t)$ of the harmonic oscillator equations must be time-independent. Indeed, according to the Bogoliubov transformation \eqref{eq:S_k to T_k}, a solution $S_{\bfk}(t)$ is a SLE if and only if the coefficient $c(\bfk)$ given in \eqref{eq:c2} vanishes. Moreover, if we impose that $S_{\bfk}(t)$ is a SLE for all smearing functions, then $\dot{S}_{\bfk}(t)^2+\omega_{\bfk}(t)^2S_{\bfk}(t)^2=0$, which is compatible with the equation of motion if and only if the frequency is constant. Note that, in the Schwinger effect, the frequency~\eqref{eq:omegaSchwinger} is constant only when the electric field vanishes. In other words, a notion of state of minimal energy does not exist when we apply an electric field.

Finally, let us note that, although the procedure to obtain SLEs in general homogeneous settings is the same as in cosmological models, we cannot assume that the construction in this case leads to Hadamard states, as we do not have such an explicit proof. Nevertheless, through the numerical computations that we will show in this chapter, we obtain states with a behavior that is compatible with the Hadamard property up to the energy scales we have probed.

In the following sections we focus our analysis on the Schwinger effect. To compute the SLEs we could take the in-solution \eqref{eq:insolution} as the fiducial solution $S_{\bfk}(t)$, and thus we could write the SLEs in terms of integrals of hypergeometric functions. However, recall that the construction of the SLEs is independent of the fiducial solution chosen, and thus one may choose any convenient one. For example, for our numerical computations we took the numerical solution with lowest-order adiabatic initial conditions \cite{Birrell1982} at $t=0$.

\section{Role of the smearing function} \label{sec:sopf}

Each SLE minimizes the energy density smeared with a certain compact support function $f(t)$. We have seen in chapter \ref{chap:SLEsinLQC} that within LQC the support of this function is essentially irrelevant as long as it is wide enough and includes the high-curvature region. We are interested in further understanding the physical interpretation of choosing different supports for the smearing function, each defining a particular notion of SLE. As the Sauter potential \eqref{eq:Sauter} is symmetric around its maximum at $t=0$, it will be useful to consider smearing functions with compact support $[-T,T]$, where $T>0$. In particular, we are going to use the same smooth window functions as in chapter \ref{chap:SLEsinLQC}, now written in terms of time $t$. Concretely, as shown in figure \ref{fig:testf}, we take
\begin{equation}\label{eq:windowf}
    f^2(t) = 
\begin{cases}
	S\left(\frac{t+T}{\delta}\pi\right) & -T\leq t < -T+\delta,\\
	1 & -T+\delta\leq t \leq T-\delta,\\
	S\left(\frac{T-t}{\delta}\pi\right) &T-\delta < t \leq T.
\end{cases}
\end{equation}
We fix a small step width of $\delta=10^{-4}\tau$ for all the figures in this chapter. For supports smaller than this width (i.e., $T<\delta$), we adapt the parameter by setting $\delta = T/2$ so that it is still smooth.

For simplicity, we are choosing to maintain the shape of the test function, considering only the effects of changing its support. In principle its shape may also be relevant to the resulting SLE. However, for large enough supports the SLEs should be fairly insensitive to the form of the test function, as long as it is reasonably behaved, as is indeed corroborated in chapter \ref{chap:SLEsinLQC}. Furthermore, even when the form of the test function may be relevant, different shapes would simply translate to more or less weight being given to specific time periods when computing the smeared energy density. Therefore, we may understand the physics behind the consequences of different shapes by understanding the physical interpretation of the support first. Besides, one may also argue that more intricate shapes are less natural choices that would require additional motivation.

The freedom in the choice of vacuum is parameterized by $D_{\bfk}(t_0)$ and $C_{\bfk}(t_0)$, which define via \eqref{eq:WY} the initial conditions of the selected basis of solutions at time $t_0$. We are going to fix $t_0=0$, the instant at which the Sauter-type electric field reaches its maximum, and consider the smearing functions \eqref{eq:windowf}, varying $T$. In addition, in this section we focus on modes whose wavevectors $\bfk$ are parallel to the direction of the electric field. Anisotropies will be analysed in detail in the following sections.

\begin{figure}[t]
    \centering
    \begin{subfigure}{0.49\textwidth}
        \centering
        \includegraphics[width=\textwidth]{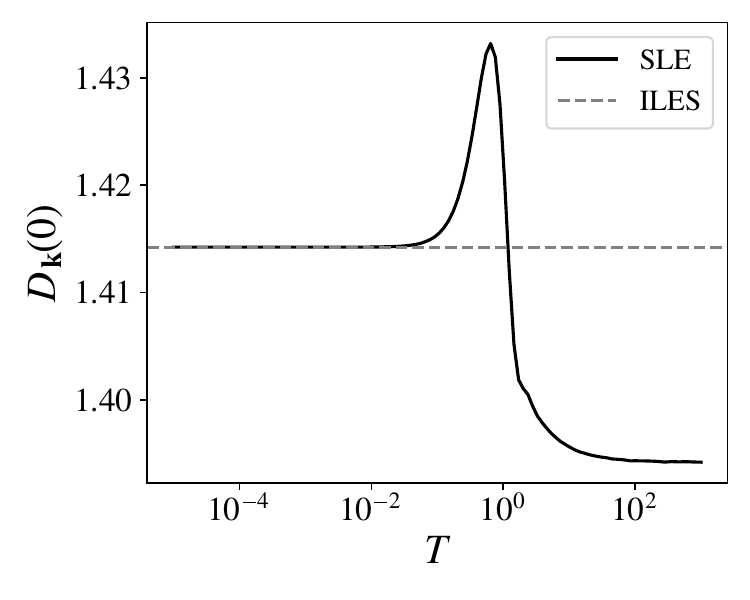}
        \caption{}
        \label{fig:WIR}
    \end{subfigure}
    \begin{subfigure}{0.49\textwidth}
        \centering
        \includegraphics[width=\textwidth]{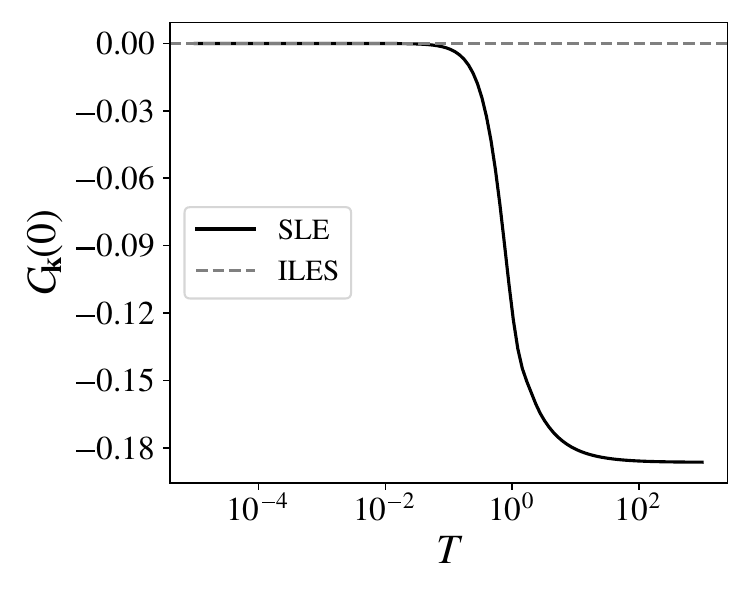}
        \caption{}
        \label{fig:YIR}
    \end{subfigure}
    \caption[Infrared behaviors of W and Y as functions of support.]{Dependence of (a) $W_{\bfk}(t_0=0)$ and (b) $Y_{\bfk}(t_0=0)$ defining the SLEs on the support $[-T,T]$ of the smearing functions \eqref{eq:windowf}. We show the infrared mode whose wavevector~$\bfk$ is parallel to the electric field and $k=10^{-5}\tau^{-1}$. 
 We use units $\tau=1$.}
    \label{fig:WYIR}
\end{figure}

In figure \ref{fig:WYIR} we show $D_{\bfk}(t_0)$ and $C_{\bfk}(t_0)$ for an infrared mode  with $k=10^{-5}\tau^{-1}$ as functions of the support of the smearing functions. We identify a transition regime around the time scale $\tau$, which is the characteristic length of the Sauter-type electric pulse, where the dependence on the support is not monotonic. It separates the behaviours of the SLEs for small and large supports. We have verified that this happens independently of the strength of the electric field.

When the support is small ($0<T\ll \tau$), the SLEs asymptotically approach the values \eqref{eq:ILES} that characterize the ILES at $t_0$. The physical justification of this fact resides in the definition of the ILES at $t_0$, which minimizes the instantaneous energy density obtained identifying the smearing function $f(t)^2$ with the Dirac delta $\delta(t-t_0)$ in \eqref{eq:Ek}. We might then say that the ILES at $t_0$ is the limit for small supports around $t_0$ of the SLEs. However, note that this limit is singular in the sense that the Dirac delta is not a smooth compact-support function, so ILESs are not a particular example of a SLE. These conclusions are also valid for other times different from $t_0$, as we have verified numerically.

On the other hand, we also find an asymptotic constant behaviour for large supports ($T \gg \tau$). This is consistent with the fact that the leading contributions to the smeared energy density are for times in the interval $[-\tau,\tau]$, and that the electric pulse decreases asymptotically. This limit defines a precise vacuum with a well-defined interpretation: the state which minimizes the energy density when it is smeared over the entire pulse. 

For other values of $k$ we also distinguish analogue behaviours of $D_{\bfk}(t_0)$ and $C_{\bfk}(t_0)$ for small and large supports. However, as we increase $k$, the dependence on the support decreases. Indeed, the limit $k\rightarrow \infty$ corresponds to local flat spacetime with no electric field, thus all vacua tend to the Minkowski vacuum defined in \eqref{eq:planewave}. Nevertheless, how fast or slow we reach the Minkowski vacuum strongly depends on each particular vacuum. We will analyse this in more detail in section \ref{sec:num} when studying the number of created particles.

Finally, one might wonder why the noticeable dependence of the SLEs on the support of the test function for supports of the order of the characteristic length of the electric pulse seems absent in the case of LQC. Indeed, in chapter \ref{chap:SLEsinLQC} we have seen that SLEs are independent of the support as long as it is large enough, which agrees with the large support convergent behaviour we observe. Let us then clarify that, although in LQC the equivalent to our potential is different, it also has a characteristic time scale (around the bounce), in which the variations of the potential are most important. This scale plays the same role as $\tau$, and there it should be less than a hundredth of a Planck second.\footnote{If we approximate the time-dependent mass in the equation of motion of cosmological perturbations in LQC \eqref{eq:s(eta)LQC} by a P{\"o}schl-Teller potential, the equivalent to $\tau$ is found approximately as the time after the bounce at which the potential reduces to half its maximum.} We find that the supports considered in chapter \ref{chap:SLEsinLQC}, though natural choices there, were already quite larger than this scale. Thus, the dependence of the SLE on them was minimal, and achieved convergence quickly. In general, the behaviour of SLEs in LQC most likely displays an intermediate regime as we observe in the Schwinger effect, though it corresponds to very small supports around the bounce, which are not physically interesting within the context of cosmology.

\section{Anisotropic power spectrum}\label{sec:anisotropies}

In this section we consider the extension of the common notion of power spectrum in cosmological scenarios to the Schwinger effect. In addition, we are interested in studying in detail the anisotropies present in this electric background. Motivated by the works in anisotropic cosmologies as Bianchi I \cite{Agullo:2020iqv}, we introduce an expansion of the power spectrum in Legendre polynomials and analyse its multipolar contributions.

Since we are working with a complex field, now the Wightman function is defined as 
\begin{equation}
    \label{eq:Wightman} W(t,\bfx;t',\bfx')=\bra{0}\hat{\phi}^{\dagger}(t,\bfx)\hat{\phi}(t',\bfx')\ket{0}=2\int \frac{d^3\bfk}{(2\pi)^3} \ e^{i\bfk\cdot (\bfx-\bfx')} \varphi_{\bfk}(t)\varphi^*_{\bfk}(t'),
\end{equation}
where we followed the notation of section \ref{sec:settings}. In the last equality we used the definition of the quantum field operator $\hat{\phi}(t,\bfx)$ in terms of the chosen basis $\varphi_{\bfk}(t)$ given by \eqref{eq:hatphi}. Note that, as in cosmology, the Wightman function only depends on the position vectors through the difference $\bfx-\bfx'$ because the electric field is spatially homogeneous. Taking the limit of coincidence $t\rightarrow t'$ yields
\begin{equation}
    \lim_{t\rightarrow t'} W(t,\bfx;t',\bfx')=\frac{1}{2\pi}\int \frac{d^3\bfk}{k^3}\ e^{i\bfk\cdot (\bfx-\bfx')} \mathcal{P}(t,\bfk),
\end{equation}
where we defined the power spectrum as
\begin{equation}
\label{eq:PS_Sch}
    \mathcal{P}(t,\bfk)=\frac{k^3}{2\pi^2}|\varphi_{\bfk}(t)|^2.
\end{equation}
Note that now this quantity depends on both $k$ and the polar coordinate $\theta$, as we take the electric field to be applied in the $z$ direction.

The power spectrum \eqref{eq:PS_Sch} depends on the solutions $\varphi_{\bfk}(t)$ that we choose to construct the quantum theory. More concretely, at the precise time $t_0$ it knows about the ambiguities in the selection of $D_{\bfk}(t_0)$ in \eqref{eq:WY}, although it is oblivious to $C_{\bfk}(t_0)$. As we are going to see in section \ref{sec:num}, the number of created particles does depend on both $W_{\bfk}(t_0)$ and $Y_{\bfk}(t_0)$. Thus, the power spectrum at a fixed time does not encode all the information about the vacuum. Furthermore, compared with the spectrum of $D_{\bfk}(t_0)$, the infrared power spectrum blurs the differences between different vacua as a consequence of the factor of $k^3$ in its definition~\eqref{eq:PS_Sch}.

\begin{figure}
    \centering
    \begin{subfigure}{0.49\textwidth}
        \includegraphics[width=\textwidth]{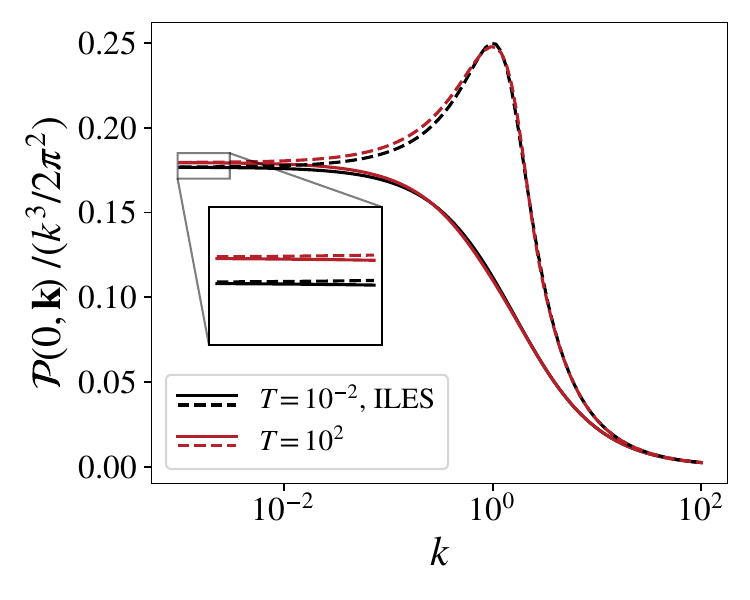}
        \caption{}
        \label{fig:PS_schw}    
    \end{subfigure}
    \begin{subfigure}{0.49\textwidth}
        \includegraphics[width=\textwidth]{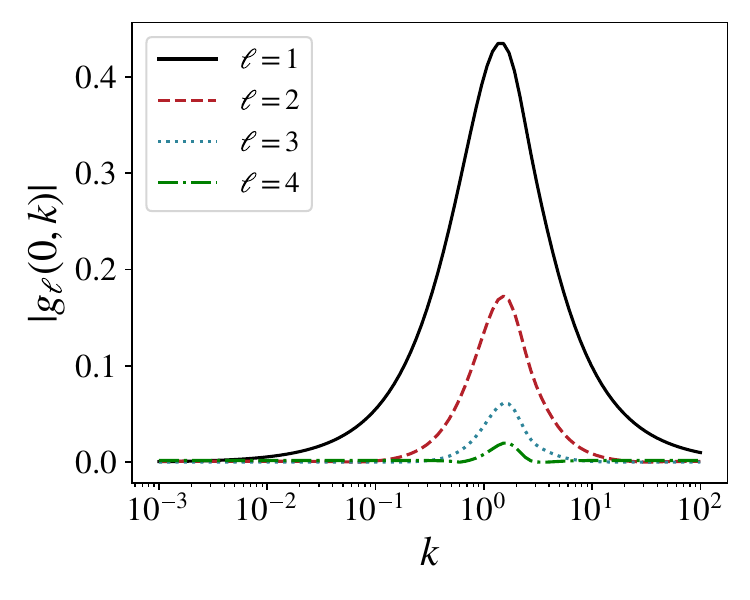}
        \caption{}
        \label{fig:gl}    
    \end{subfigure}
    \caption[Power spectrum and multipole contribution of modes parallel and anti parallel to the electric field.]{(a) Power spectrum divided by $k^3/(2\pi^2)$ at $t= 0$ for a mode parallel (solid) and antiparallel (dashed) to the electric field and for SLEs with different supports. The power spectra for ILES coincide with those for the SLE of smallest support. (b) Absolute value of the contributions $g_{\ell}$ of the multipoles $\ell$ with respect to the monopole at $t = 0$ for a SLE with support $T=10^2\tau$. Note that $g_{\ell}$ are negative for odd values of $\ell$ and positive for even~$\ell$. We use units $\tau=1$.}
    \label{fig:PSgl}
\end{figure}

We show in figure \ref{fig:PS_schw} the power spectrum $\mathcal{P}(t_0,\bfk)$ divided by the factor $k^3/2\pi^2$. This magnitude is computed for SLEs with smearing functions of the type \eqref{eq:windowf} of sufficiently small ($T=10^{-2}\tau$) and sufficiently large ($T=10^2\tau$) supports.\footnote{These are chosen according to figure \ref{fig:WYIR}. This figure refers to a particular infrared mode, but we have verified that the two supports considered here are also sufficiently small and sufficiently large for intermediate and ultraviolet modes as well.} We see that all SLEs have the same infrared behaviour except for a constant. This is in agreement with the properties found in the cosmological case, as studied in reference~\cite{Niedermaier2020}. In the ultraviolet, all vacua see a vanishing electric field at sufficiently short scales. Accordingly, they all converge to the same Minkowski vacuum at all times.

Additionally, to investigate the anisotropies we represent modes parallel and antiparallel to the direction of the electric field (i.e., $\theta=0$ and $\theta=\pi$, respectively). Both the infrared and ultraviolet behaviours are oblivious to the direction of $\bfk$. This is rooted in the angular dependence of the frequency in \eqref{eq:omegaSchwinger}, $\omega_\bfk(t)^2 = k^2 + 2q A(t) k \cos \theta + q^2 A(t)^2+ m^2$. Indeed, for $k~\ll~(q^2 A(t)^2~+~m^2)~/~|2 q A(t)|$ and $k \gg 2 |q A(t)|$ the angular contribution is negligible. Conversely, this defines an intermediate regime where the dependence on $\theta$ is important. Accordingly, in figure \ref{fig:PS_schw} the difference between parallel and antiparallel modes is significant at these intermediate scales. Note that in this regime the effects of the anisotropy are much more relevant than that of different choices of SLE. Furthermore, the curves for  $\theta = \pi$ are non-monotonic in contrast with those for $\theta = 0$. Indeed, for positive $\cos \theta$, $\omega_\bfk(t_0)^2$ grows monotonously as $k$ increases,  leading to a power spectrum that monotonously decreases. On the other hand, for negative $\cos \theta$, ${\omega_\bfk}(t_0)^2$ presents a minimum at $k=q A(t_0) |\cos \theta|$, which translates into a maximum in the power spectrum around that point (in our case, $k = \tau^{-1}$). Note that in this work we have chosen $q$ and $A(t)$ to have same sign. Had we chosen them with opposite signs, the roles of $\theta = 0$ and $\theta = \pi$ would have been interchanged. 

We now expand the power spectrum \eqref{eq:PS_Sch} in the orthonormal basis of square-integrable functions in $[-1,1]$ formed by the Legendre polynomials, $P_{\ell}(\cos{\theta})$:
\begin{equation}
    \mathcal{P}(t,\bfk)=\sum_{\ell=0}^{\infty} \mathcal{P}_{\ell}(t,k)P_{\ell}(\cos{\theta}),
\end{equation}
where the multipoles are given by
\begin{equation}
    \mathcal{P}_{\ell}(t,k)=\frac{2\ell+1}{2}\int_0^{\pi} d{(\cos{\theta})} \ \mathcal{P}(t,\bfk)P_{\ell}(\cos{\theta}).
\end{equation} 
Let us now consider the multipolar contributions $\ell\geq 1$ with respect to the isotropic monopole $\ell=0$, i.e., the coefficients
\begin{equation}
    g_{\ell}(t,k)=\mathcal{P}_{\ell}(t,k)/\mathcal{P}_0(t,k).
\end{equation}
We show in figure \ref{fig:gl} how these coefficients depend on the module $k$ of the wavevector for the SLE with large support $T=10^2\tau$. We verified that similar behaviours are obtained for smearing functions with different supports. We observe that the maximum contribution of all the multipoles with respect to the monopole happens precisely for the same scale, which is in the aforementioned intermediate regime identified also in figure \ref{fig:PS_schw}. In addition, we confirm that the contribution of multipoles decreases asymptotically in both the infrared and the ultraviolet.

\section{Number of created particles} \label{sec:num}

As mentioned in the previous section, the power spectrum at a fixed time does not fully encapsulate all the information on the vacuum. In cosmology, this is usually the only relevant quantity as it is the only one that can be related with observations of the CMB. However, in general and especially in the context of the Schwinger effect, this can be complemented with the number of created particles in one vacuum with respect to a reference one. 

We will consider as the reference vacuum $\ket{0}^{\textrm{in}}$ the one determined by the in-solution $\varphi_{\bfk}^{\textrm{in}}(t)$ in \eqref{eq:insolution}, which is a positive-frequency plane wave in the asymptotic past. Any other choice of basis of solutions $\varphi_{\bfk}(t)$ defines another quantum theory, with its corresponding annihilation and creation operators $\hat{a}^{\alpha}_{\bfk}$ and $\hat{a}^{\alpha\dagger}_{\bfk}$ according to \eqref{eq:hatphi}. The number of particles in this vacuum with respect to the in-vacuum is
\begin{equation}
\label{eq:Nkak}
    N^{\alpha}_{\bfk}= {}^{\textrm{in}}\!\bra{0}\hat{a}^{\alpha\dagger}_{\bfk}\hat{a}^{\alpha}_{\bfk}\ket{0}^{\textrm{in}}=|\beta_{\bfk}|^2,
\end{equation}
where the $\beta$-coefficients quantify the differences between the two solutions:
\begin{equation}
    \beta_{\bfk} = i\left[ \varphi_{\bfk}(t) \dot{\varphi}_{\bfk}^{\textrm{in}}(t) - \dot{\varphi}_{\bfk}(t)\varphi_{\bfk}^{\textrm{in}}(t)\right].
\end{equation}
Noticeably, at each time $t$ this depends on $\varphi_{\bfk}(t)$ as well as its derivative, and therefore encodes information on both $D_{\bfk}(t)$ and $C_{\bfk}(t)$ of the parameterization \eqref{eq:WY}. However, $N^{\alpha}_{\bfk}$ is still not fully descriptive of the vacuum, as it only depends on a combination of these two functions. As such, it may be used in addition to the power spectrum in order to characterize a given vacuum at a given time.

An interesting property of SLEs is that its predicted number of created particles per mode is proportional to the relative difference between the quantum and classical energy. More precisely,
\begin{equation}
    N_{\bfk}^{\alpha}=\frac{1}{2}\frac{{}^{\textrm{in}}\!\bra{0}E(\hat{\phi}^{\alpha}_{\bfk})\ket{0}^{\textrm{in}}-E(\varphi_{\bfk})}{E(\varphi_{\bfk})}.
\end{equation}
This is only true when $\varphi_{\bfk}$ is taken to be a SLE and $\hat{\phi}^{\alpha}_{\bfk}$ is quantized according to \eqref{eq:hatphi}. This property follows from the fact that, as it was commented in section \ref{sec:settings}, SLEs are the only states such that the constant $c(\varphi_{\bfk})$ defined in \eqref{eq:c2} vanishes.

\begin{figure}
    \centering
    \includegraphics[width=0.5\textwidth]{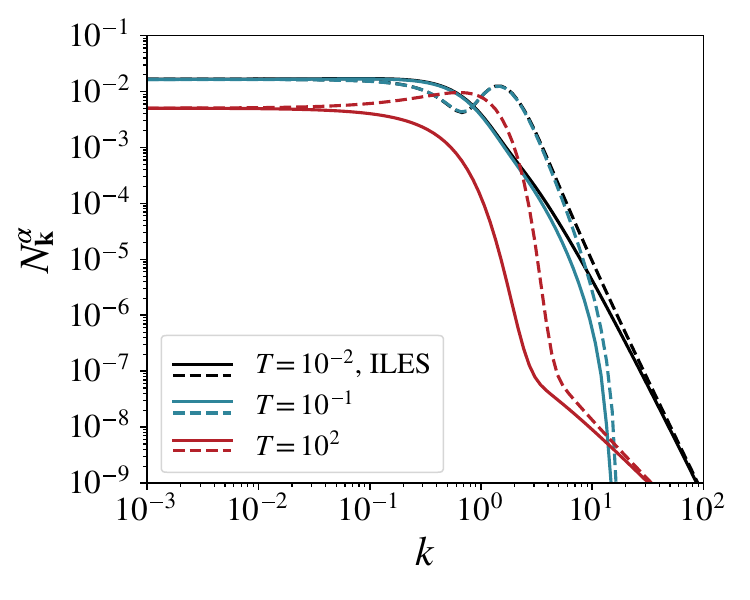}
    \caption[Number of created particles as a function of wavenumber for modes parallel and antiparallel to the electric field.]{Number of created particles $N^{\alpha}_{\bfk}$ as a function of the module $k$ of the wavevector for SLEs with small and large supports, for modes parallel (solid lines) and antiparallel (dashed lines) to the electric field. For ILES at $t_0$, $N^{\alpha}_{\bfk}$ coincides with that of the SLE with the smallest support considered. We use units $\tau=1$.}
    \label{fig:Nk}
\end{figure}

Figure \ref{fig:Nk} shows the behaviour of the number of created particles $N^{\alpha}_{\bfk}$ for modes parallel and antiparalel to the electric field, as a function of the wavenumber $k$ and for SLEs of sufficiently small ($T = 10^{-2}\tau$) and sufficiently large ($T = 10^{2}\tau$) supports around $t_0$. Again, we identify the same infrared behaviour for all vacua, which are distinguished by a constant contribution. In the ultraviolet, however, each vacuum tends to the Minkowski state at a different rate. For small supports, the spectral particle number $N^{\alpha}_{\bfk}$ of the SLE seems to agree with that of the ILES (see section \ref{sec:sopf}). However, for small enough scales, these states behave differently. To illustrate this separation, we have also represented a SLE with $T=10^{-1}\tau$, whose $N^{\alpha}_{\bfk}$ departs from that of the ILES at a lower (numerically achievable)~$k$. This behaviour is compatible with SLEs being of Hadamard type, while the ILES is not. In fact, Hadamard states are infinite-order adiabatic vacua~\cite{Pirk1993}, whose $N_{\bfk}$ decays with a power of $k$ proportional to its adiabatic order. Thus, for the ILES the $N^{\alpha}_{\bfk}$ is not exponentially suppressed, decaying more slowly than SLEs for sufficiently ultraviolet modes, not depicted in figure~\ref{fig:Nk}. Along these lines, the~$N^{\alpha}_{\bfk}$ for the SLE with large support  $T = 10^2\tau$ must also decay faster than that for the ILES, for sufficiently ultraviolet modes.\footnote{Solutions to the equation of motion are oscillatory, with increasing frequency after the maximum of the electric pulse, as well as for increasing $k$. Thus, the computation of the SLE becomes computationally demanding for large supports and large $k$, as it requires the integration of oscillations with very short periods.}

Finally, figure \ref{fig:Nk} also shows the intermediate regime where anisotropies are important. As motivated in the previous section, we verify that in the infrared and ultraviolet, the particle number is isotropic. For intermediate scales, parallel modes $\bfk$ to the electric field show a monotonic $N^{\alpha}_{\bfk}$, in contrast to antiparallel modes.

\section{Conclusions and discussion} \label{sec:conclusions_schw}

In \cite{Olbermann2007}, SLEs were introduced in general cosmological spacetimes as the states that minimize the energy density, smeared along the trajectory of an isotropic observer. They were shown to be Hadamard states, and later proven to be good candidates for the vacuum of cosmological perturbations in models with a period of kinetic dominance prior to inflation~\cite{Niedermaier2020}. In chapter \ref{chap:SLEsinLQC} we have applied SLEs in the context of LQC, where we have found that they heavily depend on the choice of smearing function only in regards to whether its support includes or excludes the bounce of LQC. Recently, they have also been applied to fermionic fields in a radiation-dominated CPT-invariant universe \cite{Nadal2023}. 

In this chapter, we have extended the construction of SLEs to general spatially homogeneous settings, with the emphasis on the Schwinger effect. To investigate the dependence of these SLEs on the choice of smearing function, we have considered regularized step-like smearing functions with a wide range of supports centered at the maximum of a Sauter-type electric pulse. We discern two asymptotic behaviours of SLEs. In the limit of small supports they behave as ILESs, which instantaneously minimize the energy density (although ILESs are not a particular case of SLEs, just a limiting behaviour). For very large supports the dependence on the support of the smearing function gradually disappears, thus determining in the limit a vacuum which minimizes the smeared energy density over the entire electric pulse. For supports of the order of the characteristic time scale of the electric pulse there is a non-trivial dependence. We have been able to draw parallels with what we have observed in chapter \ref{chap:SLEsinLQC}, and conclude that the sizes of the support considered there already corresponded to the large support regime, which is why convergence is obtained quickly and no non-trivial dependence on the smearing function is observed.

We have also calculated the power spectrum in the Schwinger effect, analogously to the usual definition in cosmology. We have shown that all SLEs have the same infrared behaviour except for a constant contribution, in agreement with what is found in cosmology \cite{Niedermaier2020}. In the ultraviolet, all vacua tend to the Minkowski vacuum although at different rates. As the power spectra only depend on the configuration of the state, they all converge for large wavenumbers. However, as the particle number encodes information not only on the configuration of the state but also on its velocity, each vacuum leads to different decay rates when approaching short scales. In particular, we observe that the particle number for all SLEs decays faster than that for the ILES. This might be an indication of SLEs being Hadamard in the Schwinger effect. In fact, after the completion of our work, SLEs have been shown to be Hadamard also in anisotropic cosmological models \cite{Banerjee:2023cxg}. Presumably, a construction along the same lines could prove that this is the case also in generic homogeneous settings.

Finally, we have analysed the anisotropy of the system. We find that in both the ultraviolet and the infrared regions, the anisotropies do not contribute to either the power spectrum or the number of created particles. An intermediate regime where they are most important has been identified.
\chapter{Conclusions}\label{chap:conclusions}

The main goal of this thesis is to contribute to a falsiable picture of quantum cosmology, particularly in the context of Loop Quantum Cosmology (LQC). Concretely, this entails the investigation of ambiguities in the theory, on one hand from a fundamental point of view in order to motivate certain proposals, and on the other from an empirical perspective, where we let observations determine the ones that best describe our Universe.

The relevant predictions to compare with observations are the primordial scalar and tensor power spectra of cosmological perturbations. In the context of LQC these are obtained by evolving the quantum corrected equations of motion over the background given by the effective dynamics of LQC. The effective dynamics are obtained through a procedure that has been proven rigorously to correctly capture the main features of the quantum dynamics of the model with massless scalar field. When the field is subject to the potential, numerical investigations of the dynamics of semiclassical states suggest that the procedure still approximates the quantum dynamics correctly.

We have applied a perturbative procedure to obtain the quantum dynamics of the model with constant scalar field potential up to first order in the potential. The potential is seen as a perturbation of the (analytically solvable) massless case. We have shown that actually the bounce of LQC in this case is generally not symmetric, as the expectation value of the volume of the massless case receives generically symmetric, anti-symmetric and asymmetric contributions. In the particular case of semiclassical states with Gaussian profiles the anti-symmetric terms vanish and the asymmetric term becomes symmetric, thus agreeing with the effective dynamics. With this we have gained insight into the nature of the bounce for generic states. However, in general we are interested in semiclassical states, for which we have further corroborated that the effective dynamics can be trusted to provide a good approximation.

Next we explored the ambiguity in the choice of vacuum by proposing States of Low Energy (SLEs) as the vacuum for cosmological perturbations in hybrid LQC. These are states that minimize the energy density when smeared along a given time window. They were introduced in cosmological settings, where they are proven to be Hadamard states \cite{Olbermann2007}. Furthermore, they had been shown to be viable candidates for vacua of cosmological perturbations in models with a period of kinetic dominance prior to inflation \cite{Niedermaier2020}. A priori, the choice of smearing function (and of time window, through its support) is free, introducing an ambiguity. We have reasoned that in LQC this window should include the Planckian regime where the bounce occurs, as this is where quantum corrections are most important. In that case, the resulting primordial power spectra are extremely insensitive to the shape and support of the smearing function, as long as it is wide enough. Thus in this case the ambiguity is effectively irrelevant. We have also shown that their primordial power spectra are essentially indistinguishable from those obtained when fixing an adiabatic vacuum of second order at the bounce. However, adiabatic states and SLEs are fundamentally different, as finite order adiabatic vacua are not Hadamard states.

On the other hand, the primordial power spectra are very sensitive to whether the Planckian regime is included in the support of the smearing function. As we move the support further away from the bounce point, the oscillations and power enhancement in the primordial power spectra gradually die out, until the power spectra become essentially near scale invariant in the ultraviolet with exponential power suppression for infrared modes. This shows that, in the context of SLEs, the bounce of LQC leaves clear imprints in the primordial power spectra: oscillations and power enhancement for intermediate scales. The exponential power suppression can be obtained simply with a period of kinetic dominance prior to inflation.

Furthermore, we have shown that the primordial power spectra of SLEs associated to smearing functions supported away from the bounce reproduce well the power spectra of the Non-Oscillatory (NO) vacuum, previously proposed in LQC. Through a direct inspection of their UV expansions we were able to prove that the NO vacuum shares the UV behavior of SLEs, and therefore is also a Hadamard state.

Then, we started the endeavor of contrasting predictions of hybrid LQC with observations with the NO vacuum, or equivalently an SLE associated to a smearing function supported away from the bounce. We chose this state to begin this section of the work because it provides the simplest power spectrum to parametrize. On one hand, the goal was to find whether such a feature in the primordial power spectra is preferred by observations, or in other words, to find constraints to the scale at which the power suppression may occur. On the other hand, we also aimed to investigate whether such a feature is helpful in alleviating the power suppression, lensing and parity anomalies identified in the data. Indeed, this has been found to be possible in certain models based on the dressed metric approach of LQC \cite{Ashtekar:2020,Ashtekar:2021,Agullo:2020b,Agullo2021}. In this spirit, we performed a Bayesian analysis of the model leaving the scale at which power suppression occurs as free parameter, as well as the other six parameters that are shared with $\Lambda$CDM. With this analysis we computed the marginalized posterior probability of the suppression scale and concluded that observations prefer it to be inside the observable window, restricted to infrared modes. The model with such a primordial power spectrum fits the observations better than the standard $\Lambda$CDM model, as evidenced by the posterior distribution of the parameter as well as the $\chi^2$ statistic.

Given the initial conditions preferred by the data, we showed that this LQC model is able to alleviate the power suppression and lensing anomalies identified in the case of the standard model. For the first we computed the distribution of the estimator according to the model and found that the p-value of the observations increases with respect to $\Lambda$CDM. This indicates that the observed power suppression is less likely to occur in a $\Lambda$CDM universe than in our model. In the case of the lensing anomaly, as had been the case in previous works in LQC, the pre-inflationary dynamics have an effect on the distribution of another parameter: the optical depth. This is a parameter that is anti-correlated with the lensing amplitude. As such, a shift in one results in a shift in the other. The anomaly is alleviated as the value of the lensing amplitude is closer to the expected one. This is not due to effects in the late-time lensing but due to shifts in the overall distribution probability of the parameters. In this sense, the inconsistencies that caused the lensing parameter to be shifted from the expected value are mitigated. On the other hand, the parity anomaly is essentially unaffected in our model. We have computed the distribution of this statistic and found the p-value of observations. Although some asymmetry is introduced for some multipoles, the variance of the distribution is affected in such a way that even in those cases the p-values are not appreciable higher, and in some cases are even lower.

In the future, we will extend this analysis to SLEs associated with smearing functions supported also on the Planckian regime. As we have shown, this introduces oscillations and power enhancement of intermediate scales in the primordial power spectra. Furthermore, we have concluded that these are a direct consequence of the bounce in the context of SLEs. Previous works \cite{Agullo:2020b,Agullo2021,Agullo:2020} have shown that with such primordial power spectra non-Gaussianities are important and affect the significance of the three anomalies investigated here, including the parity asymmetry. A natural continuation of this work is therefore to perform a Bayesian analysis of the model considering non-Gaussianities.

Finally, we have extended the construction of SLEs to generic homogeneous spacetimes. We focused particularly on the Schwinger effect: the particle/antiparticle pair production due to a very strong electric field in a flat spacetime. This introduces an anisotropy, but we have still considered a homogeneous scenario. We have investigated the role of the support of the smearing function by considering an electric pulse and identified two asymptotic regimes. In the limit of small supports the SLE coincides with the Instantaneous Lowest-Energy State (ILES), which minimizes the expectation value of the energy density at a precise time. However, rigorously, the ILES is not a particular case of SLE, only an asymptotic limit. For very large supports, the dependence of the SLE on the support gradually disappears and the SLE tends to a limiting vacuum that minimizes the energy density over the entire electric pulse. This is the regime first identified in the LQC scenario. This is why there we identified only a very week dependence on the smearing function, as long as the bounce (the equivalent of the pulse in that scenario) was included. In general, for intermediate supports, there is a non-trivial dependence on the smearing function. We have found that also in this setting the power spectra of SLEs have the same infrared behavior up to a constant and a common ultraviolet behavior. On the other hand, the number of created particles is sensitive to the chosen SLE also in the ultraviolet. Even though they all tend to the Minkowski vacuum in the limit of short wavelengths, they approach it at different rates. This is encoded in the number of created particles since it depends not only on the configuration of the vacuum but also on its velocity. In particular, we find that the number of particles of SLEs decays faster than for the ILEs. This suggests that SLEs are likely also Hadamard states in this scenario.
\appendix

\chapter{Integrals of Legendre polynomials}\label{sec:appLegendre}

The integrals $I_{\ell\ell^{\prime}}$ of \eqref{eq:S12fromCl} can be expressed as:

\begin{equation}
    I_{\ell\ell^{\prime}} = \frac{\left(2\ell+1\right)\left(2\ell^{\prime}+1\right)}{(4\pi)^2}\mathcal{I}_{\ell\ell^{\prime}}(x=1/2),
\end{equation}
where
\begin{equation}
    \mathcal{I}_{mn}(x) = \int_{-1}^x P_m(y)P_n(y) dy.
\end{equation}
Using properties of the Legendre polynomials $P_m(x)$, these are found to be \cite{Copi2008}:
\begin{equation}
    \mathcal{I}_{mn}(x) = \frac{m P_n(x)\left[P_{m-1}(x)-x P_m(x)\right]-n P_m(x)\left[P_{n-1}(x)-x P_n(x)\right]}{n(n+1)-m(m+1)}, \ \left[m\neq n\right],
\end{equation}
and when $m = n$ they can be obtained through the recursive relation:
\begin{align}
    \mathcal{I}_{nn}(x) = \frac{1}{2n+1}\bigg\{ &\left[P_{n+1}(x)-P_{n-1}(x)\right]\left[P_n(x)-P_{n-2}(x)\right]-(2n-1)\mathcal{I}_{n+1,n-1}(x)\nonumber\\
    &+(2n+1)\mathcal{I}_{n,n-2}(x)+(2n-1)\mathcal{I}_{n-1,n-1}(x)\bigg\}.
\end{align}
\chapter{Interaction picture}\label{sec:appInteractionPicture}

Let us consider that we have a system with a Hamiltonian operator that depends on time $\phi$ (for a coherent notation with the main body of this work) $\hat{H}(\phi) = \hat{H}_0 + \hat{H}_1(\phi)$. Typically $\hat{H}_0$ is the Hamiltonian of the free system, which is known and analytically under control, and $\hat{H}_I$ controls the interactions which are seen are perturbations of the free system. However, for what concerns this appendix, there is no need to impose that $\hat{H}_I$ is a perturbation of $\hat{H}_0$.
\begin{align}
    \chi(x,\phi) &= \hat{U}(x,\phi) \chi(x,\phi_0),\\
    \hat{U}(x,\phi) &= \mathcal{P}\left[\exp\left(i\int_{\phi_0}^{\phi}d\tilde{\phi} \hat{H}(\tilde{\phi})\right)\right],
\end{align}
where $\mathcal{P}$ denotes time ordering (with respect to $\phi$), whereas as operators have no time dependence (other than their possible explicit time dependencies). Expectation values can thus be calculated as:
\begin{equation}
    \langle \hat{A} \rangle_{\chi(x,\phi)} = \langle \chi(x,0)| \hat{U}^{\dagger}(x,\phi) \hat{A} \hat{U}(x,\phi) |\chi(x,0)\rangle.
\end{equation}

In the Heisenberg picture, time dependence is carried fully by the operators:
\begin{equation}
    \hat{A}_H = \hat{U}^{\dagger}(x,\phi) \hat{A} \hat{U}(x,\phi),
\end{equation}
whereas the states are fixed: $\chi_H(x) \equiv \chi(x,\phi_0)$, such that
\begin{equation}
    \langle \hat{A}_H \rangle_{\chi_H(x)} = \langle \hat{A} \rangle_{\chi(x,\phi)}.
\end{equation}

On the other hand, in the interaction picture, time dependence falls on both states and operators. The motivation is that this way we may separate the evolution due to the potential term $\hat{H}_1(\phi)$ from that of the free system given by $\hat{H}_0$. We write the states in this picture as:
\begin{align}
    \chi_I(x,\phi) &= e^{-i \hat{H}_0 \Delta\phi} \chi(x,\phi)\nonumber \\
    &= e^{-i \hat{H}_0 \Delta\phi}\mathcal{P}\left[\exp\left(i\int_{\phi_0}^{\phi}d\tilde{\phi} \hat{H}(\tilde{\phi})\right)\right] \chi(x,\phi_0),\label{eq:appchiI}
\end{align}
where $\Delta\phi=\phi-\phi_0$, and the operators as:
\begin{equation}
    \hat{A}_I(\phi) =  e^{-i \hat{H}_0 \Delta\phi} \hat{A}  e^{i \hat{H}_0 \Delta\phi},
\end{equation}
such that 
\begin{equation}
    \langle \hat{A}_I \rangle_{\chi_I(x,\phi)} = \langle \hat{A} \rangle_{\chi(x,\phi)}.
\end{equation}
Differentiating \eqref{eq:appchiI}:
\begin{align}
    \frac{d\chi_I(x,\phi)}{d\phi} &= \frac{d}{d\phi}\left\{e^{-i \hat{H}_0 \Delta\phi}\mathcal{P}\left[\exp\left(i\int_{\phi_0}^{\phi}d\tilde{\phi} \hat{H}(\tilde{\phi})\right)\right]\right\} \chi(x,\phi_0) \nonumber \\
    &= -i e^{-i \hat{H}_0 \Delta\phi} \left(\hat{H}(\phi)-\hat{H}_0\right)\mathcal{P}\left[\exp\left(i\int_{\phi_0}^{\phi}d\tilde{\phi} \hat{H}(\tilde{\phi})\right)\right]\chi(x,\phi_0) \nonumber \\
    & = -i e^{-i \hat{H}_0 \Delta\phi} \hat{H}_1(\phi) \chi(x,\phi)\nonumber \\
    & = -i e^{-i \hat{H}_0 \Delta\phi} \hat{H}_1(\phi) e^{i \hat{H}_0 \Delta\phi} e^{-i \hat{H}_0 \Delta\phi} \chi(x,\phi)\nonumber \\
    & = -i  \hat{H}_{1I}(\phi) \chi_I(x,\phi),
\end{align}
where in the second equality we have taken into account that $\hat{O_1} e^{\hat{O}_2} = e^{\hat{O}_2}\hat{O}_1$ if $\hat{O}_1$ and $\hat{O}_2$ commute. When integrating, we need to maintain the time-ordering (keeping $\hat{H}_1(\phi^{\prime})$ to the left of $\hat{H}_1(\phi^{\prime\prime})$ if $\phi^{\prime} > \phi^{\prime\prime}$), which leads to:
\begin{align}
    \chi_I(x,\phi) &= \hat{U}_I(x,\phi) \chi(x,\phi_0),\\
    \hat{U}_I(x,\phi) &= \mathcal{P}\left[\exp\left(i\int_{\phi_0}^{\phi}d\tilde{\phi} \hat{H}_{1I}(\tilde{\phi})\right)\right].
\end{align}
This is the form of the states in interaction picture that is used in chapter \ref{sec:generic potential}.

\chapter{Calculations of the volume for constant potential}\label{sec:appendix_V}

In section \ref{sec:expVgeneric}, we have laid out the necessary steps to compute $\langle\hat{V}\rangle$ for a given form of the potential $W(\phi)$. Let us do so for a constant potential $W(\phi)=W$, which is used in \ref{sec:constant potential}. With this potential, one obtains $I_1$ defined in~\eqref{eq:I1}~as:
\begin{equation}
    I_1(x,\phi) = \frac{1}{2}W\left\{\phi+\frac{1}{4\sqrt{3\pi G}}\bigg[\sinh\left(4\sqrt{3\pi G}x\right)-\sinh\left(4\sqrt{3\pi G}(x-\phi)\right)\bigg]\right\},
\end{equation}
where we have set $\phi_0=0$. This leads to
\begin{equation}
    \partial_xI_1(x,\phi) = \frac{1}{2}W\bigg[\cosh\left(4\sqrt{3\pi G}x\right)-\cosh\left(4\sqrt{3\pi G}(x-\phi)\right)\bigg].
\end{equation}

Consequently, we obtain $E(x,\phi)$ as defined in \eqref{eq:E}:
\begin{equation}
    \begin{split}
        E(x,\phi)&=\left(1+3\tilde{W}\right)\cosh\left(\sqrt{12\pi G}(x-\phi)\right)+\tilde{W}\Bigg[8\sqrt{3\pi G}\phi \sinh\left(\sqrt{12\pi G}(x-\phi)\right)\\
        &+\cosh\left(3\sqrt{12\pi G}(x-\phi)\right) -\cosh\left(\sqrt{12\pi G}(3x-\phi)\right) -3\cosh\left(\sqrt{12\pi G}(x+\phi)\right)\Bigg],
    \end{split}
\end{equation}
where, we recall, $\tilde{W}$ is defined in \eqref{eq:Wtilde}. 

Once $E(x,\phi)$ is inserted in \eqref{eq:VexpW}, the terms of the form
\begin{equation}
    \cosh\left[\sqrt{12\pi G}\left(n_x x-n_{\phi}\phi\right)\right] 
\end{equation}
lead to contributions of the form
\begin{align}
  V_{-n_x}e^{-n_{\phi}\sqrt{12\pi G} \phi}+V_{+n_x}e^{n_{\phi}\sqrt{12\pi G} \phi},
\end{align}
and similarly for the $\sinh$ terms (changing the sign of the second term in the above expression).

Finally, this leads to $\langle \hat{V} \rangle$, as defined in \eqref{eq:expv_W}.

With the same strategy, we also obtain $\langle\hat{V}^2\rangle$. Namely, we find $\hat{V}^2_J$ from \eqref{eq:Aj}, by using $\hat V_I^2$ in place of $\hat{A}_I$. Then, as in \eqref{eq:Vexp1}, we calculate
\begin{align}
    \langle \hat{V}^2 \rangle\simeq \langle \hat{V}^2_J \rangle =2i\int_{-\infty}^\infty dx \frac{dF^*(x)}{dx}\hat{V}^2_J(\phi)F(x),
\end{align}
arriving to the result
\begin{align}
\begin{split}
    \langle\hat{V}^2\rangle&\simeq\left(\frac{V_{d_{+2}}^2}{2} +\tilde{W} w_{2_+}(\phi)\right) e^{2\sqrt{12\pi G}\phi}+ \left(\frac{V_{d_{-2}}^2}{2} +\tilde{W} w_{2_-}(\phi)\right) e^{-2\sqrt{12\pi G}\phi}\\
    & + \tilde{W}\left(V_{d_{+4}}^2e^{4\sqrt{12\pi G}\phi}+ V_{d_{-4}}^2e^{-4\sqrt{12\pi G}\phi}\right)+\left(1+6\tilde{W}\right)V_{d_0}^2-4\tilde{W}\left(V_{d_{+2}}^2+V_{d_{-2}}^2\right),
\end{split}
\end{align}
where
\begin{align}
    w_{2_\pm}(\phi) &\equiv \left(4\mp 8\sqrt{3\pi G}\phi\right)V_{d_{\pm2}}^2 - V_{d_{\pm4}}^2-3V_{d_0}^2,\\
    V_{d_n} &\equiv \frac{2\pi G\gamma\sqrt{\Delta}}{3\pi G}\int_{-\infty}^{+\infty} dx \text{Im}\left[ \frac{dF^*(x)}{dx}\frac{d^2F(x)}{dx^2} \right] e^{-n\sqrt{12\pi G}\ x}.
\end{align}

This allows the computation of the relative dispersion of $\hat{V}$, ${\langle\Delta\hat{V}\rangle}/{\langle\hat{V}\rangle} = \sqrt{{\langle\hat{V}^2\rangle}/{\langle\hat{V}\rangle^2}-1}$. We~truncate to first order in $W$, getting a long expression whose form  we omit for simplicity.

\chapter{SLE and NO in kinetic dominance}\label{sec:appSLENOKD}

 In General Relativity, an early universe dominated by the kinetic energy of the inflaton field is well  described by a spacetime coupled to a perfect fluid with constant equation of state $P=\omega \rho$ with $\omega=1$, namely, stiff matter. The scale factor of the spacetime behaves as $a(\eta)=\sqrt{\eta}$ in conformal time $\eta$. The equations of motion of scalar and tensor perturbations are also well described by equations \eqref{eq:eom_uk}, with
\begin{equation}
    s(\eta)=\frac{1}{4\eta^2}.  
\end{equation}
Any complex solution can be written in terms of well-known Hankel functions $W_k(\eta)=\sqrt{\frac{\pi \eta}{4}}\bar H^{(1)}_0(k\eta)$ and its complex conjugate. 

Actually, that is the natural initial state that one adopts in a kinetically dominated classical universe \cite{Contaldi:2003zv}. Let us show that $|W_k(\eta)|^2$ has an asymptotic expansion for large $k$ of the form of the SLEs, namely, it is of the form of Eq. \eqref{eq:uFNO} --the phase of $W_k(\eta)$ is determined by $|W_k(\eta)|$ via Eq. \eqref{eq:uF} which is itself a consequence of the normalization of this basis of complex solutions. Using the asymptotic expansion of Hankel functions
\begin{equation}
H_{\nu}^{(1)}(z) \sim\left(\frac{2}{\pi z}\right)^{1 / 2} e^{i\left(z-\frac{1}{2} \nu \pi-\frac{1}{4} \pi\right)} \sum_{m=0}^{\infty}i^{m} \frac{a_{m}(\nu)}{z^{m}},
\end{equation}
with 
\begin{equation}
    a_{m}(\nu)=\frac{\left(4 \nu^{2}-1^{2}\right)\left(4 \nu^{2}-3^{2}\right) \cdots\left(4 \nu^{2}-(2 m-1)^{2}\right)}{m ! 8^{m}},
\end{equation}
one obtains
\begin{equation}
[F^{W_k}(\eta)]^{-1}=|W_{k}(\eta)|^2\sim\frac{1}{2k} \left(\sum_{m=0}^{\infty}i^{m} \frac{a_{m}(0)}{(k\eta)^{m}}\right)\left(\sum_{n=0}^{\infty} (-i)^{n} \frac{a_{n}(0)}{(k\eta)^{n}}\right),
\end{equation}
or more explicitly,
\begin{equation}
[F^{W_k}(\eta)]^{-1}\sim\frac{1}{2k} \left(1-\frac{1}{8 {k}^{2} {\eta}^{2}}+\frac{27}{128 {k}^{4} {\eta}^{4}}-\frac{1125}{1024 {k}^{6} {\eta}^{6}}+\cdots\right).
\end{equation}
On the other hand, using the recursion relation in Eq. \eqref{eq:Gn}, one obtains,
\begin{equation}
    G_1(\eta)=\frac{1}{8\eta^2}, \quad G_2(\eta)=\frac{27}{128\eta^4},\quad G_3(\eta)=\frac{1125}{1024\eta^6},\quad \cdots
\end{equation}

The asymptotic behavior at large $k$'s exactly agrees with the one of SLEs. Hence, we can claim that the state $W_k(\eta)$ is of Hadamard type. 

Now, let us see how this state is related to the SLE and the NO vacuum.  We start with the SLE prescription. Let us consider an arbitrary state $\omega$, whose modes will be denoted by $T^\omega_k$. This state will be related to the fiducial state defined by $W_k$ via a Bogoliubov transformation
 \begin{align}
 T^\omega_k=\alpha(k)W_k+\beta(k)\bar{W}_k.
 \end{align}
 Then, the smeared energy density of the mode $T^\omega_k$ is given by
\begin{align}
    E(T^{\omega}_k)=E(W_k)(1+2\beta^2(k)) +2\beta(k)\sqrt{1+\beta^2(k)}\mathcal{R}[e^{i\arg[\alpha(k)]} D(W_k)],
\end{align}
with
\begin{equation}
    E(W_k) = \frac{1}{2} \int_{\eta_0}^{\eta_f} d\eta\,f^2(\eta) \frac{k^2}{\eta} \left[\left\lvert H^{(1)}_1(k\eta)\right\rvert^2 + \left\lvert H^{(1)}_0(k\eta)\right\rvert^2\right],  
\end{equation}
and
\begin{equation}
   D(W_k) = \frac{1}{2} \int_{\eta_0}^{\eta_f} d\eta\,f^2(\eta) \frac{k^2}{\eta} \left[\left( H^{(1)}_1(k\eta) \right)^2 +\left( H^{(1)}_0(k\eta) \right)^2\right],
\end{equation} 
 Here we are calling $[\eta_0,\eta_f]$ the support of $f$, and we have used the property $\frac{d}{dx} H^{(1)}_0(x)=-H^{(1)}_1(x)$.
We obtain the  smallest value for $E(T^{\omega}_k)$ by choosing $\arg[\alpha(k)]$ such that $e^{i\arg[\alpha(k)]} D(W_k)=-|D(W_k)|$ and $\beta(k)\geq0$. If we define $\delta(k)=\beta(k)/\sqrt{1+\beta(k)^2}\in[0,1)$, then,  no matter the modes $T^\omega_k$ considered, we can write
\begin{align}
   E(T^{\omega}_k)\geq\mathcal{R}_k(\eta_f,\eta_0)E(W_k),
\end{align}
where
\begin{align}
   \mathcal{R}_k(\eta_f,\eta_0)=\left\{1+\frac{2\delta(k)}{1-\delta(k)^2}\left[\delta(k) -\frac{|D(W_k)|}{E(W_k)}\right]\right\}.
\end{align}
We have found that for test functions well approximated by a (smooth) step function with support in $[\eta_0,\eta_f]$ and in the limit $\eta_f\to\infty$, the ratio $|D(W_k)|/E(W_k)$ behaves as
\begin{equation}
\frac{|D(W_k)|}{E(W_k)} \simeq\frac{1}{2\pi(k\eta_0)^2}.
\end{equation}
Therefore,  in the limit $(k\eta_0)\gg 1$ we obtain $\mathcal{R}_k(\eta_f,\eta_0)\geq1$, and then we always have $E(T^{\omega}_k)\geq E(W_k)$. We thus conclude that the state defined by the modes $W_k$ is a SLE provided that $(k\eta_0)\gg 1$.

On the other hand, for the NO vacuum, if we follow the prescription in \cite{nonosc}, it can be implemented as the solution $v_{k}^{(NO)}(\eta)$ that minimizes the integral
\begin{equation}
    I(v_k)=\int_{\eta_0}^{\eta_f}d\eta \left|\partial_\eta|v_{k}(\eta)|^2\right|.
\end{equation}
Namely, $I(v_k)\geq I(v_k^{(NO)})$. If we write 
\begin{equation}
    v_k^{(NO)}(\eta)=\underline{\alpha}(k) W_k(\eta)+\underline{\beta}(k) \bar W_k(\eta),
\end{equation}
we have that
\begin{equation}
    |v_k^{(NO)}(\eta)|^2= J_k(\eta)|W_k(\eta)|^2,
\end{equation}
with
\begin{equation}
    J_k(\eta)=1+2|\underline{\beta}(k)|^2+2|\underline{\beta}(k)|\sqrt{1+|\underline{\beta}(k)|^2}\cos\left\{\arg[\beta(k)]-\arg[W_k(\eta)]\right\}.
\end{equation}
One can see that $J_k(\eta)\geq 0$. Now, the integral above can be written as
 \begin{equation}\label{eq:no-int}
    I(v_k^{(NO)})=\int_{\eta_0}^{\eta_f}d\eta \left|J_k(\eta)\partial_\eta|W_{k}(\eta)|^2+|W_{k}(\eta)|^2\partial_\eta J_k(\eta)\right|.
\end{equation}
In the cases in which $\partial_\eta|W_{k}(\eta)|^2=0$ one can easily see that the minimum of this integral is reached if and only if $\underline\beta(k)=0$. However, as in the SLEs, by looking at the asymptotic expansion of the solutions $W_k(\eta)$ in terms of the Hankel functions for large argument, namely, for $k\eta\gg 1$ one can see that
\begin{equation}
    |W_k(\eta)|^2=\frac{1}{2k}\left(1+{\mathcal O}\left[(k\eta)^{-2}\right]\right),\quad \arg[W_k(\eta)]=-k\eta\left(1+{\mathcal O}\left[(k\eta)^{-2}\right]\right)
\end{equation}
Therefore, $\partial_\eta|W_{k}(\eta)|^2$ decreases sufficiently fast for  $k\eta\gg1$. Besides, one can see that the leading and subleading contributions in the integrand in Eq. \eqref{eq:no-int} in the limit $k\eta\gg1$ is
\begin{equation}
    \left||\underline{\beta}(k)|\sqrt{1+|\underline{\beta}(k)|^2}\sin(\arg(\beta)+k\eta)+{\mathcal O}(k\eta)^{-3}\right|. 
\end{equation}
Therefore, in the limit $\eta_f\to\infty$, the integral $I(v_k^{(NO)})$ will diverge unless $\beta(k)=0$ for all $k$'s. 

In summary, both the SLE and NO prescriptions agree in the regimes discussed in this Appendix. Concretely, in the situations in which the smearing is carried over sufficiently large values of conformal times.

\chapter{Details of numerical computations}\label{sec:app_numerics}

In this appendix we provide some details on the numerical computations performed for the evolution of cosmological perturbations in LQC in chapters \ref{chap:SLEsinLQC} and \ref{chap:anomalies}.

\section{Initial conditions for the background}\label{sec:app_background}

In our work we have evolved numerically the cosmological perturbations, for which we have discussed several possible choices of initial conditions. We also need to evolve numerically the background of LQC. We do so through the effective Hamiltonian \eqref{eq:Ceff}, considering the quadratic and Starobinsky potentials \eqref{eq:quadV} and \eqref{eq:starV}, respectively. An agreement with observations requires $m = 1.21\times 10^{-6}$ for the quadratic potential and $V_0 = 1.77 \times 10^{-13}$ for the Starobinsky potential \cite{Bonga2015,Ashtekar:2016}. Regarding initial data, for the background we only need to specify the value of the scalar field at the bounce $\phi_B=\phi(t=t_B)$, since $H(t_B) = 0$. The kinetic energy of the scalar field is obtained from the condition $\rho(t=t_B)=\rho_c$. One may consider both $\phi_B$ and $\rho_c$ to be free parameters of the theory. In chapter \ref{chap:SLEsinLQC} we consider toy values only. As mentioned, there we take $\phi_B = 1.225$, so that the departures from near scale invariance in the power spectra are found roughly in the region where they may be allowed by observations. On the other hand, in chapter \ref{chap:anomalies}, this is indeed left as a free parameter. Finally, leaving $\rho_c$ free is equivalent to leaving $\gamma$ free. However the primordial power spectrum of the NO vacuum considered in chapter \ref{chap:anomalies} has been shown to be almost invariant under changes in the critical energy density. Thus, we fix $\gamma=0.2375$ always, a value obtained from black hole entropy calculations and that is commonly adopted in the LQC literature \cite{LQCreview_Agullo2016}.

One should also specify the value of the scale factor at some point in time. As usual in the context of LQC, we fix it to be the unit at the bounce, for convenience. However, in standard cosmology the scale factor is usually fixed to be the unit today. This choice is arbitrary, as the physical quantity is not the scale factor, but ratios of it. Nevertheless, to compare our predictions with observations we need to relate the two approaches, specifically taking into account that each one leads to different scales of the modes $k$. Ideally, one could evolve the background of LQC until today, modeling not only the early Universe but all the post inflationary history, and find the relation between the two scales today. Alternatively, one could evolve backwards the background of standard cosmology until, e.g. the end of inflation and compare the two scales there.

However, a simpler procedure is usually adopted (see for example \cite{dressedmetric_Agullo_CQG_2013}). We compute the dynamics of the background and of perturbations given our choice of scale factors, which results in a power spectrum $\mathcal{P}_{\mathcal{R}}(\tilde{k})$ for Fourier modes $\tilde{k}$. No matter the choice of vacuum, this power spectrum is the near-scale-invariant one of standard cosmology for the ultraviolet side of the spectrum. The scale at which the departures from near scale invariance occur depends on $\phi_B$. Then, to relate the scale $\tilde{k}$ with the one of standard cosmology, we resort to the pivot scale $k_{\star}$ as a reference scale. This scale is already ultraviolet with respect to the observational window, and departures from near scale invariance are severely disfavored by observations at this point. By definition, this is the scale at which the primordial power spectrum is $A_s$. For Planck data, $k_{\star} = 0.05\,{\rm Mpc}^{-1}$. Then, for a given $A_s$ we find the pivot scale in our units $\tilde{k}_{\star}$ as $\mathcal{P}_{\mathcal{R}}(\tilde{k}_{\star}) = A_s$. Finally, we can rescale $\tilde{k}$ to $k = \tilde{k}\cdot k_{\star}/\tilde{k}_{\star}$. The power spectrum in this scale may now be compared with observations from the Planck collaboration. The effect is to displace $\mathcal{P}_{\mathcal{R}}(\tilde{k})$ logarithmically in $\tilde{k}$.

\section{Parametrization of NO vacuum scalar primordial power spectrum}\label{sec:app_param}
In summary, the departures from the near-scale-invariant power spectrum occur at a scale that depends on both $\phi_B$ and $A_s$. Thus, to simplify calculations, we parametrize it with one free parameter that defines the scale at which exponential suppression occurs, encapsulating all the freedom of the power spectrum in our model. Let us define here the parametrization that we have used for our computations.

Inspired by \cite{Contaldi:2003zv}, we have parametrized the scalar primordial power spectrum obtained with the NO vacuum as:
\begin{equation}\label{eq:fkk}
    f(k,k_c) = \begin{cases}
    N f_{\textrm{sup}}(k) f_H(\frac{k_d}{2}) & \textrm{if }k < \frac{k_d}{2},\\
    N f_{\textrm{sup}}(k) f_H(k) & \textrm{if } k \geq \frac{k_d}{2},
    \end{cases}
\end{equation}
where $f_{\textrm{sup}}(k) = 1- \exp(-(k/k_d)^{\lambda_c})$ parametrizes the exponential suppression, and 
\begin{equation}\label{eq:fhk}
    f_H(k) =\frac{\pi}{2 k_d k}\Bigg| k H_1^{(2)}\left(\frac{k}{k_d}\right)\sin\left(C\frac{k}{k_d}\right)+ \frac{k_d}{2}H_0^{(2)}\left(\frac{k}{k_d}\right)\left[2\frac{k}{k_d}\cos\left(C\frac{k}{k_d}\right)-\sin\left(C\frac{k}{k_d}\right)\right]\Bigg|^2
\end{equation}
where $H_n^{(2)}(x)$ is a specific Hankel function of the second kind, and $k_d \simeq k_c/1.7$. Here, $k_d$ separates the regions of oscillation and exponential suppression, and is the parameter that encodes the ambiguities coming from LQC, $N$ is a normalization factor so that $A_s$ maintains its meaning as in standard cosmology as the amplitude of the power spectrum at the pivot scale $k_{\star}$, therefore fixed at $N = (f_{\textrm{sup}}(k_{\star}) f_H(k_{\star}))^{-1}$, $\lambda_c = 2.95$ is the slope of the suppression, and $C = 2/1.8$ parametrizes the oscillations.

It is worth commenting that this parametrization is an ad hoc choice that fits well our NO power spectrum, but only for practical purposes, namely, for the simulations that we carry out in $CLASS$ and the subsequent Bayesian analysis. However, we must remember that this parametrization has important theoretical limitations. For instance, it would correspond to a state that is not of Hadamard type, unlike the NO vacua. 

\subsection{Relation between $\phi_B$ and number of e-folds}\label{sec:app_phiB_efolds}

As mentioned in the text, we leave the value of the inflaton field at the bounce, $\phi_B$, as a free parameter. This value affects how much inflation occurs. Consequently, the range of values explored is such that the resulting background dynamics produces enough e-folds for an agreement with observations to be possible. This depends heavily on the inflaton potential.

We find numerically that the number of e-folds of inflation $N$ is related to $\phi_B$ through:
\begin{equation}
    N = A \phi_B + B,
\end{equation}
where for a quadratic potential $A \simeq 38$ and $B \simeq 29$, whereas for a starobinsky potential $A \simeq 259$ and $B \simeq 440$. The number of e-folds from the bounce until inflation is approximately constant with $k_c$, around $4.4$ e-folds for the quadratic model and $4.9$ e-folds for the Starobinsky potential.

\subsection{Relation between $k_c$ and $\phi_B$}\label{sec:app_kcefolds}

In order to relate $k_c$ of the parametrization back to $\phi_B$ we provide in this section approximate relations obtained numerically considering $A_s$ to be fixed at the $\Lambda$CDM best-fit value.
We have found numerically
\begin{equation}
	\phi_B = C \ln k_c + D,
\end{equation}
where for the quadratic potential $C = -0.027$ and $D = 0.735$, whereas for the Starobinsky potential $C = -0.004$ and $D = -1.490$.

We consider this approximation to suffice, as the variation of $A_s$ at 1-$\sigma$ we have obtained from the Bayesian analysis is of $\sim 1.7 \%$. Considering a spectral index $n_s \simeq 0.96$ this impacts the shift in the scales $\tilde{k}$ explained in \ref{sec:app_background} in $\sim 4 \%$, which would induce a variation of $< 0.1 \%$, according to the relations found above.

\addcontentsline{toc}{chapter}{References}
\bibliography{Bibliography/bib.bib}

\providecommand{\noopsort}[1]{}\providecommand{\singleletter}[1]{#1}%
\providecommand{\href}[2]{#2}\begingroup\raggedright\begin{thebibliography}{100}

\bibitem{Planck2018_parameters}
{\bfseries Planck} Collaboration, N.~Aghanim {\em et~al.}, ``{Planck 2018
  results. VI. Cosmological parameters},''
  \href{http://dx.doi.org/10.1051/0004-6361/201833910}{{\em Astron. Astrophys.}
  {\bfseries 641} (2020) A6}, \href{http://arxiv.org/abs/1807.06209}{{\ttfamily
  arXiv:1807.06209 [astro-ph.CO]}}.

\bibitem{Planck2018_inflation}
{\bfseries Planck} Collaboration, Y.~Akrami {\em et~al.}, ``{Planck 2018
  results. X. Constraints on inflation},''
  \href{http://dx.doi.org/10.1051/0004-6361/201833887}{{\em Astron. Astrophys.}
  {\bfseries 641} (2020) A10},
  \href{http://arxiv.org/abs/1807.06211}{{\ttfamily arXiv:1807.06211
  [astro-ph.CO]}}.

\bibitem{LQCreview_Bojowald2005}
M.~Bojowald, ``{Loop quantum cosmology},''
  \href{http://dx.doi.org/10.12942/lrr-2005-11}{{\em Living Rev. Rel.}
  {\bfseries 8} (2005) 11},
  \href{http://arxiv.org/abs/gr-qc/0601085}{{\ttfamily arXiv:gr-qc/0601085}}.

\bibitem{LQCreview_Ashtekar2011}
A.~Ashtekar and P.~Singh, ``{Loop Quantum Cosmology: A Status Report},'' {\em
  Class. Quant. Grav.} {\bfseries 28} (2011) 213001,
\href{http://arxiv.org/abs/1108.0893}{{\ttfamily arXiv:1108.0893 [gr-qc]}}.

\bibitem{LQCreview_Banerjee2012}
K.~Banerjee, G.~Calcagni, and M.~Martin-Benito, ``{Introduction to loop quantum
  cosmology},'' {\em SIGMA} {\bfseries 8} (2012) 016,
\href{http://arxiv.org/abs/1109.6801}{{\ttfamily arXiv:1109.6801 [gr-qc]}}.

\bibitem{LQCreview_Agullo2016}
I.~Agullo and P.~Singh, ``{Loop Quantum Cosmology},'' in {\em Loop Quantum
  Gravity: The First 30 Years}, A.~Ashtekar and J.~Pullin, eds., pp.~183--240.
\newblock WSP, 2017.
\newblock
\href{http://arxiv.org/abs/1612.01236}{{\ttfamily arXiv:1612.01236 [gr-qc]}}.
\newblock

\bibitem{APS_PRL}
A.~Ashtekar, T.~Pawlowski, and P.~Singh, ``{Quantum nature of the big bang},''
  {\em Phys. Rev. Lett.} {\bfseries 96} (2006) 141301,
\href{http://arxiv.org/abs/gr-qc/0602086}{{\ttfamily arXiv:gr-qc/0602086
  [gr-qc]}}.

\bibitem{APS_extended}
A.~Ashtekar, T.~Pawlowski, and P.~Singh, ``{Quantum Nature of the Big Bang:
  Improved dynamics},'' {\em Phys. Rev.} {\bfseries D74} (2006) 084003,
\href{http://arxiv.org/abs/gr-qc/0607039}{{\ttfamily arXiv:gr-qc/0607039
  [gr-qc]}}.

\bibitem{ACS}
A.~Ashtekar, A.~Corichi, and P.~Singh, ``{Robustness of key features of loop
  quantum cosmology},'' {\em Phys. Rev.} {\bfseries D77} (2008) 024046,
\href{http://arxiv.org/abs/0710.3565}{{\ttfamily arXiv:0710.3565 [gr-qc]}}.

\bibitem{Bentivegna2008}
E.~Bentivegna and T.~Pawlowski, ``{Anti-deSitter universe dynamics in LQC},''
  {\em Phys. Rev.} {\bfseries D77} (2008) 124025,
\href{http://arxiv.org/abs/0803.4446}{{\ttfamily arXiv:0803.4446 [gr-qc]}}.

\bibitem{Kaminski2009}
W.~Kaminski and T.~Pawlowski, ``{The LQC evolution operator of FRW universe
  with positive cosmological constant},'' {\em Phys. Rev.} {\bfseries D81}
  (2010) 024014,
\href{http://arxiv.org/abs/0912.0162}{{\ttfamily arXiv:0912.0162 [gr-qc]}}.

\bibitem{Pawlowski2012}
T.~Pawlowski and A.~Ashtekar, ``{Positive cosmological constant in loop quantum
  cosmology},'' {\em Phys. Rev.} {\bfseries D85} (2012) 064001,
\href{http://arxiv.org/abs/1112.0360}{{\ttfamily arXiv:1112.0360 [gr-qc]}}.

\bibitem{K=1Cosmologies}
A.~Ashtekar, T.~Pawlowski, P.~Singh, and K.~Vandersloot, ``{Loop quantum
  cosmology of k=1 FRW models},'' {\em Phys. Rev.} {\bfseries D75} (2007)
  024035,
\href{http://arxiv.org/abs/gr-qc/0612104}{{\ttfamily arXiv:gr-qc/0612104
  [gr-qc]}}.

\bibitem{closedFRWcosmologies}
L.~Szulc, W.~Kaminski, and J.~Lewandowski, ``{Closed FRW model in Loop Quantum
  Cosmology},'' {\em Class. Quant. Grav.} {\bfseries 24} (2007) 2621--2636,
\href{http://arxiv.org/abs/gr-qc/0612101}{{\ttfamily arXiv:gr-qc/0612101
  [gr-qc]}}.

\bibitem{Merce_bianchiI}
M.~Mart\'in-Benito, G.~A. Mena~Marug\'an, and T.~Pawlowski, ``{Loop
  Quantization of Vacuum Bianchi I Cosmology},'' {\em Phys. Rev.} {\bfseries
  D78} (2008) 064008,
\href{http://arxiv.org/abs/0804.3157}{{\ttfamily arXiv:0804.3157 [gr-qc]}}.

\bibitem{Ashtekar_bianchiI}
A.~Ashtekar and E.~Wilson-Ewing, ``{Loop quantum cosmology of Bianchi I
  models},'' \href{http://dx.doi.org/10.1103/PhysRevD.79.083535}{{\em Phys.
  Rev. D} {\bfseries 79} (2009) 083535},
  \href{http://arxiv.org/abs/0903.3397}{{\ttfamily arXiv:0903.3397 [gr-qc]}}.

\bibitem{Ashtekar_bianchiII}
A.~Ashtekar and E.~Wilson-Ewing, ``{Loop quantum cosmology of Bianchi type II
  models},'' {\em Phys. Rev.} {\bfseries D80} (2009) 123532,
\href{http://arxiv.org/abs/0910.1278}{{\ttfamily arXiv:0910.1278 [gr-qc]}}.

\bibitem{WilsonEwing_bianchiIX}
E.~Wilson-Ewing, ``{Loop quantum cosmology of Bianchi type IX models},''
  \href{http://dx.doi.org/10.1103/PhysRevD.82.043508}{{\em Phys. Rev. D}
  {\bfseries 82} (2010) 043508},
  \href{http://arxiv.org/abs/1005.5565}{{\ttfamily arXiv:1005.5565 [gr-qc]}}.

\bibitem{Taveras_2008}
V.~Taveras, ``{Corrections to the Friedmann Equations from LQG for a Universe
  with a Free Scalar Field},''
  \href{http://dx.doi.org/10.1103/PhysRevD.78.064072}{{\em Phys. Rev. D}
  {\bfseries 78} (2008) 064072},
  \href{http://arxiv.org/abs/0807.3325}{{\ttfamily arXiv:0807.3325 [gr-qc]}}.

\bibitem{FernandezMendez2012}
M.~Fernandez-Mendez, G.~A. Mena~Marug\'an, and J.~Olmedo, ``{Hybrid
  quantization of an inflationary universe},''
  \href{http://dx.doi.org/10.1103/PhysRevD.86.024003}{{\em Phys. Rev. D}
  {\bfseries 86} (2012) 024003},
  \href{http://arxiv.org/abs/1205.1917}{{\ttfamily arXiv:1205.1917 [gr-qc]}}.

\bibitem{FernandezMendez2013}
M.~Fern\'andez-M\'endez, G.~A. Mena~Marug\'an, and J.~Olmedo, ``{Hybrid
  quantization of an inflationary model: The flat case},''
  \href{http://dx.doi.org/10.1103/PhysRevD.88.044013}{{\em Phys. Rev. D}
  {\bfseries 88} no.~4, (2013) 044013},
  \href{http://arxiv.org/abs/1307.5222}{{\ttfamily arXiv:1307.5222 [gr-qc]}}.

\bibitem{FernandezMendez2014}
M.~Fern\'andez-M\'endez, G.~A. Mena~Marug\'an, and J.~Olmedo, ``{Effective
  dynamics of scalar perturbations in a flat Friedmann-Robertson-Walker
  spacetime in Loop Quantum Cosmology},''
  \href{http://dx.doi.org/10.1103/PhysRevD.89.044041}{{\em Phys. Rev. D}
  {\bfseries 89} no.~4, (2014) 044041},
  \href{http://arxiv.org/abs/1401.5256}{{\ttfamily arXiv:1401.5256 [gr-qc]}}.

\bibitem{CastelloGomar2014}
L.~Castell\'o~Gomar, M.~Fern\'andez-M\'endez, G.~A. Mena~Marug\'an, and
  J.~Olmedo, ``{Cosmological perturbations in Hybrid Loop Quantum Cosmology:
  Mukhanov-Sasaki variables},''
  \href{http://dx.doi.org/10.1103/PhysRevD.90.064015}{{\em Phys. Rev. D}
  {\bfseries 90} no.~6, (2014) 064015},
  \href{http://arxiv.org/abs/1407.0998}{{\ttfamily arXiv:1407.0998 [gr-qc]}}.

\bibitem{CastelloGomar2015}
L.~Castell\'o~Gomar, M.~Mart\'\i{}n-Benito, and G.~A. Mena~Marug\'an,
  ``{Gauge-Invariant Perturbations in Hybrid Quantum Cosmology},''
  \href{http://dx.doi.org/10.1088/1475-7516/2015/06/045}{{\em JCAP} {\bfseries
  06} (2015) 045}, \href{http://arxiv.org/abs/1503.03907}{{\ttfamily
  arXiv:1503.03907 [gr-qc]}}.

\bibitem{CastelloGomar2016}
L.~Castell\'o~Gomar, M.~Mart\'\i{}n-Benito, and G.~A. Mena~Marug\'an,
  ``{Quantum corrections to the Mukhanov-Sasaki equations},''
  \href{http://dx.doi.org/10.1103/PhysRevD.93.104025}{{\em Phys. Rev. D}
  {\bfseries 93} no.~10, (2016) 104025},
  \href{http://arxiv.org/abs/1603.08448}{{\ttfamily arXiv:1603.08448 [gr-qc]}}.

\bibitem{hybrid_Martinez2016}
F.~B. Mart\'\i{}nez and J.~Olmedo, ``{Primordial tensor modes of the early
  Universe},'' \href{http://dx.doi.org/10.1103/PhysRevD.93.124008}{{\em Phys.
  Rev. D} {\bfseries 93} no.~12, (2016) 124008},
  \href{http://arxiv.org/abs/1605.04293}{{\ttfamily arXiv:1605.04293 [gr-qc]}}.

\bibitem{hyb-vs-dress}
B.~Elizaga~Navascu\'es, D.~Martin~de Blas, and G.~A. Mena~Marug\'an,
  ``{Time-dependent mass of cosmological perturbations in the hybrid and
  dressed metric approaches to loop quantum cosmology},''
  \href{http://dx.doi.org/10.1103/PhysRevD.97.043523}{{\em Phys. Rev. D}
  {\bfseries 97} no.~4, (2018) 043523},
  \href{http://arxiv.org/abs/1711.10861}{{\ttfamily arXiv:1711.10861 [gr-qc]}}.

\bibitem{Ashtekar:2009mb}
A.~Ashtekar, W.~Kaminski, and J.~Lewandowski, ``{Quantum field theory on a
  cosmological, quantum space-time},''
  \href{http://dx.doi.org/10.1103/PhysRevD.79.064030}{{\em Phys. Rev. D}
  {\bfseries 79} (2009) 064030},
  \href{http://arxiv.org/abs/0901.0933}{{\ttfamily arXiv:0901.0933 [gr-qc]}}.

\bibitem{dressedmetric_Agullo2012}
I.~Agullo, A.~Ashtekar, and W.~Nelson, ``Quantum gravity extension of the
  inflationary scenario,''
  \href{http://dx.doi.org/10.1103/PhysRevLett.109.251301}{{\em Phys. Rev.
  Lett.} {\bfseries 109} (2012) 251301},
  \href{http://arxiv.org/abs/1209.1609}{{\ttfamily arXiv:1209.1609 [gr-qc]}}.

\bibitem{dressedmetric_Agullo_PRD_2013}
I.~Agullo, A.~Ashtekar, and W.~Nelson, ``Extension of the quantum theory of
  cosmological perturbations to the planck era,''
  \href{http://dx.doi.org/10.1103/PhysRevD.87.043507}{{\em Phys. Rev. D}
  {\bfseries 87} (2013) 043507},
  \href{http://arxiv.org/abs/1211.1354}{{\ttfamily arXiv:1211.1354 [gr-qc]}}.

\bibitem{dressedmetric_Agullo_CQG_2013}
I.~Agullo, A.~Ashtekar, and W.~Nelson, ``The pre-inflationary dynamics of loop
  quantum cosmology: confronting quantum gravity with observations,''
  \href{http://dx.doi.org/10.1088/0264-9381/30/8/085014}{{\em Classical and
  Quantum Gravity} {\bfseries 30} no.~8, (2013) 085014},
  \href{http://arxiv.org/abs/1302.0254}{{\ttfamily arXiv:1302.0254 [gr-qc]}}.

\bibitem{Merce_2008Hybrid}
M.~Mart\'{\i}n-Benito, L.~J. Garay, and G.~A. Mena~Marug\'an, ``{Hybrid quantum
  Gowdy cosmology: Combining loop and Fock quantizations},'' {\em Phys. Rev. D}
  {\bfseries 78} (2008) 083516,
  \href{http://arxiv.org/abs/0804.1098}{{\ttfamily arXiv:0804.1098 [gr-qc]}}.

\bibitem{Garay_2008hybrid}
L.~J. Garay, M.~Mart\'{\i}n-Benito, and G.~A. Mena~Marug\'an, ``{Inhomogeneous
  loop quantum cosmology: Hybrid quantization of the Gowdy model},'' {\em Phys.
  Rev. D} {\bfseries 82} (2010) 044048,
  \href{http://arxiv.org/abs/1005.5654}{{\ttfamily arXiv:1005.5654 [gr-qc]}}.

\bibitem{Merce_2010hybrid}
M.~Mart\'{\i}n-Benito, G.~A. Mena~Marug\'an, and E.~Wilson-Ewing, ``Hybrid
  quantization: From bianchi i to the gowdy model,'' {\em Phys. Rev. D}
  {\bfseries 82} (2010) 084012,
  \href{http://arxiv.org/abs/1006.2369}{{\ttfamily arXiv:1006.2369 [gr-qc]}}.

\bibitem{parker1969}
L.~Parker, ``Quantized fields and particle creation in expanding universes.
  i,'' {\em Physical Review} {\bfseries 183} no.~5, (1969) 1057.

\bibitem{fulling1979}
S.~Fulling, ``Remarks on positive frequency and hamiltonians in expanding
  universes,'' {\em General Relativity and Gravitation} {\bfseries 10} no.~10,
  (1979) 807--824.

\bibitem{Fahn:2018}
M.~J. Fahn, K.~Giesel, and M.~Kobler, ``{Dynamical Properties of the
  Mukhanov-Sasaki Hamiltonian in the context of adiabatic vacua and the
  Lewis-Riesenfeld invariant},''
  \href{http://dx.doi.org/10.3390/universe5070170}{{\em Universe} {\bfseries 5}
  no.~7, (2019) 170}, \href{http://arxiv.org/abs/1812.11122}{{\ttfamily
  arXiv:1812.11122 [gr-qc]}}.

\bibitem{Elizaga2019}
B.~Elizaga~Navascu\'es, G.~A. Mena~Marug\'an, and T.~Thiemann, ``{Hamiltonian
  diagonalization in hybrid quantum cosmology},''
  \href{http://dx.doi.org/10.1088/1361-6382/ab32af}{{\em Class. Quant. Grav.}
  {\bfseries 36} (2019) 185010},
  \href{http://arxiv.org/abs/1903.05695}{{\ttfamily arXiv:1903.05695 [gr-qc]}}.

\bibitem{agullo2015}
I.~Agullo, W.~Nelson, and A.~Ashtekar, ``Preferred instantaneous vacuum for
  linear scalar fields in cosmological space-times,'' {\em Physical Review D}
  {\bfseries 91} no.~6, (2015) 064051,
  \href{http://arxiv.org/abs/1412.3524}{{\ttfamily arXiv:1412.3524 [gr-qc]}}.

\bibitem{handley2016}
W.~Handley, A.~Lasenby, and M.~Hobson, ``Novel quantum initial conditions for
  inflation,'' {\em Physical Review D} {\bfseries 94} no.~2, (2016) 024041,
  \href{http://arxiv.org/abs/1607.04148}{{\ttfamily arXiv:1607.04148 [gr-qc]}}.

\bibitem{danielsson2002}
U.~H. Danielsson, ``Note on inflation and trans-planckian physics,'' {\em
  Physical Review D} {\bfseries 66} no.~2, (2002) 023511,
  \href{http://arxiv.org/abs/hep-th/0203198}{{\ttfamily arXiv:hep-th/0203198}}.

\bibitem{Ashtekar:2016}
A.~Ashtekar and B.~Gupt, ``{Initial conditions for cosmological
  perturbations},'' \href{http://dx.doi.org/10.1088/1361-6382/aa52d4}{{\em
  Class. Quant. Grav.} {\bfseries 34} no.~3, (2017) 035004},
  \href{http://arxiv.org/abs/1610.09424}{{\ttfamily arXiv:1610.09424 [gr-qc]}}.

\bibitem{nonosc}
D.~M. de~Blas and J.~Olmedo, ``{Primordial power spectra for scalar
  perturbations in loop quantum cosmology},''
  \href{http://dx.doi.org/10.1088/1475-7516/2016/06/029}{{\em JCAP} {\bfseries
  06} (2016) 029}, \href{http://arxiv.org/abs/1601.01716}{{\ttfamily
  arXiv:1601.01716 [gr-qc]}}.

\bibitem{hyb-obs}
L.~Castell\'o~Gomar, G.~A. Mena~Marug\'an, D.~Mart\'\i{}n De~Blas, and
  J.~Olmedo, ``{Hybrid loop quantum cosmology and predictions for the cosmic
  microwave background},''
  \href{http://dx.doi.org/10.1103/PhysRevD.96.103528}{{\em Phys. Rev. D}
  {\bfseries 96} no.~10, (2017) 103528},
  \href{http://arxiv.org/abs/1702.06036}{{\ttfamily arXiv:1702.06036 [gr-qc]}}.

\bibitem{menava}
B.~Elizaga~Navascu\'es, G.~A. Mena~Marug\'an, and S.~Prado, ``{Non-oscillating
  power spectra in Loop Quantum Cosmology},''
  \href{http://dx.doi.org/10.1088/1361-6382/abc6bb}{{\em Class. Quant. Grav.}
  {\bfseries 38} no.~3, (2020) 035001},
  \href{http://arxiv.org/abs/2005.10194}{{\ttfamily arXiv:2005.10194 [gr-qc]}}.

\bibitem{Navascues:2021}
B.~E. Navascu\'es and G.~A. Mena~Marug\'an, ``{Analytical investigation of
  pre-inflationary effects in the primordial power spectrum: from General
  Relativity to hybrid Loop Quantum Cosmology},''
  \href{http://dx.doi.org/10.1088/1475-7516/2021/09/030}{{\em JCAP} {\bfseries
  09} (2021) 030}, \href{http://arxiv.org/abs/2104.15002}{{\ttfamily
  arXiv:2104.15002 [gr-qc]}}.

\bibitem{Olbermann2007}
H.~Olbermann, ``{States of low energy on Robertson-Walker spacetimes},''
  \href{http://dx.doi.org/10.1088/0264-9381/24/20/007}{{\em Class. Quant.
  Grav.} {\bfseries 24} (2007) 5011--5030},
  \href{http://arxiv.org/abs/0704.2986}{{\ttfamily arXiv:0704.2986 [gr-qc]}}.

\bibitem{Niedermaier2020}
R.~Banerjee and M.~Niedermaier, ``{Bonus Properties of States of Low Energy},''
  \href{http://dx.doi.org/10.1063/5.0019311}{{\em J. Math. Phys.} {\bfseries
  61} (2020) 103511}, \href{http://arxiv.org/abs/2006.08685}{{\ttfamily
  arXiv:2006.08685 [math-ph]}}.

\bibitem{AshtekarSloan_probinflation}
A.~Ashtekar and D.~Sloan, ``{Probability of Inflation in Loop Quantum
  Cosmology},'' \href{http://dx.doi.org/10.1007/s10714-011-1246-y}{{\em Gen.
  Rel. Grav.} {\bfseries 43} (2011) 3619--3655},
  \href{http://arxiv.org/abs/1103.2475}{{\ttfamily arXiv:1103.2475 [gr-qc]}}.

\bibitem{Li2021}
B.-F. Li, P.~Singh, and A.~Wang, ``{Phenomenological implications of modified
  loop cosmologies: an overview},''
  \href{http://dx.doi.org/10.3389/fspas.2021.701417}{{\em Front. Astron. Space
  Sci.} {\bfseries 8} (2021) 701417},
  \href{http://arxiv.org/abs/2105.14067}{{\ttfamily arXiv:2105.14067 [gr-qc]}}.

\bibitem{Agullo2023}
I.~Agull\'o, A.~Wang, and E.~Wilson-Ewing, ``{Loop quantum cosmology: relation
  between theory and observations},''
  \href{http://arxiv.org/abs/2301.10215}{{\ttfamily arXiv:2301.10215 [gr-qc]}}.

\bibitem{Ashtekar:2020}
A.~Ashtekar, B.~Gupt, D.~Jeong, and V.~Sreenath, ``{Alleviating the Tension in
  the Cosmic Microwave Background using Planck-Scale Physics},''
  \href{http://dx.doi.org/10.1103/PhysRevLett.125.051302}{{\em Phys. Rev.
  Lett.} {\bfseries 125} no.~5, (2020) 051302},
  \href{http://arxiv.org/abs/2001.11689}{{\ttfamily arXiv:2001.11689
  [astro-ph.CO]}}.

\bibitem{Ashtekar:2021}
A.~Ashtekar, B.~Gupt, and V.~Sreenath, ``{Cosmic Tango Between the Very Small
  and the Very Large: Addressing CMB Anomalies Through Loop Quantum
  Cosmology},'' \href{http://dx.doi.org/10.3389/fspas.2021.685288}{{\em Front.
  Astron. Space Sci.} {\bfseries 8} (2021) 76},
  \href{http://arxiv.org/abs/2103.14568}{{\ttfamily arXiv:2103.14568 [gr-qc]}}.

\bibitem{Agullo:2020b}
I.~Agullo, D.~Kranas, and V.~Sreenath, ``{Anomalies in the CMB from a cosmic
  bounce},'' \href{http://dx.doi.org/10.1007/s10714-020-02778-9}{{\em Gen. Rel.
  Grav.} {\bfseries 53} no.~2, (2021) 17},
  \href{http://arxiv.org/abs/2005.01796}{{\ttfamily arXiv:2005.01796
  [astro-ph.CO]}}.

\bibitem{Agullo2021}
I.~Agullo, D.~Kranas, and V.~Sreenath, ``{Anomalies in the Cosmic Microwave
  Background and Their Non-Gaussian Origin in Loop Quantum Cosmology},''
  \href{http://dx.doi.org/10.3389/fspas.2021.703845}{{\em Front. Astron. Space
  Sci.} {\bfseries 8} (2021) 703845},
  \href{http://arxiv.org/abs/2105.12993}{{\ttfamily arXiv:2105.12993 [gr-qc]}}.

\bibitem{Copi2008}
C.~J. Copi, D.~Huterer, D.~J. Schwarz, and G.~D. Starkman, ``{No large-angle
  correlations on the non-Galactic microwave sky},''
  \href{http://dx.doi.org/10.1111/j.1365-2966.2009.15270.x}{{\em Mon. Not. Roy.
  Astron. Soc.} {\bfseries 399} (2009) 295--303},
  \href{http://arxiv.org/abs/0808.3767}{{\ttfamily arXiv:0808.3767
  [astro-ph]}}.

\bibitem{Copi2013}
C.~J. Copi, D.~Huterer, D.~J. Schwarz, and G.~D. Starkman, ``{Lack of
  large-angle TT correlations persists in WMAP and Planck},''
  \href{http://dx.doi.org/10.1093/mnras/stv1143}{{\em Mon. Not. Roy. Astron.
  Soc.} {\bfseries 451} no.~3, (2015) 2978--2985},
  \href{http://arxiv.org/abs/1310.3831}{{\ttfamily arXiv:1310.3831
  [astro-ph.CO]}}.

\bibitem{Copi2018}
C.~J. Copi, J.~Gurian, A.~Kosowsky, G.~D. Starkman, and H.~Zhang, ``{Exploring
  suppressed long-distance correlations as the cause of suppressed large-angle
  correlations},'' \href{http://dx.doi.org/10.1093/mnras/stz2962}{{\em Mon.
  Not. Roy. Astron. Soc.} {\bfseries 490} no.~4, (2019) 5174--5181},
  \href{http://arxiv.org/abs/1812.03946}{{\ttfamily arXiv:1812.03946
  [astro-ph.CO]}}.

\bibitem{Weinberg2008}
S.~Weinberg, {\em Cosmology}.
\newblock Oxford University Press, Oxford, 2008.

\bibitem{Bastero-Gil:2009sdq}
M.~Bastero-Gil and A.~Berera, ``{Warm inflation model building},''
  \href{http://dx.doi.org/10.1142/S0217751X09044206}{{\em Int. J. Mod. Phys. A}
  {\bfseries 24} (2009) 2207--2240},
  \href{http://arxiv.org/abs/0902.0521}{{\ttfamily arXiv:0902.0521 [hep-ph]}}.

\bibitem{Riotto:2002yw}
A.~Riotto, ``{Inflation and the theory of cosmological perturbations},'' {\em
  ICTP Lect. Notes Ser.} {\bfseries 14} (2003) 317--413,
  \href{http://arxiv.org/abs/hep-ph/0210162}{{\ttfamily arXiv:hep-ph/0210162}}.

\bibitem{Langlois:2010xc}
D.~Langlois, ``{Lectures on inflation and cosmological perturbations},''
  \href{http://dx.doi.org/10.1007/978-3-642-10598-2_1}{{\em Lect. Notes Phys.}
  {\bfseries 800} (2010) 1--57},
  \href{http://arxiv.org/abs/1001.5259}{{\ttfamily arXiv:1001.5259
  [astro-ph.CO]}}.

\bibitem{Mukhanov:1988jd}
V.~F. Mukhanov, ``{Quantum Theory of Gauge Invariant Cosmological
  Perturbations},'' {\em Sov. Phys. JETP} {\bfseries 67} (1988) 1297--1302.

\bibitem{Allen:1985ux}
B.~Allen, ``{Vacuum States in de Sitter Space},''
  \href{http://dx.doi.org/10.1103/PhysRevD.32.3136}{{\em Phys. Rev. D}
  {\bfseries 32} (1985) 3136}.

\bibitem{Mottola2000}
S.~Habib, C.~Molina-Paris, and E.~Mottola, ``{Energy momentum tensor of
  particles created in an expanding universe},''
  \href{http://dx.doi.org/10.1103/PhysRevD.61.024010}{{\em Phys. Rev. D}
  {\bfseries 61} (2000) 024010},
  \href{http://arxiv.org/abs/gr-qc/9906120}{{\ttfamily arXiv:gr-qc/9906120}}.

\bibitem{Fewster2000}
C.~J. Fewster, ``A general worldline quantum inequality,''
  \href{http://dx.doi.org/10.1088/0264-9381/17/9/302}{{\em Classical and
  Quantum Gravity} {\bfseries 17} no.~9, (Apr, 2000) 1897–1911},
  \href{http://arxiv.org/abs/gr-qc/9910060}{{\ttfamily arXiv:gr-qc/9910060}}.

\bibitem{DegnerPhD2013}
A.~Degner and DESY, \href{http://dx.doi.org/10.3204/DESY-THESIS-2013-002}{{\em
  {P}roperties of {S}tates of {L}ow {E}nergy on {C}osmological {S}pacetimes}}.
\newblock Dr., Universit\"at Hamburg, 2013.
\newblock \url{https://bib-pubdb1.desy.de/record/145520}.
\newblock Universit\"at Hamburg, Diss., 2013.

\bibitem{Tristram:2021tvh}
M.~Tristram {\em et~al.}, ``{Improved limits on the tensor-to-scalar ratio
  using BICEP and Planck data},''
  \href{http://dx.doi.org/10.1103/PhysRevD.105.083524}{{\em Phys. Rev. D}
  {\bfseries 105} no.~8, (2022) 083524},
  \href{http://arxiv.org/abs/2112.07961}{{\ttfamily arXiv:2112.07961
  [astro-ph.CO]}}.

\bibitem{Fixsen:1996nj}
D.~J. Fixsen, E.~S. Cheng, J.~M. Gales, J.~C. Mather, R.~A. Shafer, and E.~L.
  Wright, ``{The Cosmic Microwave Background spectrum from the full COBE FIRAS
  data set},'' \href{http://dx.doi.org/10.1086/178173}{{\em Astrophys. J.}
  {\bfseries 473} (1996) 576},
  \href{http://arxiv.org/abs/astro-ph/9605054}{{\ttfamily
  arXiv:astro-ph/9605054}}.

\bibitem{Blas_2011}
D.~Blas, J.~Lesgourgues, and T.~Tram, ``{The Cosmic Linear Anisotropy Solving
  System ({CLASS}). Part {II}: Approximation schemes},''
  \href{http://dx.doi.org/10.1088/1475-7516/2011/07/034}{{\em JCAP} {\bfseries
  2011} no.~07, (2011) 034--034},
  \href{http://arxiv.org/abs/1104.2933}{{\ttfamily 1104.2933}}.

\bibitem{COBE}
G.~Hinshaw, A.~J. Banday, C.~L. Bennett, K.~M. Gorski, A.~Kogut, C.~H.
  Lineweaver, G.~F. Smoot, and E.~L. Wright, ``{2-point correlations in the
  COBE DMR 4-year anisotropy maps},''
  \href{http://dx.doi.org/10.1086/310076}{{\em Astrophys. J. Lett.} {\bfseries
  464} (1996) L25--L28},
  \href{http://arxiv.org/abs/astro-ph/9601061}{{\ttfamily
  arXiv:astro-ph/9601061}}.

\bibitem{PlanckVII}
{\bfseries Planck} Collaboration, Y.~Akrami {\em et~al.}, ``{Planck 2018
  results. VII. Isotropy and Statistics of the CMB},''
  \href{http://dx.doi.org/10.1051/0004-6361/201935201}{{\em Astron. Astrophys.}
  {\bfseries 641} (2020) A7}, \href{http://arxiv.org/abs/1906.02552}{{\ttfamily
  arXiv:1906.02552 [astro-ph.CO]}}.

\bibitem{DiValentino2019}
E.~Di~Valentino, A.~Melchiorri, and J.~Silk, ``{Planck evidence for a closed
  Universe and a possible crisis for cosmology},''
  \href{http://dx.doi.org/10.1038/s41550-019-0906-9}{{\em Nature Astron.}
  {\bfseries 4} no.~2, (2019) 196--203},
  \href{http://arxiv.org/abs/1911.02087}{{\ttfamily arXiv:1911.02087
  [astro-ph.CO]}}.

\bibitem{Arnowitt:1962hi}
R.~L. Arnowitt, S.~Deser, and C.~W. Misner, ``{The Dynamics of general
  relativity},'' \href{http://dx.doi.org/10.1007/s10714-008-0661-1}{{\em Gen.
  Rel. Grav.} {\bfseries 40} (2008) 1997--2027},
  \href{http://arxiv.org/abs/gr-qc/0405109}{{\ttfamily arXiv:gr-qc/0405109}}.

\bibitem{Thiemann_notes}
T.~Thiemann, {\em {Modern canonical quantum general relativity}}.
\newblock Cambridge University Press,
2008.
\newblock

\bibitem{Ashtekar_Bojowald_Lewandowski}
A.~Ashtekar, M.~Bojowald, and J.~Lewandowski, ``{Mathematical structure of loop
  quantum cosmology},''
  \href{http://dx.doi.org/10.4310/ATMP.2003.v7.n2.a2}{{\em Adv. Theor. Math.
  Phys.} {\bfseries 7} no.~2, (2003) 233--268},
  \href{http://arxiv.org/abs/gr-qc/0304074}{{\ttfamily arXiv:gr-qc/0304074}}.

\bibitem{Merce_Guillermo_Javier}
M.~Martin-Benito, G.~A. Mena~Marug\'an, and J.~Olmedo, ``{Further Improvements
  in the Understanding of Isotropic Loop Quantum Cosmology},''
  \href{http://dx.doi.org/10.1103/PhysRevD.80.104015}{{\em Phys. Rev. D}
  {\bfseries 80} (2009) 104015},
  \href{http://arxiv.org/abs/0909.2829}{{\ttfamily arXiv:0909.2829 [gr-qc]}}.

\bibitem{Singh:2012zc}
P.~Singh, ``{Numerical loop quantum cosmology: an overview},''
  \href{http://dx.doi.org/10.1088/0264-9381/29/24/244002}{{\em Class. Quant.
  Grav.} {\bfseries 29} (2012) 244002},
  \href{http://arxiv.org/abs/1208.5456}{{\ttfamily arXiv:1208.5456 [gr-qc]}}.

\bibitem{Diener:2014mia}
P.~Diener, B.~Gupt, and P.~Singh, ``{Numerical simulations of a loop quantum
  cosmos: robustness of the quantum bounce and the validity of effective
  dynamics},'' \href{http://dx.doi.org/10.1088/0264-9381/31/10/105015}{{\em
  Class. Quant. Grav.} {\bfseries 31} (2014) 105015},
  \href{http://arxiv.org/abs/1402.6613}{{\ttfamily arXiv:1402.6613 [gr-qc]}}.

\bibitem{Liddle:2009xe}
A.~R. Liddle, ``{Statistical methods for cosmological parameter selection and
  estimation},''
  \href{http://dx.doi.org/10.1146/annurev.nucl.010909.083706}{{\em Ann. Rev.
  Nucl. Part. Sci.} {\bfseries 59} (2009) 95--114},
  \href{http://arxiv.org/abs/0903.4210}{{\ttfamily arXiv:0903.4210 [hep-th]}}.

\bibitem{hobson_2009}
M.~P. {Hobson}, A.~H. {Jaffe}, A.~R. {Liddle}, P.~{Mukherjee}, and
  D.~{Parkinson}, \href{http://dx.doi.org/10.1017/CBO9780511802461}{{\em
  {Bayesian Methods in Cosmology}}}.
\newblock Cambridge University Press, 2009.

\bibitem{ours_Lambda}
M.~Mart\'\i{}n-Benito and R.~B. Neves, ``{The Effect of a positive cosmological
  constant on the bounce of Loop Quantum Cosmology},''
  \href{http://dx.doi.org/10.3390/math8020186}{{\em Mathematics} {\bfseries 8}
  no.~2, (2020) 186}, \href{http://arxiv.org/abs/2001.02918}{{\ttfamily
  arXiv:2001.02918 [gr-qc]}}.

\bibitem{paper1_domainV}
M.~Mart\'\i{}n-Benito and R.~B. Neves, ``{Solvable Loop Quantum Cosmology:
  domain of the volume observable and semiclassical states},''
  \href{http://dx.doi.org/10.1103/PhysRevD.99.043525}{{\em Phys. Rev. D}
  {\bfseries 99} no.~4, (2019) 043525},
  \href{http://arxiv.org/abs/1901.05091}{{\ttfamily arXiv:1901.05091 [gr-qc]}}.

\bibitem{Assanioussi:2018}
M.~Assanioussi, A.~Dapor, K.~Liegener, and T.~Paw\l{}owski, ``{Emergent de
  Sitter Epoch of the Quantum Cosmos from Loop Quantum Cosmology},''
  \href{http://dx.doi.org/10.1103/PhysRevLett.121.081303}{{\em Phys. Rev.
  Lett.} {\bfseries 121} no.~8, (2018) 081303},
  \href{http://arxiv.org/abs/1801.00768}{{\ttfamily arXiv:1801.00768 [gr-qc]}}.

\bibitem{Liegener:2019}
K.~Liegener and P.~Singh, ``{Some physical implications of regularization
  ambiguities in SU(2) gauge-invariant loop quantum cosmology},''
  \href{http://dx.doi.org/10.1103/PhysRevD.100.124049}{{\em Phys. Rev. D}
  {\bfseries 100} no.~12, (2019) 124049},
  \href{http://arxiv.org/abs/1908.07543}{{\ttfamily arXiv:1908.07543 [gr-qc]}}.

\bibitem{Liegener:2019_2}
K.~Liegener and P.~Singh, ``{New Loop Quantum Cosmology Modifications from
  Gauge-covariant Fluxes},''
  \href{http://dx.doi.org/10.1103/PhysRevD.100.124048}{{\em Phys. Rev. D}
  {\bfseries 100} no.~12, (2019) 124048},
  \href{http://arxiv.org/abs/1908.07001}{{\ttfamily arXiv:1908.07001 [gr-qc]}}.

\bibitem{Li:2018}
B.-F. Li, P.~Singh, and A.~Wang, ``{Towards Cosmological Dynamics from Loop
  Quantum Gravity},'' \href{http://dx.doi.org/10.1103/PhysRevD.97.084029}{{\em
  Phys. Rev. D} {\bfseries 97} no.~8, (2018) 084029},
  \href{http://arxiv.org/abs/1801.07313}{{\ttfamily arXiv:1801.07313 [gr-qc]}}.

\bibitem{Kaminski:2008td}
W.~Kaminski, J.~Lewandowski, and T.~Pawlowski, ``{Physical time and other
  conceptual issues of QG on the example of LQC},''
  \href{http://dx.doi.org/10.1088/0264-9381/26/3/035012}{{\em Class. Quant.
  Grav.} {\bfseries 26} (2009) 035012},
  \href{http://arxiv.org/abs/0809.2590}{{\ttfamily arXiv:0809.2590 [gr-qc]}}.

\bibitem{SLEs_ours}
M.~Mart\'\i{}n-Benito, R.~B. Neves, and J.~Olmedo, ``{States of Low Energy in
  bouncing inflationary scenarios in Loop Quantum Cosmology},''
  \href{http://dx.doi.org/10.1103/PhysRevD.103.123524}{{\em Phys. Rev. D}
  {\bfseries 103} (2021) 123524},
  \href{http://arxiv.org/abs/2104.03035}{{\ttfamily arXiv:2104.03035 [gr-qc]}}.

\bibitem{mno-sles}
M.~Mart\'\i{}n-Benito, R.~B. Neves, and J.~Olmedo, ``{Non-Oscillatory Power
  Spectrum From States of Low Energy in Kinetically Dominated Early
  Universes},'' \href{http://dx.doi.org/10.3389/fspas.2021.702543}{{\em Front.
  Astron. Space Sci.} {\bfseries 0} (2021) 133},
  \href{http://arxiv.org/abs/2104.14850}{{\ttfamily arXiv:2104.14850 [gr-qc]}}.

\bibitem{Dapor:2017}
A.~Dapor and K.~Liegener, ``{Cosmological Effective Hamiltonian from full Loop
  Quantum Gravity Dynamics},''
  \href{http://dx.doi.org/10.1016/j.physletb.2018.09.005}{{\em Phys. Lett. B}
  {\bfseries 785} (2018) 506--510},
  \href{http://arxiv.org/abs/1706.09833}{{\ttfamily arXiv:1706.09833 [gr-qc]}}.

\bibitem{Assanioussi:2019}
M.~Assanioussi, A.~Dapor, K.~Liegener, and T.~Paw\l{}owski, ``{Emergent de
  Sitter epoch of the Loop Quantum Cosmos: a detailed analysis},''
  \href{http://dx.doi.org/10.1103/PhysRevD.100.084003}{{\em Phys. Rev. D}
  {\bfseries 100} no.~8, (2019) 084003},
  \href{http://arxiv.org/abs/1906.05315}{{\ttfamily arXiv:1906.05315 [gr-qc]}}.

\bibitem{quismondo:2019}
A.~Garc\'\i{}a-Quismondo and G.~A. Mena~Marug\'an, ``{The Martin-Benito-Mena
  Marugan-Olmedo prescription for the Dapor-Liegener model of Loop Quantum
  Cosmology},'' \href{http://dx.doi.org/10.1103/PhysRevD.99.083505}{{\em Phys.
  Rev. D} {\bfseries 99} no.~8, (2019) 083505},
  \href{http://arxiv.org/abs/1903.00265}{{\ttfamily arXiv:1903.00265 [gr-qc]}}.

\bibitem{Agullo:2018}
I.~Agullo, ``{Primordial power spectrum from the Dapor-Liegener model of loop
  quantum cosmology},'' \href{http://dx.doi.org/10.1007/s10714-018-2413-1}{{\em
  Gen. Rel. Grav.} {\bfseries 50} no.~7, (2018) 91},
  \href{http://arxiv.org/abs/1805.11356}{{\ttfamily arXiv:1805.11356 [gr-qc]}}.

\bibitem{Li:2019}
B.-F. Li, P.~Singh, and A.~Wang, ``{Primordial power spectrum from the dressed
  metric approach in loop cosmologies},''
  \href{http://dx.doi.org/10.1103/PhysRevD.101.086004}{{\em Phys. Rev. D}
  {\bfseries 101} no.~8, (2020) 086004},
  \href{http://arxiv.org/abs/1912.08225}{{\ttfamily arXiv:1912.08225 [gr-qc]}}.

\bibitem{Li:2020}
B.-F. Li, J.~Olmedo, P.~Singh, and A.~Wang, ``{Primordial scalar power spectrum
  from the hybrid approach in loop cosmologies},''
  \href{http://dx.doi.org/10.1103/PhysRevD.102.126025}{{\em Phys. Rev. D}
  {\bfseries 102} (2020) 126025},
  \href{http://arxiv.org/abs/2008.09135}{{\ttfamily arXiv:2008.09135 [gr-qc]}}.

\bibitem{Olmedo:2018}
J.~Olmedo and E.~Alesci, ``{Power spectrum of primordial perturbations for an
  emergent universe in quantum reduced loop gravity},''
  \href{http://dx.doi.org/10.1088/1475-7516/2019/04/030}{{\em JCAP} {\bfseries
  04} (2019) 030}, \href{http://arxiv.org/abs/1811.04327}{{\ttfamily
  arXiv:1811.04327 [gr-qc]}}.

\bibitem{BICEP:2021xfz}
{\bfseries BICEP, Keck} Collaboration, P.~A.~R. Ade {\em et~al.}, ``{Improved
  Constraints on Primordial Gravitational Waves using Planck, WMAP, and
  BICEP/Keck Observations through the 2018 Observing Season},''
  \href{http://dx.doi.org/10.1103/PhysRevLett.127.151301}{{\em Phys. Rev.
  Lett.} {\bfseries 127} no.~15, (2021) 151301},
  \href{http://arxiv.org/abs/2110.00483}{{\ttfamily arXiv:2110.00483
  [astro-ph.CO]}}.

\bibitem{Contaldi:2003zv}
C.~R. Contaldi, M.~Peloso, L.~Kofman, and A.~D. Linde, ``{Suppressing the lower
  multipoles in the CMB anisotropies},''
  \href{http://dx.doi.org/10.1088/1475-7516/2003/07/002}{{\em JCAP} {\bfseries
  07} (2003) 002}, \href{http://arxiv.org/abs/astro-ph/0303636}{{\ttfamily
  arXiv:astro-ph/0303636}}.

\bibitem{nonosc_BeaDaniGuillermo}
B.~Elizaga~Navascu\'es, D.~M. de~Blas, and G.~A. Mena~Marug\'an, ``{The Vacuum
  State of Primordial Fluctuations in Hybrid Loop Quantum Cosmology},''
  \href{http://dx.doi.org/10.3390/universe4100098}{{\em Universe} {\bfseries 4}
  no.~10, (2018) 98}, \href{http://arxiv.org/abs/1809.09874}{{\ttfamily
  arXiv:1809.09874 [gr-qc]}}.

\bibitem{Agullo:2020}
I.~Agullo, D.~Kranas, and V.~Sreenath, ``{Large scale anomalies in the CMB and
  non-Gaussianity in bouncing cosmologies},''
  \href{http://dx.doi.org/10.1088/1361-6382/abc521}{{\em Class. Quant. Grav.}
  {\bfseries 38} no.~6, (2021) 065010},
  \href{http://arxiv.org/abs/2006.09605}{{\ttfamily arXiv:2006.09605
  [astro-ph.CO]}}.

\bibitem{Iteanu2022}
S.~Iteanu and G.~A. Mena~Marug\'an, ``{Mass of Cosmological Perturbations in
  the Hybrid and Dressed Metric Formalisms of Loop Quantum Cosmology for the
  Starobinsky and Exponential Potentials},''
  \href{http://dx.doi.org/10.3390/universe8090463}{{\em Universe} {\bfseries 8}
  no.~9, (2022) 463}, \href{http://arxiv.org/abs/2208.01987}{{\ttfamily
  arXiv:2208.01987 [gr-qc]}}.

\bibitem{ElizagaNavascues2021}
B.~Elizaga~Navascu\'es, G.~A. Mena~Marug\'an, and S.~Prado, ``{Nonoscillating
  vacuum states and the quantum homogeneity and isotropy hypothesis in loop
  quantum cosmology},''
  \href{http://dx.doi.org/10.1103/PhysRevD.104.083541}{{\em Phys. Rev. D}
  {\bfseries 104} no.~8, (2021) 083541},
  \href{http://arxiv.org/abs/2107.11054}{{\ttfamily arXiv:2107.11054 [gr-qc]}}.

\bibitem{Brinckmann2018}
T.~Brinckmann and J.~Lesgourgues, ``{MontePython 3: boosted MCMC sampler and
  other features},'' \href{http://dx.doi.org/10.1016/j.dark.2018.100260}{{\em
  Phys. Dark Univ.} {\bfseries 24} (2019) 100260},
\href{http://arxiv.org/abs/1804.07261}{{\ttfamily arXiv:1804.07261
  [astro-ph.CO]}}.

\bibitem{Audren2012}
B.~Audren, J.~Lesgourgues, K.~Benabed, and S.~Prunet, ``{Conservative
  Constraints on Early Cosmology: an illustration of the Monte Python
  cosmological parameter inference code},''
  \href{http://dx.doi.org/10.1088/1475-7516/2013/02/001}{{\em JCAP} {\bfseries
  1302} (2013) 001},
\href{http://arxiv.org/abs/1210.7183}{{\ttfamily arXiv:1210.7183
  [astro-ph.CO]}}.

\bibitem{PlanckV}
{\bfseries Planck} Collaboration, N.~Aghanim {\em et~al.}, ``{Planck 2018
  results. V. CMB power spectra and likelihoods},''
  \href{http://dx.doi.org/10.1051/0004-6361/201936386}{{\em Astron. Astrophys.}
  {\bfseries 641} (2020) A5}, \href{http://arxiv.org/abs/1907.12875}{{\ttfamily
  arXiv:1907.12875 [astro-ph.CO]}}.

\bibitem{PlanckVIII}
{\bfseries Planck} Collaboration, N.~Aghanim {\em et~al.}, ``{Planck 2018
  results. VIII. Gravitational lensing},''
  \href{http://dx.doi.org/10.1051/0004-6361/201833886}{{\em Astron. Astrophys.}
  {\bfseries 641} (2020) A8}, \href{http://arxiv.org/abs/1807.06210}{{\ttfamily
  arXiv:1807.06210 [astro-ph.CO]}}.

\bibitem{Gupt2017}
B.~Gupt, ``{Constraining Planck scale physics with CMB and Reionization Optical
  Depth},'' \href{http://arxiv.org/abs/1710.00759}{{\ttfamily arXiv:1710.00759
  [astro-ph.CO]}}.

\bibitem{Ford1987}
L.~H. Ford, ``Gravitational particle creation and inflation,''
  \href{http://dx.doi.org/10.1103/PhysRevD.35.2955}{{\em Phys. Rev. D}
  {\bfseries 35} (1987) 2955}.

\bibitem{Sauter1931}
F.~Sauter, ``{\"Uber das Verhalten eines Elektrons im homogenen elektrischen
  Feld nach der relativistischen Theorie Diracs},''
  \href{http://dx.doi.org/10.1007/BF01339461}{{\em Z. Phys.} {\bfseries 69}
  (1931) 742}.

\bibitem{Schwinger1951}
J.~S. Schwinger, ``On gauge invariance and vacuum polarization,''
  \href{http://dx.doi.org/10.1103/PhysRev.82.664}{{\em Phys. Rev.} {\bfseries
  82} (1951) 664}.

\bibitem{Pla2021}
S.~Pla, I.~M. Newsome, R.~S. Link, P.~R. Anderson, and J.~Navarro-Salas, ``Pair
  production due to an electric field in $1+1$ dimensions and the validity of
  the semiclassical approximation,''
  \href{http://dx.doi.org/10.1103/PhysRevD.103.105003}{{\em Phys. Rev. D}
  {\bfseries 103} (2021) 105003},
  \href{http://arxiv.org/abs/2010.09811}{{\ttfamily arXiv:2010.09811}}.

\bibitem{Pla2022}
J.~Navarro-Salas and S.~Pla, ``{Particle Creation and the Schwinger Model},''
  \href{http://dx.doi.org/10.3390/sym14112435}{{\em Symmetry} {\bfseries 14}
  no.~11, (2022) 2435}, \href{http://arxiv.org/abs/2211.10414}{{\ttfamily
  arXiv:2211.10414 [hep-th]}}.

\bibitem{Yakimenko2018}
V.~Yakimenko {\em et~al.}, ``{On the prospect of studying nonperturbative QED
  with beam-beam collisions},''
  \href{http://dx.doi.org/10.1103/PhysRevLett.122.190404}{{\em Phys. Rev.
  Lett.} {\bfseries 122} (2019) 190404},
\href{http://arxiv.org/abs/1807.09271}{{\ttfamily arXiv:1807.09271
  [physics.plasm-ph]}}.

\bibitem{Fedotov2010}
A.~M. Fedotov, N.~B. Narozhny, G.~Mourou, and G.~Korn, ``{Limitations on the
  Attainable Intensity of High Power Lasers},''
  \href{http://dx.doi.org/10.1103/PhysRevLett.105.080402}{{\em Phys. Rev.
  Lett.} {\bfseries 105} (2010) 080402},
  \href{http://arxiv.org/abs/1004.5398}{{\ttfamily arXiv:1004.5398}}.

\bibitem{Bulanov2010}
S.~S. Bulanov, T.~Z. Esirkepov, A.~G.~R. Thomas, J.~K. Koga, and S.~V. Bulanov,
  ``{Schwinger Limit Attainability with Extreme Power Lasers},''
  \href{http://dx.doi.org/10.1103/PhysRevLett.105.220407}{{\em Phys. Rev.
  Lett.} {\bfseries 105} (2010) 220407},
  \href{http://arxiv.org/abs/1007.4306}{{\ttfamily arXiv:1007.4306}}.

\bibitem{Gonoskov2013}
A.~Gonoskov, I.~Gonoskov, C.~Harvey, A.~Ilderton, A.~Kim, M.~Marklund,
  G.~Mourou, and A.~Sergeev, ``{Probing Nonperturbative {QED} with Optimally
  Focused Laser Pulses},''
  \href{http://dx.doi.org/10.1103/physrevlett.111.060404}{{\em Phys. Rev.
  Lett.} {\bfseries 111} (2013) },
  \href{http://arxiv.org/abs/1302.4653}{{\ttfamily arXiv:1302.4653}}.

\bibitem{Ilderton2023}
M.~Marklund, T.~G. Blackburn, A.~Gonoskov, J.~Magnusson, S.~S. Bulanov, and
  A.~Ilderton, ``Towards critical and supercritical electromagnetic fields,''
  \href{http://dx.doi.org/10.1017/hpl.2022.46}{{\em High Power Laser Sci. Eng.}
  (2023) 1}, \href{http://arxiv.org/abs/2209.11720}{{\ttfamily
  arXiv:2209.11720}}.

\bibitem{Vincenti2019}
H.~Vincenti, ``Achieving extreme light intensities using optically curved
  relativistic plasma mirrors,''
  \href{http://dx.doi.org/10.1103/PhysRevLett.123.105001}{{\em Phys. Rev.
  Lett.} {\bfseries 123} (2019) 105001},
  \href{http://arxiv.org/abs/1812.05357}{{\ttfamily arXiv:1812.05357}}.

\bibitem{Fedotov2023}
A.~Fedotov, A.~Ilderton, F.~Karbstein, B.~King, D.~Seipt, H.~Taya, and
  G.~Torgrimsson, ``{Advances in QED with intense background fields},''
  \href{http://dx.doi.org/https://doi.org/10.1016/j.physrep.2023.01.003}{{\em
  Phys. Rep.} {\bfseries 1010} (2023) 1},
  \href{http://arxiv.org/abs/2203.00019}{{\ttfamily arXiv:2203.00019}}.

\bibitem{Schmitt2023}
A.~Schmitt {\em et~al.}, ``{Mesoscopic Klein-Schwinger effect in graphene},''
  \href{http://dx.doi.org/10.1038/s41567-023-01978-9}{{\em Nature Phys.}
  {\bfseries 19} no.~6, (2023) 830--835},
  \href{http://arxiv.org/abs/2207.13400}{{\ttfamily arXiv:2207.13400
  [cond-mat.mes-hall]}}.

\bibitem{Alvarez2023}
A.~\'Alvarez-Dom\'\i{}nguez, J.~A.~R. Cembranos, L.~J. Garay,
  M.~Mart\'\i{}n-Benito, A.~Parra-L\'opez, and J.~M. S\'anchez~Vel\'azquez,
  ``{Operational realization of quantum vacuum ambiguities},''
  \href{http://arxiv.org/abs/2303.07436}{{\ttfamily arXiv:2303.07436
  [hep-th]}}.

\bibitem{Ilderton2022}
A.~Ilderton, ``{Physics of adiabatic particle number in the Schwinger
  effect},'' \href{http://dx.doi.org/10.1103/PhysRevD.105.016021}{{\em Phys.
  Rev. D} {\bfseries 105} (2022) 016021},
  \href{http://arxiv.org/abs/2108.13885}{{\ttfamily arXiv:2108.13885
  [hep-th]}}.

\bibitem{Yamada2021}
Y.~Yamada, ``Superadiabatic basis in cosmological particle production:
  application to preheating,''
  \href{http://dx.doi.org/10.1088/1475-7516/2021/09/009}{{\em JCAP} {\bfseries
  2021} (2021) 009}, \href{http://arxiv.org/abs/2106.06111}{{\ttfamily
  arXiv:2106.06111 [hep-th]}}.

\bibitem{Fewster2013}
C.~J. Fewster and R.~Verch, ``{The Necessity of the Hadamard Condition},''
  \href{http://dx.doi.org/10.1088/0264-9381/30/23/235027}{{\em Class. Quant.
  Grav.} {\bfseries 30} (2013) 235027},
  \href{http://arxiv.org/abs/1307.5242}{{\ttfamily arXiv:1307.5242 [gr-qc]}}.

\bibitem{ours_Schwinger}
A.~\'Alvarez-Dom\'\i{}nguez, L.~J. Garay, M.~Mart\'\i{}n-Benito, and R.~B.
  Neves, ``{States of low energy in the Schwinger effect},''
  \href{http://dx.doi.org/10.1007/JHEP06(2023)093}{{\em JHEP} {\bfseries 06}
  (2023) 093}, \href{http://arxiv.org/abs/2303.15294}{{\ttfamily
  arXiv:2303.15294 [hep-th]}}.

\bibitem{Poschl1933}
G.~P{\"o}schl and E.~Teller, ``{Bemerkungen zur Quantenmechanik des
  anharmonischen Oszillators},'' {\em Z. Phys.} {\bfseries 83} (1933) 143.

\bibitem{Ilderton2013}
A.~Gonoskov, I.~Gonoskov, C.~Harvey, A.~Ilderton, A.~Kim, M.~Marklund,
  G.~Mourou, and A.~M. Sergeev, ``{Probing nonperturbative QED with optimally
  focused laser pulses},''
  \href{http://dx.doi.org/10.1103/PhysRevLett.111.060404}{{\em Phys. Rev.
  Lett.} {\bfseries 111} (2013) 060404},
  \href{http://arxiv.org/abs/1302.4653}{{\ttfamily arXiv:1302.4653 [hep-ph]}}.

\bibitem{BeltranPalau2019}
P.~Beltr\'an-Palau, A.~Ferreiro, J.~Navarro-Salas, and S.~Pla, ``{Breaking of
  adiabatic invariance in the creation of particles by electromagnetic
  fields},'' \href{http://dx.doi.org/10.1103/PhysRevD.100.085014}{{\em Phys.
  Rev. D} {\bfseries 100} (2019) 085014},
  \href{http://arxiv.org/abs/1905.07215}{{\ttfamily arXiv:1905.07215
  [hep-th]}}.

\bibitem{Abramowitz1965}
M.~Abramowitz and I.~Stegun, {\em Handbook of Mathematical Functions: With
  Formulas, Graphs, and Mathematical Tables}.
\newblock Applied mathematics series. Dover Publications, 1965.

\bibitem{Mukhanov2007}
V.~Mukhanov and S.~Winitzki,
  \href{http://dx.doi.org/10.1017/CBO9780511809149}{{\em {Introduction to
  Quantum Effects in Gravity}}}.
\newblock Cambridge University Press, 2007.

\bibitem{Birrell1982}
N.~D. Birrell and P.~C.~W. Davies,
  \href{http://dx.doi.org/10.1017/CBO9780511622632}{{\em Quantum Fields in
  Curved Space}}.
\newblock Cambridge Monographs on Mathematical Physics. Cambridge University
  Press, 1982.

\bibitem{Agullo:2020iqv}
I.~Agullo, J.~Olmedo, and V.~Sreenath, ``{Observational consequences of Bianchi
  I spacetimes in loop quantum cosmology},''
  \href{http://dx.doi.org/10.1103/PhysRevD.102.043523}{{\em Phys. Rev. D}
  {\bfseries 102} no.~4, (2020) 043523},
  \href{http://arxiv.org/abs/2006.01883}{{\ttfamily arXiv:2006.01883 [gr-qc]}}.

\bibitem{Pirk1993}
K.-T. Pirk, ``Hadamard states and adiabatic vacua,''
  \href{http://dx.doi.org/10.1103/PhysRevD.48.3779}{{\em Phys. Rev. D}
  {\bfseries 48} (1993) 3779},
  \href{http://arxiv.org/abs/gr-qc/9211003}{{\ttfamily arXiv:gr-qc/9211003}}.

\bibitem{Nadal2023}
S.~Nadal-Gisbert, J.~Navarro-Salas, and S.~Pla, ``{Low-energy states and CPT
  invariance at the big bang},''
  \href{http://dx.doi.org/10.1103/PhysRevD.107.085018}{{\em Phys. Rev. D}
  {\bfseries 107} no.~8, (2023) 085018},
  \href{http://arxiv.org/abs/2302.08812}{{\ttfamily arXiv:2302.08812 [gr-qc]}}.

\bibitem{Banerjee:2023cxg}
R.~Banerjee and M.~Niedermaier, ``{States of Low Energy on Bianchi I
  spacetimes},'' \href{http://arxiv.org/abs/2305.11388}{{\ttfamily
  arXiv:2305.11388 [gr-qc]}}.

\bibitem{Bonga2015}
B.~Bonga and B.~Gupt, ``{Inflation with the Starobinsky potential in Loop
  Quantum Cosmology},'' \href{http://dx.doi.org/10.1007/s10714-016-2071-0}{{\em
  Gen. Rel. Grav.} {\bfseries 48} no.~6, (2016) 71},
  \href{http://arxiv.org/abs/1510.00680}{{\ttfamily arXiv:1510.00680 [gr-qc]}}.

\end{thebibliography}\endgroup
\bibliographystyle{utphys}

\end{document}